\documentclass[showpacs,amsmath,amssymb,prl]{revtex4}

\usepackage{graphicx}
\usepackage{dcolumn}
\usepackage{amsmath}
\usepackage{multirow}
\usepackage{appendix}
\usepackage{booktabs}
\usepackage{url}

\newcommand{\be}{\begin{equation}}
\newcommand{\ee}{\end{equation}}
\newcommand{\bey}{\begin{eqnarray}}
\newcommand{\eey}{\end{eqnarray}}
\newcommand{\bw}{\begin{widetext}}
\newcommand{\ew}{\end{widetext}}

\newcommand{\ra}{\rangle}
\newcommand{\la}{\langle}
\newcommand{\br}{ {\bf r} }
\newcommand{\bp}{ {\bf p} }

\begin{document}
 %\draft

\title {Loschmidt Echo for edge of chaos and chaotic sea in Standard Map:Cases Study}

\author{Wen-jun Shi\textsuperscript{1,2}}

 %\address{
 \affiliation{$^1$Department of Physics, University at Albany, United States of America.   $^2$College of Mathematical and Physical Sciences,Ningde Normal university,Ningde,Fujian,352100,China(Part of the work was done in this organization.) }

\date{1 December 2021}

\begin{abstract}
Using the model as Standard Map,I study the decay laws of quantum Loschimedt Echo(LE) with the classical limit as edge of chaos and chaotic sea which all belong to mixed-type phase space.Here I propose there is the existence of common decay law as $M(t)\approx e^{-c_{0}(t)\sigma^{\nu(t)}t^{\alpha(\sigma,t)}}$ for typical decay process of LE which is testified by numerical study,and $\alpha$ only can be the indicator to characterize the decay law if $c_{0},\nu$ fixed.Then the variations of related parameters to describe the decay of LE are heavily studied and the statistical semi-classical method is developed to understand the results with the probability of $P(s)$ evolved with time as the essential variable,where $s$ is classical action.From the study of process of LE,there are typical three decay processes as the initial decay with same decay law,transitive decay,and followed decay.Although there is some different expressions for edge of chaos and chaotic sea,but the basic decay features are all hold.For chaotic sea,variation of $\alpha$ with $\sigma$ can taken as logarithm law if Levy distribution for describing $P(s)$ can be found in a high-level approximation.Applying the decay law as $M(t)\approx e^{-c_{0}(t)\sigma^{\nu(t)}t^{\alpha(\sigma,t)}}$,the critical perturbation for comparing the decay extent can be found commonly with the referent decay of LE as the strong chaos.Further more,time scale is studied and find there is the rule commonly existing as $\tau\propto{\sigma}^{-\gamma}$,where $\gamma$ generally is not $1$ or $2$ corresponding to rule of time scale of stable classical dynamics or strong chaos in the classical limit.To understand the accuracy of semi-classical method in terms of Levy distribution,carefully mathematical analysis is carried out although it has not very tight connection with study of decay law as the focus in this research.
\end{abstract}

\pacs{05.45.Mt, 05.45.Pq, 03.65.Sq }

\maketitle

%\begin{multicols}{2}

\section{1. Introduction}

   Reversibility and sensibility to perturbations of quantum systems are at the heart of fields of research as vast as quantum phase transition(such as BEC),quantum information processing and coherent transport\cite{Nielsen,Datta,Gorin,Jacquod,Casabone}.The interest in these subjects has greatly increased due to the development of experimental techniques that enable the manipulation of a great number of quantum systems from photons to mesoscopic devices\cite{Raimond,Sohn}.

   The suitable magnitude for measuring the stability of the quantum motion,as well as its  irreversibility,is the Loschmidt echo(LE)or fidelity which put forward originally by Peres\cite{peres}to investigate the so called sensitivity in the quantum systems corresponding to the dynamical chaos in the classical system.The LE is explicitly expressed using $ M(t) = |m(t)|^2 $, measures the overlap of two states started from the same initial state and evolved under slightly different Hamiltonians in the classic limit,$H_0$ and $ H=H_0 + \epsilon V $, \be m(t)=\la\Psi_0|{\rm exp}(iHt/\hbar ){\rm exp}(-iH_0t /\hbar)|\Psi_0\ra\label{mat}\ee.The key point here is the perturbation for the Hamiltonian rather than the initial condition as the Hamiltonian evolution is unitary for quantum systems,ie,the scalar product of any two states is invariant,therefore two initially neighboring states will always remain neighbors in Hilbert space.

    Although the importance of measurement to the stability of quantum systems with LE,but taken little attention for the decay behavior of LE for a long time.Only recently,joined with the application in the quantum computation(especially after the paper\cite{Jalabert}using semiclassical method to obtain the relationship between the Lyapunov exponent and the decay of LE) and the likely characterizing of critical points for quantum phase transitions\cite{Quan,Wang,Zheng},the LE has been extensively studied.The different time and perturbation strength regimes were shown to exist\cite{Jacquod_01,Gorin,Jacquod,Wisniacki}.As a function of time,this magnitude has three well-known regimes.For very short times,it is parabolic or Gaussian,as the perturbation theory is valid to first order\cite{Wisniacki_03}.This transient regime is followed by a decay exponential in chaotic systems\cite{Jacquod_01}.Finally,the LE finds a long-time saturation at values inversely proportional to the effective size of effective size of the Hilbert space\cite{Jacquod,Guti}.

   As a function of the strength of the perturbation,the decay of the LE has mainly three different
behaviors\cite{Jalabert,Jacquod_01}.When the perturbation is very small,in which a typical matrix element $W$ of the perturbation is smaller than the mean level spacing $\bigtriangleup$ ,the decay is
Gaussian until $M$ reaches its asymptotic values.If $W>{\bigtriangleup}$ ,this regime has an exponential decay,with decay rate given by the width $\sigma$ of the local density of states$(\rm
LDOS)$.This is usually called Fermi golden rule regime(FGR).Finally,when $\sigma>\lambda$,with $\lambda$ the mean Lyapunov exponent of the classical system,the regime becomes independent of the perturbation and the decay rate is given by $\lambda$.Although this Lyapunov regimes seems to be universal as the intensive numerical studies have shown in the literature\cite{Gorin,Jacquod},recent works have found a nonuniform behaviors of the decay rate as a function of the perturbation
strength.This was observed in an echo spectroscopy experiment on ultracold atoms in optical billiards and in theoretical studies of the kicked rotator,the sawtooth map,and Josephson flux qubits\cite{Andersen,WgWang_1,WgWang_2,Pozzo}.Similar qualitative behavior was shown in Ref.\cite{Goussev}for local perturbations.In this case the authors found an oscillating regime of the decay of the LE around the value of the classical escape rate.

    Now the above picture for the fidelity decay is far from satisfaction,and the paper\cite{WgWang_2}gave a explicit description of the abnormal expression of the fidelity decay to the common understanding.Our work here has its focus on the two question,the one is the decay laws of LE for the systems with classical counterpart as chaotic sea but also having regular torus which is quite different from those in systems with strong chaos or with regular motion\cite{WgWang_1,WgWang_2,WgWang_3,Zheng}.The stretched exponential decay are obtained but the condition for emerging the decay behavior is not clear.Another interesting subject is the decay behavior with classical counterpart as edge of chaos.As we know,the edge of chaos is an important issue in the study of quantum chaos.In classical chaos,the edge of chaos is fractal boundary separating the regular and chaotic regions.The behavior of LE in the border region between regular and chaotic components might be particularly tricky,the decay law in such region has been
numerically studied in Ref.\cite{WLT02},where we can see a universal situation for decay behavior for LE,ie,a initial power law decay is followed by an exponential decay but only showed from naked eyes although the paper used the non-extensive statistical mechanics to give an theoretical fit for the numerical results.

The motivation of this paper here is to investigate the LE decay of these two regions discussed above to a classical counterpart with mixed phase space,as these situations are very common but having very few careful study.We want to give explicit observation of the LE decay for the two regions,and make an contrast with the decay laws for them.Beside to the extensive and careful numerical study,we also want to find likely analytical understanding from semi-classical method from the angle of the correspondence between the classical and quantum expression.Further more,we try to find some connection of our study with the previous work,in particular for the likely differences as the main consideration.

The paper is organized as follows.In Sec.2,we study the LE decay corresponding to the chaotic sea and edge of chaos,directly using the numerical technique with the focus in the decay laws.In this section,we try to find the differences from the previous study for the decay laws in particular for the established exponential decay in strong chaos. In the Sec.3,we want to develop the statistical-type semi-classical analytical method \cite{WgWang_2} combined with related numerical technique to treat the numerical study above and try to find some central causes for the new decay laws.In this study,we want to seriously consider the comparison of the evaluation of semi-classical method approaching Loschimidt echo in a very careful way,surely including the comparison of direct decay process.Meanwhile we also want to study some dynamics process with the correspondence of decay features of Loschimidt echo.In the Sec.4,we will summarize the study results in a clear logic way and try to consider the likely new research direction.In the Sec.5,we will give some appendix for some theoretical analysis in detail.For simplicity,we call Loschmidt echo as LE.

\section{2.Study model and research method}

In this paper, we take the both condition of the edge of chaos and weak chaos into our consideration.The model we study is a very famous model-standard Map(or Chirikov-Taylor Model)\cite{RS96,WLT02},the Hamiltonian of this model \be H=\frac{1}{2}p^2+V(r)\sum_{n}\delta(t-nT)\label{Hamilton} \ee with $V(r)=K{\rm cos}\hspace{0.2em}r$.For simplicity,the period T is set to be unit,$T=1$.Kicks are switched on at $t=n$, $n=0,1,2,3...$. So we have the map,\be r_{n+1}=r_n+p_n, p_{n+1}=p_n+K {\rm sin}(r_{n+1})\ee The $r$ and $p$ are all take the up bound as $2\pi$ and this model is described very clearly by Ott in the book\cite{Ott} and we recommend it to anyone who have a interest in the model in detail.Actually the map we use is the same style for the seminar paper by the authors J.Van\'{i}\v{c}ek and Eric J.~Heller\cite{vanicek} to treat the fidelity decay with dephasing representation firstly which is our starting point here for the semi-classical analysis described latter and there is another equivalent style\cite{WgWang_2} with the map as:
   \be p_{n+1}=p_n+K {\rm sin}(r_n), r_{n+1}=r_{n}+p_{n+1}\ee
We notice that the seeming two different kind of formulations for Standard Map actually show the same dynamics for a initial condition of $(r,p)$ if we just do a transformation to $r \to 2\pi-r$ as well as changing the order of sequence for $r$ and $p$.For convenience,we use the latter form for computation in terms of quantum part.Based on the requirement of making an classical contrast,we choose the coherent state for the initial state.We fixed the $r$ and changed the $p$ as the variation of the position of wave packet we use. Now we firstly give the direct numerical simulation.The arrangement of the numerical simulation is to probe into the decay of LE with the perturbation of typical strength to different regions corresponding to the classical limitation.

Then we can consider the numerical method for calculating LE and the classical system is quantized on a torus\cite{Hannay,Ford,Wilkie,Haake}.As to have a good corresponding for the wave packet with the classical phase space's point,we take the large Hilbert space with $2^{21}$ so the edge of chaos for the quantum can be studied\cite{WLT02} as the very thin layer existing between the chaotic and regular regions.The quantum computation for
the algorithm is as follows:to $1D$ finite configuration space for $0<r<r_m$,and $1D$ finite momentum space for $0<p<p_m$,the effective Planck constant $h_{eff}$ and the dimension $N$ of Hilbert space has the relationship $Nh_{eff}=r_{m}p_{m}$. $r_{m}=p_{m}=2\pi$ and hence $h_{eff}=2\pi/N$.Floquet operators in the quantized systems have the form $U=exp[-i{\hat{p}}^2/(2\hbar)]exp[-iV(\hat{r})/\hbar]$.Eigenstates of $\hat{r}$ are denoted by $|j\ra$,$\hat{r}{|j\ra}=j\hbar|j\ra$,with $j=0,1,2,...,N-1$.The same formulation for eigenstates of $\hat{p}$. The evolution of states,$\psi(t)=U^{t}\psi_0$,is calculated numerically by the fast Fourier transform(FFT)method.For the perturbation of system,we take $K=K_0+\epsilon$ and $k=k_0+\sigma$.where $\sigma=\epsilon/\hbar$
and $\epsilon<<K_0$.Actually this is the very basic numerical method,we can use alternative one FFT method\cite{WgWang_2} to increase the computational speed when we can take into the matrix representation of Floquet operator in a analytical way.

   Now we want to choose the specific cases for edge of chaos and chaotic sea as our research objective and try to find more decay features as possible as we could.For edge of chaos,we consider a specific line as $r=2.2$ with corresponding system parameter as $K=3$.Alongside with this line,we can observe a typical edge of chaos for the transition from stable field to chaotic sea.To clearly determine this field of edge of chaos is the first thing we should consider,and we can track the evolution of related ensemble to find whether there is the typical situation of edge of chaos for sticking in the stable field for a limited time scale.The simplest method to construct a classical ensemble is to choose samples uniformed from a little circle using Monte Carlo technique(united probability distribution).The method is the same to the work by G.Casati et al\cite{Casati05} with a ensemble with fixed in a very little place and track the evolution of the whole phase points of the ensemble.But here we do a some change inspired directly from the correspondence between the classical ensemble and quantum state and the the center of ensemble $r_{center}$ is the quantum average for $\overline{r}=2.2$ we fix and the variation of $p_{center}$ is to find the edge of chaos.Thus we use the symbol as $r_{center}$ and $p_{center}$ to point out the location of classical ensemble precisely and the extending field of classical ensemble is set by the uncertainty relation of quantum Gaussian wave-packet we use.To probe into the edge of chaos from the viewpoint of quantum world,Large Hilbert space should be used as the field is very thin.For chaotic sea,we fix the location of quantum wave-packet as $\overline{r}=2.2,\overline{p}=3$ but with the variation of $K$ from $K=1.5$ to $K=3$ in terms of the mixed phase space.The study of edge of chaos strongly depends on the fine structure of phase space and we depict it in the Figure 1.

\begin{center}
\begin{figure}
\includegraphics[width=18cm,height=8cm]{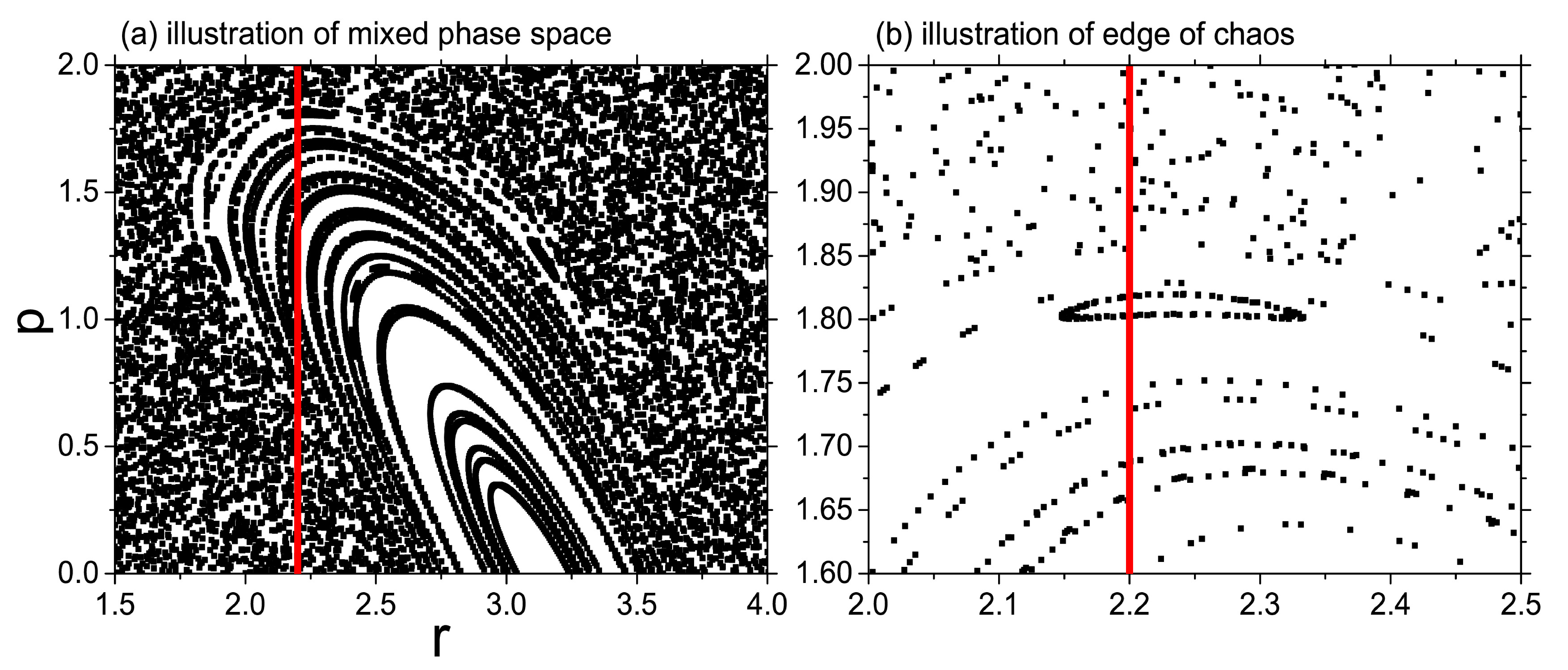}
\vspace{-0.5cm}\caption{Illustration of mixed phase space for Standard Map with $K=3$,the red line in the figure shows the constant of $r_{centre}=2.2$,the local enlargement is to show the edge of chaos. } \label{fig1}
\end{figure}
\end{center}

Now we should precisely show the field of edge of chaos and tracking process can be quantified to consider time scale for sticking in the stale field.Further more,we also can use the Lyapunov exponent which is the indicator of chaos to express the variation of dynamical feature in terms of the field of edge of chaos.The sticking situation is commonly understood as the initial point of phase space can not escape the stable field or at least for some limited time,then for the purpose to have the correspondence for the classical ensemble and quantum state,here we also consider the sticking situation for the classical ensemble with quite a lot points of phase space included,we hope this study can help to distinguish the likely different decay features for LE.Based on this consideration,we consider three cases respectively as the single initial condition,classical ensemble with the points of phase space distributed evenly in a little circle with the radius as $10^{-3}$ and the classical ensemble based on the quantum Gaussian wave-packet with the extending field from quantum uncertainty relation.We set a sufficient large number as $10^{5}$ for time evolution and the sticking time smaller than this number means the escaping situation just do happen.The classical ensemble is constructed with the statistical sampling as the Monte Carlo method,and we take the number of points of phase space as $10^4$ which is enough to do the numerical work.For calculating the sticking time,firstly we should set a feasible procedure to catch the time beginning to enter the chaotic sea in terms of mixed phase space.As what we want to investigate is around the edge between the chaotic sea and stable field illustrated in the figure 1,it is reasonable to set a condition that we can consider the escaping situation to chaotic sea do happen if $\left|r-r_{center}\right|<0.1$ as well as $p>2$.We want to observe the time to enter the chaotic sea without caring about what is the specific point of phase space to escape and also have a large points to use,thus the sticking time we obtain is reliable.For a more convincing evidence to find the fine structure of phase space of edge of chaos,we also calculate the Lyapunov exponents using the technique of tangent space,and the variations of largest exponent and second exponent are symmetric as it is the feature of Hamitonian map\cite{Ott}.

In the figure 2,we illustrate numerical result for the variation of sticking time and Lyapunov exponents with $p_{center}$ which is taken as the center of $p$ for a classical ensemble and also can be represented as the initial value of $p$ for a given single initial condition.we can find the variation of sticking time is not a monotonic process and can be stuck in the stable field again,it is reflected as the multi-fractal structure of edge of chaos\cite{Ott} numerically verified here.With the size of classical system increasing,we can find the re-sticky situation is weakened conforming to our expectation because there is the situation for sticking location and non-sticking location closing to each other.If the size is increased to across the sticking location and non-sticking location,the original sticking situation can be changed to the non-sticking situation.There is a very good correspondence between the sticking time of single initial condition and Lyapunov exponents as the value around zero is the manifestation of stable field and we also can find the escaping situation with the Largest Lyapunov exponent changed to far larger than zero in terms of increasing time steps for calculating the exponent.Quantitatively to describe,we can find there is a first escaping situation with the center close to $1.86$ and then undergoes some kind of process for going back to the sticking location.The second escaping situation appears in different $p_{center}$ based on the different initial ensemble or condition,but we can find it do happen with $p_{center}>1.872$ except the isolated $p_{center}=1.882$ of single initial condition with corresponding Lyapunov exponents as zero.The expression of classical ensemble constructed with quantum Gaussian wave-packet actually is our main consideration,and then we can select the represented $p_{center}$ as the quantum average of $p$ to study the decay of LE with the connection of classical expression have been showed here.As the second escaping situation can be taken as the main part of edge of chaos,we choose the typical cases as $\overline{p}=1.86,1.868,1.872,1.874,1.876,1.878,1.880,1.882,1.884,1.886,1.888,1.89$ to study LE carefully.To testify the previous study of LE in stable field,we also consider the case as $\overline{p}=1.8$ for a small regular torus encircled by a much larger torus which has not been studied before yet.

\begin{figure}
\includegraphics[width=12cm,height=14cm]{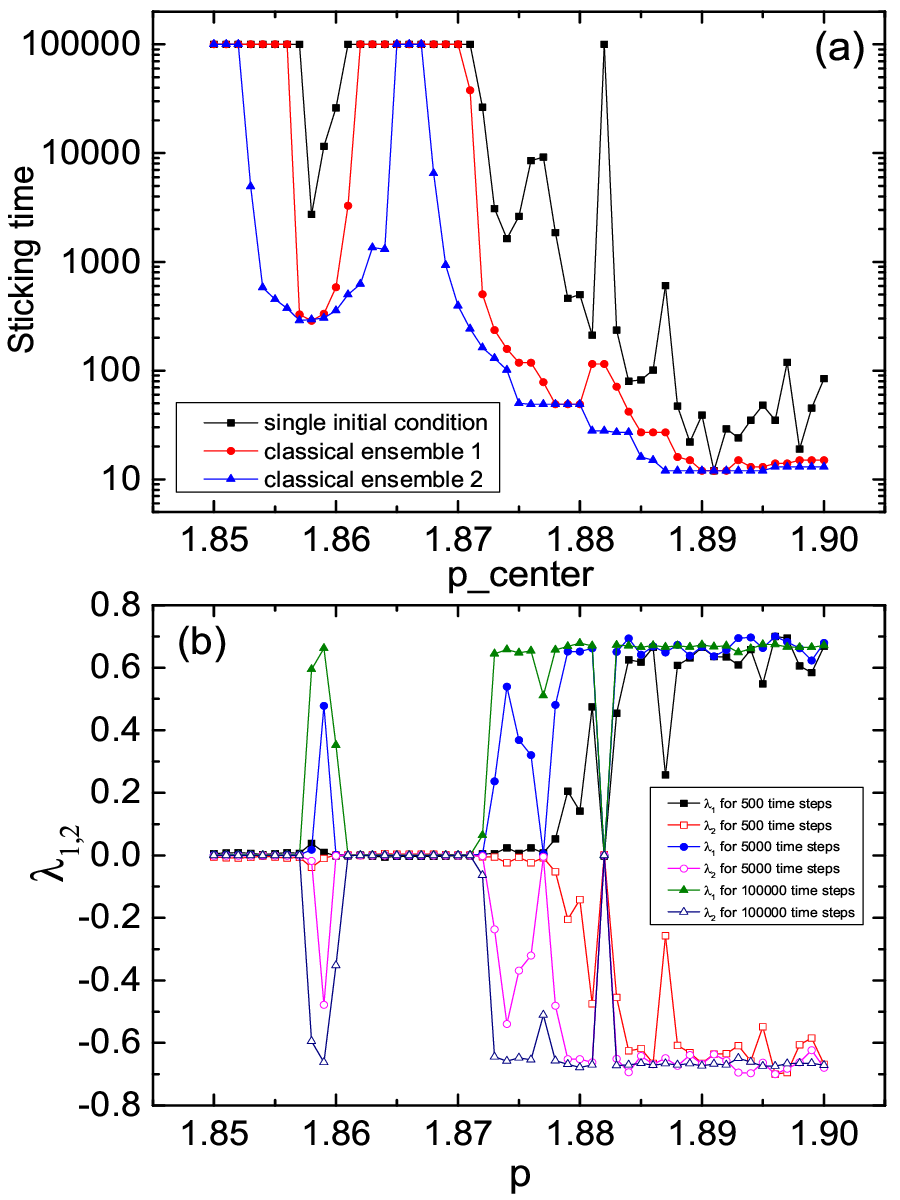}
\vspace{0cm} \caption{Illustration of feature of edge of chaos with variables as sticking time and Lyapunov exponents depicted respectively in the figure(a)and(b),here $K=3$ and $r_{center}=2.2$ fixed.In the figure(a),the mark of single initial condition means to change $p$ of initial condition corresponding to one point in the phase space,and
so-called classical ensemble 1 is realized from statistical uniform sampling taken from a little circle with the radius as $10^{-3}$.The classical ensemble 2 is constructed with quantum Gaussian wave-packet with uncertainty relation for the extending field slightly larger than the size of classical ensemble 1.The symbol $p_{center}$ in the figure(a) also represents the location of $p$ for single initial condition.$10^{5}$ is the largest time for evolution of ensemble and we concentrate on the field around the edge of chaos,and set a feasible procedure to judge the time beginning to escape to the chaotic sea if $p>2$ as well as the discrepancy between $r$ and $r_{center}=2.2$ smaller than $0.1$.The points of phase space in the classical ensemble is taken as $10^{4}$ which is enough.The sticking time shows a non-monotonic variation although has a whole tendency to decrease which reflects the complicated structure of the edge of chaos.With the size of classical ensemble increasing,the re-bonding situation to the stable field with the sticking time again for $10^{5}$ becomes less appear as the manifestation of fine structure of phase space using sticking time should be blurred.In the figure(b),we can find the expression of Lyapunov exponents has a good correspondence for the variation of sticking time of single initial condition,further more,the enlargement of time for calculating Lyapunov exponents shows clearly the escaping situation as the largest Lyapunov exponent can be changed to be far larger than zero.} \label{fig2}
\end{figure}

To study LE with the classical limit for chaotic sea of mixed phase space,we let $\overline{r}=2.2$ and $\overline{p}=3$ fixed together and change the system parameter $K$.The selection of the field of $K$ should guarantee the dynamics property is not strong chaos,then we choose the field from $K=1.5$ to $K=3$ with the corresponding largest Lyapunov exponent below 1.We use the figure 3 to illustrate the numerical result with $500$ time steps to calculate the exponents which is enough to get the stable value with the accuracy more than $10^{-2}$.

\begin{center}
\begin{figure}
\includegraphics[width=12cm,height=10cm]{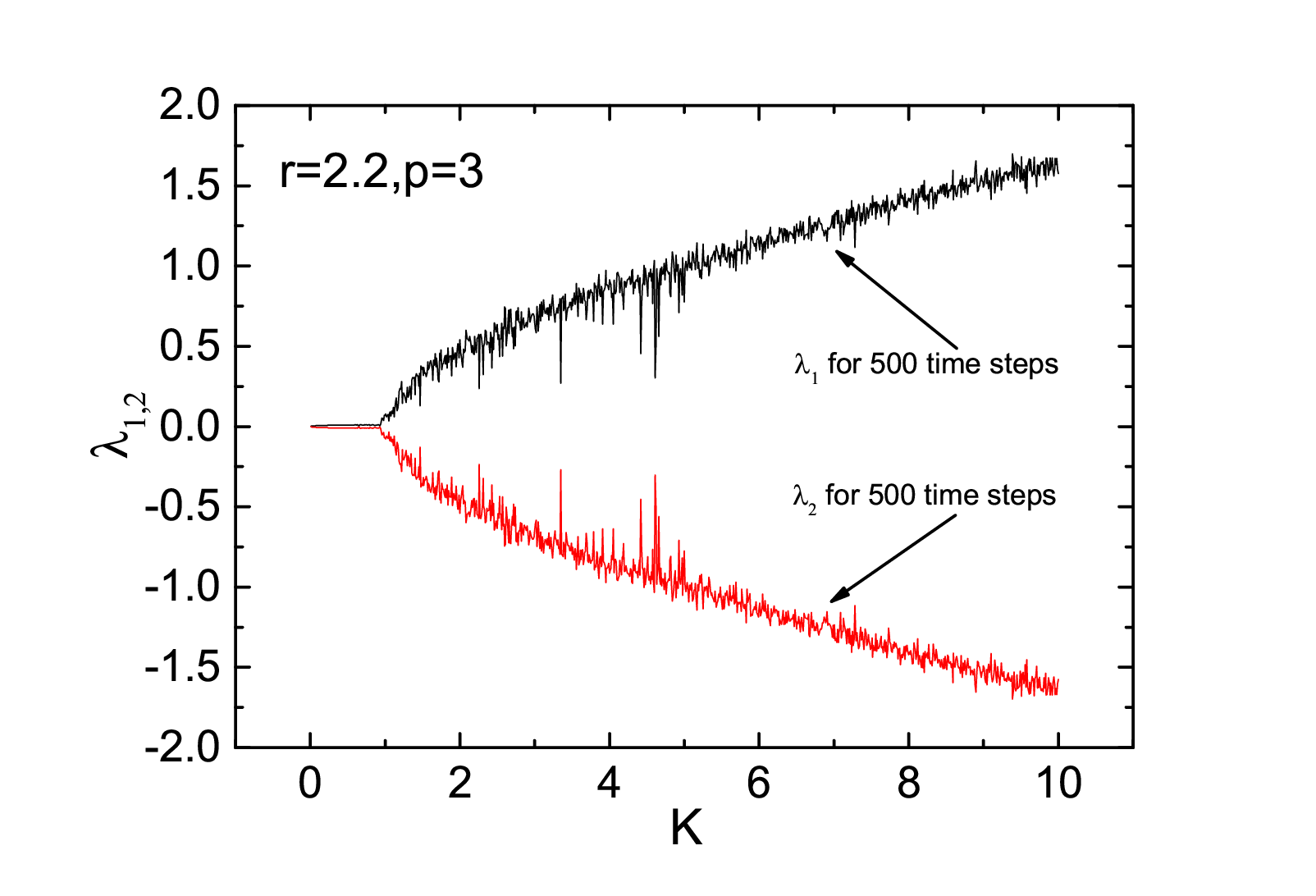}
\vspace{-0.5cm}\caption{Variation of Lyapunov exponents with $K$,the initial condition is taken as $r=2.2$ and $p=3$.It is reasonable to choose the field of $K=1.5$ to $K=3$ to study LE for the chaotic sea of mixed phase space as the Largest Lyapunov exponent for them are all well below $1$ which do not show strong chaos.}\label{fig3}
\end{figure}
\end{center}

  Then we want to show our semi-classical method to treat LE,and it has a direct connection with classical ensemble which can not clearly be point out in previous work.In this paper, we want to develop the basic method with
semi-classical treatment, and unify as possible as we could the decay expressions in our numerical study.Here we should make something clear for using uniform semi-classical approach to address the fidelity decay,that is the affect of width of Gaussian wave-packet\cite{WgWang_2}.
    For the quantum Gaussian wave-packet centered at $\tilde \br_0$,with dispersion $\xi$ and mean momentum $\tilde\bp_0$,  \be \label{wave-packet} \psi_{0}(\br_0)=\left(\frac{1}{\pi \xi^{2}}\right)^{d/4}\rm exp\left [\frac {i}{\hbar}{\tilde \bp}_0 \cdot \br_0-\frac {(\br_0-{\tilde \br}_0)^2} {2\xi^2}\right ] \ee    
    If the variable as $k=\hbar/{\xi^2}>>1$,then we can have the first order formula obtained in the seminal paper\cite{vanicek}as     
       \be \label{first_order} m_{sc1}(t)=\left(\frac {\xi^2}{\pi \hbar^2}\right)^{d/2}\int d\rm \bp_0 \rm exp\left [\frac {i}{\hbar}\Delta S(\bp_0,{\tilde \br}_0;t)-\frac {(\bp_0-{\tilde \bp}_0)^2} {(\hbar /\xi)^2}\right ]\ee
       where $\Delta S(\bp_0,{\tilde \br}_0;t)$ is the action difference alongside one trajectory based on the shadowing argument that there is always a perturbed trajectory very closer to the original trajectory.To have a good semi-classical approximation for $k=\hbar/{\xi^2}\sim 1$ of one dimensional kicked system,using the trick of shadowing theory described above and initial momentum representation\cite{WgWang_2},we obtain the second formula as
          \be \label{second_order} m_{sc2}(t)=\int dp_0  \frac {\xi} {\sqrt{\pi}\hbar D} \rm exp\left [\frac {i}{\hbar}\Delta S(p_0,{\tilde r}_0;t)-\frac {(p_0-{\tilde p}_0)^2} {(\hbar D /\xi)^2}\right ]\ee
where
           \be \label{D} D=\sqrt{1+\frac {1}{k^2}\left(\frac {\partial p_s}{\partial {{\tilde r}_0}}\right)^2}\ee

Note $D$ as the function of $p_0$,${\tilde r}_0$,and t,further more,$p_s$ come from the second order expansion as:
     \be \label{second_expansion} S_{s}(r,r_0;t)\simeq S_{s}(r,{\tilde r}_0;t)-(r_0-{\tilde r}_0)p_s-\frac {1}{2}\frac {\partial {p_s}}{\partial {\tilde r}_0 }(r_0-{\tilde r}_0)^2\ee

  If the width is sufficient small with the requirement of $k=\hbar/{\xi^2}>>1$,$D$ is changed to be $1$,but the focus we study in this paper is $k=1$ with the same widths of wave-packet in terms of position representation and momentum representation for the best contrast to the classical point particle.The paper\cite{WgWang_2} shows that the difference between the first order and second order could just in the short time and have the similar decay behaviors in terms of long time.From the previous study,it seems that we could use the first order semi-classical approximation to the LE with $k=1$\cite{Zheng},and here we give a argument like this:The different effect should be related to the function of $D$. When the value of $D$ change very little in terms of different $p_0$ for some given time,the function of $D$ just change the integral window with Gaussian weight,collaborated with the factor $1/D$ to give the basic same result for the situation without $D$.But this condition can be changed and if $D$ is changed sharply in terms of the variation of $p_0$ for some given time,we can not reasonably expect the same result.So in this paper,we use the first order to treat LE and also consider the likely effect from the second order.

   From the application of probability idea,it is seen that $M_{sc}(t)$ can be expressed in  terms of the distribution $P(\Delta S)$ of the action difference $\Delta S$,
     \be\label{Ma} M_{sc}(t) \simeq \left | \int d\Delta S e^{i\Delta S/ \hbar } P(\Delta S)\right |^2,\ee for point sources
     \be\label{PdS} P(\Delta S) = \frac 1{ \int d\br_0 d{\bf p}_0 } \int d\br_0 d{\bf p}_0 \delta \left [\Delta S - \Delta S(\bp_0 , \br_0;t)\right ],\ee 
      and for the Gaussian wave-packet
     \be \label{guass-PRO} P(\Delta  S)=\int {\frac {d{\bf p}_0} {(\pi (\hbar/\xi)^2)^{-1/2}}} e^{{-({\bf p}_0-\tilde{\bf p}_0)^2}/{(\hbar/\xi)^2}} \delta \left [\Delta S - \Delta S_{{\bf p}_0}
 \right ].\ee

   Here $\tilde{\bf p}_0$ is the mean momentum for a Gaussian wave-packet and the factor $ 1/{(\pi (\hbar/\xi)^2)^{-1/2}}$ is the integral factor with the requirement of normalization for probability theoretically,but we can ignore it in the numerical simulation just because this relationship can be satisfied spontaneously with integral of the density of probability for one.Now it is the time we should point out something easy to misunderstand in the previous work. In the previous work of J.Van\'{i}\v{c}ek and Eric J.~Heller,the one of key points to get the celebrated semi-classical formula Eq.~(\ref{first_order})is the argument with shadowing
theorem,which means action difference $\Delta S(\bp_0,\br_0;t)$ is not the general on the two different trajectories but a same trajectory.So we should consider the action along the one trajectory,for action definition as $S=\int_0^t dt (T-V_p)$ with $V_p$ as the potential for a given system and here $\Delta S(\bp_0,\br_0;t)=S_0(\bp_0,\br_0;t)-S_V(\bp_0,\br_0;t)$,$S_0(\bp_0,\br_0;t)$ for the original systems and $S_V(\bp_0,\br_0;t)$ for the perturbed one.
Therefore it is very easy to see 
   \be\label{delts}\Delta S(\bp_0,\br_0 ;t)= \epsilon \int_0^t dt' V[{\bf r}(t')]\ee with $V$
is not the potential but the potential $V_p$ divided by $K$ in terms of the perturbation $K=K_0+\epsilon$. Generally to say,the $V$ is different based on the perturbation in detail,so one should be very careful for the expression of $V$.

The key point of our interpretation is to use the statistical formula Eq.~(\ref{Ma}) and we want to use this formula as our theory to explain the result.The behavior of fidelity in the border between regular and chaotic components might be particularly tricky,it is far from clear understanding to these fields\cite{RS96,WLT02,Robledo}.Although we pay attention to the very recent work related to non-Markovian behavior of the chaos border using some measure means with information flow\cite{Mata},but systematical investigation of fidelity decay laws is the first time in our work.On the other side,from the previous work on the $P(\Delta S)$ related specific models\cite{WgWang_2}, the Levy
distribution\cite{Levydistr} can be used as an approximation to $P(\Delta S)$ with the classical limit of weak chaos.The formula Eq.~(\ref{Ma})is the starting point for us to make the semi-classical approximation,and we show the technique here.

The strategy to consider the fidelity decay theoretically in detail can be separated into two procedures,the first one is taken as the so called "seed".For it seems a totally new name here,We want to describe this idea in some detail,in terms of the part of probability formula Eq.~(\ref{Ma}),$P(\Delta S)d\Delta S$ can be written also as $P(s)ds$,hence $s=\int_0^t dt' V[{\bf r}(t')$,and pay attention $s$ itself is not the action.So what we really care
about is the information of $P(s)$ or even further more to consider the assumption of Levy distribution for $P(s)$. Once we obtain the information of $P(s)$,the quantum fidelity of different perturbation can be evaluated just like a seed.Hence the second procedure is to evaluate the fidelity decay using the seed $P(s)$ for all different perturbations within effective field,therefore what we use actually is the revised formula Eq.~(\ref{Ma})as:
   \be\label{Ma_new}M_{sc}(t) \simeq \left | \int ds e^{i\epsilon s/ \hbar } P(s)\right|^2,\ee
   obviously $\epsilon/\hbar$ gives the strength of perturbation $\sigma$.If Eq.~(\ref{Ma_new})give the decay process
reconciled with quantum fidelity,then our theory is valid which means the fluctuation term can be neglected and it is also important to observe what kind of situation can lead to the fluctuation term giving significant contribution.

In terms of Levy distribution for the consideration,the Levy distribution can be written as:
   \be\label{Levy}L(x,\eta,\beta)=\frac {1}{2\pi}\int_{-\infty}^{\infty}F_L(z)e^{izx}dz.\ee 
   Here the Fourier transform of the Levy distribution is 
   \be\label{Flz}F_L(z)={\rm exp}\{-igz-D_L|z|^{\eta}\left[1+i\beta sgn(z)\omega(z,\eta)\right]\}\ee
Where 
   \be\label{Omega1}\omega(z,\eta)=tan(\pi\eta/2)~ \rm{for}~\eta\neq1,\ee
   \be\label{Omega2}\omega(z,\eta)=(2/\pi)ln|z|~\rm{for} ~\eta=1.\ee 
The parameter $\eta$, with $0<{\eta}<2$,determines the decay of long tails,ie.,
    $L(x)\sim |x|^{-(1+\eta)}$ for large $|x|$;the parameter $\beta$ has the domain $[-1,1]$,with $\beta=0$ giving the symmetric distribution;the parameter $g$ gives a shift along the $x$ direction; and $D_L$ is related to the width of the distribution.What we need really here is the Fourier transform of $P(s)$,denoted by $F(z)$.If the Levy
distribution can be used to as an approximation of $P(s)$,with $s=\Delta S/\epsilon$,we can get the formulation by Fourier transform 
   \be\label{semiclass}M_{sc}(t)=\rm exp(-2(\epsilon/\hbar)^\eta D_L).\ee 
   Joined the force with the direct numerical integral of Eq.~(\ref{Ma_new}),they constitute the basic theoretical tool to evaluate the quantum fidelity decay in this paper.

To give the direct illustration for the assumption of Levy distribution to $P(s)$, we need to do a numerical integral of characteristic function to give the contrast between the Levy distribution and the probability density of $P(s)$ from the map.Obviously the parameter $g$ can be deleted only by re-arranging the distribution for the variable $s$ to $s-\la s \ra$,and $\la s \ra$ is the statistical average value for $s$.Then we can unify the formula Eq.~(\ref{Levy})and Eq.~(\ref{Flz})together with the different expressions for $\eta$,hence we can make a transform from the original integral to entirely real variable integral.For $\eta\neq1$ and $g=0$, 
    \be\label{levycompute1}L(x,\eta,\beta)=\frac {1}{\pi}\int_{0}^{\infty}e^{-{D_L}z^{\eta}}{\rm cos}[{D_L}z^{\eta}{\beta}tan(\pi\eta/2)-zx]dz.\ee For $\eta=1$ and $g=0$,
    \be\label{levycompute2}L(x,\eta,\beta)=\frac {1}{\pi}\int_{0}^{\infty}e^{-{D_L}z^{\eta}}{\rm cos}[{D_L}z^{\eta}{\beta}(2/\pi)lnz-zx]dz.\ee 
    
    Actually the original integral can surely be used numerically and give the exact same result with our revised real integral.From general viewpoint,$\eta$ and $D_L$ are all changed with time from the semiclassical viewpoint with the Levy distribution assumption applied.As the general stochastic process,there exists a gradual attractor for a specific probability density distribution as the limit mathematically.A fixed $\eta$ used in physical system may
be taken as some approximation under the condition for the variation of $\eta$ in terms of time quite slowly,so it also help to understand the paper\cite{Zheng}just considering the linear relation of $lnt$ via $ln(-ln D_L)$ for explaining some stretched exponential fidelity decay in a specific atom system.But we will show it is not a general situation studied in this paper.

   Now one can find the $P(s)$ can be generated by the classical ensemble based on the Quantum initial state,and we can track the evolution of the classical ensemble to get a useful information to understand the quantum decay of LE.Further more,the evolution of classical ensemble can be characterized with abnormal diffusion.The expression of the abnormal diffusion should have some connection with dynamical properties we observe before.What we care about is the relationship for $ln(\la s^2 \ra)$ versus $ln{t}$ and the normal diffusion we learn from standard statistical physics have the simplest formula as $\la s^2\ra_t=D t^1$,$D$ is called the coefficient of normal diffusion.Thus we want to study the slope by fitting $ln(\la s^2 \ra)$ versus $ln{t}$.For a general situation,we can write the formula as $\la s^2\ra_t \sim D t^\alpha$ and $\alpha$ characterize the feature of diffusion,for $\alpha<1$ corresponding to sub-diffusion and $\alpha>1$ corresponding to super-diffusion with the limitation $\alpha=2$ taken as ballistic diffusion.We also fit the diffusion exponents to show the diffusion type really as super-diffusion in the figure 4.

We use the relation for $ln\la s^2 \ra$ versus $lnt$ to get the fitted slope and also fitted intercept numerically using different fitting time to obtain more information about the variation of diffusion process although actually we just want to study the asymptotic stable process.From the figure 5(a),we can find the dropping situation of the fitted slope in the edge of chaos happened repeatedly,it is easy to understand for the first two dropping situations as there are two separated escaping processes from the stable field to chaotic sea for the time evolution of initial classical ensemble described previously.It is non-trivial for the third dropping situation as the escaping process is gradually strengthened but the fitted slope still undergo a non-monotonic variation which could be attributed to the different dynamic process in term of different $p_{center}$ of the initial ensemble.It also can be observed from the corresponding expression of fitted intercept in the figure(b)and one can find there is entirely the opposite variation for the fitted slope and intercept until $p_{center}=1.88$ illustrated in the figure(a)and(b).For the figure(c)and(d)with the initial ensemble set into the chaotic sea,the fitted slope in (c) and corresponding fitted intercept in (d) still have the non-monotonic variations which show opposite tendency basically except the case of $K=2.4$ with fitting initial time as $3001$ and the case of $K=2.9$ with the fitting initial time as $6001$. A big drop around the parameter $K=2.4$ for the fitted slope deserves much attention as the slope actually represent the diffusion exponent which shows the classical dynamic undergoes a sudden change.Therefore,we try to find out some important expression of the classical ensemble and want to make some connection with quantum LE to help us to understand some decay processes.

\begin{center}  
\begin{figure}
\includegraphics[width=18cm,height=20cm]{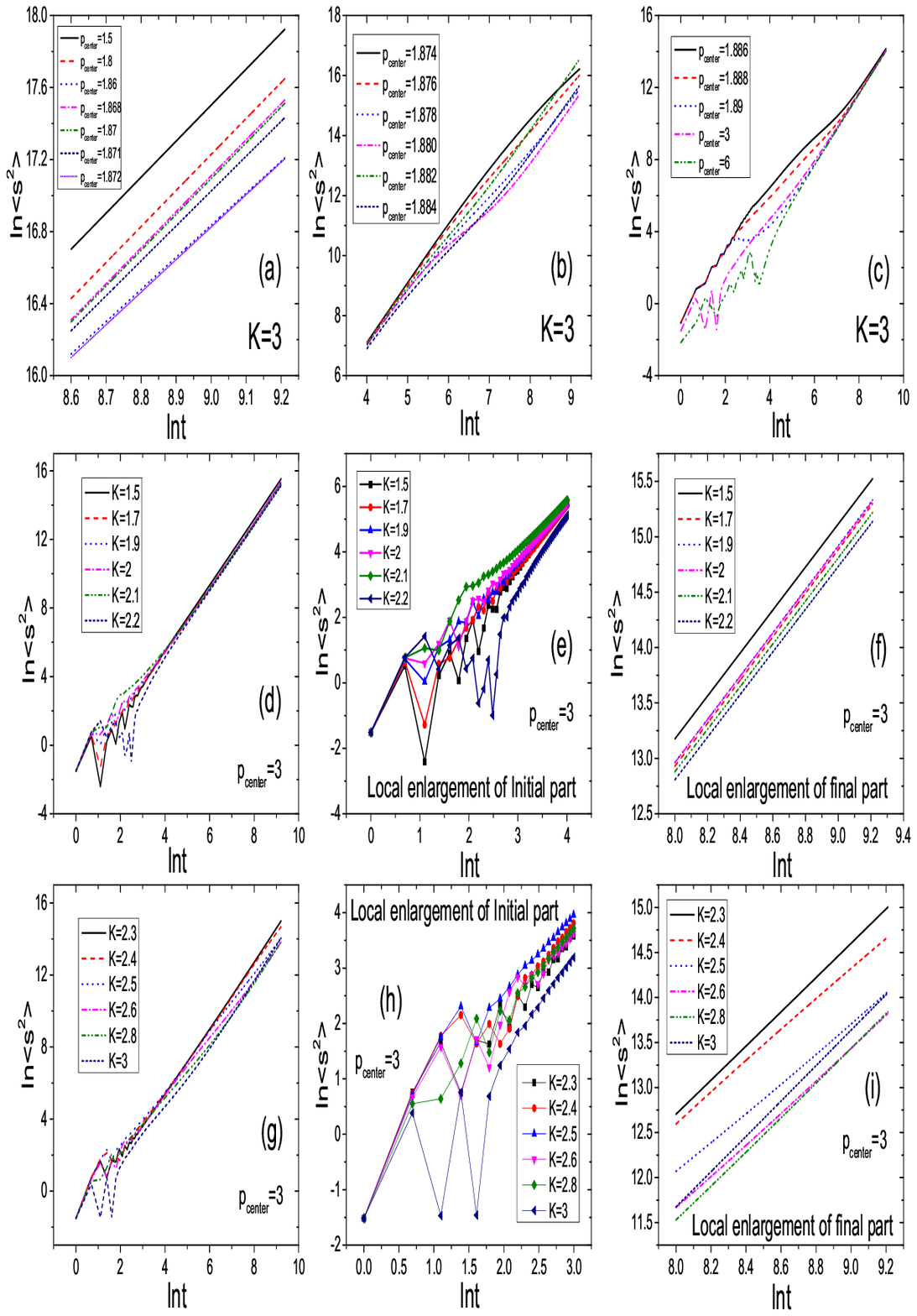}
\vspace{0cm} \caption{Classical diffusion behaviors of $s$ for the situations of edge of chaos and chaotic sea are showed with the relation as $ln\la s^2 \ra$ versus $lnt$.Using the same classical ensemble applied to the semi-classical analysis,we show the changing pattern with the variations of $p_{center}$ for the edge of chaos and system parameter $K$ for the chaotic sea.For figure(a),we can find the cases of $p_{center}=1.86$ and $p_{center}=1.872$ share basically the same line with the situation of obviously escaping to the chaotic sea with evolution time increased.For(b),we show the diffusion process which actually can enter a final process with a straight line although there is not the common converging gradual straight line.For figure(c),basically after the case of $p_{center}=1.886$,we can find there have a common tendency to enter a common straight line with the slope smaller than 2.To show the main variation clearly,we omit the initial very short transitive process with some fluctuation in the figure(a)and(b)which can be seen in the figure(c).For the figure(d),(e)and(f),we show the similar diffusion process mainly as super-diffusion for the chaotic sea until actually to the case of $K=2.3$ which is showed in the figure(g),(h)and(i)to strengthen the comparison.Further more,one can find there is a transitive pattern.} \label{fig4}
\end{figure}
\end{center}

\begin{center}  
\begin{figure}
\includegraphics[width=16cm,height=14cm]{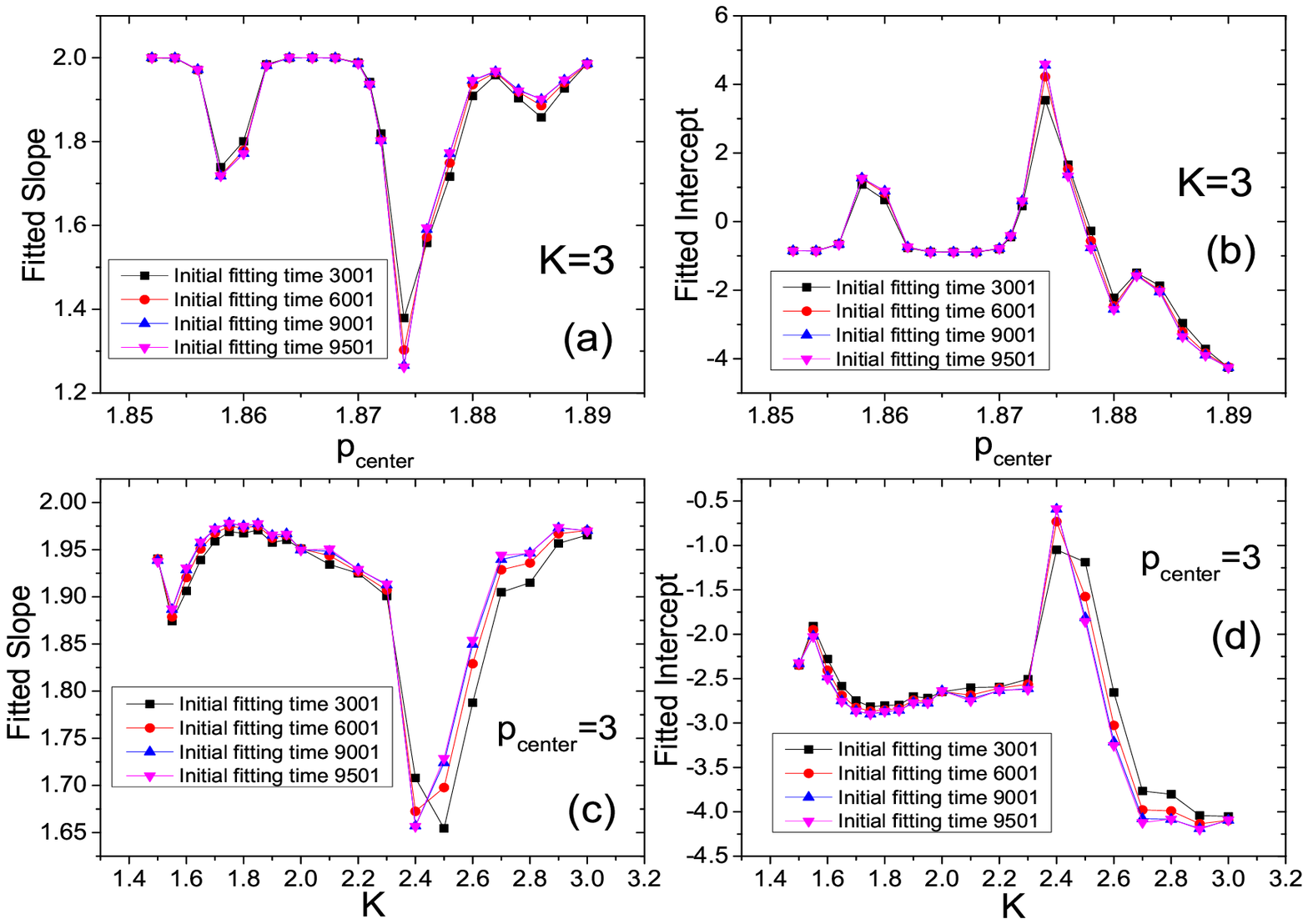}
\vspace{0cm} \caption{Fitted slope and intercept for the relation as $ln\la s^2 \ra$ versus $lnt$ corresponding to the edge of chaos and chaotic sea.(a)and(b)correspond to the variations of fitted slope and fitted intercept accordingly with the different $p_{center}$ we select in terms of the edge of chaos for the system parameter $K$ fixed as $3$.(c)and(d)correspond to the variations of fitted slope and fitted intercept accordingly with the different system parameter $K$ we select in terms of the chaotic sea for the $p_{center}$ fixed as $3$.} \label{fig5}
\end{figure}
\end{center}

\section{3. Study of LE for the edge of chaos between the chaotic sea and regular torus}

By checking the variation $ln(-ln|F(z)|)$ with $ln{z}$ in terms of Fourier transform of $P(s)$,we can find it is not the Levy distribution but generally is a localized-type distribution without a obvious long tail as a typical feature of Levy distribution.Thus the main research method here is the direct numerical study with the comparison of the semi-classical approximation for $M_{sc}(t) \simeq \left | \int ds e^{i\sigma s} P(s)\right|^2$ that is given the name as semi-classical integral for simplicity.The study procedure in detail is firstly to use some typical perturbations to directly observe the decay processes with the comparison of the semi-classical approximation to see the effectiveness of our semi-classical theory.Then we want to find some decay features as a whole in the edge of chaos with different perturbation,during this study,we want to obtain some variables to characterize the decay features which is the focus in the study such as the decay rate in the strong chaos in the previous work.Here we will consider the connection for different location of quantum wave-packet as well the likely variation of decay laws with perturbation increasing.Further more,we want to consider the likely regular for the time scale and concentrate the relation for the time scale and perturbation in terms of different location of quantum state.Based on this study procedure,we want to figure out the decay laws in the edge of chaos.

Then we show our numerical results in the very first time to see the typical distributions and find the variations of the contribution from the long tail of P (s) show the transition of the situation gradually escaping to chaotic sea from the edge of chaos,meanwhile maybe the most important feature here is the peak-like shape of P (s) for the typical distribution for the edge of chaos clearly depicted in the figure 6.

\begin{center}  
\begin{figure}
\includegraphics[width=18cm,height=20cm]{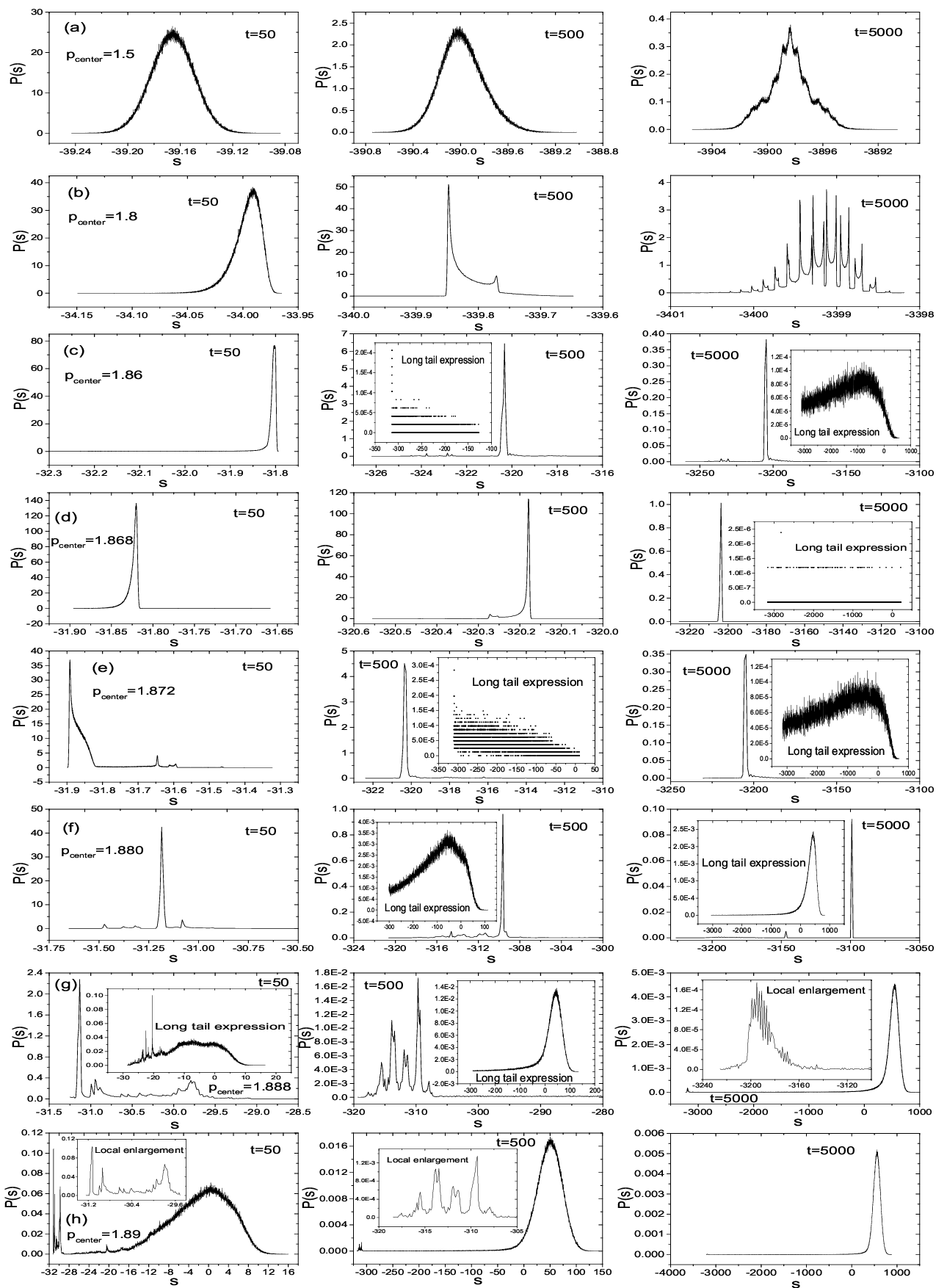}
\vspace{-0.5cm} \caption{The typical evolution of the probability density distribution $P(s)$ are showed for the different $P_{center}$ as $1.5,1.8,1.868.1.872,1.880,1.888,1.89$ and we find there have obvious basic laws for the evolution.For the entirely regular field in phase space as $p_{center}=1.5$,the distribution is very extended although it is not a common Gaussian or Levy shape from checking the frequency relation.For $p_{center}=1.8$ with the situation of meeting the regular ellipse insulaire and we can find the distribution shape finally turn to be similar to fractal type distribution which is highly non-trivial.For the initial region to escaping to chaotic sea in terms of the cases for $p_{center}=1.86$ and $p_{center}=1.872$,we can find similar expressions in the figure(c)and (d),the distributions basically show peak shape and there gradually have a small but increased contribution from the long tail with time increased.From the case of $p_{center}=1.868$ to prevent from escaping to the chaotic sea again,the distributions always show peak shape more localised than before and the contribution from the long tail is extreamly small that can be ignored.From the cases of $p_{center}=1.880,1.888,1.89$,the contribution of the long tail increases gradually with the time increased and it is harder to hold the  peak shape gradually replaced by the similar Levy shape corresponding to the original long tail part as the process of leaving the edge.} \label{fig6}
\end{figure}
\end{center}

Then we want to study the effectiveness of our semi-classical approximation in the edge of chaos and respectively use the two figures as figure 7 and 8 to illustrate the comparison between the direct numerical computation and semi-classical integral with perturbations selected as $\sigma=0.01,0.05,0.1,1.5,10$.The variation of the accuracy of semi-classical approximation is strongly depend on the location of quantum state,specifically $p_{center}$ from the corresponding classical ensemble which is the base to do the semi-classical approximation.Our finding is that the accuracy is good for the stable field that can be seen from the case of $p_{center}=1.5$ and $p_{center}=1.8$,and then it can be some kind of complicated for the edge of chaos.For the perturbation is not large,one can find the accuracy is not good for the value of $p_{center}$ initially set in the field corresponding to have the escaping situation and then gradually it can become better with increasing $p_{center}$ without the sticking situation strengthened as the case of $p_{enter}=1.868$ having a large fluctuation.For the perturbation is sufficient large,we can find the accuracy can be good universally for the edge of chaos.Further more,we can find a quite unusual expression for the revival of semi-classical evaluation of LE,numerically find that the revival to the maximum value is basically insensitive for specific $p_{center}$ emergent for $\sigma>4$ within the time scale as $10^{4}$ which is universal for the edge of chaos.The revival is periodic and the period can be changed to smaller with the perturbation increasing to some extent.This universal expression shall has the root in the common feature of $P(s,t)$ although we can not give a analytical formula accounting for it.Qualitative argument is the revival is happened for the large perturbation as well as the long time with the order as $10^{3}$ and the reason for it is that the large value of $\sigma s$ gives a very big fluctuation for $e^{i\sigma s}$ joined with some specific $P(s)$ which leads to the numerical calculation abnormal.More careful study could be required in the future.

\begin{center}  
\begin{figure}
\includegraphics[width=18cm,height=20cm]{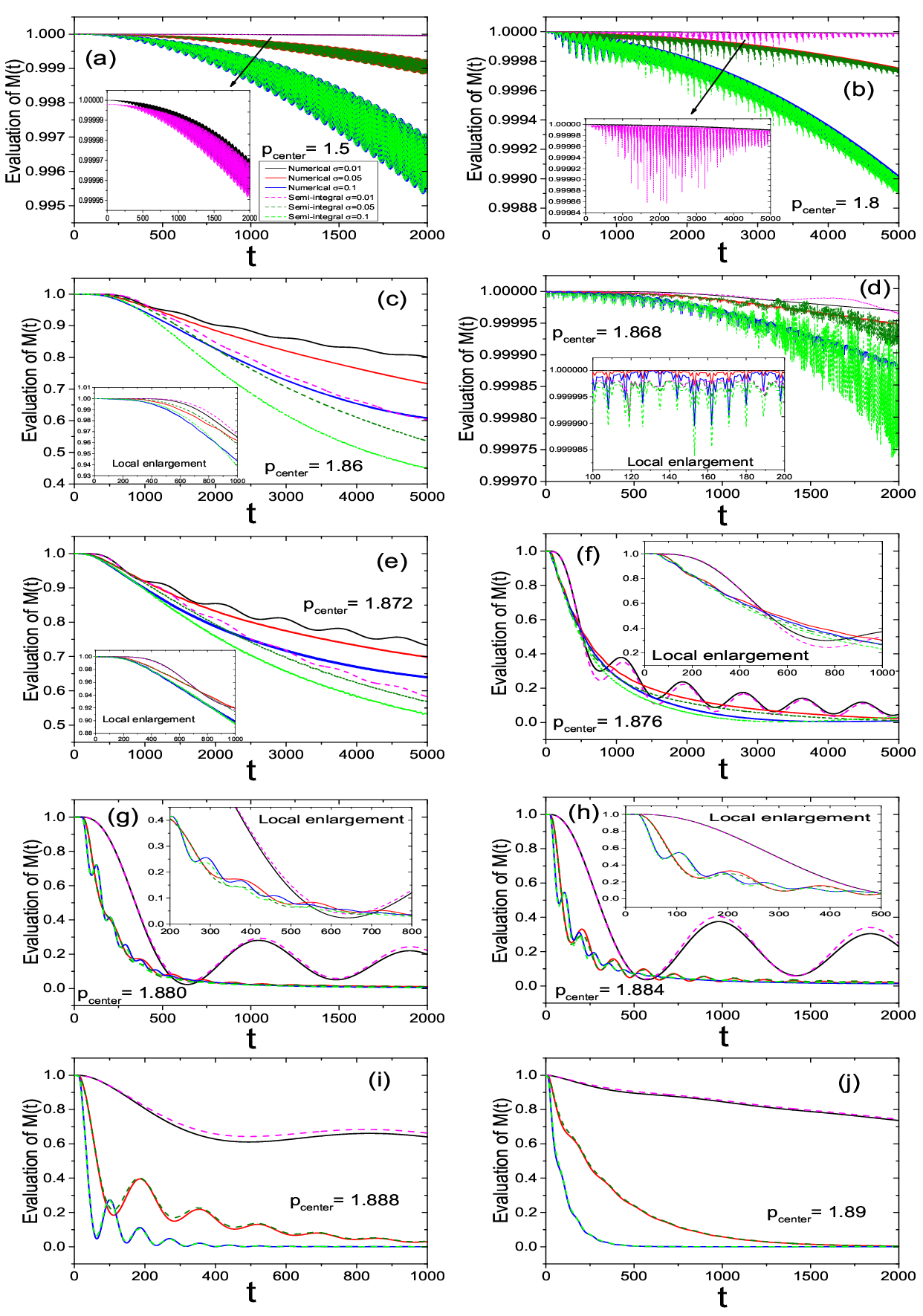}
\vspace{-0.5cm} \caption{The contrast of Loschmidt echo between the direct numerical computation and corresponding semi-integral one is showed for typical positions of initial quantum wave-packet in the edge chaos in terms of three typical weak perturbations $\sigma=0.01,0.05,0.1$.Here we can find there have obvious correspondence for the accuracy of our semi-classical theory to evaluate Loschmidt echo from the dynamics of classical ensemble.For inside of regular field,we can find basically good agreement with numerical result illustrated with the case of $p_{center}=1.5$,and then enter a small regular ellipse insulaire the agreement seems still good but find larger fluctuation from our semi-classical evaluation with the illustration about $p_{center}=1.8$.For the first situation of escaping to the chaotic sea with the case of $p_{center}=1.86$,the agreement can not hold for a long time but can be better for the escaping situation ceased afterwards illustrated with the case of $p_{center}=1.868$ that yet have a very large fluctuation in terms of the perturbation and time increased.Then we can find the second time for escaping situation illustrated with the case of $p_{center}=1.872$ sharing the similarity with the case of $p_{center}=1.868$,and afterwards the agreement tend to be better basically during the process of quantum wave-packet leaving the edge of chaos and finally we can find a very good agreement for the case of $p_{center}=1.89$ corresponding to the critical field of going out the edge of chaos.} \label{fig7}
\end{figure}
\end{center}

\begin{center}  
\begin{figure}
\includegraphics[width=18cm,height=20cm]{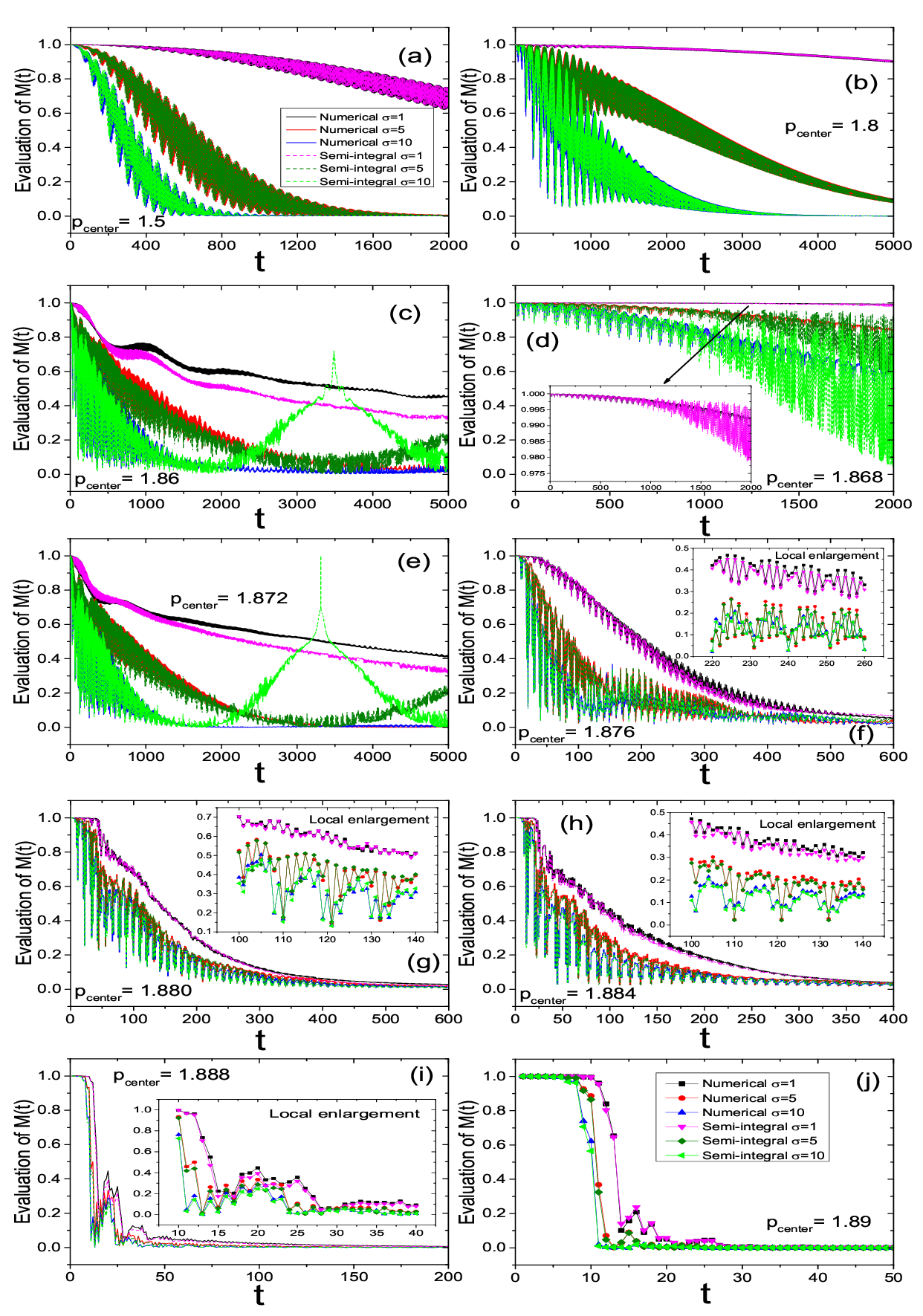}
\vspace{-0.5cm} \caption{The same consideration of comparison of Loschmidt echo for large perturbation $\sigma=1,5,10$ in terms of edge of chaos in the classical limit.For large perturbation,the accuracy for the comparison show better than the small perturbation in the same time scale and the case of $p_{center}=1.868$ still have the larger fluctuation with the semi-integral than the numerical result.One can find there is a very particular expression for the revival of semi-classical evaluation after sufficient decay process and repeat again and again,the field of range of this kind of revival process show the periodicity universally insensitive about the location of $p_{center}$ seen clearly for the cases of $p_{center}=1.86$ and $p_{center}=1.872$ and this situation is actually emergent for all classical ensemble for semi-classical calculation escaping to chaotic sea in terms of time increasing.} \label{fig8}
\end{figure}
\end{center}

Now we want to directly observe the decay features as a whole for different location as $p_{center}$,for clarity,we just use the direct numerical computation to illustrate the decay features and then we obtain some variables to characterize the decay features with the semi-classical approximation to see the effectiveness of the theory.Thus there are two factors as $p_{center}$ and $\sigma$ that should be considered together to give a whole likely regular governing the decay process.As the previous study of edge of chaos in the Kicked top,it claims that a initial power decay is the signature of the edge of chaos,but in our study we find the universally initial decay in the edge of chaos is the Cubic-exponential decay theoretically proposed by J.Van\'{i}\v{c}ek \cite{vanicek_arxiv} in quasi-integrable field.Thus in this paper,we give a typical case of edge of chaos from the quasi-integrable field to chaotic sea.To enhance the convincingness,we take the decay expression of different $p_{center}$ corresponding typical perturbations together to show the decay features.On the one hand,we could fix the $p_{center}$ to observe the decay features in terms of different perturbation,and on the other hand,we could fix the perturbation $\sigma$ to observe the decay features in terms of different $p_{center}$.As LE is a slow variable compared to variation of time except the very large perturbation,thus we should pay attention to the illusory decay law illustrated with the so-called linear relation for sufficient limited time steps,such as the variation of $lnM$ with $lnt$ for assumed power law decay but actually is not the case.

From our previous study\cite{WgWang_2,WgWang_3,Zheng},it seems the stretched exponential decay can be expected as a important decay for the field out of strong chaos,thus we assume the likely decay with the formula as $M(t)\sim e^{-ct^{\alpha}}$ and $c$ is the decay rate in the study of strong chaos\cite{Scholarpedia},thus we also name $\alpha$ as decay exponent which is $1$ for exact exponential decay,$2$ for Gaussian decay and $3$ for Cubic-exponential decay.Based on this idea,we numerically study the decay features in the edge of chaos.Now we select the $p_{center}=1.5,1.8,1.86,1.868,1.872,1.876,1.880,1.884,1.888,1.89$ fixed respectfully to study the decay features of LE in terms of different perturbations as $\sigma=0.01,0.1,1,10$ using the relation as the variation of $ln[-ln(M(t))]$ with $ln(t)$ which can be called as the decay relation for simplicity in this paper.As there are two escaping parts separated by a sticking part that can be measured by the sticking time of corresponding classical ensemble studied above,thus we can use two groups with the $p_{center}=1.5,1.8,1.86,1.868$ and $p_{center}=1.872,1.876,1.880,1.884,1.888,1.89$ to study the likely decay features and show the results respectfully in the figure 9 and 10.

we can find there is basically Gaussian decay for the cases of $p_{center}=1.5,1.8$ which are set in the stable field of phase space using the reference line as the slope for $2$.As the initial quantum wave-packet for $p_{center}=1.8$ actually reaches out to the small regular torus,so we can find there is some difference illustrated in the figure 9(a)and(b).For the case of $p_{center}=1.86$ belonging to the first escaping situation happened for the corresponding classical ensemble illustrated in the figure 2,the initial decay can be seen basically as the Cubic-exponential decay showed with the straight line as $3$ and then later continued decay undergo some obvious deviation from Cubic-exponential decay for small perturbation illustrated with $\sigma=0.01$ in the figure 9(c) and then afterwards we can find two decay processes typically for the small perturbation with the averaged slope of decay relation larger than $3$ and smaller than $3$ accordingly.With the perturbation increasing,the decay process with the average slope of decay relation larger than $3$ can be smoothed out and finally the whole decay process can be divided into two decay processes as the first Cubic-exponential decay and then approximately stretched exponential decay with averaged $\alpha<1$ having some large fluctuation.The initial Cubic-exponential decay can become important with the perturbation increasing and it is a fast decay compared with the much slower decay afterwards which have been described as the approximately stretched exponential decay.For simplicity,we call the whole decay processes can be mainly divided into three likely decay processes as the first decay as Cubic-exponential decay,the second decay as the transitive decay depending on the perturbation,and the third decay approximately as the stretched exponential decay at least for the large perturbation which can be observed in the figure 9(c) for $\sigma=1,10$.Actually the comparison of the fast or slow extent for two given decay process is simple if we can consider the difference between the beginning time and ending time of LE divided by all the decay time for a given decay process,thus we can quantify it just like the average velocity.Based on this very simple method,we can find the second decay is faster than the third decay for the small perturbation $\sigma=0.01$.We also can study the time scale to decide the degree of decline and will be studied in detail later.For the case of $p_{center}=1.868$ as the escaping situation has been heavily hindered again,thus LE decay shows the mixed type sharing the decay feature of $p_{center}=1.86$ for $\sigma=0.01$ as well as the decay feature of $p_{center}=1.8$ for $\sigma=0.1,1,10$.This expression has a good correspondence for the sticking time we illustrate in the figure 2,and the corresponding classical ensemble can finally has the escaping situation but for a long time with the order as $10^{3}$,so we can observe the mixed features.

The decay features for the case of $p_{center}=1.86$ is the basic pattern for the different $p_{center}$ of edge of chaos which has a obvious escaping situation for the corresponding classical ensemble illustrated in the figure 10.Although the basic decay pattern is similar for each other but the variation of the second decay as the transitive decay is deserved us to notice.The slope of this decay process becomes smaller monotonically with $p_{center}$ increasing and this pattern of variation can be seen very clearly compared with the reference line as the slope for $3$ used in the figure 10,thus the difference for the slope between second decay and the third decay tends to decrease.Meanwhile we also find the degree of decline of transitive decay undergoes non-monotonous variation with $p_{cener}$,it can become to decay to the small value that can be taken as the main decay process for $p_{center}=1.878,1.880,1.882,1.884$ and then begin to degrade for $p_{center}=1.886,1.888,1.89$ illustrated in the figure 10(d),(e),(f),(g),(h),(i),(j).It implicates that even a small perturbation still can make LE decreasing heavily in a limited time scale for some selection of $p_{center}$,and this feature is quite counter-intuitive as the assumption of positive correlation between the perturbation and degree of decline particularly taken it as ground in the study of LE for strong chaos.The decay relation of transitive decay process can not show a good straight line basically which means it is not a strict stretched exponential decay but we can loosely accept it from the average viewpoint,so-called slope can be seen as the numerical fitting of the decay process.The initial decay can be seen approximately as the Cubic-exponential decay with the duration of time having a good correspondence for the sticking time of corresponding classical ensemble although the decay become less alike for Cubic-exponential decay gradually with the enlargement of $p_{enter}$.Meanwhile the time of duration of the transitive decay for small perturbation is also deserved us to notice,and one can carefully find the time scale of transitive decay for $\sigma=0.01$ is similar around $6.5$ with the logarithm scale for most $p_{center}$ in the edge of chaos,therefore the study for the comparison of the dependence of time scale on $p_{center}$ is the essential point in this paper.

\begin{center}  
\begin{figure}
\includegraphics[width=18cm,height=10cm]{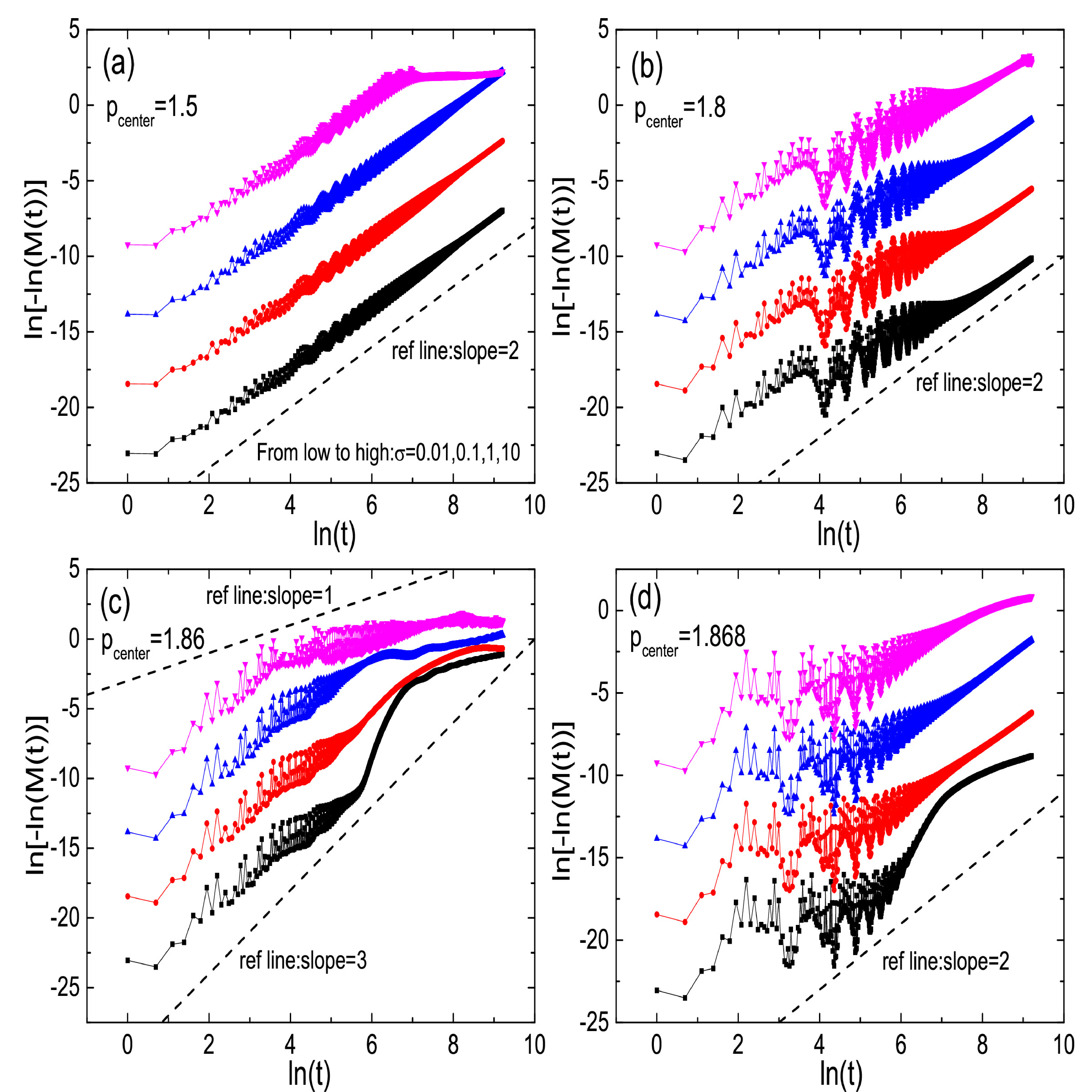}
\vspace{-0.5cm} \caption{Variation of $ln[-ln(M(t))]$ with $ln(t)$ for typical perturbation $\sigma=0.01,0.1,1,10$ with different $p_{center}$ respectfully as $1.5,1.8,1.86,1.868$.The reference lines used as the slopes for $1$,$2$ and $3$ are to show the likely exponential decay,Gaussian decay and  Cubic-exponential decay in terms of guiding eyes.In the figure(a)for $p_{center}=1.5$,Gaussian decay can be found for all the perturbations illustrated with the slope as $2$.In the figure(b)for $p_{center}=1.8$,the Gaussian decay for LE as a whole also can be found illustrated with the slope as $2$ but there is some large oscillation which is quite different from the case of $p_{center}=1.5$ as the corresponding initial classical ensemble actually reaches out to the small regular torus.In the figure(c)for $p_{center}=1.86$ having the escaping situation happened,there are typical three decay processes of LE for small perturbation $\sigma=0.01$ which can be transformed to two main decay processes as the second transitive decay is gradually smoothed out with the perturbation increasing.In the figure(d)for $p_{center}=1.868$,the escaping situation to chaotic sea for corresponding classical ensemble is hindered heavily,thus LE decay shows the mixed type sharing the decay feature of $p_{center}=1.86$ for $\sigma=0.01$ as well as the decay feature of $p_{center}=1.8$ for $\sigma=0.1,1,10$.}\label{fig9}
\end{figure}
\end{center}

\begin{center}  
\begin{figure}
\includegraphics[width=18cm,height=20cm]{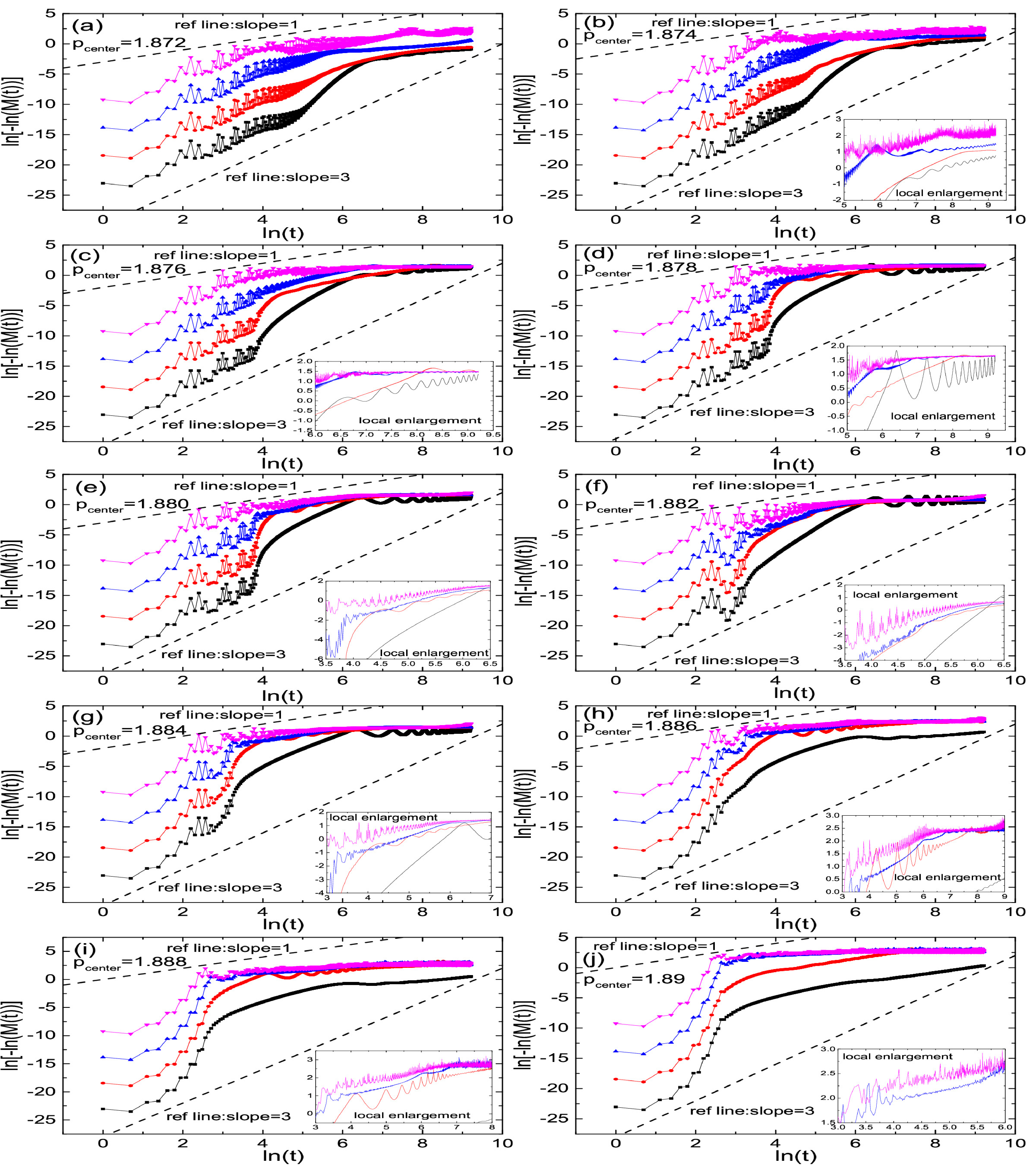}
\vspace{-0.5cm} \caption{The same consideration to figure 9 with different $p_{center}$ respectfully from $1.872$ to $1.89$ taking the interval as $0.002$.The decay features as a whole for different $p_{center}$ are similar for each other,the initial decay can be seen approximately as the Cubic-exponential decay with the duration of time having a good correspondence for the sticking time of corresponding classical ensemble although the decay become less alike for Cubic-exponential decay gradually with the enlargement of $p_{enter}$.The second transitive decay can become dropping to the minimum value observed in the figure(d)for $p_{center}=1.878$ taken as the leading decay process which is quite similar to the cases of $p_{center}=1.880,1.882,1.884$ and afterwards LE shows some oscillation having the decreasing tendency or not for the case of $p_{center}=1.882$.The slope of second transitive decay tends to decrease with the enlargement of $p_{enter}$ compared clearly with the reference line for exact slope as $3$.With perturbation increasing,two main decay processes can be observed as the first initial decay dominate more and more.}\label{fig10}
\end{figure}
\end{center}

With a clear understanding about the decay law in the stable field above,now what we concentrate is the edge of chaos having the obvious escaping situation which means the corresponding classical ensemble just can be stuck for a limited time scale within order as $10^{2}$ in our numerical study.So we put together $p_{center}=1.86,1.872,1.876,1.880,1.884,1.888,1.89$ to show the comparison for the decay processes in terms of the typical perturbation as $\sigma=0.01,0.1,1,10$.We depict our numerical result in the figure 11 and there are some important properties we find.For small perturbation as $\sigma=0.01$,after the initial non-observed decay,there is a transitive fast decay followed by a slow decay easily observed except the cases of $p_{center}=1.86,1.872$ which actually is but just for the small decay.Further more,we can find the time scales for the transitive decay are close to each other and the basically same frequency of oscillation after the transitive decay can be found for the cases of $p_{center}=1.876,1.880,1.884$ which is much faster than the decay of the cases of $p_{center}=1.888,1.89$ very close to entering the chaotic sea.The degree of decline as a whole actually undergoes a non-monotonous variation from a increasing tendency to a decreasing tendency with the enlargement of $p_{center}$ for perturbation as $\sigma=0.01$.With the perturbation increasing,the pattern of the variation can hold except the cases of $p_{center}=1.888,1.89$.For $\sigma=0.1$,we can find the decay of $p_{center}=1.888$ is the fast,but the status is changed for $\sigma=1$ as the decay of $p_{center}=1.89$ is the fast remaining for the perturbation as $\sigma=10$.Our finding here shows there is some critical perturbation determining the slower decay to the faster decay for the comparison of degree of decline,this similar situation was firstly discovered by Prosen\cite{Gorin} in the study of comparison of time scales for LE in the stable field and strong chaos.In terms of time scale as a whole for the decay,we can find there are three groups as $p_{center}=1.86,1.872$,$p_{center}=1.876,1.880,1.884$ and $p_{center}=1.888,1.89$ respectfully having the close time scale although the decay process is different for each other.The short-term fluctuation before a relatively smoothed decay is deserved to notice clearly observed in the figure 11(c)for the cases of $p_{center}=1.880,1.884,1.888,1.89$ with $\sigma=1$ which also can be illustrated by corresponding Variation of $ln[-ln(M(t))]$ with $ln(t)$ in the figure 10 and the decay of LE for $p_{center}=1.886$ also typically has this kind of process as well.For a very careful check,we can find this process just happen between the initial decay and second relative smoothed decay for the revival of LE emergent following a considerable but not too heavy initial decay in a observable way.As the big fluctuation,it make us some kind of hard to observe that short process in the figure 11(d) with these decays put together but it does exist for $p_{center}=1.878,1.880,1.884,1.886$ with a single check.

\begin{center}  
\begin{figure}
\includegraphics[width=18cm,height=16cm]{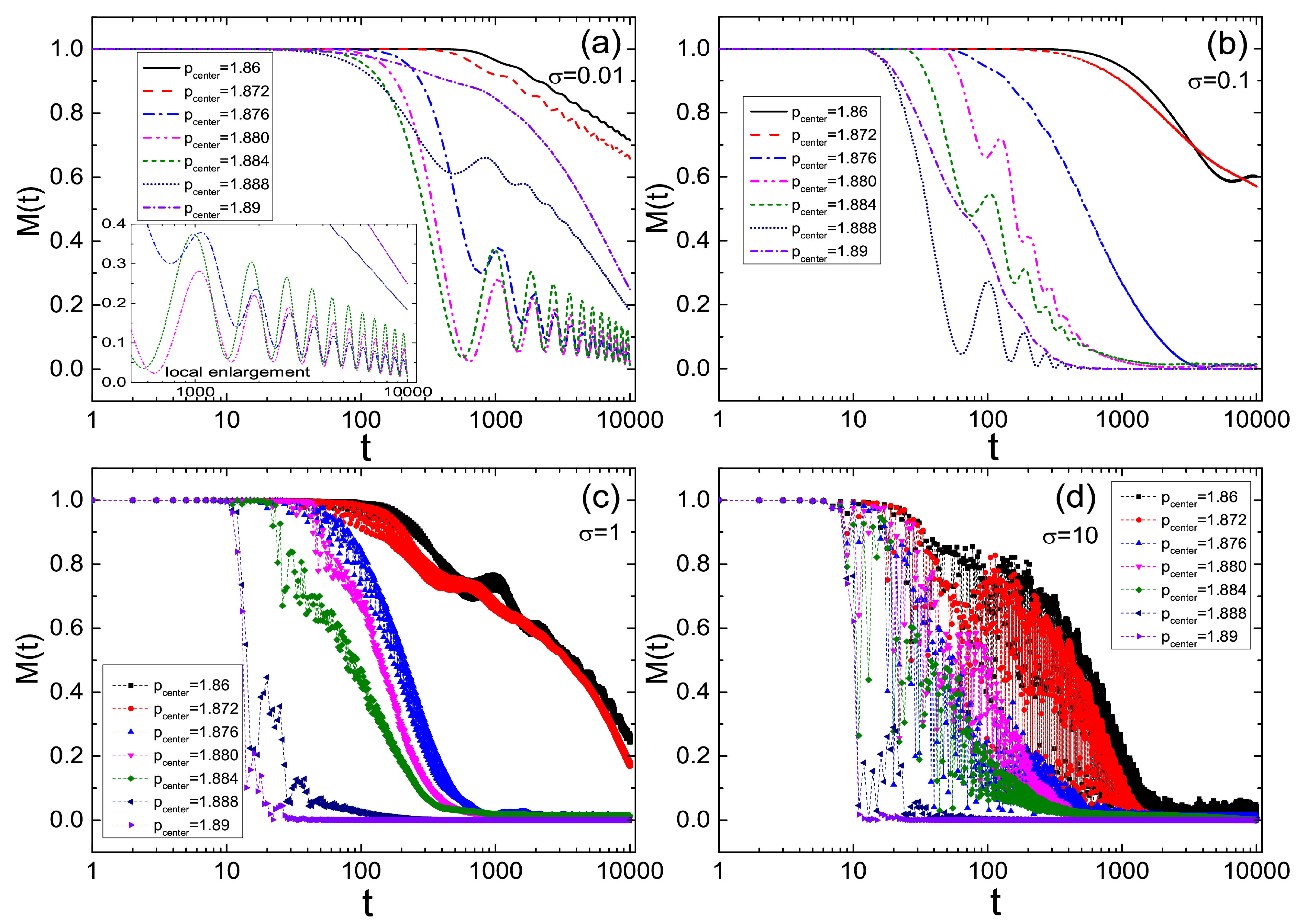}
\vspace{-0.5cm} \caption{Variation of $M(t)$ with $t$ for different $p_{center}$ with typical perturbation fixed respectfully as $0.01,0.1,1,10$.The $p_{center}$ as $1.86,1.872,1.876,1.880,1.884,1.888,1.89$ are selected as the corresponding classical ensembles all have the escaping situation within the order of sticking time for $10^{2}$.In the figure(a)for small perturbation as $\sigma=0.01$,there is a fast decay followed by a slow decay clearly observed for most $p_{center}$ except for the cases of $p_{center}=1.86,1.872$ as the degree of decline is small.There is also a common feature for the cases of $p_{center}=1.876,1.880,1.884$ that LE can decay to some value followed by the oscillation with frequencies very close to each other as well as the tendency to decline quite slowly.The degree of decline for LE with the enlargement of $p_{center}$ undergoes non-monotonous variation,it is monotonically enhanced by increasing $p_{center}$ to $1.884$,and then monotonically retrieved afterwards.In the figures(b)and(c),the variation of degree of decline changes the order for the cases of $p_{center}=1.888,1.89$ as the fast decay of LE is for $p_{center}=1.888$ with $\sigma=0.1$ but for $p_{center}=1.89$ with $\sigma=1$.In the figure(d),the fluctuations of LE have been changed to be quite large but the order of variation of degree of decline remain the same to the situation of $\sigma=1$.In terms of logarithm scale for describing the variation of $t$,we can find there are three groups as $p center = 1.86,1.872$,$p center = 1.876,1.880,1.884$ and $p center = 1.888,1.89$ respectfully have the similar time scale although actually the considerable differences exist from the common scale of $t$.The processes without obvious decline for LE can be observed clearly in the figure(c)for $p_{center}=1.880,1.882,1.884,1.888,1.89$ before a continued decay that also can be illustrated by variation of $ln[-ln(M(t))]$ with $ln(t)$ as the intermediate process between initial decay and the second decay for large perturbation.}\label{fig11}
\end{figure}
\end{center}

To explicitly determine the time scale and decay law in detail,we need to extract some variables to characterize the basic feature we directly observe,meanwhile we also want to do some comparison for our effectiveness of semi-classical method.Firstly,we want to get the variation $c$ and $\alpha$ with time $t$ in terms of different $p_{center}$ and $\sigma$.Then we can fix the typical time to study the variation of $c$ and $\alpha$ with $p_{center}$ and $\sigma$ respectfully.At last,we want to study the time scale based on the information we get from the variation of $c$ and $\alpha$.As the decay law is expected as:
    \be \label{decay_law}\ln(-\ln{M(t)})\approx \ln{c_0}+{\nu}ln{\sigma}+{\alpha}\ln{t}\ee,
then we want to directly check the expected linear relation as $ln(-ln(M(\sigma)))$ versus $ln(\sigma)$ for a given time $t$ with numerical observation.To show the decay law we find in terms of edge of chaos is common,we study the center of wave-packet spanning from $p_{center}=1.85$ to $p_{center}=1.89$ with actually two escaping situation happened for a corresponding classical ensemble.

From the direct numerical observation,we can find the expected linear relation $ln(-ln(M(\sigma)))$ versus $ln(\sigma)$ does exist for all the field of perturbation $\sigma$ within the order of time as $10^{1}$ independent of different $p_{center}$,and the process of distortion of this linear relationship deserve us attention,thus the typical cases as $p_{center}=1.85,1.86,1.87,1.88,1.89$ are used to illustrate it in terms of different time for $t=10,50,100,200,400,600,800,1000$.We can find the linear dependence of $ln(-ln(M))$ on  $ln(\sigma)$ for the direct quantum computation in the figure 12(a) as $t=10$,and the comparison of the numerical results with semi-classical method show the obvious deviation just happen in the initial small field of perturbation $\sigma$.With time increasing,we can find the destructions of linear relationship begin for most cases of $p_{c}$ except the case of $p_{c}=1.85$ from the near field of largest perturbation and then extend gradually to the smaller field of perturbation.Linear relationship for the case of $p_{c}=1.85$ can hold until the field of perturbation with saturation happened for a given time and this expression is due to the Gaussian decay for the case of $p_{c}=1.85$ inside the regular region without the escaping situation happened in terms of classical correspondence.For the cases of $p_{c}=1.88,1.89$ in terms of some given time with clear observation,after approximately foregoing linear relationship,a quite slow variation for $ln(-ln(M))$ can be typically found in the figure 12(b),(c)and(d).Further more this field of perturbation with a slow variation of $ln(-ln(M))$ do not simply correspond to saturation by checking different time together in a given $p_{c}$ and we can find this expression is similar to fidelity decay in strong chaos having the independent decay without much effect from the perturbation beyond a threshold.We also can find the linear increasing relationship can not be hold even for the very initial field of perturbation in terms of the case of $p_{c}=1.88$ with $t=600$ depicted in the figure 12(f) and it means this kind of process of decay have the fluctuated feature after some monotonic decay with the evidence clearly seen in the figure 11(a).The change of relative distance of values of $ln(-ln(M))$ among different $p_{c}$ during time increasing deservers attention and a typical and noticeable change exists for the pairs of $p_{c}=1.88,1.89$ as the value of $ln(-ln(M))$ for the case of $p_{c}=1.88$ can become smaller than the corresponding value of the case of $p_{c}=1.89$ below almost a common threshold of perturbation as $\sigma=0.06$ with time increasing.Actually we also find the similar expression for the pairs as $p_{c}=1.85,1.86$ and $p_{c}=1.85,1.87$ and the key point here is the emergence of of intersection of perturbation that can be taken as the indicator for the comparison of faster or slower of fidelity decay related to existence of critical perturbation directly observed for the decay process in the figure 11.

\begin{center}  
\begin{figure}
\includegraphics[width=16cm,height=20cm]{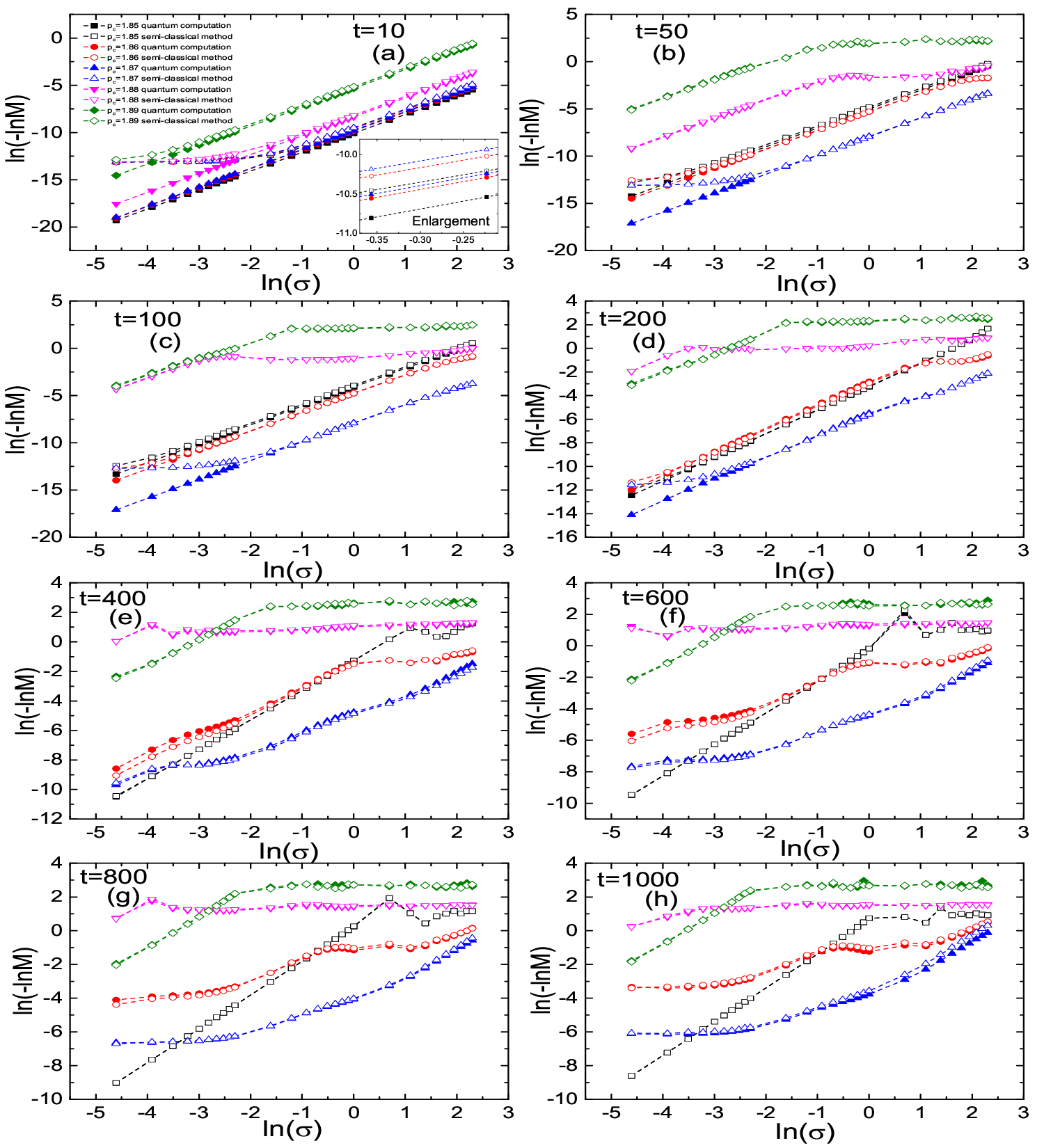}
\vspace{-0.2cm}\caption{Variation of $ln(-ln(M))$ with $ln(\sigma)$ for different $p_{center}=1.85,1.86,1.87,1.88,1.89$ evenly selected from $1.85$ to $1.89$ corresponding to fixed different time as $t=10,50,100,200,400,600$.To simplify the expression,the symbol $p_{center}$ is substituted by $p_{c}$.The numerical results are calculated by the direct quantum computation and semi-classical method for comparison respectfully illustrated with the solid and hollow interior.For the very initial time as $t=10$ in the figure(a),the expected linear relationship can be hold independent of different $p_{c}$.With time increasing,the distortion of linear relationship happens initially in the field of large perturbation and gradually extends to the small perturbation field except the expression of $p_{c}=1.85$.In terms of the cases of $p_{c}=1.88,1.89$,after approximately linear part,a quite slow variation for $ln(-ln(M))$ typically is found in the figure(b),(c)and(d).It deserves us to notice that the linear increasing relationship can be basically effaced for the case of $p_{c}=1.88$ seen clearly in the figure(f).Further more,one also can notice the change of relative distance of values of $ln(-ln(M))$ among different $p_{c}$ during time increasing,which typically can be seen for the pairs of $p_{c}=1.88,1.89$ and $p_{c}=1.85,1.87$.Therefore the intersection of perturbation implicates the change of relative decay speed related to time scale.}\label{fig12}
\end{figure}
\end{center}

To get a whole understanding about the rule of variation of the dependence of $ln(-ln(M))$ on $ln(\sigma)$ with different time,we have to put some selected time as $t=50,100,200,500,1000,2000,5000$ together to study the dependent relationship carefully for every $p_{c}$ illustrated in the figure 13.For the very initial time such as $t=10$,we can observe the universal linear relationship and also illustrate typically in the figure 12 which do not deliberately show it for every $p_{c}$ here.The distinction of the variation of $ln(-ln(M))$ versus $ln(\sigma)$ for different $p_{c}$ in terms of some given time we study can be understood well from the classical correspondence using the relation between the classical ensemble and quantum wave-packet illustrated in the figure 2.To make our work more convincing,we choose $p_{c}$ from $1.85$ to $1.89$ evenly selected by the interval as $0.02$,thus we can get enough information to support our basic research idea for the classical correspondence.For the sticking situation happened as $p_{c}=1.85,1.852,1.866$,we can find the obvious linear relationship in the corresponding figure(a),(b)and(i) almost for all the time with the field of perturbation from $\sigma=0.01$ to $\sigma=10$.For a carefully observation for $p_{c}=1.866$ in the figure(i),the whole linear relationship have some deviation for a long time as $t=5000$ originating from the field of large perturbation.We fit the slopes of the linear dependence about the variations in terms of $p_{c}=1.85,1.852,1.866$ for different time and they are all very close to $2$ which means the corresponding quantum decay can be seen as Gaussian conforming to the classical correspondence.Further more we can find the proximate values to $p_{c}=1.852,1.866$ as $p_{c}=1.854,1.864,1.868$ can hold the basic linear relationship except the field of small perturbation and it is a quite non-trivial also supported by the relationship as $ln(-ln(M))$ versus $ln(t)$ for different typical perturbations in the figure 9 for $p_{c}=1.868$.Thus we can not call this kind of situation as the common distortion of linear relationship pointed out above for the distortion initially happened in the field near largest perturbation.For the variations of $ln(-ln(M))$ with $ln(\sigma)$ in terms of $p_{c}=1.856,1.868,1.86,1.862$,they basically share the similar distortion of linear relationship for a given time with the value of $ln(-ln(M))$ continuously increasing to the likely saturation.For the $p_{c}$ we choose from $1.87$ to $1.89$ which is our main consideration in the study of edge of chaos,we can find a pattern of the variations for different given time is formed gradually with $p_{c}$ increasing as the initial parts corresponding to different time all show the approximately linear relationship and then a quite slow variation can be found that could be called the platform but not just as saturation and can be obtained after some threshold of perturbation for a given time.During the process of forming this pattern which becomes notably after the variations for $p_{c}=1.878$,we can find a common tendency that $ln(-ln(M))$ versus $ln(\sigma)$ still can increase more or less after some slow variation for a given time even that is quite long obviously seen in the figure(q)and(r)corresponding to $p_{c}=1.882,1.884$.In terms of a given time,we can find the slow variation gradually comes into being with the $p_{c}$ increasing and gradually become notably and finally extends to all the field of perturbation above the threshold.As the perturbation we choose have a large span,thus what we find here is a basic rule.Further more the threshold of perturbation can not be necessarily same for different time in terms of a fixed $p_{c}$ which is different from the feature of fidelity decay of strong chaos to differentiate the FGR decay and independent exponential decay with a explicit and universal threshold of perturbation independent of a given time.\cite{Scholarpedia}.

\begin{center}  
\begin{figure}
\includegraphics[width=18cm,height=20cm]{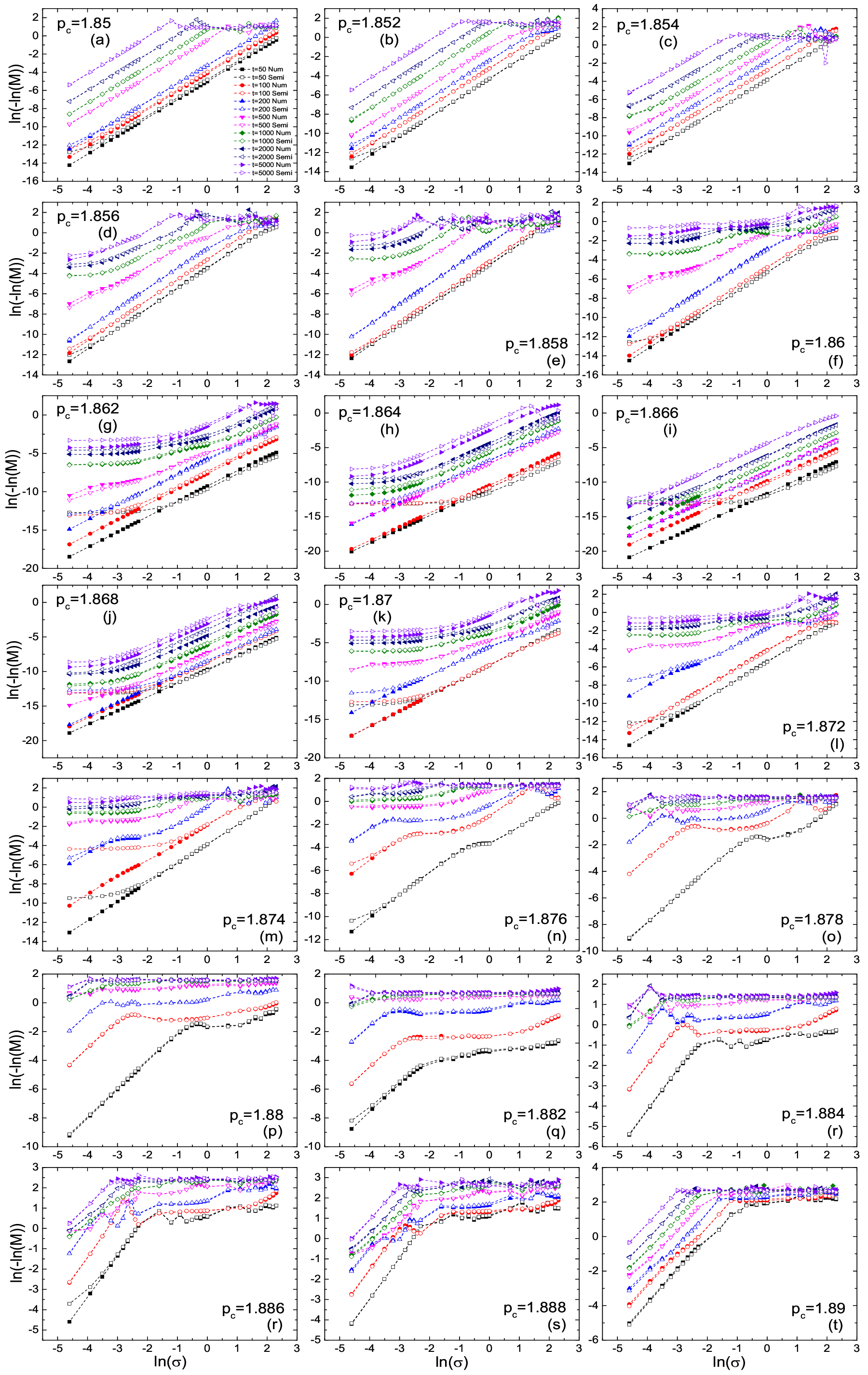}
\vspace{-0.5cm} \caption{Variation of $ln(-ln(M))$ with $ln(\sigma)$ for a given $p_{c}$ with different time as $t=50,100,200,500,1000,2000,5000$.The solid and open shapes still show respectively the results from direct numerical computation and semi-classical method.The obvious linear relationships can be found in the figure(a),(b)and(i)corresponding to $p_{c}=1.85,1.852,1.866$ for almost all the time we study without the saturation happened.Further more,we can find the proximate values to $p_{c}=1.852,1.866$ as $p_{c}=1.854,1.864,1.868$ can hold the basic linear relationship except the field of small perturbation.For the figure(d),(e),(f),and(g)corresponding to $p_{c}=1.856,1.868,1.86,1.862$,the variations share the similar distortion of linear relationship with the value of $ln(-ln(M))$ basically increasing to the likely saturation if applied for some given time.From the figure(k)to(u),a patter of the variation can be formed gradually as the initial parts corresponding to different times all show the approximately linear relationship and then a platform can be found but not for saturation obtained after some threshold of perturbation for a given time.Actually the universal linear relationship can exist in the very initial time selected to be illustrated in the figure 12 and do not deliberately show it in this figure.}\label{fig13}
\end{figure}
\end{center}

Based on the numerical study of the variations of $ln(-ln(M))$ with $ln(\sigma)$,it is not reasonable to fit all the field of perturbation to get $\nu$ as a important variable to characterize the decay law.So we can consider two ways to fit,one way is to fit with just initial three perturbations as $\sigma=0.01,0.02,0.03$ for finding the local variation of relationship as $ln(-ln(M))$ versus $ln(\sigma)$ and the other way is to fit with the perturbations
corresponding to the increasing value of $ln(-ln(M))$ which can be taken as the averaged effect.For simplicity,the methods to get the values of $\nu$ can be called as $Num1,Semi1,Num2,Semi2$ corresponding to two different fitting procedures in terms of respectfully the direct computation and semi-classical integral.To clearly show the variation,here we depict them separately with two figures as figure 14 and 15 although combined to show the typical features of variation of $\nu$.Meanwhile we also show the fitted value from semi-classical integral for the comparison,and here we mainly show the feature of variations of fitted $\nu$ from $Num1$ and $Num2$ and then the comparisons with the fitted $\nu$ from $Semi1$ and $Semi2$ are also investigated.

 The first notable expression is that the common value as $2$ of $\nu$ fitted from $Num1$ and $Num2$ is for initial time at least for the order $10^1$ independent of different $p_{c}$,which is illustrated clearly in the figure 14 and 15 with logarithmic coordinate.For the variation of value of $\nu$ fitted from $Num1$ besides $p_{c}=1.852,1.866$ with the classical correspondence as sticking situation,we can find there is always some obvious decreasing process after the initial time for the frozen value as $2$ and then it has the rising tendency with some fluctuation weakened to the very small value close to zero clearly seen for $p_{c}=1.87,1.872$ in the figure 14(j)and the figure 15(a).The situation about the quite small value of $\nu$ is highly non-trivial,it means that approximately independent decay can exist in the field of small perturbation which is out of previous study in the strong chaos that considers this kind of decay only can exist for large perturbation.Thus we can divided the variation of $\nu$ fitted from $Num1$ with time increasing as three basic parts illustrated with logarithmic expression,the value of $\nu$ for third part still can undergo some complicated change with notable oscillation as $p_{c}$ is increased from $1.874$ to $1.884$ with interval as $0.02$,and then it shows the increasing tendency again for $p_{c}=1.886$,and finally the increasing tendency can be ceased to show some asymptotically stability for $p_{c}=1.89$.For the corresponding value of fitted $\nu$ from $Num2$,we still can find some clear decreasing process after the initial frozen value as $2$,and then gradually the value of $\nu$ can vary small which also can be called some stable field commonly found for $p_{c}$ from $1.85$ to $1.87$ but except $p_{c}=1.85,1.866$ close to $2$ belonging to classical sticking situation,this kind of variations have been illustrated in the figure 14.In terms of the variation of $\nu$ for $p_{c}=1.872$,we can find a quite special expression that its value can decrease again before a quite slow variation and then the typical three parts for the variation pointed out above can be seen for $p_{c}=1.874,1.876$.For $p_{c}$ from $1.878$ to $1.884$,the variations of $\nu$ fitted by $Num1$ and $Num2$ are quite similar with some notably large oscillation and then they also share the common tendencies for $p_{c}=1.886,1.888,1.89$ but there are some large fluctuation for the value of $\nu$ from $Num2$.These features of variation can be seen clearly in the figure 15.

For the comparison between $\nu$ from $Num1$ and $Semi1$ as well as from $Num2$ and $Semi2$ besides $p_{c}=1.85,1.866$ corresponding to classical sticking situation,the degree of consistency for the initial time is not good independent of different $p_{c}$ at least corresponding to $\nu=2$ and the degrees of consistency after the initial stage can be much better at least for the decreasing process as a common feature.For $p_{c}$ from $1.86$ to $1.868$ except $p_{c}=1.866$,we can find the occupancy of the part of good consistency for the decreasing process is gradually reduced as well as degree of consistency for the part of rebounding process is increased although the fluctuation becomes notably larger seen in the figure 14(g)and(i)as $p_{c}=1.864$ and $p_{c}=1.868$.The occupancy of the part of good consistency for the decreasing process is gradually increased for $p_{c}$ from $1.87$ to $1.878$,and degree of consistency can be seen as good without initial stage after $p_{c}=1.878$.The degree of consistency after some threshold become poor for $p_{c}=1.85$ that can be directly connected with the relationship as $ln(-ln(M))$ versus $ln(\sigma)$ for quite a long time that is not showed yet in the figure 13(b),and they are all poor for $p_{c}=1.866$ as the large difference of evaluation of fidelity between the direct computation and semi-integral for small perturbation.It is easy to find these expressions in the figure 14 and 15.

\begin{center}  
\begin{figure}
\includegraphics[width=18cm,height=18cm]{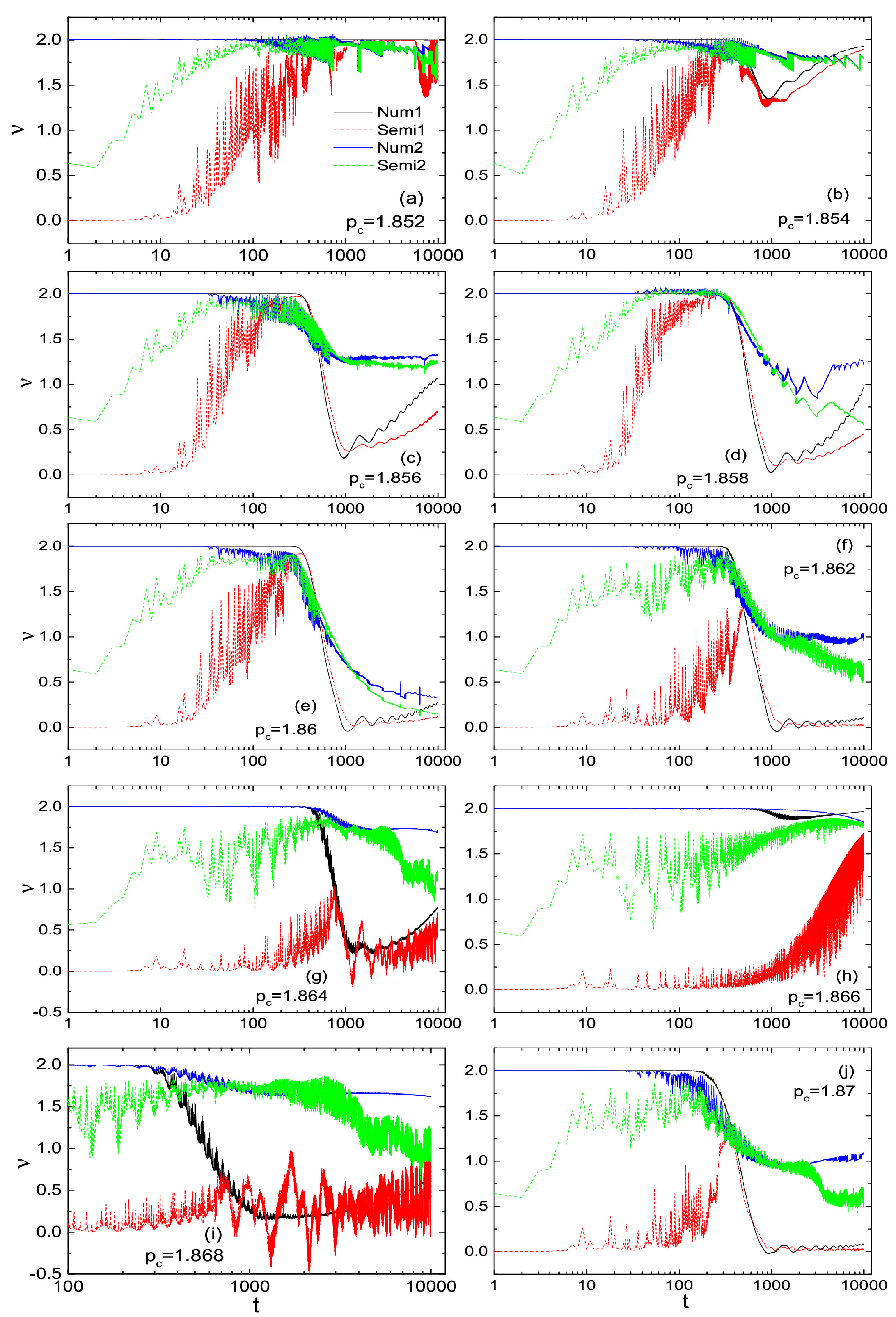}
\vspace{-0.5cm} \caption{Variation of fitted $\nu$ with time $t$ for $p_{c}$ from $1.85$ to $1.87$.$Num1,Semi1,Num2,Semi2$ are the methods for fitting $\nu$ through the relationship as $ln(-ln(M))$ versus $ln(\sigma)$ by the selecting perturbations as $0.01,0.02,0.03$ and the restriction of increasing function in terms of the value of $ln(-ln(M))$ calculated from direct computation and semi-classical integral respectfully.There are obviously three basic processes for the variation of fitted $\nu$ from $Num1,Num2$ as the initial frozen value for $2$,decreasing process and rebounding process besides for $p_{c}=1.85,1.866$ with the values of $\nu$ all close to $2$ with some large fluctuation for $p_{c}=1.85$.The variation of $\nu$ from $Num2$ for so-called rebounding process is some different from corresponding expression from $Num1$ as some quite slow variation can be commonly found.Meanwhile,we can find the quite small value of $\nu$ from $Num1$ in the figure(j) for $p_{c}=1.87$ which means independent decay can exist approximately for the field of small perturbation.For the comparison between $\nu$ from $Num1$ and $Semi1$ as well as from $Num2$ and $Semi2$,the degree of consistency for the initial time is not good at least corresponding to $\nu=2$ and the following degrees of consistency for different $p_{c}$ are all much better.For $p_{c}$ from $1.86$ to $1.868$ except $p_{c}=1.866$,we can find the occupancy of the part of good consistency for the decreasing process is gradually reduced as well as degree of consistency for the part of rebounding process is increased although the fluctuation becomes notably larger for these two parts clearly seen in the figure(g)and(i)as $p_{c}=1.864$ and $p_{c}=1.868$.The degrees of consistency after some threshold become obviously worse for $p_{c}=1.85$,and they are not good for $p_{c}=1.886$ as the large difference of evaluation of fidelity between the direct computation and semi-integral for small perturbation seen clearly in the figure 13(i).}\label{fig14}
\end{figure}
\end{center}

\begin{center}  
\begin{figure}
\includegraphics[width=18cm,height=20cm]{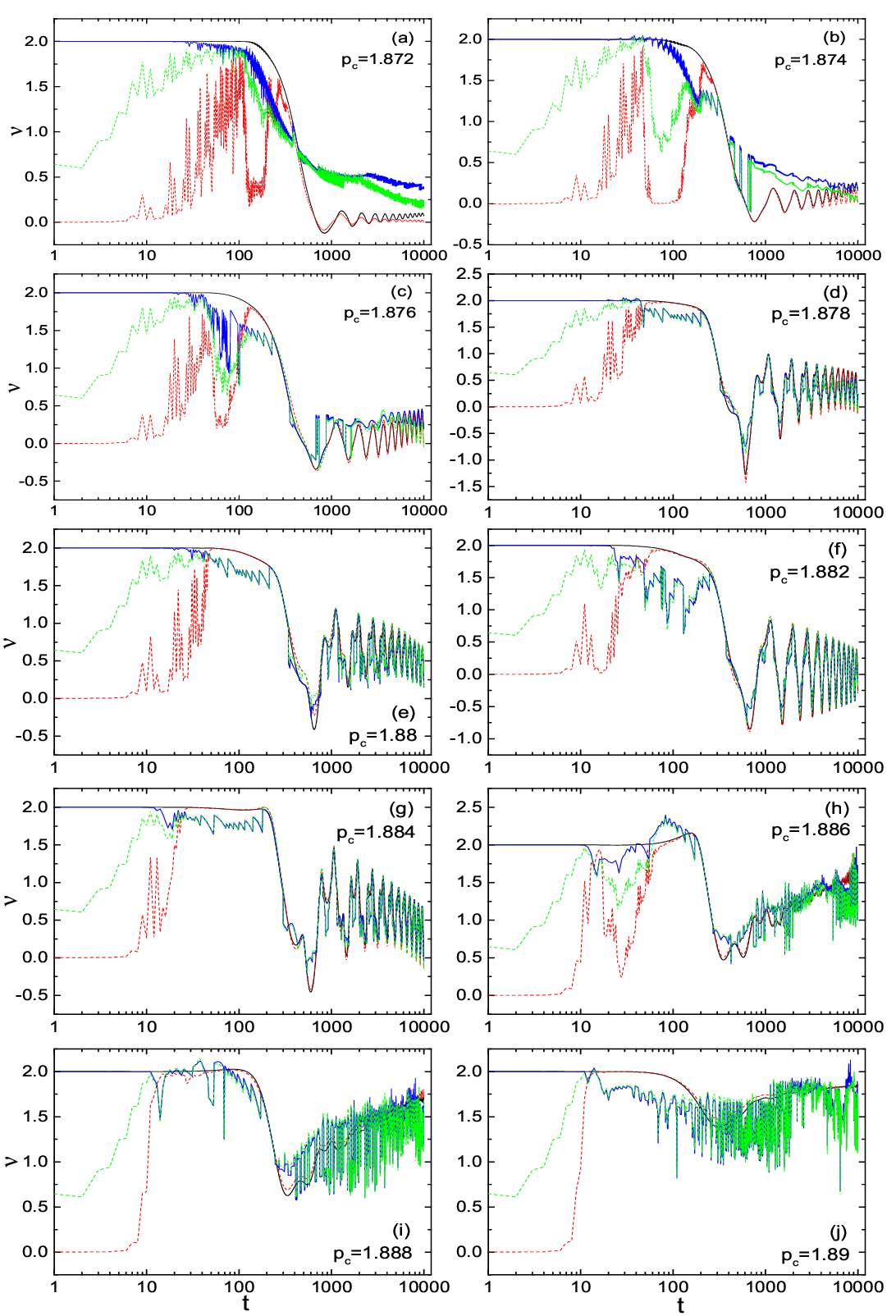}
\vspace{-0.5cm} \caption{Variation of fitted $\nu$ with time $t$ for $p_{c}$ from $1.872$ to $1.89$.The basic three parts of variation can be found similar to the expression in the figure 14 and we also can find the situation of quite small value of $\nu$ from $Num1$ in the figure(a) for $p_{c}=1.872$ is similar to $p_{c}=1.87$ in the figure 14(j).For $p_{c}$ from $1.878$ to $1.884$,the variations of $\nu$ fitted by $Num1$ and $Num2$ are quite similar for each other with some notably large oscillation and then they also share the common tendencies for $p_{c}=1.886,1.888,1.89$ but there are some large fluctuations for the values of $\nu$ from $Num2$.The rebounding process can undergo some non-monotonic change for overall trend with notable oscillation as $p_{c}$ is increased from $1.874$ to $1.884$ with the interval as 0.02,and then it shows the increasing tendency again for $p_{c}=1.886$,and finally the increasing tendency can be ceased to show some asymptotically stability for $p_{c}=1.89$.In terms of the variation of $\nu$ from $Num2$ for $p_{c}=1.872$,its value can decrease again before a quite slow variation.For the comparisons between $\nu$ from $Num1$ and $Semi1$ as well as from $Num2$ and $Semi2$,the degree of consistency is not good still at least corresponding to $\nu=2$,and the following degrees of consistency can be seen as good after $p_{c}=1.878$,further more,the occupancy of the part of good consistency for the decreasing process is notably increased more and more for $p_{c}$ from $1.872$ to $1.878$ corresponding to the figure(a),(b),(c),(d). }\label{fig15}
\end{figure}
\end{center}

Actually we still want to find the variations of $\nu$ with different $p_{c}$ for different time,and the study results are illustrated in the figure 16.For the very initial time,we can find the values of fitted $\nu$ from $Num1,Num2$ are seen as $2$ with high precision independent of different $p_{c}$,and this expression is conformed to  the previous study in short time with the relationship as the variation of $ln(-ln(M))$ versus $ln(\sigma)$ in the figure 14 and 15 and here we use the time as $t=10$ to show it in the figure 16(a).Then we can find some deviations from $2$ do happen firstly in the latter part of $p_{c}$ with time increasing,and gradually this kind of deviations can continue to extend to adjacently smaller $p_{c}$ which can be seen in the figure(b),(c),(d).But then we can observe the new deviation can bypass the intermediate part of $p_{c}$ to the initial part for the time increasing further more clearly seen in the figure 16(e),(f),(g).In terms of this kind deviations,we can find the values of $\nu$ fitted by $Num1$ are more robust than corresponding values by $Num2$ and these expressions implicate that the linear relationship for $ln(-ln(M))$ versus $ln(\sigma)$ can be expected in the field of small perturbation within short time as $10^2$ independent of different $p_{c}$.For the time scale spanning from $10^2$ to $10^3$,we can find the variations of $\nu$ with different $p_{c}$ can evolve into a pattern with the values of $\nu$ corresponding to $p _{c}=1.85,1.852,1.866,1.89$ as the backbones to form a non-monotonic variation similar to $W$ shape which can be seen clearly in the figure 16(h),(i),(j),(k).Then we can find this basic pattern for the variation can hold without much change in terms of time scales from $10^3$ to $10^4$ illustrated by the figure 16(l)to(p).For the comparisons between $\nu$ from $Num1$ and $Semi1$ as well as from $Num2$ and $Semi2$,we can find the degrees of consistency are not good for the very initial time illustrated in the figure 16(a) for $t=10$,and then they tend to be better for the field of later part of $p_{c}$ with time initially increasing that can be clearly seen from the figure 16(b)to(e),meanwhile we also can find this kind of better consistencies can gradually move to the field of initial part of $p_{c}$ but not for the field of intermediate part which can be seen clearly in the figure 16(f),(g).With time increasing more with the time scale from $10^{2}$ to $10^{3}$,we can find some good degrees of consistency are obtained gradually for the field of intermediate part of $p_{c}$ distributed around $p_{c}=1.866$ which can be seen clearly in the figure(h),(i),(j),(k).For the time scale from $10^{3}$ to $10^{4}$,the basic good degrees of consistency for most $p_{c}$ can hold which can be seen for the figure 16 from(l)to(p)although the field of intermediate part pointed out before can not have a good consistency for a long time.

\begin{center}  
\begin{figure}
\includegraphics[width=18cm,height=18cm]{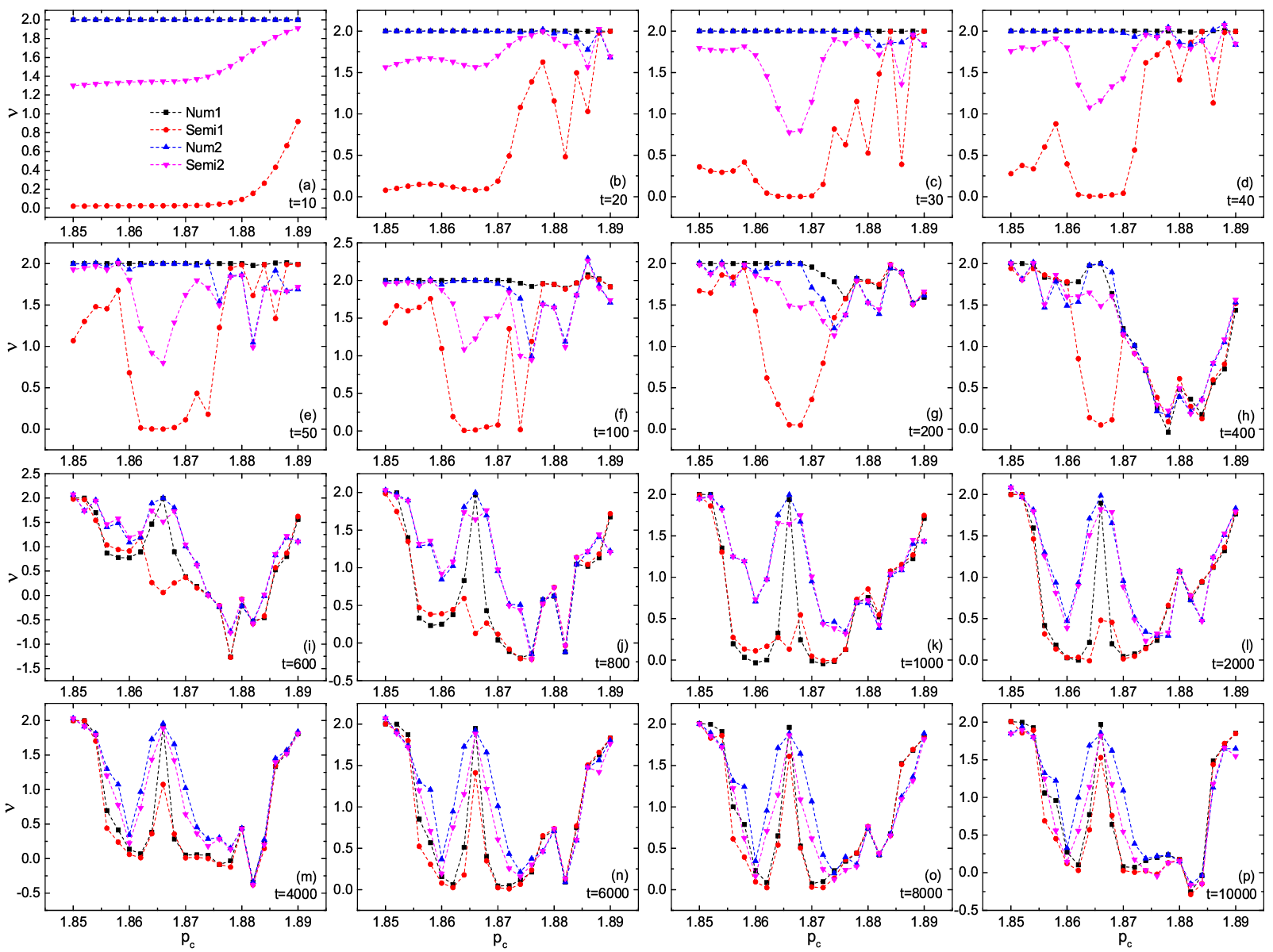}
\vspace{-0.5cm} \caption{Variation of $\nu$ fitted from $Num1,Semi1,Num2,Semi2$ with $p_{c}$ for different given time $t$.For the very initial time,the values of fitted $\nu$ from $Num1,Num2$ are seen for $2$ with high precision independent of different $p_{c}$ seen clearly in the figure(a).The deviations from $2$ can be seen firstly for the field between $1.88$ and $1.89$ found in the figure(b),and gradually the deviations can extend to adjacently smaller $p_{c}$ as well as the field of initial part of $p_{c}$ we consider leaving the intermediate field untouched which can be seen clearly from the figure(c)to(g).In terms of this kind deviations,we can find the values of $\nu$ fitted by $Num1$ are more robust than corresponding values by $Num2$ and these expressions implicate that the linear relationship for $ln(-ln(M))$ versus $ln(\sigma)$ can be expected in the field of small perturbation within short time as $10^2$ independent of different $p_{c}$.From the figure(h)to(k),the variation of $\nu$ with $p_{c}$ evolves a pattern with the values of $\nu$ corresponding to $p_{c}=1.85,1.852,1.866,1.89$ as the backbone to form a non-monotonic variation similar to $W$ shape.From the figure(l)to(p),the basic pattern of variation can hold without much change.For the comparisons between $\nu$ from $Num1$ and $Semi1$ as well as from $Num2$ and $Semi2$,we can find the degrees of consistency are not good for the very initial time clearly seen in the figure(a),and then they tend to be better firstly for the field of later part of $p_{c}$ typically seen in the figure(e) and this kind of better consistencies move on for the field of initial part of $p_{c}$ typically seen in the figure(h),and the field of intermediate part of $p_{c}$ can also obtain some good degrees of consistency afterwards in a gradual process until for $t=1000$ corresponding to figure(k).For the figure(l)to(p),the basic good degrees of consistency for most $p_{c}$ can hold as a whole although the field of intermediate part pointed out before can not have a good consistency for a long time.}\label{fig16}
\end{figure}
\end{center}

Then we want to study the variation of $\alpha$ with time as $t$ which is a key point to differentiate the decay laws.Through the carefully numerical investigation,$50$ can be taken as the time step to do the fitting to show the basic variation of $\alpha$.The fidelity decay can be taken as exponential decay for $\alpha=1$,Gaussian decay for $\alpha=2$,cubic-exponential decay for $\alpha=3$,and stretched exponential decay for $\alpha$ between $0$ and $2$ but not for $1$.Now we need to show the variation versus time in terms of some typical perturbations,and choose
$\sigma=0.01,0.1,1,10$ to observe the decay laws and the time used to show the variation is not necessarily same as the consideration here is the basic pattern for the variation.As the span for studying $p_{c}$ is wide from $1.85$ to $1.89$ including the edge of chaos,it is reasonable in practice to use three figures as figure 17,18,19 to show the variation versus time clearly but we should take them as a whole to understand likely rules for the variation in the field of edge of chaos.In practice,We can recognize the expected decay laws if the values of fitted $\alpha$ can approximately dwell on or oscillate unattenuated around some fixed value for some time and thus we use some reference lines likely as $0,1,2,3$ to guide the eyes.Here we can judge the ending of effective fidelity decay by finding the value of $\alpha$ beginning to below $0$ and fluctuating evenly around $0$ afterwards.

Firstly we consider the variation of $\alpha$ for $p_{c}$ with span from $1.85$ to $1.868$ as there are related to stable dynamics in terms of classical ensemble for $p_{c}=1.85$ and $p_{c}=1.866$.For $p_{c}=1.85,1.852,1.866$ corresponding to classical stable field,$\alpha$ can almost fix or oscillate unattenuated around the value as 2 even for very large perturbation as $\sigma=10$ without consideration of the saturation leading to its value dropping to zero seen clearly in the figure 17(a),(b),(c),(d) and 18(c),(d)which is reconciled with previous study of fidelity for stable dynamics as Gaussian decay\cite{Gorin},then we just study the variation of $\alpha$ for the rest of $p_{c}$.For perturbation as $\sigma=0.01$,the common variation of $\alpha$ for $p_{c}=1.854,1.864,1.868$ seen in the figure 17(e),18(a),(e)can show chronologically as the very initial non-monotonic variation,gradually increasing process,and then decreasing process evolving into the stable variation with its value between $0$ and $2$.For $p_{c}=1.856,1.858,1.86,1.862$ in terms of $\sigma=0.01$,the variations of $\alpha$ are quite similar to the situations of $p_{c}=1.854,1.864,1.868$ but the obvious differences consist in that the increasing processes after very initial non-monotonic variation are more steep and emergent stable oscillation after the transient decreasing process seen clearly in the figure 17(g),(i),(k),(m).For perturbation as $0.1$,there are some oscillations around $2$ for $p_{c}=1.864,1.868$ or at least close to $2$ for $p_{c}=1.854$ besides the considerable decreasing process within initial non-monotonic variation.There is a change for $\sigma=0.1$ from parabola-like process for $p_{c}=1.856,1.858$ to gradual decreasing process as the typical feature for $p_{c}=1.86,1.862$ although the common expression for the initial non-monotonic variation.For $\sigma=1$,$\alpha$ in terms of $p_{c}=1.856,1.86$ can undergo some non-monotonic variations going through some ups and downs with the gradual process to below $0$ or enter the field basically between $0$ and $2$ respectfully seen in the figure 17(h),(l),and some minor variation of $\alpha$ in terms of $p_{c}=1.858$ basically between $2$ and $3$ can be seen in the figure 17(j),basic unattenuated oscillation balanced below $2$ in terms of $p_{c}=1.862$ can be seen in the figure 17(n),$\alpha$ in terms of $p_{c}=1.854$ can show the initial non-monotonic variation and afterwards decreasing process gradually crossing $2$ to the field below $0$ seen in the figure 17(f),and there are the similar oscillations for $\alpha$ in terms of $p_{c}=1.864,1.868$ around the value near $2$ seen in the figure 18(b),(f).For $\sigma=10$,$\alpha$ in terms of $p_{c}=1.854,1.856,1.858$ can show quickly drops to negative values and some fluctuations around $0$ can be followed and $\alpha$ in terms of $p_{c}=1.86$ can also show some quickly drop but the irregular fluctuation afterwards with net fluctuation range above $0$ can be found as a whole,and $\alpha$ in terms of $p_{c}=1.862,1.864,1.868$ can oscillate with the balanced values almost around $2$ from direct observation.For comparison with the fitted $\alpha$ in terms of semi-classical method,the consistency can be seen as good for $p_{c}=1.85$ and tends to be better with perturbation increased for $p_{c}=1.852$,and the consistencies show the large deviations for $p_{c}=1.854,1.862$ in terms of $\sigma=0.01,0.1$ and can be seen as basic good for $p_{c}=1.856,1.858,1.86$,tend to be better with perturbation increased in the initial time scale around $10^3$ for $p_{c}=1.864,1.866,1.868$.Combining the study results,variations of $\alpha$ for $p_{c}=1.854,1.864,1.868$ closest to $p_{c}=1.852,1.866$ respectfully can show the obvious deviations from $2$ for $\sigma=0.01$ but are more close to $2$ at least for balanced values in terms of $\sigma=0.1,1,10$.For $p_{c}=1.856,1.858,1.86,1.862$,the variations of $\alpha$ can show obvious deviations from $2$,it could be seen to be more vulnerable for smaller perturbation from the effect of change of $p_{c}$.We also notice the very first values of $\alpha$ in terms of $p_{c}=1.85,1.852,1.854,1.856,1.858,1.86$ are between $2$ and $3$,and the very first values of $\alpha$ in terms of $p_{c}=1.862,1.864,1.866,1.868$ are between $0$ and $2$.

Then we can move to the study the variation of $\alpha$ for $p_{c}$ from $1.87$ to $1.89$ corresponding to different typical perturbations,and also want to find the basic rules.For perturbation as $\sigma=0.01$,we can find the variations of $\alpha$ seen in the figure 18(g),(i),(k),(m) can show the initial steep climbing processes after the very first drops and afterwards decreasing processes until some emergent large oscillations unattenuated for $p_{c}=1.87,1.872$ and gradually increased for $p_{c}=1.874,1.876$ during time going.For more careful observation,we can find the time for ending the decreasing process tends to be shorter with $p_{c}$ increased from $1.87$ to $1.876$,and the balanced values for these oscillations can be seen between 0 and 1 corresponding to stretched exponential decay actually.Then the variations of $\alpha$ for $p_{c}=1.878,1.88,1.882,1.884$ can become to dwell on the values around $3$ at least larger than $2$ continuing for some time towards basically beyond $t=500$ after the very initial alternations seen in the figure 19(a),(c),(e),(g)and then they express large oscillations with the amplitudes increasing more and more that can be found with the balanced values near to 0 which implicate quite slow stretched exponential decay.For $p_{c}$ increased as $1.886,1.888$ in terms of $\sigma=0.01$,the variation of $\alpha$ can change again to have the transition from a continuously decreasing process to progressively stable oscillation with the balanced values between $0$ and $1$ as stretched exponential decay seen clearly in the figure 19(i),(k),and finally the value of $\alpha$ for $p_{c}=1.89$ in terms of $\sigma=0.01$ can vary closely around $1$ as approximately exponential decay seen in the figure 19(m).For perturbation as $\sigma=0.1$ with $p_{c}$ increased,variations of $\alpha$ can show the main slowly deceasing processes crossing the value $1$ for $p_{c}=1.87,1.872,1.874$ seen in the figure 18(g),(i),(k)and then $\alpha$ for $p_{c}=1.876$ can dwell on the value obviously around $1$ for quite a long time with the order as $10^3$ seen in the figure 18(m),afterwards the values of $\alpha$ can quickly enter the field between $0$ and $1$ for $p_{c}=1.878,1.88,1.882,1.884$ seen in the figure 19(a),(c),(e),(g)and then the variation of $\alpha$ can be divided into two basic parts with the fluctuation of first part much larger seen in the figure 19(i),(k),finally $\alpha$ can basically vary around $1$ without the very first sharp drop showing the exponential decay seen in the figure 19(m) which is similar to the situation about $\sigma=0.01$.For variation of $\alpha$ for $\sigma=0.1$ in terms of initial time,we can find the variation as the initial raise after the very first drop can still exist for $p_{c}=1.87,1.872,1.874$ and then the purely initial raise can be found for $p_{c}=1.876,1.878,1.88$ also for $\sigma=0.01$,and then the very initial raise can disappear to let the very initial dramatic drop emergent firstly happened for $p_{c}=1.882$ and hold afterwards for $p_{c}=1.884,1.886,1.888,1.89$.Actually the first emergence of initial dramatic drop is postpone for $p_{c}=1.884$ in terms of $\sigma=0.01$,thus the very initial dramatic drops for $\sigma=0.01$ are related to $p_{c}=1.884,1.886,1.888,1.89$.Meanwhile we also notice the very first values of $\alpha$ are very close each other for $\sigma=0.01,0.1$ corresponding to $p_{c}$ from $1.87$ to $1.886$ and these values which can be near to $3$ are very common for $p_{c}$ from $1.872$ to $1.88$.

For perturbations as $\sigma=1$ with $p_{c}$ increased,the variation of $\alpha$ can undergo some quite different change.Firstly we can find some continues oscillation for $p_{c}=1.87$ with gradually balanced value between $0$ and $2$ seen in the figure 18(h) and it can change to be similar variation for $p_{c}=1.1.872,1.874$ seen in the figure 18 (h),(j),(l)respectfully as there are three parts for the initial values basically varying around $2$ or even dwelling close to $2$,the transitive decreasing process and the stable variation slightly above than $0$ or obviously oscillating around the value larger than $0$.Then the situation of variation of $\alpha$ as sticking around $2$ can not be maintained for $p_{c}=1.876$ but instead initially continuous decreasing process which can be ceased to rebound crossing $1$ and then the decreasing process can be found again until the emergent fluctuation around $0$ seen clearly in the figure 18(n).In terms of $p_{c}=1.878,1.88$,the similar pattern of variations of $\alpha$ to the situation of $p_{c}=1.876$ can be found but the bottoms of rebound are quite different smaller than $1$ seen in the figure 19(b),(d).In terms of $p_{c}=1.882$,there is basically the smooth decreasing process which can gradually enter into the field slightly above $0$ without rebounding process seen clearly in the figure 19(f).Then we can find  a very first steep drop to the values around $1$ before the later decreasing process to the values around $0$ for $p_{c}=1.884$ seen in the figure 19(h) and the variations of $\alpha$ for $p_{c}=1.886,1.888,1.89$ are similar as 
the very first steep drops to the values showing some rebounding zigzag-like variations until the field fluctuating around $0$.For perturbation as $\sigma=10$,the tendency of variations of $\alpha$ is similar to the situation of $\sigma=1$,but there are more likely abrupt drops in the initial time and also some large fluctuations happened which can be seen obviously in the figure 18(j),(l),(n),and figure 19(b),(d)corresponding to $p_{c}=1.872,1.874,1.876,1.878,1.88$.For $p_{c}=1.884$ particularly,the dwelling situation for $\sigma=1$ can still remain for $\sigma=10$ although weakened with the sticking value obviously smaller than $1$,and we even particularly find two dwelling situations for $\sigma=2$ not depicted yet as it is not a common situation.For the consistency of
comparison with the fitted $\alpha$ in terms of semi-classical method,it can be taken as basic good except for the time scale basically above $10^{3}$ for $\sigma=10$ in terms of $p_{c}=1.87,1.872,1.874,1.882,1.884$,here the large deviation from the satiation of fidelity decay arrived do not take into account.Meanwhile we also notice the very first values of $\alpha$ are very close each other for $\sigma=1,10$ corresponding to $p_{c}=1.87,1.872,1.874$,and these values which can vary around $3$ at least larger than $2$ are very common for $p_{c}$ from $1.872$ to $1.88$.

\begin{center}  
\begin{figure}
\includegraphics[width=18cm,height=17cm]{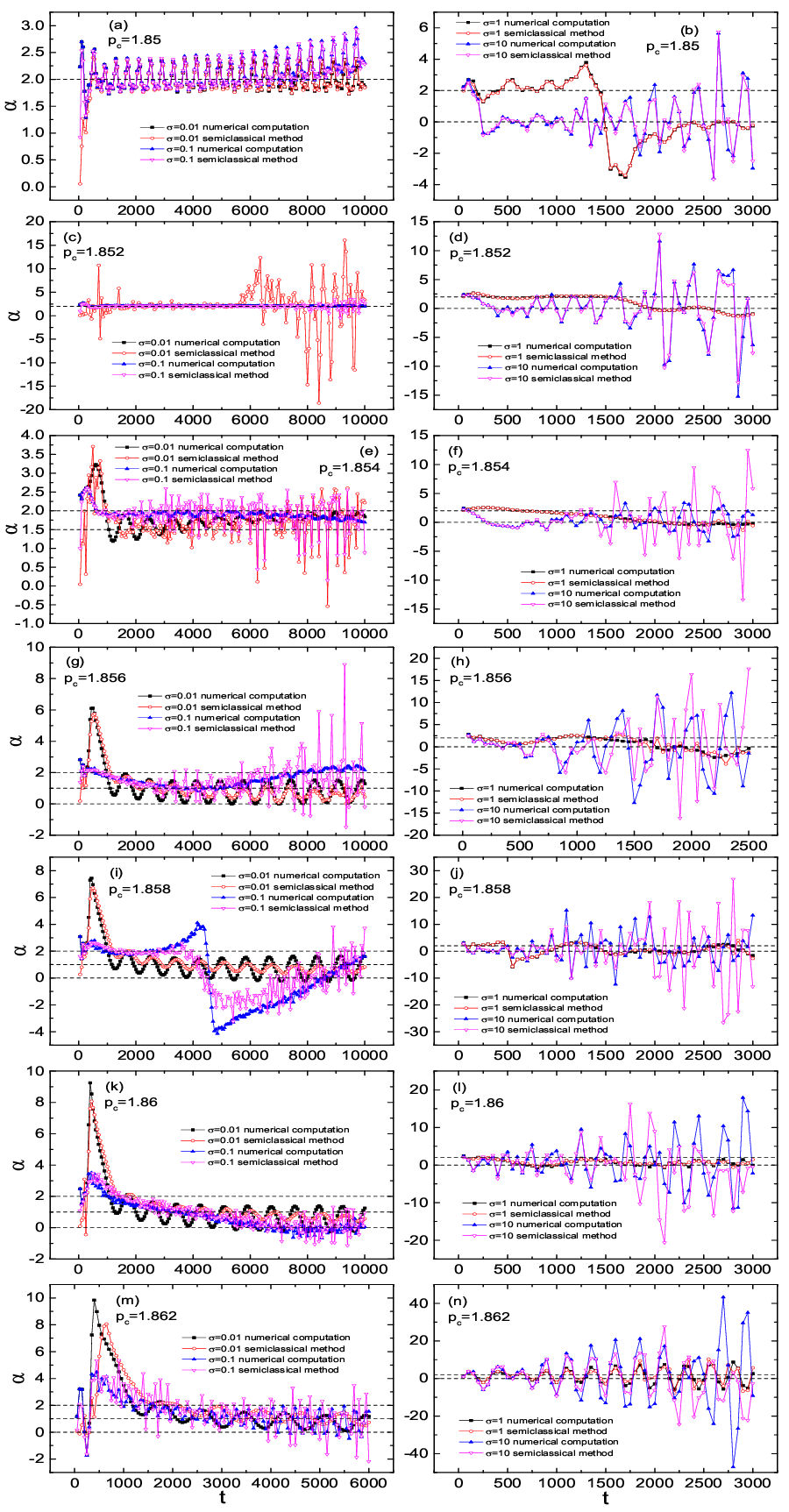}
\vspace{-0.2cm} \caption{Variations of $\alpha$ fitted by numerical computation and semi-classical method with time $t$ for $\sigma=0.01,0.1,1,10$ in terms of $p_{c}$ from $1.85$ to $1.862$ with the interval as $0.002$.In the figure(a),(b),(c),(d)for $p_{c}=1.85,1.852$,we can find $\alpha$ can dwell on the value as $2$ or around in some oscillated way except the gradual transitions to $0$ for $\sigma=1$ as well as some quickly drops to $0$ for $\sigma=10$.In the figure(e),(f)for $p_{c}=1.854$,the variation of $\alpha$ is similar to the situations of $p_{c}=1.85,1.852$ except obvious deviation from $2$ for $\sigma=0.01$ as the initial increasing process before the very beginning non-monotonic variation and then decreasing process evolving into oscillation finally increasing but bounded with $2$.In the figure(g),(i),(k),(m)for $p_{c}=1.856,1.858,1.86,1.862$ in terms of $\sigma=0.01,0.1$,we can find the variations of $\alpha$ for $\sigma=0.01$ are all quite similar to the corresponding expression of same perturbation for $p_{c}=1.854$,and actually the differences are the more steep increasing process for initial time and more stable oscillation after a transient decreasing process.There is a change for $\sigma=0.1$ from parabola-like process for $p_{c}=1.856,1.858$ to gradual decreasing process as the typical feature for $p_{c}=1.86,1.862$ although the common expression for the initial non-monotonic variation.For $\sigma=1$,the values of $\alpha$ in terms of $p_{c}=1.856,1.86$ can undergo some non-monotonic variations going through some ups and downs with the gradual process to below $0$ or enter the field basically between $0$ and $2$ respectfully seen in the figure(h),(l),and some minor variation of $\alpha$ in terms of $p_{c}=1.858$ basically between $2$ and $3$ can be seen in the figure(j)as well as basic unattenuated oscillation balanced near $2$ in terms of $p_{c}=1.862$ can be seen in the figure(n).For $\sigma=10$,some similar expressions in terms of $p_{c}=1.856,1.858$ can be found as quickly drops to negative values and some fluctuations around $0$ followed in the figure(h),(j),and also a quickly drop for $p_{c}=1.86$ but the irregular fluctuation afterwards with net fluctuation range above $0$ as a whole can be seen in the figure(l),and the unattenuated oscillation for $p_{c}=1.862$ balanced near $2$ can be found in the figure(n).For the comparison with the fitted $\alpha$ in terms of semi-classical method,the consistency can be seen as good for $p_{c}=1.85$ and tends to be better with perturbation increased for $p_{c}=1.852$,the consistencies can show some similar large deviations for $p_{c}=1.854,1.862$ in terms of $\sigma=0.01,0.1$,and can be seen as basic good for $p_{c}=1.856,1.858,1.86$.}\label{fig17}
\end{figure}
\end{center}

\begin{center}  
\begin{figure}
\includegraphics[width=18cm,height=18cm]{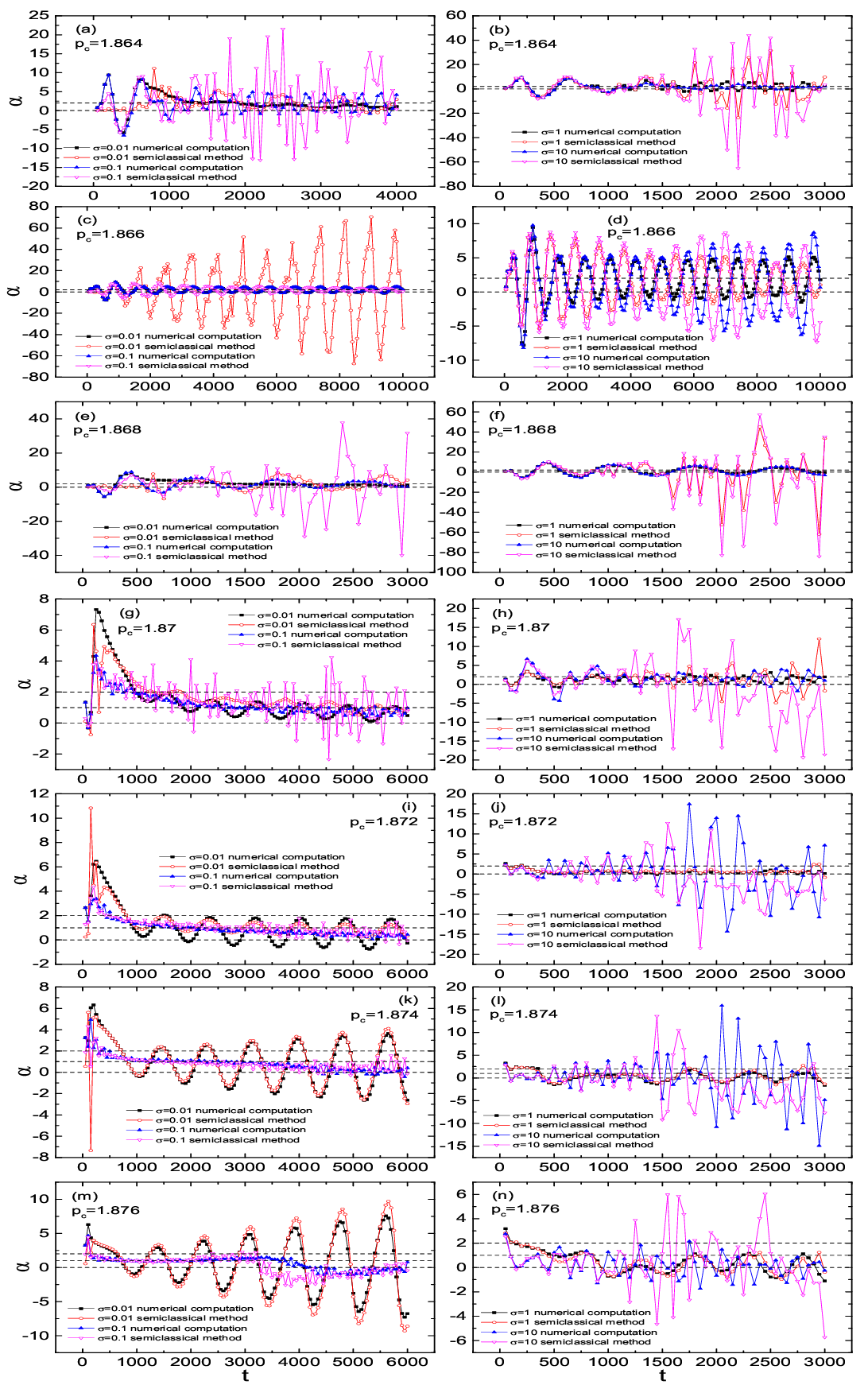}
\vspace{-0.2cm} \caption{The same consideration to the figure 17 for $p_{c}$ from $1.864$ to $1.876$ with the interval as $0.002$.The variations of $\alpha$ are similar for $p_{c}=1.864,1.868$ showed specifically in the figure(a),(b),(e),(f)as they can gradually evolve into some values between $0$ and $2$ for $\sigma=0.01$ and can have basic stable oscillations with the balanced values at least larger than $0$ for $\sigma=0.1,1,10$.For $p_{c}=1.866$ corresponding to the figure(c)and(d),we can find the variations of $\alpha$ are just around $2$ with some unattenuated oscillation which means mainly for Gaussian decay.In the figure(g),(i),(k),(m),we can find variations of $\alpha$ for $\sigma=0.01$ can mainly show the initial steep climbing process and afterwards decreasing process until some emergent large oscillations unattenuated for $p_{c}=1.87,1.872$ and increased for $p_{c}=1.874,1.876$ as time goes and actually there are the very first drops for $p_{c}=1.87,1.872,1.874$.For more carefully observation,we can find the time for ending the decreasing process becomes shorter with $p_{c}$ increased from $1.87$ to $1.876$,and the balanced values for these oscillations can be seen between $0$ and $1$.Meanwhile slowly decreasing processes as the main variation of $\alpha$ in terms of $\sigma=0.1$ tend to be ceased to dwell on the value as $1$ with $p_{c}$ increased from $1.87$ to $1.876$.In the figure(h),(j),(l),(n),we can find the variations of $\alpha$ for $\sigma=1$ can undergo some change from single continues oscillation for $p_{c}=1.87$ to mainly two parts for $p_{c}=1.872,1.874,1.876$ expressed specifically as the basically initial decreasing process with the very first value around 3 and then gradually stable field obtained with some oscillation more clear but the balanced value tending to be ceased to around $0$ for $p_{c}$ increased.Meanwhile the peculiarities of variation of $\alpha$ for $\sigma=10$ which are different from the corresponding expression for $\sigma=1$ are the initial abrupt drops and large fluctuations.The consistencies for the comparison with the fitted $\alpha$ in terms of semi-classical method tend to be better with perturbation increased in the order of initial time scale as $10^{3}$ for $p_{c}=1.864,1.866,1.868$,and this kind of consistency can quickly become basic good as a whole with $p_{c}$ increased from $1.87$ to $1.876$ except for the time after entering satiation of fidelity decay.}\label{fig18}
\end{figure}
\end{center}

\begin{center}  
\begin{figure}
\includegraphics[width=18cm,height=18cm]{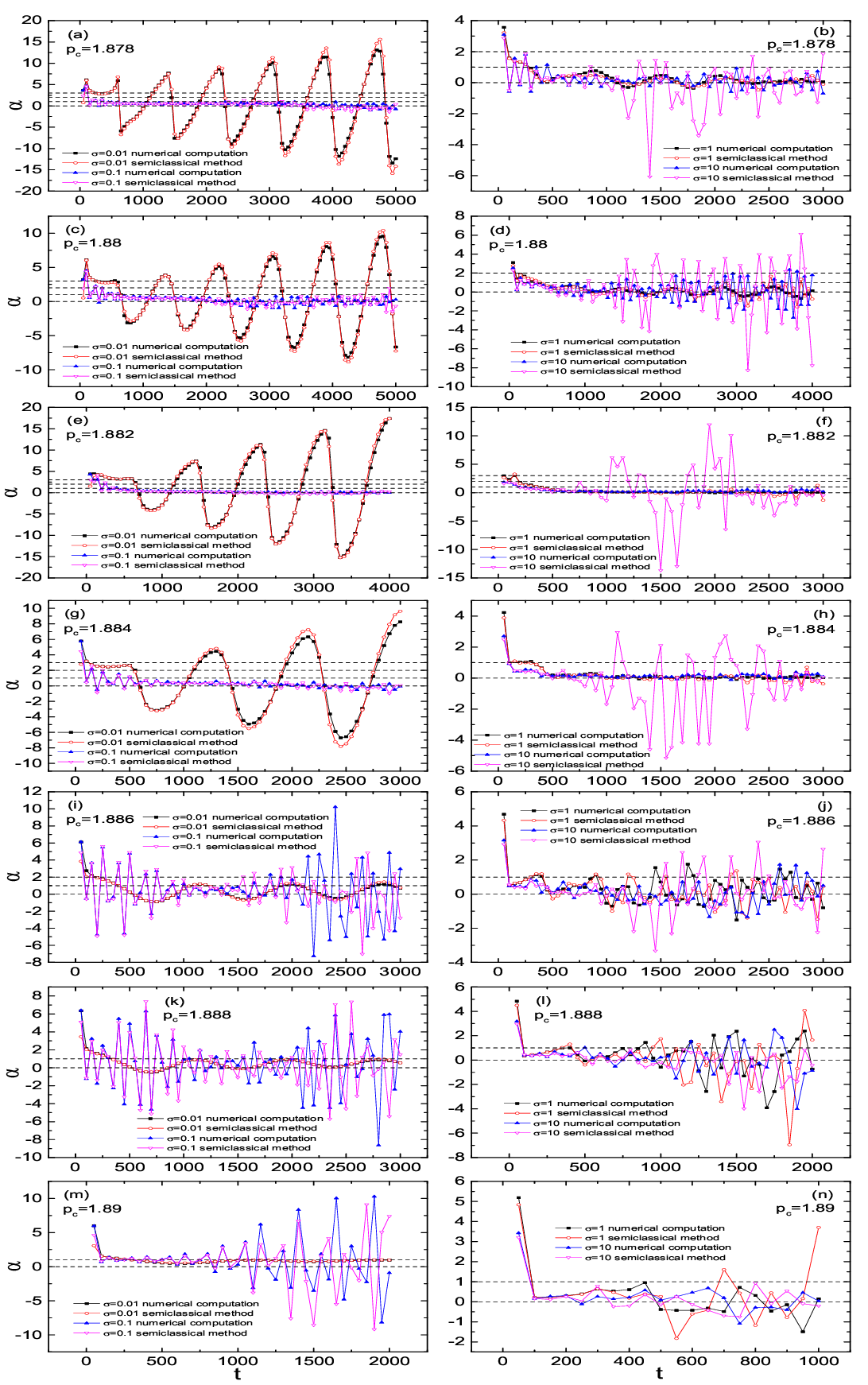}
\vspace{-0.2cm} \caption{The same consideration to the figures 17,18 for $p_{c}$ from $1.878$ to $1.89$ with the interval as $0.002$.There are similar expressions for $p_{c}=1.878,1.88,1.882,1.884$ in the figure(a),(c),(e),(g) as $\alpha$ for $\sigma=0.01$ can dwell on the values typically around $3$ as basically cubic-exponential decay with the time scale as $10^2$ after initially short non-monotonic variation and then large oscillation with the amplitude increasing more and more can be found with the balanced values near to $0$ which implicate quite slow stretched exponential decay.Meanwhile the values of $\alpha$ for $\sigma=0.1$ in terms of $p_{c}=1.878,1.88,1.882,1.884$ can decrease as a whole with two parts as the initial non-monotonic variations with the basic same first values to the corresponding values for $\sigma=0.01$ and gradually slow variations between $0$ and $1$.In the figure(b),(d),(f)as $\sigma=1,10$,the values of $\alpha$ can also decrease as a whole to the field as some obvious fluctuations around $0$ for $p_{c}=1.878,1.88$,basically as $0$ for $p_{c}=1.882$.In the figure(h) as $\sigma=1,10$ in terms of $p_{c}=1.884$,there is a first large drop to the platform around $1$ or smaller corresponding to $\sigma=1$ or $\sigma=10$ before the last variation slightly larger than $0$.In the figure(i),(k),the values of $\alpha$ for $\sigma=0.01$ in terms of $p_{c}=1.886,1.888$ can continuously decrease until some unattenuated oscillation with balanced values between $0$ and $1$ showing stretched exponential decay and meanwhile the expressions for $\sigma=0.1$ can be basically divided into two parts with the fluctuation of first part much larger.For large perturbations as $\sigma=1,10$ in terms of $p_{c}=1.886,1.888$ in the figure(j)and(l),$\alpha$ can show zigzag-like variations in the efficient time without the very first much larger values.The values of $\alpha$ without its first large values as well for $p_{c}=1.89$ can basically vary around $1$ taken as approximately exponential decay for $\sigma=0.01,0.1$ and also can show zigzag-like variations for $\sigma=1,10$.The comparison with the fitted $\alpha$ from semi-classical method here can show some good consistency except mainly for the time for entering the satiation of decay reflected as the fluctuation around $0$ basically evenly.}\label{fig19}
\end{figure}
\end{center}

From the study of variations of $\alpha$ versus time $t$ for different $p_{c}$,we can find the typical three decay processes for small perturbation as the initial same decay,transitive fast decay and afterwards slow decay,then the transitive fast decay can be gradually shortened to neglect with the perturbation increased.Originally for the small perturbation,we can define three characteristic time as $t_{1},t_{2},t_{3}$ according to the ending time of typical three decay processes pointed above and then $t_{2},t_{3}$ can alternately be considered as the ending time of the second decay process and the maximum time we consider here for $10^{4}$ with perturbation increased to some extent.Based on our assumed decay law as $M(t)\approx e^{-ct^{\alpha}}$,we can get the fitted $\alpha$ with the time interval as $\Delta t_{1},\Delta t_{2},\Delta t_{3}$ corresponding to the length of time respectively for $t_{1},t_{2}-t_{1},t_{3}-t_{2}$.Therefore we need to investigate the variations of $\alpha$ versus $\sigma$ for different $p_{c}$ and it seems easy to treat but here how to get the time for fitting is a essential point.

Firstly the initial time is related to the same decay law expressed with the similar relation for $ln(-ln(M))$ versus $ln(\sigma)$ pointed it out clearly in the previous study.But we can find the initial time can be shortened gradually after perturbation is sufficiently increased for some $p_{c}$ which have been found numerically.Here we define the original initial time $t_0$ corresponding the measurement from the smallest perturbation we choose as $\sigma=0.01$ and thus the diminishing situation for original initial time can be understood by two elements,one is the basic linear relation as $ln(-ln(M))$ versus $ln(t)$ and the other is the saturation.As the perturbation is increased,the basic linear line can be considered for shifting with the same slope,and there will be a intersection point emergent as the saturation value of $ln(-ln(M))$ can be seen as a horizontal line in terms of independent variable $ln(\sigma)$.The intersection point determines the largest perturbation $\sigma_{h}$ that can hold $t_{0}$  as the initial time and we can get the analytical formula to describe initial time for perturbation larger than $\sigma_{h}$
     \be\label{initial_edge}t_{1}=t_{0}(\frac{{\sigma}_{h}}{\sigma})^{\frac{2}{\alpha_{0}}}.\ee 
Here the number $2$ comes from the value of $\nu$ that is set as $2$ for $t_0$ supported by our previous numerical study in terms of the assumed decay law as $\ln(-\ln{M(t)})\approx \ln{c_0}+{\nu}ln{\sigma}+{\alpha}\ln{t}$,and $\alpha_{0}$ is the value of $\alpha$ fitted with $t_{0}$.Once the values of $t_0$,$\alpha_{0}$,$\sigma_{h}$ are set,we can get the corresponding initial time diminished as power law with the characteristic exponent $2/\alpha
_{0}$ if the perturbation surpasses $\sigma_{h}$.

Then secondly we use a flexible method to get $t_2$ and $t_3$ and the key point is to use fitted local $\alpha$ in terms of some fixed time interval to find typical transitive time,here we use the name as local $\alpha$ to distinguish it from the $\alpha$ fitted with the time that is not fixed as well as can change a lot.As the variations of local $\alpha$ for small perturbation can typically undergo the processes from rise to decline which have been illustrated in our previous study for the variation of $\alpha$ with time,and the ending time of this decreasing process related to the local minimum value of $\alpha$ could be obtained as the transitive value of time.This transitive time can be set as $t_{2}$ if it can be found,afterwards we can consider the time corresponding local $\alpha$ closed to zero which means saturation just comes,so we can set this time as $t_3$.But we should include all the perturbation put into consideration for different $p_{c}$ and also there are some very large fluctuation for the local $\alpha$ with perturbation increased to some extent,thus the technique in detail is more complicated.The priority for us is to do the smoothing procedure for the variation of local $\alpha$,and we select the fixed time to get the local $\alpha$ as $20$ for keeping it small enough to express sufficient continuity with time as well as reducing too much fluctuations.Then there still exists fluctuation need to be smoothed out for getting the transitive time as accurate as possible and we use the moving average method with the span as $5$ for smoothing.Further more,based on a lot of comparisons between the smoothing curves and the original variations of local $\alpha$,we use moving average repeatedly for $20$ times to get the sufficient smooth curve but also to hold the basic feature of the original variation.Actually this method is quite effective which also have been found in the study of eigenstate thermalization and quantum chaos very recently\cite{moving_average}.Now we use the smooth curve stemmed from original $\alpha$ to get $t_2,t_3$ for all the $p_{c}$ we consider,the procedure actually is separately treated with the condition as whether there is any value can surpass the value of smooth $\alpha$ corresponding to $t_1$ in a given sufficient time,here we take the time as $2000$.If this condition can be satisfied,the minimum point need to be found after a maximum point,otherwise directly find the minimum point along the decline.For simplicity,we use the procedure one and procedure two to describe them when we need to point them out.The minimum point is related to the transitive time but we also consider the situation as saturation emergent or revival happened,here we can set a small value as $0.2$ for judging saturation or likely revival,then we give the name respectively as $t_{tr}$ and $t_{s}$,and if $t_{tr}>t_{s}$ we set $t_2=t_{tr}$,$t_3=t_{s}$,otherwise we set $t_2=t_{s}$,$t_3=10000$,the number $10000$ is the largest scale of time for studying fidelity.Through very careful observation,we can find the condition as minimum point can be softened to be the inflection point for getting the transitive time as accurate as possible if they are quite similar.Thus we can get a value from fitting with two values of smooth $\alpha$,then continue to do this fitting and a new curve can be obtained.Along this new curve but without consideration of $t_1$,we can count it saved as the likely transitive time if the corresponding value of the new curve also is the minimum value for procedure one or the maximum value for procedure two and meanwhile it is quite small but positive or the absolute value is also quite small but negative,here we set the value for judgement as $0.0005$ based on a lot of numerical observations for whether this setting is effective to catch the transitive time.

Now we show the study result of $t_1$ in the figure 20 and want to give a comprehensive explanation.Firstly we should find the seed of $t_1$ as $t_0$ which is the original initial time we have described above.Based on the very similar expression for the relation as $ln(-ln(M))$ via $ln(t)$ corresponding to all the perturbation not too large,we can make a comparison for local $\alpha$ between $\sigma=0.01$ and $\sigma=0.1$ alongside with time increasing,then we can consider the absolute difference and set a small value to judge whether they are still very close,so basically we can count the time as $t_0$ corresponding to breakdown happened.Here we set the value for judging as $0.05$ supported by the direct observation of the similarity for the relation as $ln(-ln(M))$ via $ln(t)$ and actually we use the condition with the threshold as $0.05$ twice to double check that it is indeed a real time for breakdown.To obtain the power law of the variation of $t_1$ with perturbation $\sigma$,we have assumed $\nu=2$ valid for the original initial time and actually we also can use the condition as absolute difference between $\nu$ and $2$ not more than $0.1$ to get the corresponding time if breakdown happens,here we just use the first three perturbations to get the $\nu$ and we still can give the name as original initial time to make a comparison with the time obtained from local $\alpha$.In the figure 20(a),we can find the time obtained from $\nu$ is always larger than the original initial time we actually use obtained from $\alpha$,so the comparisons also support to use $\nu=2$ to get the Eq.~(\ref{initial_edge}).In the figure 20(b),we show the variation of $\sigma_{h}$ for different $p_{c}$ and there is not decline for some $p_{c}$ as the value of $\sigma_{h}$ larger than $10$ which has already been the largest perturbation we use.Further more,as the fixed time for fitting is $20$,then the decline only can be observed for $p_c$ below $1.86$.So in the figure 20(c),we show the variation of $t_1$ with perturbation $\sigma$ obtained from the formula Eq.~(\ref{initial_edge}) for $p_c=1.85$ to $p_c=1.86$ with interval as $0.002$ and corresponding characteristic exponents $\gamma=2/\alpha_{0}$ for the power-law decline are showed in the figure 20(d).As we use the discrete time with the base as $20$,thus there is some differences between the analytical evaluation as $2/\alpha_{0}$ and direct fitting result from the variation of $t_1$ which has been showed in the figure 20(e).Further more,we need to know the values of $\alpha_{0}$ for different $p_c$ which is very important for showing the decay laws and the direct numerical result as well as semi-classical evaluation are illustrated in the figure 20(f),the comparison shows the poor agreement with each other not surprised as the poor agreement for $M(t)$,and we can find the variations of $\gamma$ and $\alpha_{0}$ are opposite for each other as $\gamma \propto 1/\alpha_{0}$.Here we can find the variation of $\alpha_{0}$ with different $p_c$ undergoes a non-monotonic process from the value close to $2$ to the value around $3$ and afterwards the steep rise comes.The value for $2$ represents the Gaussian decay and $3$ means the Cubic-exponential decay theoretically predicted by J.Van\'{i}\v{c}ek applying to quasi-integrable field in short time studied by the correlation of action differences in terms of dephasing representation\cite{vanicek_arxiv}.Obviously the decay laws we find here is quite novel but also have a tight connection with the previous research.

\begin{center}  
\begin{figure}
\includegraphics[width=14cm,height=14cm]{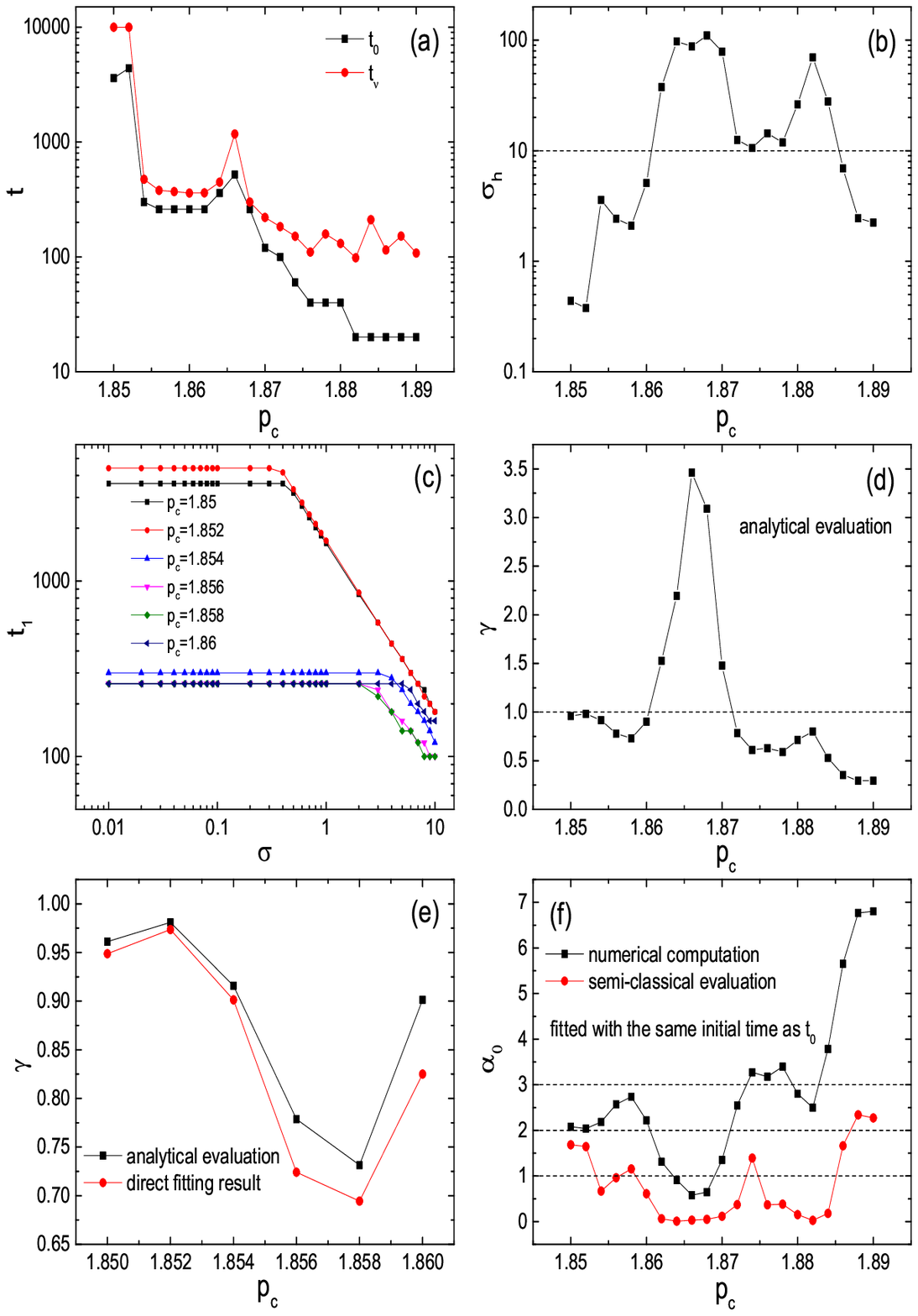}
\vspace{-0.2cm} \caption{(a)variations of $t_0$ and $t_{\nu}$ with $p_c$,$t_0$ and $t_{\nu}$ are obtained respectively from $\alpha$ and $\nu$.$t_{\nu}$ shows larger than $t_0$,and the variations are non-monotonic with $1.866$ corresponding to common local maximum points.(b)variation of $\sigma_{h}$ with $p_c$,$\sigma_{h}$ is the perturbation as up bound for holding $t_0$ as the initial time $t_1$.The line is to guide the eyes as $\sigma=10$ is the largest perturbation we consider.(c)variation of $t_1$ with perturbation $\sigma$ in terms of $p_c$ from $1.85$ to $1.86$ with interval as $0.002$,the log-log plot is to show the power-law decline somewhat not very accurate as the discrete time with the base for $20$.(d)variation of $\gamma$ with $p_c$,and $\gamma$ is the characteristic exponent as $\gamma=2/\alpha_{0}$ for the variation of $t_1$ with $\sigma$.(e)comparison of $\gamma$ for analytical evaluation as $\gamma=2/\alpha_{0}$ and direct fitting result as the slope of variation of $t_1$ with $\sigma$ without the unchanged part in the log-log scale illustrated in the (c).As the expression of $\sigma_{h}$,we only have the comparison for $p_c$ from $1.85$ to $1.86$ with interval as $0.002$.(f)variations of $\alpha_{0}$ with $p_c$ obtained from direct numerical computation and semi-classical evaluation,there is some large difference between them.
} \label{edge_inifin}
\end{figure}
\end{center}

Now we show the variations of $t_2$ and $t_3$ with $\sigma$ in the figure 21 with log-log scale as we have found the power-law decline is quite common.As we know,the time scale for stable field in terms of quantum fidelity is $\tau \propto \sigma^{-1}$ in the previous research,the characteristic exponent is $-1$ which can be taken as a important reference for our study here.For simplicity,we will elaborate the variation of $t_2$ and $t_3$ for different $p_c$ in the related sub-figure without referring to the name as figure 21 and try to connect with the feature of decay of $M(t)$.To get the characteristic exponent $\gamma$ technically for a part showing basic power-law decline in terms of $\tau \propto \sigma^{-\gamma}$,we can use the initial point and end point to decide a line and get the slope of this line neglecting interior points.

For the variation of $t_2$ for $p_c=1.85$ illustrated in the (a),a initial unchanged part exists until $\sigma=0.1$,and then a basic power-law decline can be found until $\sigma=2$,and then another approximated power-law decline follows until $\sigma=8$ and afterwards rising situation is happened for $\sigma=9,10$.There is the similar qualitative expression of variation of $t_2$ for $p_c=1.852$ illustrated in the (b)and the obvious difference here is the value of $\sigma$ to end the unchanged part is $0.2$.$t_3$ for $p_c=1.85$ is unchanged and $t_3$ for $p_c=1.852$ can decline from $\sigma=0.3$ to $\sigma=1$ and a power-law decline can be found clearly although there is a exceptional case as $\sigma=0.7$ accidentally jumping back to $10000$.The reason for the transition for $t_2$ commonly for $p_c=1.85,1.852$ from unchanged part to decline is that the saturation appears with perturbation increasing to some extent and the unchanged part obviously is related to the so called transitive time we define but not typical that can be hold same.If $t_1$ is more than the time corresponding to the value of smooth $\alpha$ as $0.2$,then the value of $t_2$ is only over $20$ than $t_1$ based on our procedure,such as the part of $t_2$ that begins at $\sigma=0.4$ and ends at $\sigma=2$ for $p_c=1.85$.Here we can find $\gamma$ for the first power-law of $t_2$ is closed to $1$ for $p_c=1.85$ or just can be seen as $1$ for $p_c=1.852$,$\gamma$ for second power-law of $t_2$ are all obviously larger than $1$ respectively as $1.77311,1.63011$,further more,$\gamma$ of the part of $t_3$ having the power-law decline for $p_c=1.852$ can also be found as $1$ with high accuracy.To set the value as $0.2$ of smooth $\alpha$ to judge the saturation or likely revival is basically reasonable but also leads to some likely oscillation for $t_3$ connected directly with the changeable role of $t_2$ as the ending time of transitive process or main decay process,this kind of situation just happens for the closeness for $t_{tr}$ and $t_s$ with perturbation relatively large and then this function of $t_3$ is basically trivial that does not affect the validity of study results.

For simplicity,we give the name for the largest time $10^4$ as $T$,and basic straight line with log-log scale means power law variation,thus we also simply point it out below the likely power law without pointing out corresponding straight line.There are two basic patterns for the variations of $t_2$ and $t_3$,the first one is that that $t_2$ can change but $t_3$ can hold for $T$ and the second one is that $t_2$ and $t_3$ all can change,we give the names as type 1 and type 2.For the investigation further more,we can find there is the difference for the type 1 existing before type 2 or after with the new names respectively as patter 1 and patter 3,and the original type 2 can be called patter 2.Then the explicit patterns for the variations of $t_2$ and $t_3$ as a whole respectively mean the transitive decay process and afterwards continual decay process without emergent saturation,still the transitive decay process and afterwards decay process ceased by emergent saturation,and the decay process after initial decay process determined by $t_1$.We can find there are not all the variations of $t_2$ and $t_3$ showing the clearly patterns we described above,but have the function to remind the corresponding decay features of LE,and below we concentrate on the elaboration of variations of $t_2$ and $t_3$ supplemented by the corresponding expressions of LE if needed.

For $p_c=1.854$ related to graph(c),$t_2$ can show initial twisted variation,and then enter stable field with little change,and afterwards has a steep rise and then show some decline but with twisted situation around $\sigma=3$.Meanwhile $t_3$ for $p_c=1.854$ can hold the value as $T$ until for $\sigma=0.2$,and then show a power law decline until for $\sigma=0.6$,afterwards rebound to $T$ although the occasional drop for $\sigma=3$.Actually we notice the variations of $t_2$ and $t_3$ can link together to form a obvious common power law decline with the joined perturbation as $\sigma=0.6$ corresponding to the transition from $t_2=t_{tr}$ to $t_2=t_s$ and then we specifically call it as linked power law decline once meeting this situation again.In terms of graph(d),$t_2$ for $p_c=1.856$ can show initial twisted rise and then have two parts showing the basic same power law declines,afterwards enter the field belonging to the linked power law decline from $\sigma=0.8$ to $\sigma=4$ and then variate irregularly.Meanwhile there is the similarity for variations of $t_3$ between $p_c=1.854$ and $p_c=1.856$ all having three parts as unchanged part,power law decline and unchanged part again as the rebound but with some drops happened.In terms of graph(e),$t_2$ for $p_c=1.858$ can show initial twisted variation and then enter the field belonging to the linked power law decline from $\sigma=0.2$ to $\sigma=1$ and then variate with some fluctuation.$t_3$ for $p_c=1.858$ share the similar variation as typical three parts compared with the variations of $t_3$ for $p_c=1.854$ and $p_c=1.856$.For $p_c=1.854,1.856,1.858$ taking a whole,we can easily find the three patterns for the variations of $t_2$ and $t_3$ can be observed.

For $p_c=1.86,1.862,1.864$ related to the graphs $(f,g,h)$,there are only pattern 1,and patter 2 and the variation of $t_2$ in terms of patter 1 can have approximately piecewise power laws for the rising part and declining part that are quite clearly for $p_c=1.86$,less clearly for $p_c=1.862$,and entirely can not be applied for $p_c=1.862$.It is noteworthy that the variation of $t_3$ of the part as patters 2 shows the opposite tendency compared with variation of $t_2$ of the part as pattern 1 for the degree of accepting power law in terms of the increasing order of $p_c$ as $1.86,1.862,1.864$ and accordingly the power law can not be applied for $p_c=1.86$,roughly can be applied for $p_c=1.862$,and then can be well applied for $p_c=1.864$.Meanwhile we notice the part of pattern 2 for $p_c=1.86,1.862,1.864$ gradually decrease and this tendency has the tight connection with approaching $p_c=1.866$ belonging to classically stable dynamics.For $p_c=1.866,1.868$ related to the graphs $(i,j)$,there just exists pattern 1 and $t_3$ for $p_c=1.866$ and $p_c=1.868$ all are $T$ and $t_2$ for $p_c=1.866$ basically do not change,then $t_2$ for $p_c=1.868$ can show initial part having some tiny decline and then have a roughly power law decline from $\sigma=0.06$ to $\sigma=0.2$ and afterwards have the part with small variation,lastly show a sudden drop happened for $\sigma=8$ and hold the closed value afterwards.Then we move on to $p_c=1.87,1.872$ related to the graphs $(k,l)$ showing the patter 1 and patter 2,approximately piecewise power laws are also found similar to the variation of $t_2$ for $p_c=1.86,1.862$,and the variation of $t_3$ in terms of patters 2 can show power law decline that is basically suitable for $p_c=1.872$ and particularly well for $p_c=1.87$.There is the similarity for the variations of $t_2$ and $t_3$ between $p_c=1.874$ and $p_c=1.858$ illustrated in the graphs $(m,e)$ all having the integrated patterns and particularly a clear linked power law decline can be found with the same starting perturbation as $\sigma=0.05$ in terms of $t_3$ and ending perturbation as $\sigma=1$ in terms of $t_2$.Further more,it is the same distance from $p_c=1.866$ for $p_c=1.858$ and $p_c=1.874$,and the similar variations then are observed.It is not a coincidence and someone want to observe the variations of $t_2$ and $t_3$ in the figure 21 very carefully,thus this similarity also can be found in the pairs for $(p_c=1.864,p_c=1.868)$,$(p_c=1.862,p_c=1.87)$,and $(p_c=1.86,p_c=1.872)$ and there is the same distance for two $p_c$ belonging to every single pair taking $p_c=1.866$ as the common reference.Here we also notice the twisted rise for the variation of $t_2$ from decline to rise corresponding to $\sigma=0.01,0.02,0.03$,this kind of variation is common since $p_c=1.874$.

The similarity that is based on the same distance from the common reference as $p_c=1.866$ even can be roughly suitable for the pair of $p_c$ as $(1.856,1.876)$ illustrated in the graphs $(d,n)$,and the common feature is that there is the existence of power law declines for the variations of $t_2$ and $t_3$ happened together corresponding to related perturbations.We also notice the initial twisted variation of $t_3$ for $p_c=1.876$ which makes the part of pattern 1 quite short seriously just for $\sigma=0.01$.Although we can not find a clear pattern 3 in the graph (n) but we can expand this characterization to include the field of fluctuation of $t_3$ after $\sigma=4$ for the first rebound to $t_3=T$ as afterwards the main decay time without saturation then can be represented by $t_2$.For $p_c=1.878$ related to $(o)$,power law declines for the variations of $t_2$ and $t_3$ also can happen almost together after the initial twisted variations and we can find a common power law decline can be formed together by the variation of $t_2$ and $t_3$ with the linked perturbation as $\sigma=0.4$ corresponding to the first rebound to $t_3=T$,and the rest of field of perturbation can be seen as the pattern 3 but a clear power law decline also can be found in the field of perturbations from $\sigma=0.5$ to $\sigma=1$ for variation of $t_3$ showing the existence of specific decay processes of LE near saturation with serious check.For $p_c=1.88$ related to the graph $(p)$ showing integrated ordered patterns,there is a obvious power law decline for variation of $t_2$ until $\sigma=0.5$ after initial twisted variation and $t_3$ can not hold for $T$ beginning as $\sigma=0.6$,then a power law decline for $t_3$ can be formed by $\sigma=0.6,0.7,0.8$ that is also connected with the existence of specific decay processes of LE closed to saturation.For $p_c=1.882,1.884$ related to the graphs $(q,r)$,variations of $t_2$ all undergo some twisted rises and then two clear power law declines can be found and $t_3$ can hold for $T$ except the initial twisted variations.For $p_c=1.886,1.888$ related to the graphs $(s,t)$,variations of $t_2$ do not change a lot and only can show basic power law declines belonging to intermediate part of the perturbations within $0.1$ and $1$.Variations of $t_3$ for $p_c=1.886,1.888$ share the common features as the heavy drops for small perturbations not more than $\sigma=0.1$ and irregular variations for large perturbations marked with rebounds of $t_3$ to $T$,a rough power law decline can be found for $p_c=1.886$ seen clearly in the graph $(s)$.For $p_c=1.89$ related to the graph (u),the variation of $t_2$ can initially have a twisted rise within the intermediate part showing basic power law,then a power law decline can be found for perturbations $\sigma$ belonging to $(0.1,0.4)$,and afterwards reach the field of fluctuation.In the same graph (u),$t_3$ can undergo unchanged part for $\sigma=0.01,0.02,0.03$,rough power law decline below $\sigma=0.1$ and the part of fluctuation with rebounds to $t_3=T$ before $\sigma=0.9$,and afterwards basically unchanged part again for $t_3=T$ although having the drop for $\sigma=5$.

Here we actually need to know $\gamma$ as the characteristic exponent for the power law,so we need to get the slope for the corresponding linear relationship in terms of log-log scale for $t$ versus $\sigma$,and the technique we use is very simple for only considering the starting point and ending point and then connect them to form a line and then we get the slope of this line to represent $\gamma$ as we try to delete some effect of likely wild points.Based on the expressions of variations of $t_2$ and $t_3$ described above,it is reasonable to observe $\gamma$ for different $p_c$ classified by the groups as $p_c=1.85,1.852,1.854,1.856$,$p_c$ from $1.858$ to $1.874$ symmetrical for $1.866$, $p_c$ from $1.876$ to $1.89$.For the first group,fitted $\gamma$ are all closed to $1$ except the obvious deviation related to the large perturbations for $p_c=1.85,1.852,1.854$ seen clearly for graphs $(a,b,c)$.For the second group ,$\gamma$ for variations of $t_2$ for $p_c=1.862,1.864,1.868,1.87$ adjacent symmetrically to $p_c=1.866$ are obviously smaller than $1$ and $\gamma$ for other $p_c$ are all closed to $1$.For the third group,expressions of $\gamma$ are some kind of complicated but all can obviously deviate from $1$.To see the effectiveness of our semi-classical method,we also use the semi-classical evaluation to compare with the direct numerical result.In terms of comparison,there are some large deviations all for $t_2$ and $t_3$ related to $p_c$ adjacent to $p_c=1.866$ as $p_c=1.862,1.864,1.868,1.87$ and otherwise some basic agreement can exist for the variations of $t_2$ related to the rest of $p_c$.

\begin{center}  
\begin{figure}
\includegraphics[width=18cm,height=20cm]{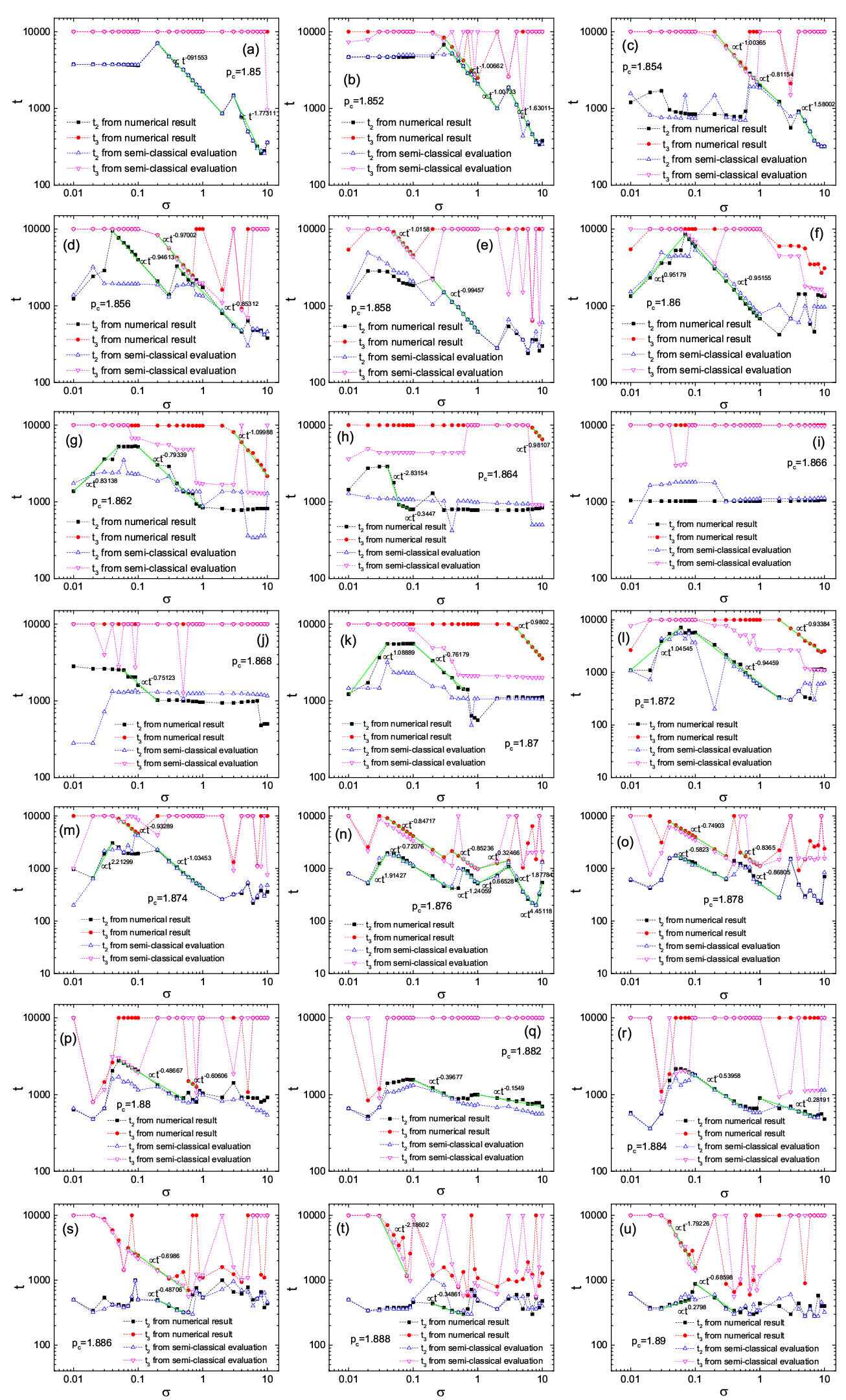}
\vspace{-0.5cm} \caption{$t_2$ and $t_3$ versus perturbation $\sigma$ for $p_c$ from $1.85$ to $1.89$ with the interval as $0.002$,$t_2$ and $t_3$ are obtained from numerical result and semi-classical evaluation for comparison.Power law declines can be found for most $p_c$ corresponding to related fields of perturbations and there are clearly similar expressions for the pairs of $p_c$ as $(1.864,1.868),(1.862,1.87),(1.86,1.872),(1.858,1.874)$ symmetrically distributed with the reference as $p_c=1.866$.For $p_c$ from $1.876$ to $1.89$,the values of $\gamma$ are commonly and obviously deviate from $1$ in terms of $\tau \propto \sigma^{-\gamma}$ with $\tau$ as $t_2$ or $t_3$.In terms of comparison between the numerical result and semi-classical evaluation,there are some large deviations all for $t_2$ and $t_3$ related to $p_c$ adjacent to $p_c=1.866$ as $p_c=1.862,1.864,1.868,1.87$ and otherwise some basic agreement can exist for the variations of $t_2$ related to the rest of $p_c$.Some large alternations exist for the variations of $t_3$ in terms of some $p_c$ correlated with the procedure corresponding to the small value as $0.2$ to judge the saturation or likely revival of LE affecting the value of $t_3$ chosen.}\label{global_t_edge}
\end{figure}
\end{center}

Now we want to obtain the exponents $\alpha$ from direct numerical computation corresponding to time interval as $\Delta t_2$ and $\Delta t_3$,it is very important to distinguish the different decay laws of LE.As we know,$\alpha$ as $1,2$ is related to classical strong chaos and regular dynamics respectively,then we want to substantiate the new decay laws by showing the values of $\alpha$ with the obvious deviation from $1$ or $2$.We show the variations of $\alpha$ with different $p_c$ in the figure 22 and here neglect the figure name but only use related sub-figure name to illustrate the variation of $\alpha$ for some specific $p_c$.For simplicity,we can call $\alpha$ obtained from $\Delta t_2$ and $\Delta t_3$ respectively as $\alpha_{2}$ and $\alpha_{3}$ although we do not have the names in the figure 22,meanwhile the value of $\alpha_{1}$ can be seen as $\alpha_{0}$ for the basically same decay law but with different decay time.Then,$\alpha_{2}$ and $\alpha_{3}$ for $p_c=1.85$ showed in graph $(a)$ can hold basically around $2$ not more than $\sigma=0.1$ and then $\alpha_{2}$ can rise steeply for $\sigma=0.2,0.3$ showing the obvious deviation from the Gaussian decay.After dropping to the value below $0$,$\alpha_{2}$ can rise again to have the value quite larger than $0$ from $\sigma=3$ to $\sigma=8$,and this kind of situation implicates there is a decay process of LE connected with main decay part decided by $t_1$ and saturation.It is not a isolated case but can be found for some other $p_c$,and we will point it out once meeting this situation.For $p_c=1.852$ related to graph (b),$\alpha_{2}$ and $\alpha_{3}$ can dwell around the value as $2$ not more than $\sigma=0.2$.Afterwards $\alpha_{2}$ can drop to the field with the value basically around $0.6$ from $\sigma=0.4$ to $\sigma=3$,and then rise to another field with the value significantly larger than previous one.Meanwhile $\alpha_{3}$ then can have basically stable variation with the value close to $0.4$ from $\sigma=0.3$ to $\sigma=1$ although there is a exceptional case for $\sigma=0.7$ dropping below $0$,and then can reach the field quite close to $0$ although having the interrupted situation for $\sigma=3$.Through very serious check,actually the variation of $\alpha_{3}$ is trivial after $\sigma=0.3$ as the fitting time has already put into the saturation.

Based on the expressions of $\alpha_{2}$ and $\alpha_{3}$ for $p_c=1.85,1.852$,we can find they are similar as they all can stay around the value $2$ which is the indicator of Gaussian decay of LE corresponding to classical stable dynamics,and $p_c=1.85,1.852$ are related to the classical stable field proved by sticking time and Lyapunov exponents illustrated in the figure 2.In addition,$\alpha_{2}$ after the heavy drops for $p_c=1.85,1.852$ described above actually all show there are processes connected with the main decay process related to $t_1$ and saturation and the larger of the value show the more significant of this process,thus we can not neglect them.Therefore our assumption as the decay of LE is entirely dominated by $t_1$ if the turning perturbation $\sigma_{h}$ is surpassed is not very accurate for existence of the variation of $\alpha_{2}$ not close to $0$.For $p_c=1.854$ related to graph (c),we can find $\alpha_{2}$ can basically dwell on the value as $2$ until for $\sigma=0.6$ except the initial large deviation for $\sigma=0.01$,meanwhile $\alpha_{3}$ can undergo initial rise to gradual stable variation close to $2$ and afterwards decline until for $\sigma=0.6$.There are drops steeply all for $\alpha_{2}$ and $\alpha_{3}$ corresponding to $\sigma=0.7$,and afterwards $\alpha_{2}$ can show a decline until for $\sigma=4$ followed by the variation with some considerable values that can not be neglected but it is trivial for the variation of $\alpha_{3}$ with the value below $0$ or close.Observing all the heavy drops happened for $p_c=1.85,1.852,1.854$,we can understand they are directly connected with $\sigma_{h}$ and anyone who has interest can see the illustration of figure 21 and this situation obviously can be suitable for $p_c=1.856,1.858$ but not for $p_c=1.86$ as related $\sigma_{h}$ is not accurate,so same decay law for large perturbations can not be suitable very well for $p_c=1.86$.

For $p_c$ from $1.856$ to $1.876$,we can find the similar expressions for the pair of $p_c$ with the same distance from $p_c=1.866$,and have showed the similarity in the study of variations of $t_2$ and $t_3$ for the pairs of $p_c$ as $(1.856,1.876),(1.858,1.874),(1.86,1.872),(1.862,1.87),(1.864,1.868)$.Then $\alpha_{2}$ for $p_c=1.856$ related to graph (d) can undergo the initial sharply drop,stable variation basically above the value $1.5$ lasting to $\sigma=2$,afterwards the variation with the value mostly quite smaller than $1$.Meanwhile $\alpha_{3}$ can show a basically rise and then gradually become closing to $\alpha_{2}$ and then have a large drop happened for $\sigma=0.8$ leading to the value below $0$ but can re-rise to the value obviously larger than $0$ for $\sigma=2,3,4$ showing the existence of two distinguished decay processes after the main decay process determined by $t_1$.Then the variations of $\alpha_{2}$ for $p_c=1.858,1.86$ related to graph (e) and (f) also can be seen as three parts but the intermediate parts are not stable showing some obvious deviations having the strong tendency for rising or declining.Meanwhile $\alpha_{3}$ for $p_c=1.858$ can have a obvious rise gradually closed to $\alpha_{2}$,and then have a sharply drop for $\sigma=0.2$,and afterwards vary near to $0$.For $p_c=1.86$,$\alpha_{3}$ can have a initial twisted variation and then have a drop towards $0$ happened for $\sigma=0.7$,afterwards tend to close to $0$,and then return to the value quite larger than $0$ since $\sigma=2$.To have a attention,the values of $\alpha_{2}$ and $\alpha_{3}$ for $p_c=1.86$ are all quite larger than $0$ from $\sigma=3$ to $\sigma=10$ showing existence of two distinguished decay processes after the main decay process determined by $t_1$.For simplicity,we will call them as non-trivial variations related to two distinguished decay processes once meeting the same situation.

For $p_c=1.862$ related to graph (g),we still can find there are three parts for the variation of $\alpha_{2}$ as the initial drop until for $\sigma=0.05$ and then stable variation around the value for $2$ gradually changing into another stable variation with the value a little bit above $1$.Meanwhile $\alpha_{3}$ for $p_c=1.862$ can undergo initial stable variation with the values below $1$ until for $\sigma=0.1$ and then the transitive part and lastly the new stable variation with the values closed to $2$.For $p_c=1.864$ related to graph (h),$\alpha_{2}$ can undergo just two parts as the first decline until for $\sigma=0.3$ and then stable variation quite closed to $1.5$.Meanwhile $\alpha_{3}$ for $p_c=1.864$ can rise initially from the value above $1$ and then can reach the stable variation closed to $2$ from $\sigma=0.2$ to $\sigma=1$,and afterwards show the obvious decline.For $p_c=1.866$ related to graph (i),$\alpha_{2}$ can show initial slight decline and then quickly enter the stable variation with the value a little bit above $1.5$.Meanwhile $\alpha_{3}$ for $p_c=1.866$ can undergo the very initial slight rise,then stable variation emergent quite closed to $2$ lasting to $\sigma=1$ and then the obvious decline afterwards.Based on the symmetrical expressions with the reference as $p_c=1.866$,one can indeed find the basic similar processes for $p_c$ coming from the same pair,thus we strongly encourage one to observe these similarities from the corresponding graphs of pairs as $(d,n),(e,m),(f,l),(g,k),(h,j)$.In terms of two $p_c$ from the same pair,there certainly can be some differences obvious or not for the values of $\alpha_{2}$ and $\alpha_{3}$ related to the corresponding processes.

For $p_c=1.878$ related to graph (o),$\alpha_{2}$ can drop initially until for $\sigma=0.05$ and then have the stable variation with the main values near to $1$ lasting to $\sigma=2$,and then have a sharply drop to another variation with the values mainly below $0.5$.Meanwhile $\alpha_{3}$ for $p_c=1.878$ can show the irregular variation with the values basically below $0.5$.Here we still can find there is the existence of roughly two decay processes since $\sigma=4$ as the main decay process is dominated by $t_1$ based on the sign of sharp drop of $\alpha_{2}$.There are quite similar expressions for $p_c=1.88,1.882,1.884$ corresponding to the variations of $\alpha_{2}$ and $\alpha_{3}$ depicted in the graphs $(p,q,r)$,and thus we elaborate them together.$\alpha_{2}$ can firstly have the basic drop basically until for $\sigma=0.05$ although having the very initial rise for $p_c=1.884$ and then enter the twisted variations with the first rising part and second declining part.Meanwhile $\alpha_{3}$ can have the initial twisted variations with the extent decreasing in terms of increasing $p_c$ and then quickly show the value very close to $0$ for perturbation not more than $5$ but interrupted for $\sigma=0.6,0.7,0.8$ related to $p_c=1.88$,and afterwards show some rise.Here we notice the final rise of $\alpha_3$ means there is a clearly decay emergent after a quite slow decay that even can be seen as saturation but not true.Further more this kind of new emergent decay can be strengthened with perturbation increased and is common for $p_c=1.88,1.882,1.884$.This decay is quite unusual and need more to investigate the physical mechanism in the future.

For $p_c=1.886$ related to graph (s),$\alpha_{2}$ can show the basic decline although having some interrupted rise including the very initial rise,and afterwards vary around about $0.75$ from $\sigma=0.2$ to $\sigma=1$,then drop to the field varying around $0.5$.Meanwhile the variation of $\alpha_{3}$ for $p_c=1.886$ is some kind of complicated but can be divided into two parts with the turning perturbation as $\sigma=1$ and the values of latter part are smaller than $0.5$.For $p_c=1.888$ related to graph (t),$\alpha_{2}$ can also show the first gradual decline until for $\sigma=0.1$ and then vary with some oscillation as the values basically between $0.4$ and $0.6$.Meanwhile $\alpha_{3}$ for $p_c=1.888$ can initially rise to the value quite close to $\alpha_{2}$ and then have irregular variation afterwards,actually the variation of $\alpha_{3}$ also can be divided into two parts with the turning perturbation taken as $\sigma=1$ based on the extent of fluctuation as well as the value itself.For $p_c=1.89$ related to graph (u),$\alpha_{2}$ can undergo the initial decline and then quickly have parabolic variation with the value above $1$ until for $\sigma=0.1$,then show a long decline until for $\sigma=0.9$,and afterwards vary with some oscillation as the values basically between $0.2$ and $0.4$.Meanwhile $\alpha_{3}$ for $p_c=1.89$ can have a rise that gradually close to $\alpha_{2}$ and afterwards show a basic drop beginning for $\sigma=0.09$ and then vary irregularly with some oscillation basically below $0.4$ lasting to $\sigma=0.8$,and lastly be quite close to $0$ except small rise for $\sigma=5$.Here we also want to have the comparison with $\alpha$ obtained from semi-classical method,and there are some large differences for $\alpha_{2}$ and $\alpha_{3}$ corresponding to $p_c$ adjacent to $p_c=1.866$ as $p_c=1.862,1.864,1.868,1.87$ and otherwise some basic agreement can exist for the variations of $\alpha_{2}$ and $\alpha_{3}$ corresponding to the rest of $p_c$.

\begin{center}  
\begin{figure}
\includegraphics[width=18cm,height=17cm]{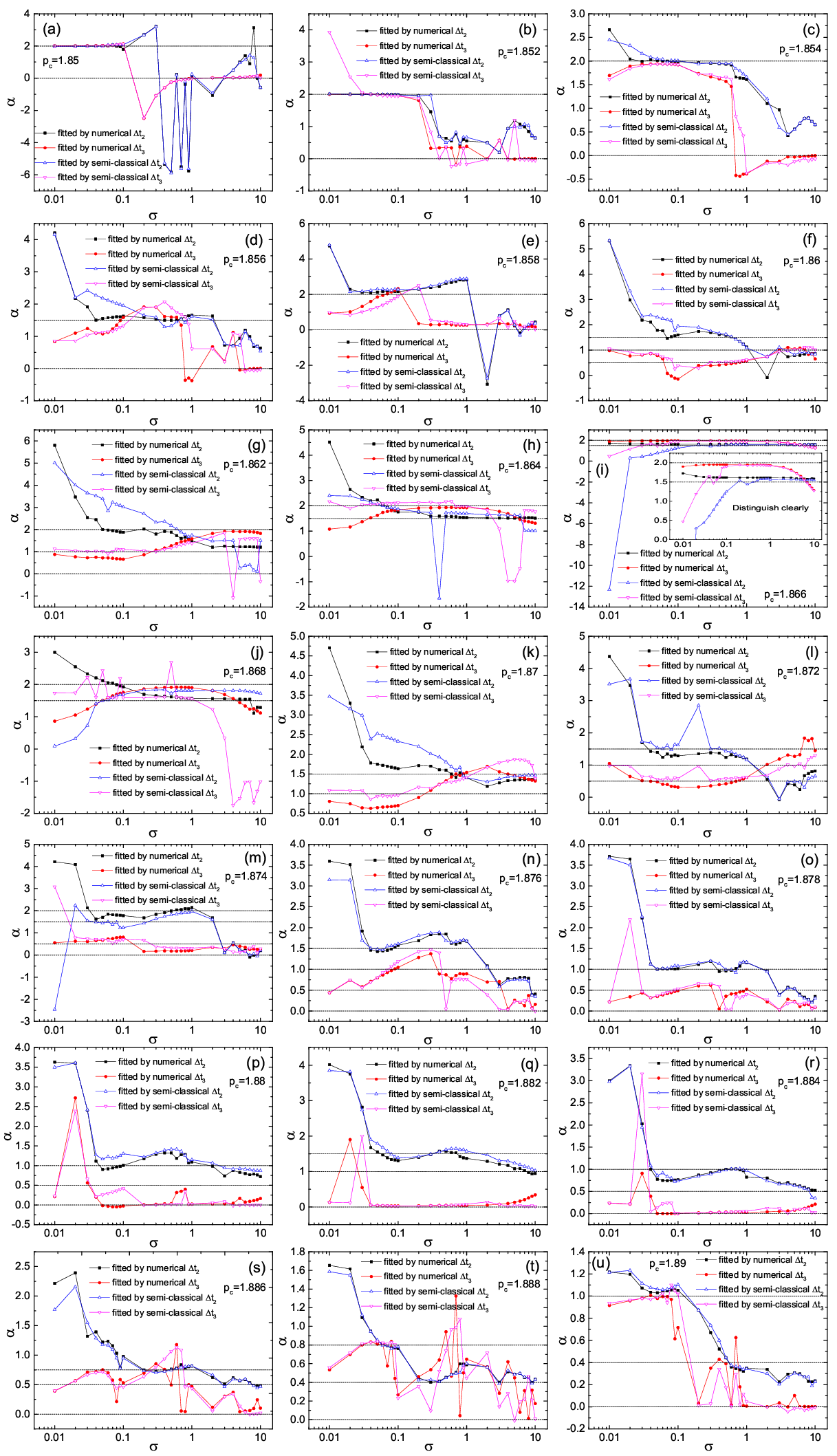}
\vspace{-0.5cm} \caption{variation of $\alpha$ with $\sigma$ for $p_c$ from $1.85$ to $1.89$ with the interval as $0.002$,and related $\alpha$ is fitted respectively from $\Delta{t_2}$ or $\Delta{t_3}$ of LE stemming from numerical calculation or semi-classical evaluation.For simplicity,we call them as $\alpha_{2q},\alpha_{3q},\alpha_{2sc},\alpha_{3sc}$.For $p_c=1.85,1.852$ related to graphs $(a,b)$,the common feature of $\alpha_{2q}$ and $\alpha_{3q}$ is to dwell on the value as $2$ for corresponding field of perturbation respectively,and then $\alpha_{2q}$ can show some clearly re-rise after $\sigma=3$.For $p_c=1.854$ related to graph $(c)$,$\alpha_{2q}$ can also basically dwell on the value as $2$ for corresponding field of perturbation except the large deviation for $\sigma=0.01$ and meanwhile $\alpha_{3q}$ can undergo initial rise,stable variation close to $2$ and gradual decline,afterwards we can find the obvious drops in common very sharply for $\alpha_{3q}$.For $p_c$ from $1.856$ to $1.876$ related to graphs $(d,e,f,g,h,i,j,k,n)$,we can find the similar expressions for the pair of $p_c$ with the same distance to $p_c=1.866$ taking as $(1.856,1.876),(1.858,1.874),(1.86,1.872),(1.862,1.87),(1.864,1.868)$.Two-part processes of variation of $\alpha_{2q}$ can be quickly transformed to typical three-part processes with alternating $p_c$ related to increase the distance from the reference $p_c=1.866$,and meanwhile $\alpha_{3q}$ can undergo some complicated change from parabolic variation to two-part variation with the first basic rise as the main feature.Typical three part variation are the initial decline,intermediated stable variation followed by a sharply drop,and final variation with obvious smaller values that can not be observed for two-part variation.For $p_c$ from $1.878$ to $1.89$ related to graphs $(o,p,q,r,s,t,u)$,typical three-part variation of $\alpha_{2q}$ can hold for $p_c=1.878$,and the second part and third part can be smoothed out showing the twist variation for $p_c=1.88,1.882,1.884$,and then three-part variation can be regained for $p_c=1.886$ but approximated two-part variation can be found for $p_c=1.888$,and lastly four-part variation combining two typical two-part variation can be found for $p_c=1.89$.Meanwhile $\alpha_{3q}$ can have some irregular variation with the tendency to decrease for $p_c=1.878$,and show the initial variation with up and down as the main feature for $p_c=1.88,1.882,1.884$.$\alpha_{3q}$ can have initial gradual rise for $p_c=1.886,1.888,1.89$ but only show the evolved closeness to $\alpha_{2q}$ for $p_c=1.888,1.89$,and afterwards can undergo the limited drops followed by irregular oscillations tending to decrease finally for $p_c=1.886,1.888$ as well as the sharply drop followed by some limited oscillation tending to zero for $p_c=1.89$.There are some large differences compared with $\alpha_{2sc}$ and $\alpha_{3sc}$ corresponding to $p_c$ adjacent to $p_c=1.866$ as $p_c=1.862,1.864,1.868,1.87$ and otherwise some basic agreements can exist for the rest of $p_c$.}\label{global_alp_t_edge}
\end{figure}
\end{center}

%\begin{center} % as used for the smoothed data and thus it seems not very good but could be used latter.
%\begin{figure}
%\includegraphics[width=18cm,height=22cm]{global_devia_edge.eps}
%\vspace{-0.2cm} \caption{111}\label{global_devia_edge}
%\end{figure}
%\end{center}

Now we want to show the variations of local $\alpha$ with $\sigma$ for different time we select in terms of $p_c$ from $1.87$ to $1.89$ and thus draw three figures to carefully show the study results and suggest to read the captions all together to know the connection among them.Firstly we use the fitting time step as $50$ as it can reflect the feature of decay process finely but also is not small used to suppress the likely fluctuation to some extent as the previous study of variation of $\alpha$ with time.Besides for $p_c=1.85,1.852,1.89$,we can find three typical patterns of variation of $\alpha$ as follows.Firstly we can find the situation to maintain the same value of $\alpha$ for short time as $t=50$ is universal at least for the related efficient field of perturbation,and then this kind of expression can be weakened by showing smaller valid field of perturbation.The reason for this variation of $\alpha$ is the same decay law have been described before and the deviation from same value of $\alpha$ with the perturbation increased to some extent can be understood by the maximum perturbation to hold the same decay law in terms of a given time.The maximum perturbation can be given the symbol as $\sigma_{h}$ has been used previously corresponding to fixed $t_0$ but here we do not need this limitation.For a given time,we can have the related $\sigma_{h}$ with the formula as 
     \be\label{sigmah_givent}\sigma_{h}=(-\frac{\ln{F_{\infty}}}{c_0})^{\frac{1}{2}}t^{-\frac{\alpha_{0}}{2}}.\ee 
and $F_{\infty}$ is the symbol for the value of saturation.

Thus we can find $\sigma_{h}$ is decided by $t$ with other variables fixed and the unchanged value of $\alpha$ can be found for whole field of perturbation with the up bound we used as $\sigma=10$ if $\sigma_{h}>10$.From Eq.~(\ref{sigmah_givent}),we can find the efficient field of perturbation can be shortened through decreasing $\sigma_{h}$ with the fixed time $t$ increased.Although the same decay law showing the same $\alpha$ within the time scale as $t_0$ without considering saturation can not hold rigorously for large perturbation typically as $\sigma>1$ in terms of most $p_c$ not near $1.866$ and this situation can lead to some deviation from our theoretical prediction such as for the case of $p_c=1.862$ analytically showing unchanged value of $\alpha$ for $t=200$ as $t_0>200$ and related $\sigma_{h}>10$ based on the analytical result in the figure 20,but our analysis of this tendency for the variation of $\alpha$ is still reliable.Therefore variation of $\alpha$ can not hold the same value even for the smallest perturbation as $\sigma=0.01$ if $t>t_0$,and then we can get the transitive pattern as the obvious decline of $\alpha$ from the very initial perturbation for $\sigma=0.01$ caused by the part of obvious faster decay of LE typically for $\sigma<0.1$ which is the result of direct numerical observation.Then this kind of transitive pattern can be gradually vanished for the time to surpass the part of faster decay of LE and the relatively stable variation of $\alpha$ can come with the feature as the values of $\alpha$ alter not too much.But we notice the effect of the fluctuation of LE before saturation or just for saturation,thus we give the specific names accordingly as non-trivial fluctuation and trivial fluctuation.As the fitting time step as $50$,the effect of fluctuation could not be smoothed out and the related expression should be have some abnormal feature.Therefore we will summarize the study result depicted in the figure 23,24,25 for variation of local $\alpha$ with $\sigma$ for different $p_c$ based on our analysis and trivial fluctuation will not be pointed out particularly.

For $p_c=1.85,1.852$ corresponding to classical stable dynamics,$\alpha$ can hold the basically same value for the effective field of perturbation before some obvious change happened for every fixed time and particularly non-zero changed value of $\alpha$ for $t=200$ shows that there is not rigorously same decay law for large perturbation as $t=200$ is relatively short time which is not related to saturation for $\sigma=4$ and $\sigma=3$ respectively corresponding to first obvious change.With fixed time we choose becomes large,the effective field of perturbation for holding basically same value can become smaller as the effect of saturation.For $p_c=1.866$,there is the basically same value of $\alpha$ for every give time respectively as $t=50,200,500,2000,5000$ showing the saturation is not yet coming even for the largest perturbation as $\sigma=10$ in terms of the fixed time we investigate.Meanwhile the discrepancy of values related to different time is quite large as the effect of non-trivial fluctuation,and thus we can not observe the expected value as $2$.For $p_c=1.854,1.856$ corresponding to initially leaving classical stable field,the three patterns of variation still can be found very clearly as the first pattern characterized by unchanged value related to $t=50,200$ in terms of effective field of perturbation,the transitive pattern characterized by the obvious decline from the very initial small perturbation as $\sigma=0.01$ for $t=500$,the relatively stable variation for $t=2000,5000$.

We can notice that most values of $\alpha$ for $p_c=1.854$ are still close to $2$ for $t=50,2000,5000$ with the fluctuation diminished heavily but the values of $\alpha$ for $p_c=1.856$ related to the third pattern as $t=2000,5000$ are set basically between $1$ and $2$ showing the effect of increasing distance from $p_c=1.85,1.852$.Enlightened by this expression,we can find the similar variations for the pairs of $p_c$ as $(1.858,1.874),(1.86,1.872),(1.862,1.87),(1.864,1.868)$ symmetrically distributed with the reference as $p_c=1.866$ sticking to classical stable field and the basic three pattens all can hold for them.To investigate the time to firstly have the transitive pattern of variation of $\alpha$,we can find it is $t=500$ for $p_c$ before $p_c=1.87$ and then it is $t=200$ for rest of $p_c$.This expression is agree with our analysis to firstly have the situation of finding transitive pattern if the condition as $t>t_0$ can be found and the figure 20(a) showing the variation of $t_0$ with $p_c$ can easily give the information that $p_c=1.87$ is the watershed.Here the effect of non-trivial fluctuation can make some seeming unreasonable variation of $\alpha$ based on the previous study of global $\alpha$ which actually reflect the basic decay laws,and they are related to $t=2000,5000$ for $p_c=1.86$,$t=200,500,2000,5000$ for $p_c=1.862$,$t=2000,5000$ for $p_c=1.872$ and $t=500,2000$ for $p_c=1.87$ judged by the continuous decline even with the negative value or sharply rise happened for long time that actually can be related to relatively stable variation.Meanwhile the large fluctuations also can make the notable discrepancies among the values of local $\alpha$ of different fixed time respectively for $p_c=1.864,1.866,1.868$.For relatively short time as $t=500$ for $p_c=1.874,1.876$,the non-monotonic large alternations of $\alpha$ after firstly passing zero can not be taken as the effect of saturation but as the effect of slow decay of LE with large fluctuation.

For $p_c$ from $1.878$ but not including $1.89$,the irregular oscillation within the field of perturbation as $0.2$ can be found commonly quite notably for $t=200,500$,and actually this kind of irregular oscillation has been found for $p_c=1.876$ in terms of $t=500$ implicating the obvious fluctuation of LE for the field of small perturbation.Further more,this large fluctuation of LE for small perturbation can have the affect on the variation of $\alpha$ in terms of long time as $t=2000$ and $t=5000$ showing sharp alteration like up and down for small perturbation that can be seen notably for $p_c=1.878,1.88,1.882,1.884$.For $p_c=1.89$,the typical initial decline of $\alpha$ can not be found as it means the transient fast decay related to small perturbation is vanished,and alternatively we can find some stable oscillation of $\alpha$ for $t=200$ within the field of perturbation as $0.2$ followed by the variation with small value implicating the existence of quite slow decay of LE.With careful comparisons of values of $\alpha$ in quantity for different $p_c$ in terms of different fixed time,we find higher value is directly related to the closer distance from the referenced $p_c$ corresponding to classical stable field as $1.852$ or $1.866$.Without the consideration of abnormal expressions of $\alpha$ caused by large fluctuation of LE,the alternation of the values of $\alpha$ in quantity with $p_c$ increasing do undergo an non-monotonic change notably showed respectively for $t=500$,$t=2000$ and $t=5000$ as gradual decreasing tendency for $p_c=1.854,1.856,1.858,1.86$,gradual increasing tendency for $p_c=1.86,1.862,1.864,1.866$,and gradual decreasing tendency for $p_c=1.866,1.868,1.87,1.872,1.874,1.876,1.878$.For $p_c$ after $1.878$ but not including $1.89$,such clear tendency can not be found and the values related to relative stable variation corresponding to $t=500$ are basically between $0$ and $1$.In terms of $p_c=1.89$,the values of $\alpha$ can be around or quite close to $1$ respectively for $t=500,2000,5000$ within the effective field of perturbation showing exponential decay of LE.We also want to have the comparison with local $\alpha$ obtained from semi-classical integral,and there are some large discrepancies of $\alpha$ for short time typically manifested as $t=50$ independent of $p_c$ we study,for $p_c$ distributed around $p_c=1.866$ as $p_c=1.862,1.864,1.866,1.868,1.87$,and for $p_c=1.88,1.882,1.884$ with the long time manifested as $t=2000,5000$ in terms of large perturbation basically larger than $1$.The discrepancies of $\alpha$ for $p_c=1.88,1.882,1.884$ pointed out above under the conditions of the long time and large perturbation together deserve our attention as they are related to the situation for showing the continuous although quite slow decay of LE after some part seeming for saturation.

\begin{center} 
\begin{figure}
\includegraphics[width=18cm,height=18cm]{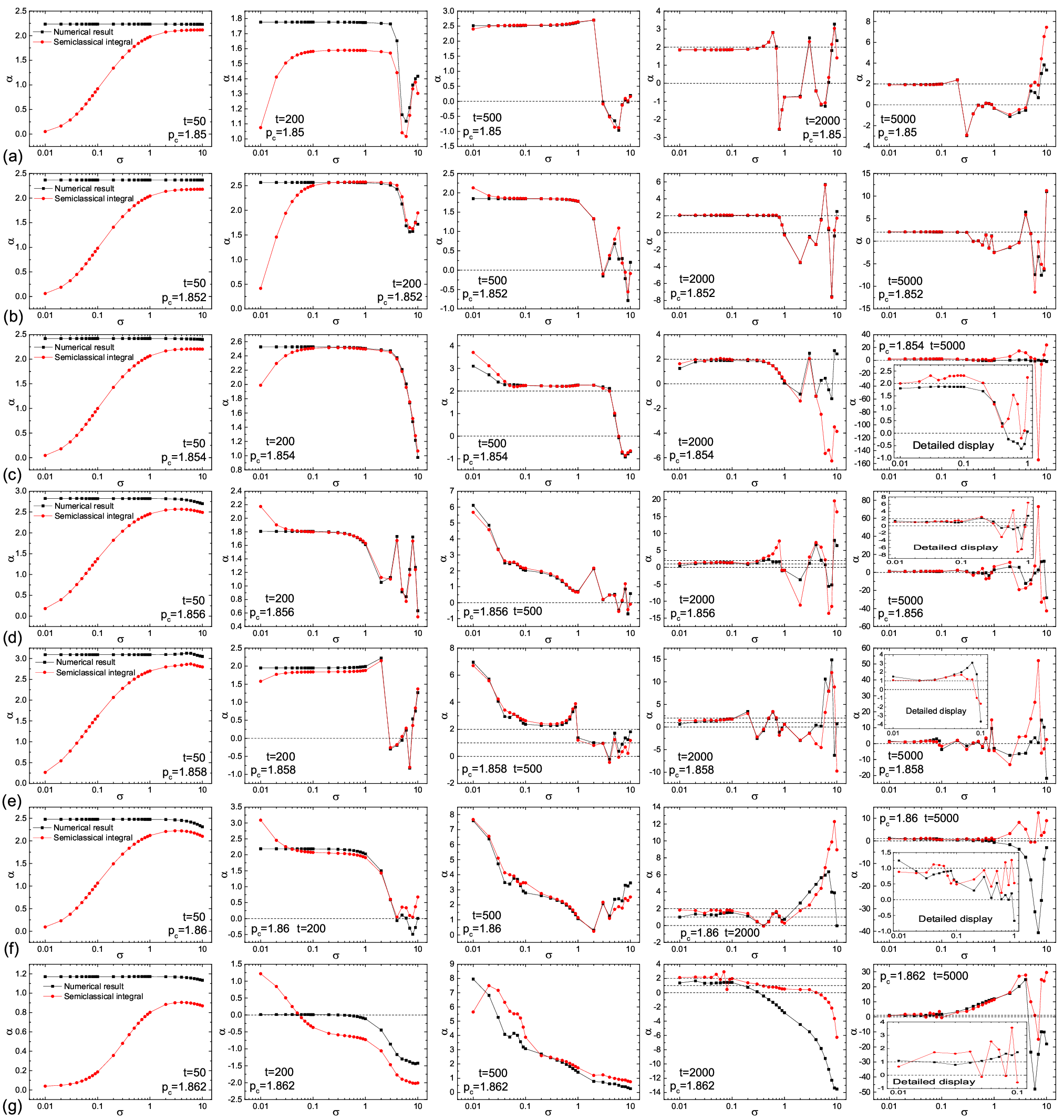}
\vspace{-0.2cm} \caption{Variation of local fitted $\alpha$ with $\sigma$ corresponding to different selected time in terms of $p_c$ from $1.85$ to $1.862$ with the interval as $0.002$,the fitted time is fixed as $50$.For $p_c=1.85,1.852$ respectively,$\alpha$ can show the main parts which are around expected $2$ for $t=2000,5000$ and are deviated from $2$ for $t=50,200,500$ due to the effect of fluctuations of LE but also with basically same values respectively.For the rest of $p_c$,the situation to maintain the same value of $\alpha$ for short time as $t=50$ can be found in terms of respectively valid field and it also can be available for $t=200$ corresponding to smaller valid field of perturbation.Meanwhile $\alpha$ can show the obvious declines for the rest of $p_c$ from the very initial perturbation as $\sigma=0.01$ for $t=500$ and the non-monotonic large alternation of $\alpha$ after firstly passing zero can be found notably for $t=2000,5000$ and mainly saturation is accountable for this expression.But the local $\alpha$ can also show some large deviation from the normal expectation based on the study result of global $\alpha$ caused by the effect of notable fluctuation of LE that can not be smoothed out for the fitting time we use as $50$ and the related obvious expressions can be observed as $t=2000,5000$ for $p_c=1.86$ and $t=200,500,2000,5000$ for $p_c=1.862$ judged by the continuous decline even with the negative value or sharply rise for long time with the order of $10^{3}$ that actually can be related to relatively stable variation.From the comparison of local $\alpha$ obtained from semi-classical integral,basic agreement can be found except for short time commonly related to $t=50$ as well as $p_c=1.862$ showed in different time.}\label{local_alp_edge1}
\end{figure}
\end{center}

\begin{center} 
\begin{figure}
\includegraphics[width=18cm,height=18cm]{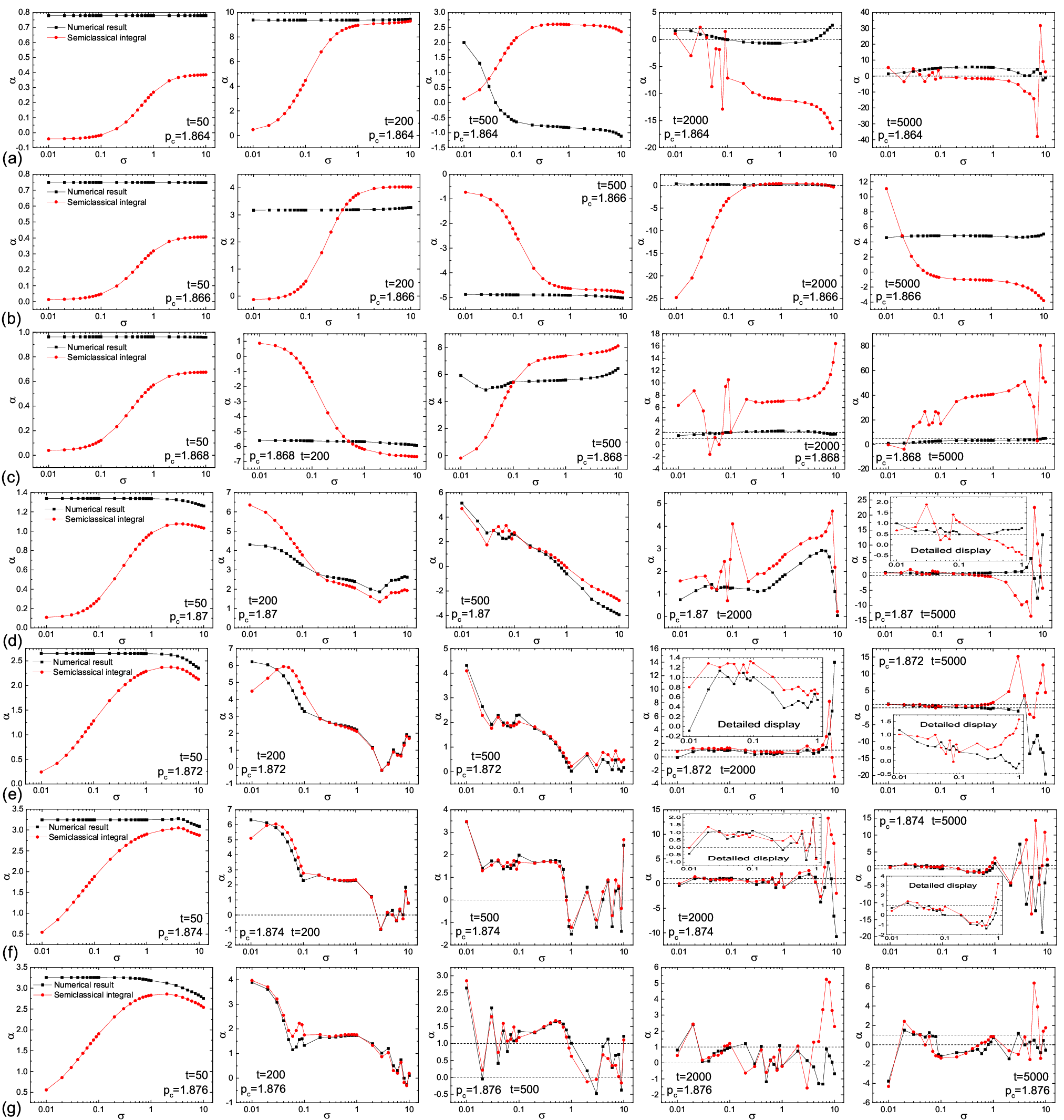}
\vspace{-0.5cm} \caption{Variation of local fitted $\alpha$ with $\sigma$ corresponding to different selected time in terms of $p_c$ from $1.864$ to $1.876$ with the interval as $0.002$,the fitted time is fixed as $50$.Just like the illustration in the last figure,the variations of $\alpha$ with $\sigma$ except for $p_c=1.866$ still can show the three typical patterns with time increasing as the situation to maintain same value within valid field of perturbation in terms of short time as $t=50,200$ for $p_c=1.864,1.868$ and $t=50$ for $p_c=1.87,1.872,1.874,1.876$,then the obviously monotonous decline beginning with the very initial perturbation as $\sigma=0.01$ emergent as $t=500$ for $p_c=1.864,1.868$ and $t=200$ for $p_c=1.87,1.872,1.874,1.876$,and afterwards relatively stable variation that can be seen clearly for $t=2000,5000$.Non-monotonic large alternation of $\alpha$ after firstly passing zero also can be found mainly as the effect of saturation or slow decay of LE after main decay related to $t=500$ for $p_c=1.874,1.876$.Besides the expressions of the large deviation from the normal expectation based on the study result of global $\alpha$ as $t=2000,5000$ for $p_c=1.872$ and $t=500,2000$ for $p_c=1.87$,we also notice the values of local $\alpha$ related to $p_c=1.864,1.866,1.868$ can differ quite largely compared with different time particularly for $t=50,200,500$ caused by the large fluctuation of LE as well but we still can differentiate the typical three patterns of variation of $\alpha$ with $\sigma$.Further more,we can find the similar variations for the pairs of $p_c$ as $(1.858,1.874),(1.86,1.872),(1.862,1.87),(1.864,1.868)$ symmetrically distributed with the reference as $p_c=1.866$ sticking to classical stable field,and the change of $\alpha$ with $\sigma$ for $p_c=1.866$ is basically small for given time which is similar to the variations of $\alpha$ for $p_c= 1.85,1.852$ in terms of valid fields of perturbation.From the comparison of local $\alpha$ obtained from semi-classical integral,there are some large discrepancies for short time commonly related to $t=50$ as well as for $p_c=1.864,1.866,1.868,1.87$ showed in different time.}\label{local_alp_edge2}
\end{figure}
\end{center}

\begin{center}
\begin{figure}
\includegraphics[width=18cm,height=18cm]{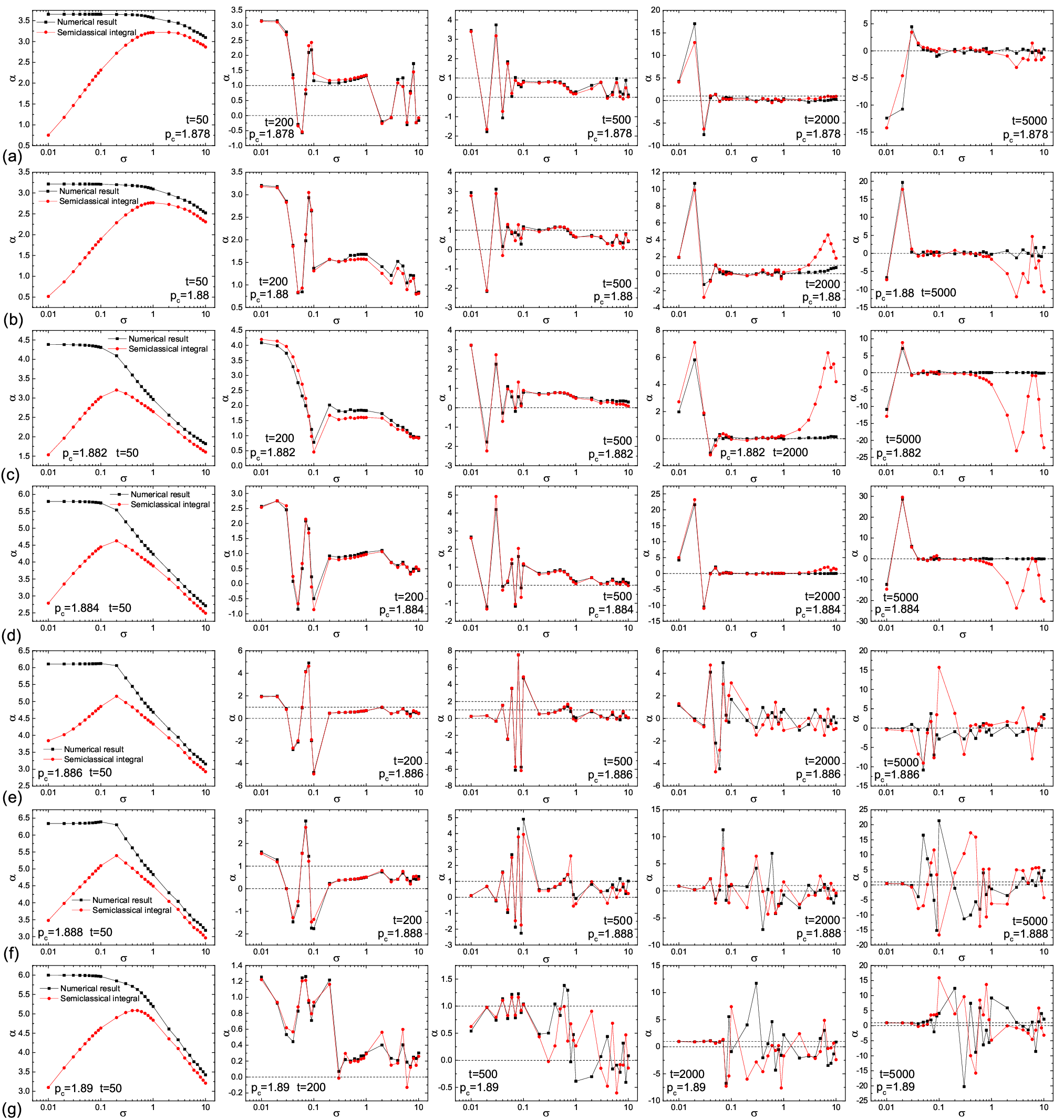}
\vspace{-0.2cm} \caption{Variation of local fitted $\alpha$ with $\sigma$ corresponding to different selected time in terms of $p_c$ from $1.878$ to $1.89$ with the interval as $0.002$,the fitted time is fixed as $50$.For $p_c$ which is not $1.89$,the three basic patterns for the variation of $\alpha$ still can be basically valid as the situation to maintain same value within valid field of perturbation for $t=50$,the decline as a whole commonly found for $t=200$ in terms of the field of perturbation not more than $0.2$ having some large alternation,relatively stable variation as a whole for $t=500$ but with the initial part having some intensely irregular oscillation in terms of the field of perturbation not more than $0.2$ as well.Actually this kind of irregular oscillation has been found for $p_c=1.876$ in terms of $t=500$ implicating the obvious fluctuation of LE for the field of small perturbation and it can have the affect for variation of $\alpha$ in terms of long time as $t=2000$ and $t=5000$ showing sharp alteration like up and down for small perturbation that can be seen notably in the figure for $p_c=1.878,1.88,1.882,1.884$.Meanwhile non-monotonic large fluctuations of $\alpha$ after passing zero also can be found have been showed in the last two figures and saturations are accountable for them.For $p_c=1.89$,the typical initial decline of $\alpha$ can not be found as it means the transient fast decay related to small perturbation is vanished,and alternatively we can find some stable oscillation of $\alpha$ for $t=200$ within the field of perturbation as $0.2$ followed by the variation with small value implicating the existence of quite slow decay of LE.Further more,$\alpha$ can be around or quite close to $1$ for $t=500,2000,5000$ in terms of the effective field of perturbation showing exponential decay of LE.From the comparison of local $\alpha$ obtained from semi-classical integral,there are some obvious discrepancies for short time commonly related to $t=50$ as well as for long time as $t=2000,5000$ related to $p_c=1.88,1.882,1.884$ in terms of strong perturbation basically larger than $1$.}\label{local_alp_edge3}
\end{figure}
\end{center}

Then we want to study $c_0$ and our expectation is to find the values basically independent of time or at least show stable oscillations around some center values that validate our assumed decay law as $M(t)\approx e^{-c_{0}(t)\sigma^{\nu(t)}t^{\alpha(\sigma,t)}}$.Here we use $\sigma=0.01$ to extract $c_0$ from the decay law we assume in terms of the fitting time as $50$ to reduce the likely large fluctuation as well as show the variation finely.Actually we still find the fluctuations can be large commonly,and thus $\ln(c_{0})$ is used to replace the original $c_0$.The variations of $\ln(c_{0})$ are clearly showed in the figure 26 and thus encourage to investigate the study results firstly.Here we also want to connect the study results to previous results in terms of correspondence for the classical dynamics.Firstly there are similar variations of $\ln(c_{0})$ with $t$ for $p_c=1.85,1.852,1.866$ related to classical stable dynamics as the basic feature for having stable oscillations,and further more $\ln(c_{0})$ can oscillate around the values largely for $p_c=1.85$ but relatively small for $p_c=1.852,1.866$ and the orders of center values are similar for $p_c=1.85,1.852$ but not for $p_c=1.866$ illustrated with reference lines commonly used in the figure 26.For $p_c=1.854,1.856,1.858,1.86,1.862,1.864,1.868,1.87,1872$,the basic stable oscillations of $\ln(c_{0})$ are also quite common after the previous large alternations and the amplitudes of oscillation for $p_c=1.864,1.868$ closest to $p_c=1.866$ are much smaller compared with other $p_c$.Here we still can find the similar variations of $\ln(c_{0})$ for the pairs of $p_c$ as $(1.86,1.872),(1.862,1.87),(1.864,1.868)$ which are all symmetrically distributed with the reference as $p_c=1.866$.This kind of similarity is also found in the previous study but is not applied to the pair as $1.858,1.874$ here as the obvious discrepancies for the gradually increasing amplitude for $p_c=1.874$ as well as the considerable different center values.For $p_c=1.874,1.876,1.878,1.88,1.882,1.884$,the common feature for the variations of $\ln(c_{0})$ is the gradually increasing amplitude and the related center values begin to around $0$ from $p_c=1.878$.For $p_c=1.886$,$\ln(c_{0})$ can show the slow transition to basic stable oscillation with the very initial large alternations happened.The variation of $\ln(c_{0})$ for $p_c=1.888$ is quite similar to the case of $p_c=1.886$ pointed above,but the stable oscillation can not be observed clearly.For $p_c=1.89$,$\ln(c_{0})$ mainly can show the limited fluctuation irregularly with the center value clearly smaller than other $p_c$ since $1.878$ and the initial large alternation is similar to the cases of $p_c=1.886,1.888$.From the comparison of $\ln(c_{0})$ obtained from semi-classical evaluation,there are some obviously large discrepancies for $p_c=1.862,1.864,1.866,1.868,1.87$ symmetrically distributed with the reference $p_c=1.866$.Therefore we can find the variations of $\ln(c_{0})$ basically are agree with our expectation and at least the order of center value can be used to study time scale.

\begin{center}
\begin{figure}
\includegraphics[width=18cm,height=18cm]{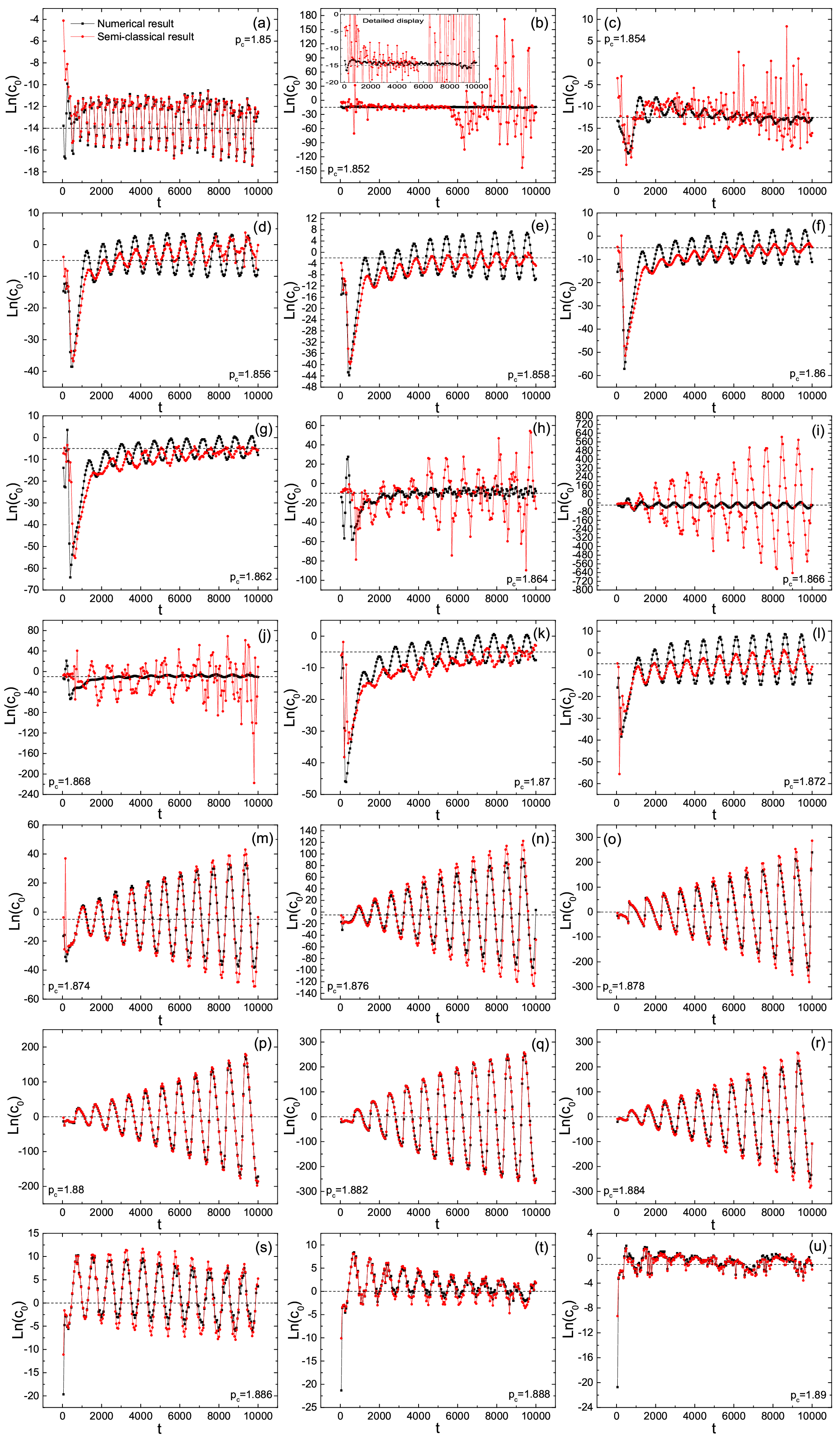}
\vspace{-0.5cm} \caption{Variation of $\ln(c_{0})$ with $t$ for $p_c$ from $1.85$ to $1.89$ with the interval as $0.002$,$c_{0}$ is extracted from the assumed decay law as $M(t)\approx e^{-c_{0}(t)\sigma^{\nu(t)}t^{\alpha(\sigma,t)}}$ with the perturbation as $\sigma=0.01$ under the condition for the fitting time as $50$ to reduce the likely large fluctuation as well as show the variation finely.In terms of graphs as $(a,b,i)$,$\ln(c_{0})$ can oscillate around the values largely for $p_c=1.85$ but relatively small for $p_c=1.852,1.866$ and the orders of center values are similar for $p_c=1.85,1.852$ but not for $p_c=1.866$ illustrated with reference lines also commonly used in all graphs.For $p_c=1.854,1.856,1.858,1.86,1.862,1.864,1.868,1.87,1872$ related to graphs as $(c,d,e,f,g,h,j,k,l)$,the basic stable oscillations of $\ln(c_{0})$ are also quite common after the previous large alternations and the amplitudes of oscillation for $p_c=1.864,1.868$ closest to $p_c=1.866$ are much smaller compared with other $p_c$.The similar variations of $\ln(c_{0})$ for the pairs of $p_c$ as $(1.86,1.872),(1.862,1.87),(1.864,1.868)$ can be found which are all symmetrically distributed with the reference as $p_c=1.866$.For $p_c=1.874,1.876,1.878,1.88,1.882,1.884$ related to graphs as $(m,n,o,p,q,r)$,the common feature for the variations of $\ln(c_{0})$ is the gradually increasing amplitude and the related center values begin to around $0$ from $p_c=1.878$.For $p_c=1.886$ related to graph $(s)$,$\ln(c_{0})$ can show the slow transition to basic stable oscillation with the very initial large alternations happened.The variation of $\ln(c_{0})$ for $p_c=1.888$ related to graph(t)is quite similar to the case of $p_c=1.886$ pointed above,but it seems that the stable oscillation can not be observed yet.For $p_c=1.89$ related to graph $(u)$,$\ln(c_{0})$ mainly can show the limited fluctuation irregularly with the center value clearly smaller than other $p_c$ since $1.878$ and the initial large alternation is similar to the cases of $p_c=1.886,1.888$.From the comparison of $\ln(c_{0})$ obtained from semi-classical evaluation,there are some obviously large discrepancies for $p_c=1.862,1.864,1.866,1.868,1.87$ symmetrically distributed with the reference $p_c=1.866$.}\label{c0_edge}
\end{figure}
\end{center}

Now we want to use the study result of decay laws to consider the issue for time scale particularly for the comparisons for decay speed compared with LE of strong chaos.As our assumed decay law supported extensively by numerical results is $M(t)\approx e^{-c_{0}(t)\sigma^{\nu(t)}t^{\alpha(\sigma,t)}}$,so the corresponding time scale can be written as $\tau=(c_0\sigma^{\nu})^{-\frac{1}{\alpha}}$.For the comparison of decay degree of LE related to strong chaos,we can choose $K=7$ commonly used for expressing strong chaos and meanwhile set the averaged momentum of initial quantum wave packet as $p_c=3$.Thus we can have a simple mathematical inequality as $\sigma^{2-\frac{\bar{\nu}}{\bar \alpha}}>\frac{\overline{c_{0}}^{\frac{1}{\bar{\alpha}}}}{2K(E)}$ from the equivalent concept has been used in the study of LE for chaotic sea in the classic limit,and the rest of work is to consider the condition definitely with related values as $\overline{c_{0}},\bar{\nu},\bar \alpha$ put into.Actually we can set the value of $\bar \alpha$ selected from $\alpha_{2},\alpha_{3}$ preferentially for $\alpha_{3}$ not close to $0$ in terms of fixed perturbation as $\sigma=0.1$,$\bar{\nu}$ from the mid-value during the decreasing process for the variation of $\nu$ with $t$,and $\overline{c_{0}}$ from the averaged value of $c_0$ extracted by $\sigma=0.1$ within time as $10^3$,thus indeed find the universal existence of critical perturbations typically with the order of magnitude as $10^{-1}$.For the specific case as $p_c=1.874$,the values of $\overline{c_{0}},\bar{\nu},\bar \alpha$ can be reasonably set as $e^{-5},0.5,1$ and we get the condition as $\sigma>0.0514$ with a quite straightforward calculation.Therefore we want to check out our expectation although it is a quite rough estimate.

For directly observing this existence of critical perturbations,we firstly show the comparisons for decay degree between the edge of chaos and strong chaos in the classical limit with different perturbation selected as $\sigma =0.01,0.05,0.1.0.2,0.3,0.4,0.5,0.6,1$.Meanwhile we choose $p_c=1.874,1.878,1.882,1.886,1.89$ with $K$ fixed as $3$ to represent typical edge of chaos and meanwhile choose $K=7,10$ with $p_c$ fixed as $3$ to represent strong chaos,thus want to observe the transition from slower to faster decay for LE of strong chaos compared with the decay of LE of edge of chaos.Such transitions indeed can be found for all $p_c$ we choose,it means the critical perturbations do commonly exist.Further more,this kind of transition for the comparisons of decay degree also can be found among the decay of LE for different $p_c$ of edge of chaos and the expressions are some complicated particularly for small perturbation seen in the graph(a),(b).After directly showing the transitions with decay of LE,we also can compare the decay time of LE to show critical perturbation by intersection.Here decay time $\tau$ can be numerically determined by the condition for the value of LE as $e^{-1}$ and we can consider the situations belonging to edge of chaos for $p_c=1.874,1.876,1.878,1.88,1.882,1.884,1.886,1.888,1.89$ with $K$ fixed as $3$,and choose the strong chaos for $K=7,10$ with $p_c$ fixed as $3$.The intersections for the variations of decay time between strong chaos and edge of chaos can be found commonly and these expressions are reconciled with the theoretical expectation of existence of critical perturbation with the order of magnitude as $10^{-1}$.Among the comparisons for decay time of LE related to cases of edge of chaos,the intersections also can be found but show some kind of complicated expressions in particular for the field of small perturbation.As the established rules for stable dynamics and strong chaos are $\tau\propto\sigma^{-\gamma}$ with respectively $\gamma=1,2$,we can use the logarithmic coordinates to find likely similar rules.Actually these clear rules of time scale can not be found commonly except for beginning part of perturbation related to $p_c=1.89$,further more we notice there is a dramatic change from $p_c=1.888$ to $p_c=1.89$ as the slope of $\ln{\tau}$ versus $\ln{\sigma}$ can suddenly shift roughly from $1$ to $2$ at least for the main part of small perturbation within $0.1$.For $p_c=1.878,1.88,1.882,1.884$,we have known the small perturbation such as $\sigma=0.01$ can lead to heavy decay of LE and thus the relatively small values of $\tau$ with the order as $10^{2}$ show a quite sensitive field but the variations of $\tau$ with $\sigma$ as a whole are mild.

\begin{center}
\begin{figure}
\includegraphics[width=18cm,height=16cm]{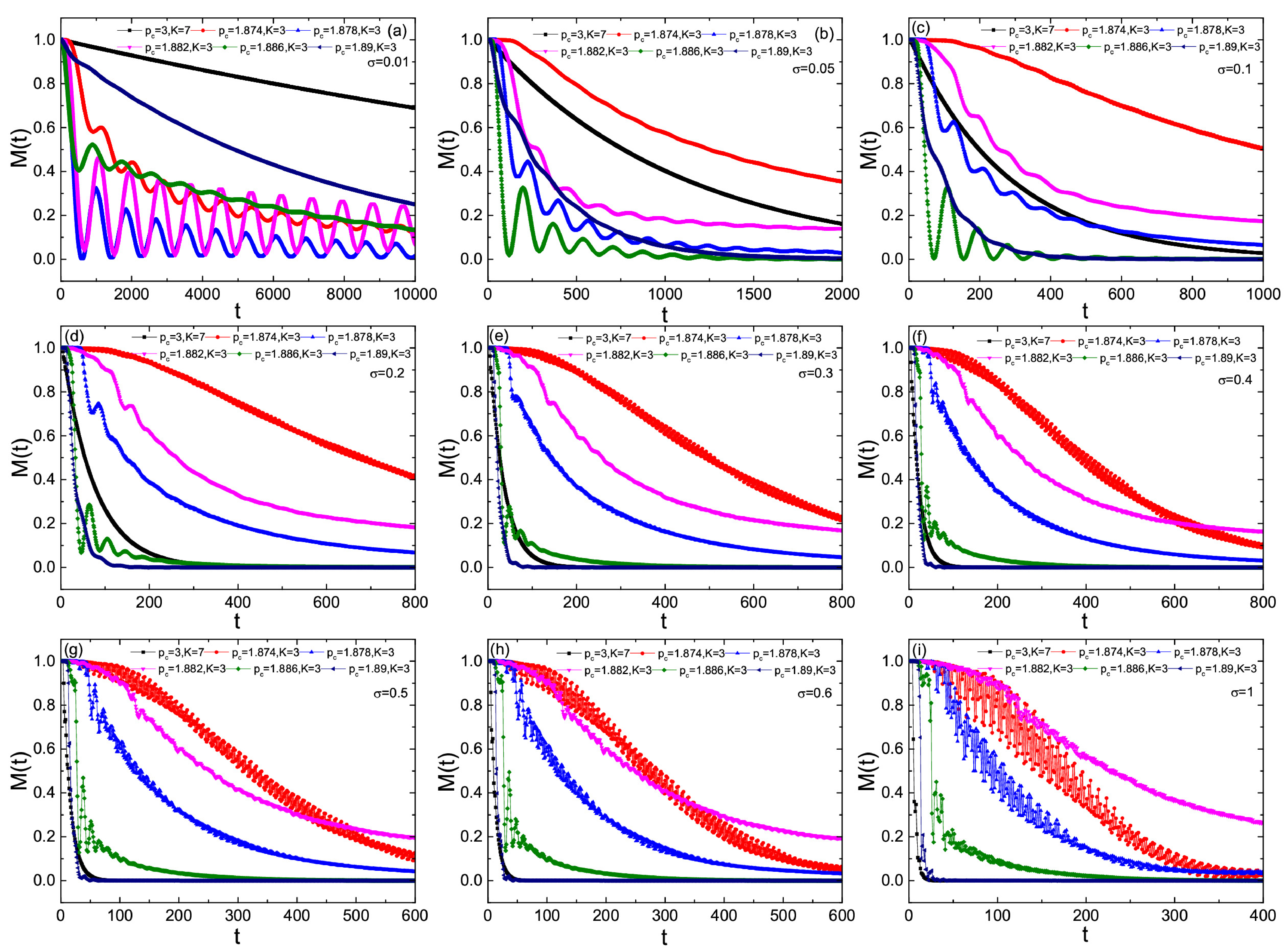}
\vspace{-0.5cm} \caption{Comparison of decay speed of LE between the edge of chaos and strong chaos in the classical limit with different perturbation selected as $\sigma=0.01,0.05,0.1.0.2,0.3,0.4,0.5,0.6,1$.Here we choose the typical cases of edge of chaos for $p_c=1.874,1.878,1.882,1.886,1.89$ with $K$ fixed as $3$,and choose the strong chaos for $K=7,10$ with $p_c$ fixed as $3$.There is a clear transition from slower to faster decay for LE of strong chaos compared with the decay of LE of edge of chaos,and this transition is conform to theoretical analysis for showing the existence of critical perturbation.There also are some transitions for comparisons of decay speed of LE among edge of chaos which are particularly complicated for small perturbation seen in the graph(a),(b).}\label{fidelity_edge_transition}
\end{figure}
\end{center}

\begin{center}
\begin{figure}
\includegraphics[width=14cm,height=12cm]{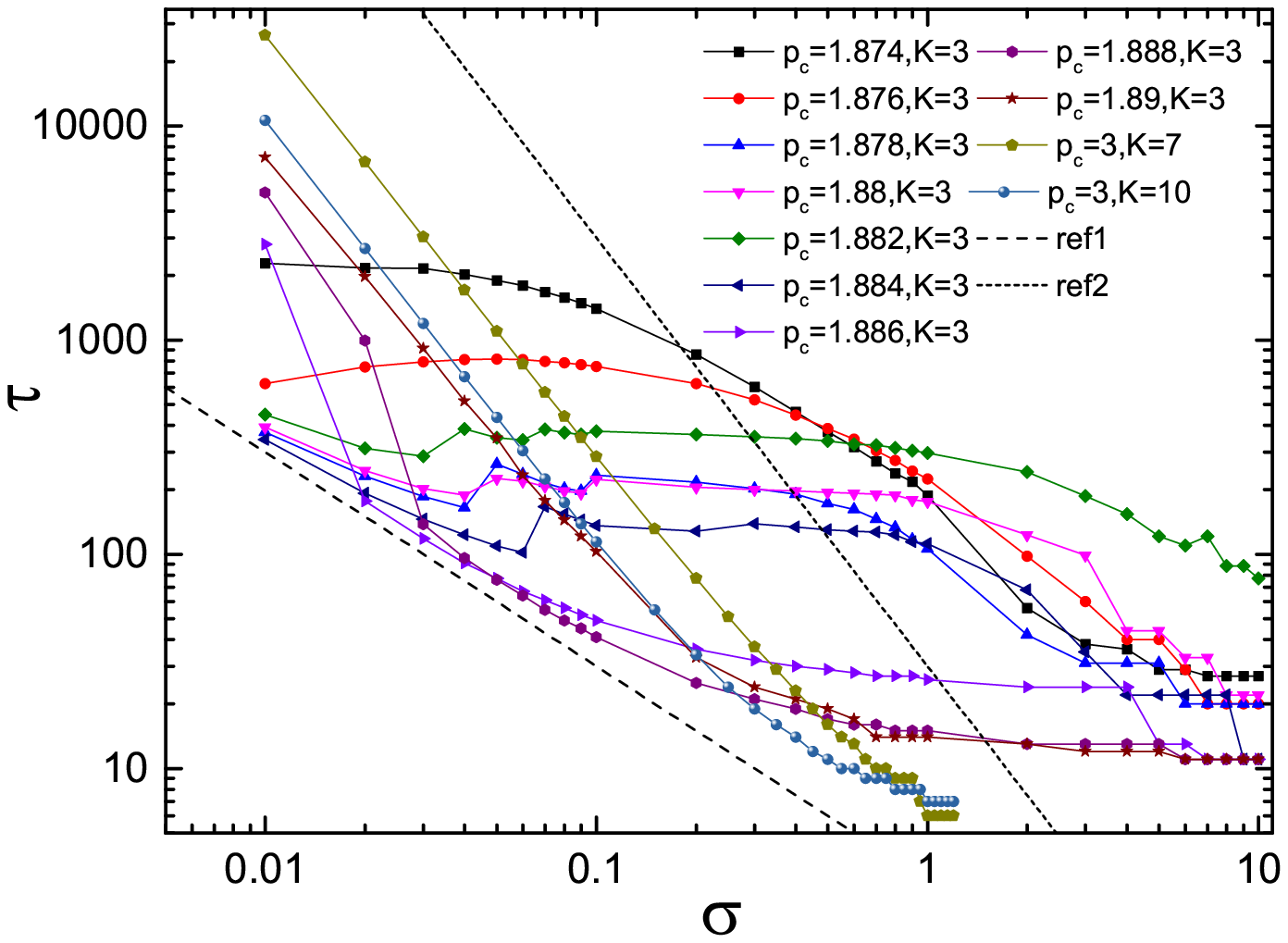}
\vspace{-1.2cm} \caption{Comparison of decay time $\tau$ between the edge of chaos and strong chaos in the classical limit in terms of logarithmic coordinates.$\tau$ can be numerically determined by the decay process within the value of quantum fidelity as $e^{-1}$ and two lines with the names as $ref{1}$ and $ref{2}$ are illustrated to guide the eyes for respective slopes as $1$ and $2$ corresponding to the rules of governing stable dynamics and strong chaos.Here we consider the situations belonging to edge of chaos for $p_c=1.874,1.876,1.878,1.88,1.882,1.884,1.886,1.888,1.89$ with $K$ fixed as $3$,and choose the strong chaos for $K=7,10$ with $p_c$ fixed as $3$.The intersections for the variations of decay time between strong chaos and edge of chaos can be found commonly and these expressions are reconciled with the theoretical prediction of existence of critical perturbation to determine the transition from the slower decay speed to faster decay speed for LE of strong chaos compared with the decay process of LE for edge of chaos we consider here.Without considering strong chaos,the good linear dependence of $\ln{\tau}$ on $\ln{\sigma}$ basically can not be found except the beginning part of perturbation for $p_c=1.89$,so the rule as $\tau\propto\sigma^{-\gamma}$ can not be applied commonly.Among the comparisons for decay time of LE related to cases of edge of chaos in the figure,the intersections also can be found but are quite complicated.For $p_c=1.878,1.88,1.882,1.884$,the small perturbation such as $\sigma=0.01$ can lead to heavy decay of LE reflecting the relatively small value of $\tau$ but the variations as a whole are mild.}\label{timescale_critial_edge}
\end{figure}
\end{center}

Although we can not find the expected rules as $\tau\propto\sigma^{-\gamma}$ for the time numerically related to $M(t)=e^{-1}$,but we consider the likely rules still could exist for the decay time $T$ corresponding to main decay process of LE.Therefore,we consider two ways to get the time for studying likely rules of time scale.The one way is to use the fitting technique to get the main time of decay of LE which has been used before to get global $\alpha$,and the time we can get here should beyond the first initial time corresponding to same decay law pointed out in the previous study.The time step to do the fitting of LE is $20$ and the condition to finish the searching is the fitting local $\alpha$ smaller than $0.2$,this procedure is same to the previous study to get global $\alpha$ and the purpose is for catching the expected time as accurate as we could which means trial-and-error method.The other way to get the time is through the value of saturation numerically obtained from the mean value of LE related to the time period without clear decay but just show purely very small fluctuation,thus we can get the expected time when the value of LE just below it based on the fluctuating feature of saturation.In detail,we consider the time period for averaging as the last $20000$ time steps with the whole evolved time as $50000$ except for the cases of $p_c=1.852,1.854,1.866,1.868$ requiring more evolved time as $100000$ with the last $20000$ time steps for averaging.The study result is showed in the figure below and we give some clarification of the results for understanding.

With the expectation of likely rule of time scale as $T\propto\sigma^{-\gamma}$,we use the referent dash line and short dash line to show the slop respectively as $1$ and $2$ for $ln{T}$ versus $ln{\sigma}$ actually representing the rules of time scale related to stable field and strong chaos in the classical limit as $T\propto\sigma^{-1}$ and $T\propto\sigma^{-2}$.Considering the two ways to get $T$ from direct numerical computation and semi-classical evaluation,so we can have four type of study results that can be compared together.If we just show the result from which way to get,we just call them separately as study result 1 and study result 2 for convenience.From the graph(a)to(m),we find the rule of time scale as $T\propto\sigma^{-1}$ is common for $p_c$ from $1.85$ to $1.874$ in terms of perturbation within $1$ and thus the rule of time scale is not just for the pure stable dynamics in the classical limit.According to $p_c=1.85,1.852,1.854$,we can find the rule of time scale can shift sharply towards $T\propto\sigma^{-2}$ for some field of perturbation larger than $2$.For other $p_c$ except $1.856$,$T\propto\sigma^{-1}$ still can be the dominated rule governing time scale for large perturbation which can be applied for having the effective decline of $T$.For $p_c$ from $1.876$ to $1.884$ corresponding to the graph from(n)to(r),the beginning discernible deviation from $T\propto\sigma^{-1}$ can be found for $p_c=1.876$ and then this deviation becomes notable for $p_c=1.88,1.882,1.884$.According to the declining part of $T$,the basic good linear relation for $\ln T$ versus $\ln{\sigma}$ can be commonly observed showing the new rule as $T\propto\sigma^{-\gamma}$.Further more for the numerical result 2 and semi-classical result 2,the rule as $T\propto\sigma^{-1}$ still can be found for $p_c=1.88,1.882,1.884$ in terms of large perturbation larger than $1$,and these results show the whole decay process rather than the dominantly main decay still obey the rule for stable field in the classical limit.With $p_c$ increased further more from $p_c=1.886$ to $1.89$,we can find originally explicit rule of time scale seems to break but meanwhile the new rule as $T\propto\sigma^{-2}$ tends to form gradually for the part of perturbation with clear decline of $T$ observed in the graph $(s),(t),(u)$.This expression do agree with our expectation as $p_c$ is approaching chaotic sea and $T\propto\sigma^{-2}$ is related to strong chaos.

It is worth noting that the variations of $T$ for semi-classical result 1 in terms of $p_c=1.862,1.864,1.866,1.868,1.87,1.872$ are quite irregular compared with corresponding numerical result 1,and actually this expression is reconciled with the poor agreement between numerical computation of LE and semi-classical evaluation for $p_c$ basically symmetrically distributed with the reference $1.866$.Thus this kind of semi-classical result 1 can not be considered for showing the rule of time scale and we should set the direct numerical result as priority.Further more we notice the so-called abnormal expression for the basic rise of $T$ in terms of small perturbation and we call the existence of rise of $T$ joined by the numerical result 1 and semi-classical result 1 as pattern $A1$ and the existence of rise of $T$ joined by all the study results as pattern $A2$.Thus firstly we can find the very beginning abnormal rise of $T$ for $p_c=1.858$ as $A1$ pattern as well and then the situation can be observed for $p_c=1.86$ but the study result 1 can form a separate part to show the explicit rule of time scale as $T\propto\sigma^{-1}$ for perturbation within $\sigma=1$ besides the latter part joined with study result 2.The cause to have this kind expression of $A1$ is the large oscillation of LE for quite small perturbation here as $\sigma=0.01$,and the condition we set for smoothed local $\alpha$ smaller than $0.2$ can met although actually the decay process can not yet end at all.The reason for special expression of time scale for $p_c=1.86$ in terms of two distinct parts is that the twisted situation of LE can happen after some perturbation,actually this kind twisted situation is quite similar to oscillated situation with the key feature as there always is the first heavy decay followed by the next relatively slow decay connected with strong rebound.The first dramatic decay of LE can not be seen as the main decay but manifested by the first way to get $T$ and this kind of dramatic decay here is also the transitive decay process showed in the previous study.Therefore we can conclude that the strong rebound leading to smoothed local $\alpha$ smaller than $0.2$ is the reason for having single decline of $T$ for study result 1 if the value of saturation can not meet yet.With $p_c$ increased,we also can find pattern $A1$ for $p_c=1.872,1.874,1.876$,and then pattern $A2$ for $p_c=1.878,1.88,1.882,1.884$,then find $A1$ again for $p_c=1.886,1.888$.If there is the other situation as the first continued decay of LE is so heavy that the smallest value of this kind of decay process could even below the value of saturation,then we can find $A2$ joined with the large oscillation of LE for small perturbation as well.Finally as these two situations just disappear,thus the normal expression of $T$ can be found for $p_c=1.89$.Here we meet again the very tricky situation as small perturbation can lead to heavy decay of LE for $p_c=1.878,1.88,1.882,1.884$.For the comparison for the two ways to get $T$,we can find the big difference observed for $p_c=1.852,1.854,1.874,1.878,1.88,1.882,1.884,1.886$ to show the typical expression as the decline of $T$ for study result 1 but unchanged value of $T$ for study result 2.The cause is the very slow and small decay process after the main decay process,and we can not observe the decline of $T$ just because the time to meet the saturation value is quite longer than $10^4$ which is largest time we set although we notice the irregularity of $p_c=1.886$ which is not a typical case.Here the values of saturation for different $p_c$ deserve attention,and we also show the variation in the related figure below.

\begin{center}
\begin{figure}
\includegraphics[width=18cm,height=20cm]{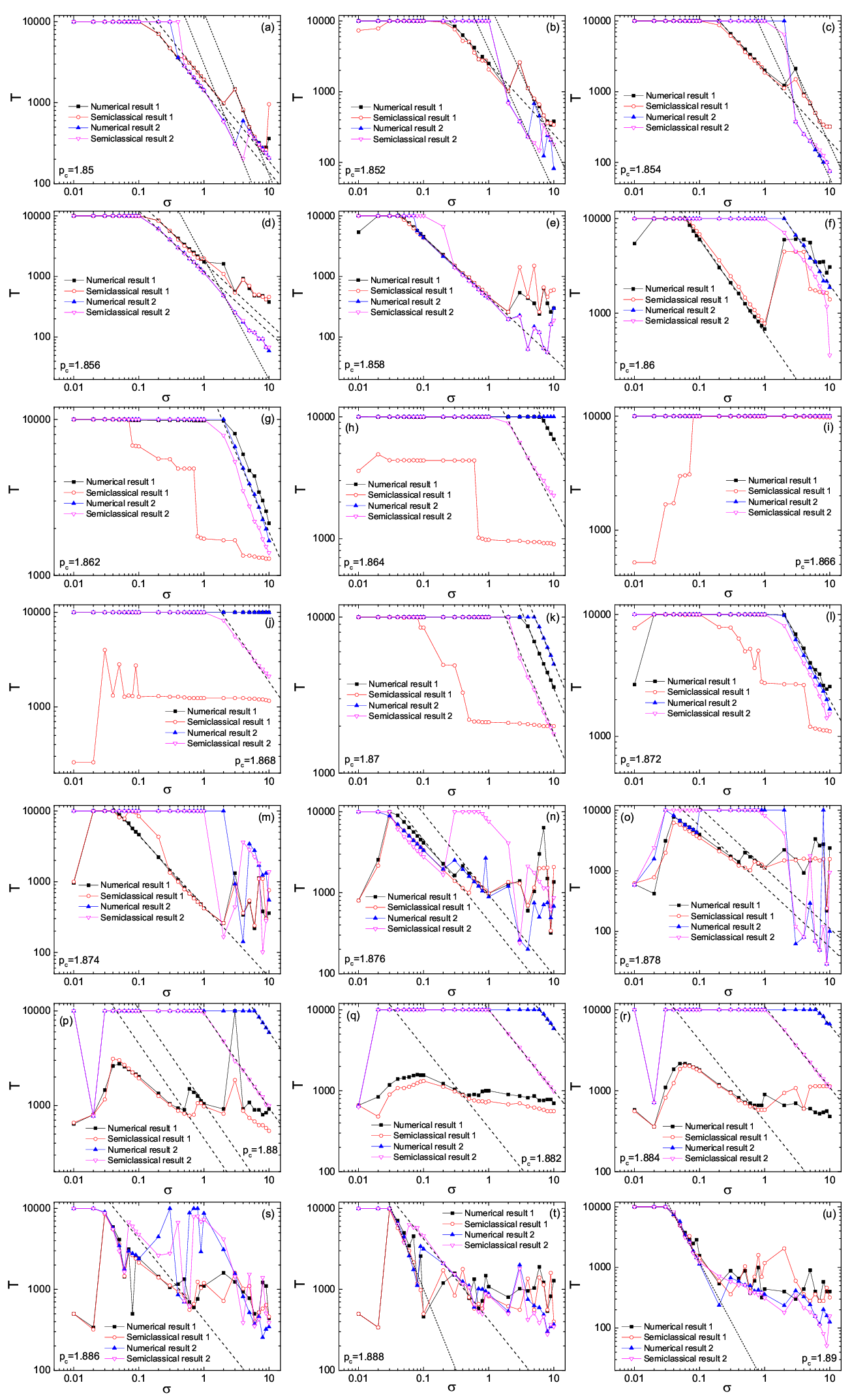}
\vspace{-0.5cm} \caption{Variation of decay time $T$ with $\sigma$ for $p_c$ from $1.85$ to $1.89$ with the interval as $0.002$ and there are two ways to get $T$ labelled as $1$ and $2$.For comparison,the direct numerical result and semi-classical result are all showed.The first way of getting $T$ is to set the condition as local $\alpha$ smaller than $0.2$ with the fitting step as $20$ as well as beyond the first period of time having the same decay law.The second way is to set the condition as the value of LE smaller than the value of saturation numerically obtained.The dash line and short-dash line are related to slope respectively as $1$ and $2$ corresponding to stable field and strong chaos in the classical limit and can help for discerning the likely rule of time scale here.Until for $p_c=1.876$ from the graph(a)to(m),the rule of time scale can be showed commonly as $T\propto\sigma^{-1}$ but it shows the obvious change towards $T\propto\sigma^{-2}$ for some field of large perturbation larger than $2$ in terms of $p_c=1.85,1.852,1.854$.For $p_c$ from $1.876$ to $1.884$ corresponding to the graph from(n)to(r),the beginning discernible deviation from $T\propto\sigma^{-1}$ can be found for $p_c=1.876$ and then the clear deviation follows. Meanwhile basically good linear relation for $\ln T$ versus $\ln{\sigma}$ can be commonly observed showing the rule as $T\propto\sigma^{-\gamma}$ and the rule as $T\propto\sigma^{-1}$ for $p_c=1.88,1.882,1.884$ still can be found from the numerical result 2 and semi-classical result 2 for large perturbation larger than $1$ and thus there is another rule of time scale for the slow decay process beside main decay process.With $p_c$ increased further more from $p_c=1.886$ to $1.89$,the rule as $T\propto\sigma^{-2}$ tends to form gradually for the part of perturbation with clear decline observed in the graph $(s),(t),(u)$.It is worth noting that variations of $T$ for semi-classical result 1 are quite different from numerical result 1 in terms of $p_c=1.862,1.864,1.866,1.868,1.87,1.872$.}\label{whole_decay_time_edgeofchaos}
\end{figure}
\end{center}

\begin{center}
\begin{figure}
\includegraphics[width=10cm,height=8cm]{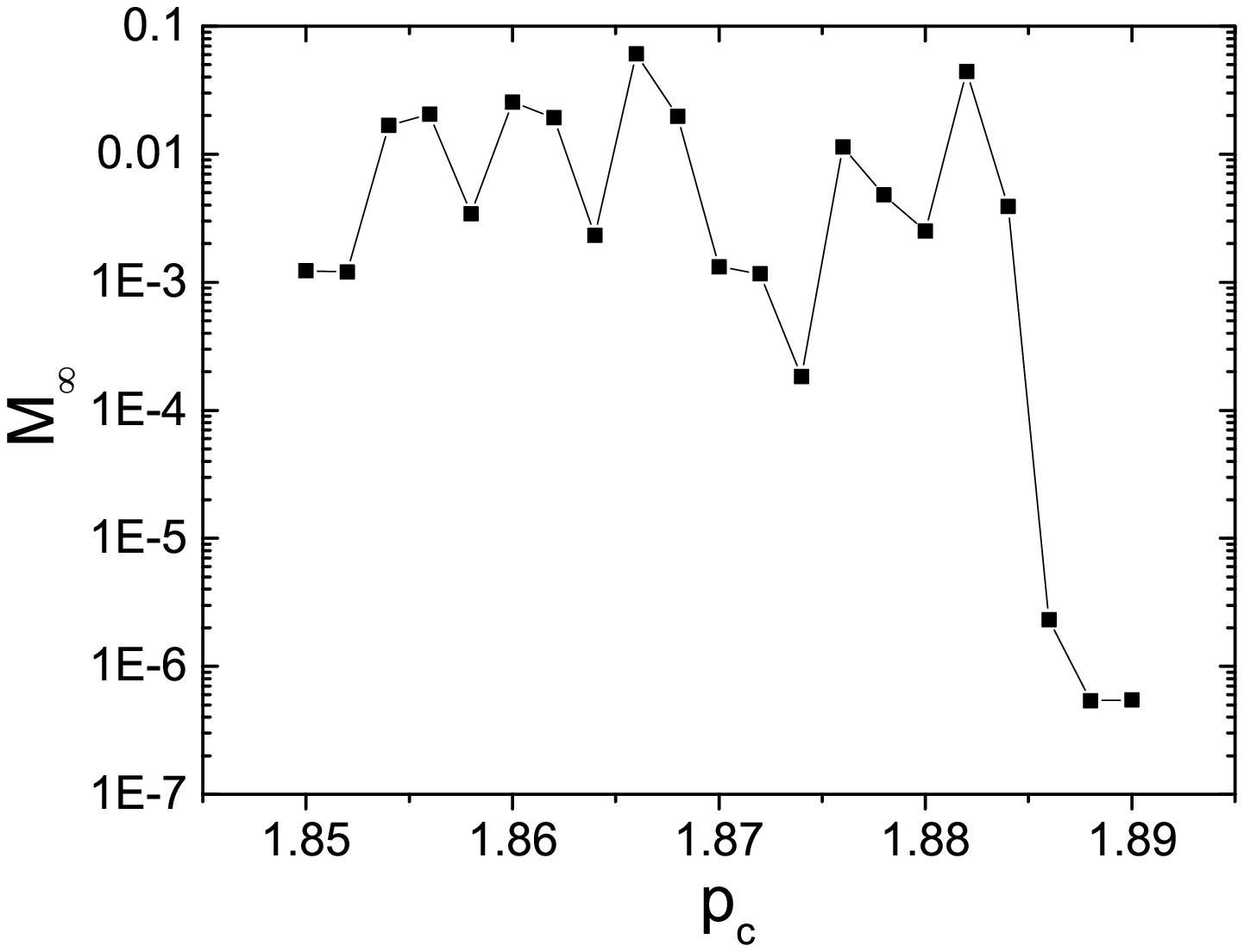}
\vspace{-0.5cm} \caption{Variation of $M_{\infty}$ with $p_c$}\label{edge_saturation}
\end{figure}
\end{center}

\section{4. Study of LE for chaotic sea in terms of mixed phase space}

Here the key point study is not to find the decay process in a very detailed way,but we want to find 
the general features in terms of decay process.Further more,as there are quite a few different positions for the studied quantum wave-packet,we also want to find a likely transition of decay laws as possible as we could.In one word,we do not use some way of a dairy of events to study the decay features for every single quantum wave-packet rather than to find some universal expression with the transition rule.Meanwhile we also consider the previous work related to the edge of chaos and the typical power decay law and exponential decay law are showed which are also the decay laws we want to investigate.Based on this concept,we use some typical perturbations to study decay features with different wave-packets and try to find the likely rules in terms of the assumed power decay and exponential decay respectively.We show our numerical observation in the figure 7.

\subsection{A.Semi-classical method}

If the width is sufficient small with the requirement of
$k=\hbar/{\xi^2}>>1$,$D$ can not be considered,but the focus we
study in this paper is $k=1$ with the same widths of wave-packet in
terms of position representation and momentum representation for the
best contrast to the classical point particle.The
paper\cite{WgWang_2} shows that the difference between the first
order and second order could just in the short time and have the
similar decay behaviors in terms of long time. As we basically have
not the sufficient theoretical information about $D$,so use the
testing perturbation $\sigma=0.1$ with some chosen parameters $K$ to
see the likely differences.

\begin{figure}
\includegraphics[width=16cm,height=16cm]{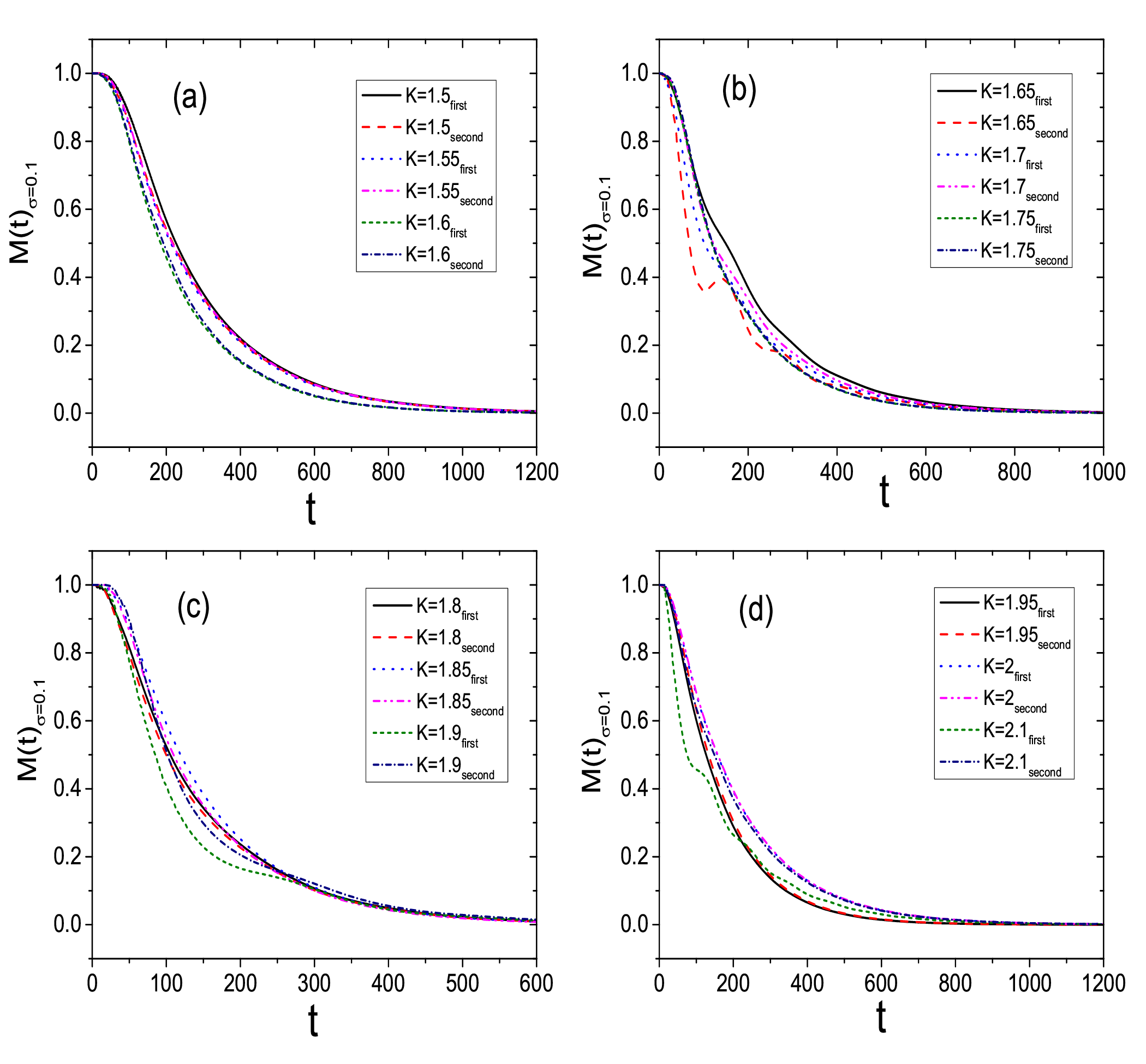}
\vspace{-0.5cm} \caption{ Quantum fidelity decay in terms of the
first order type $k=\hbar/{\xi^2}=50$ and second order type
$k=\hbar/{\xi^2}=1$ for chosen typical nonlinear parameters
$K=1.5,1.55,1.6,1.65,1.7,1.75,1.8,1.85,1.9,1.95,2,2.1$.The
perturbation $\sigma=0.1$ is used to see the difference
numerically.}\label{quantum_fir_sec1}
\end{figure}

\begin{figure}
\includegraphics[width=18cm,height=8cm]{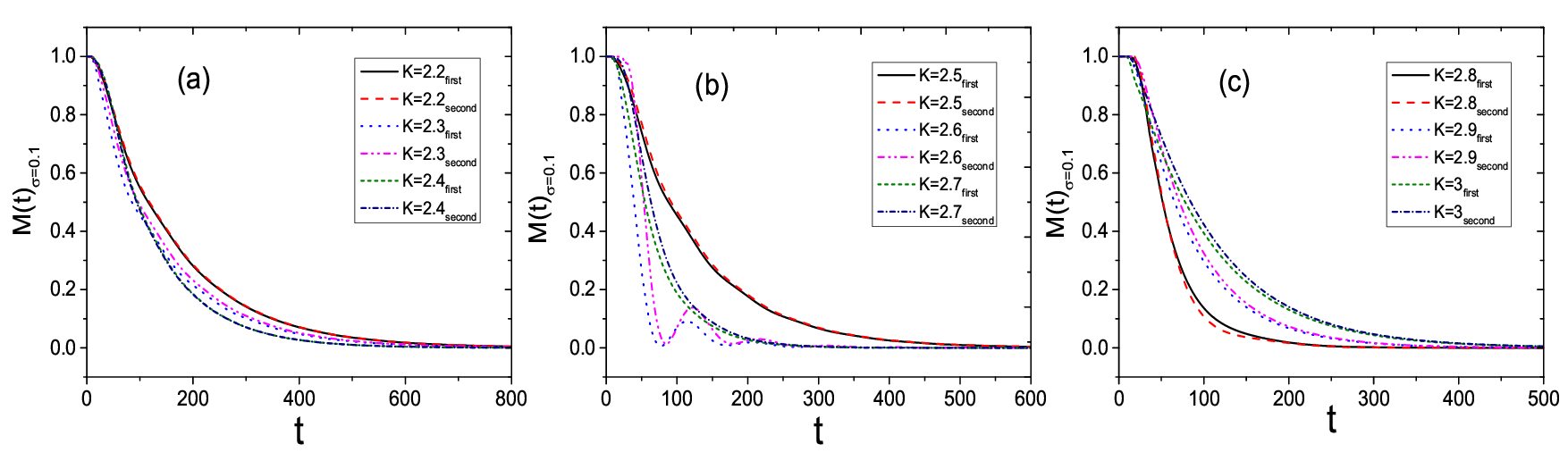}
\vspace{-0.5cm} \caption{Quantum fidelity decay in terms of the
first order type $k=\hbar/{\xi^2}=50$ and second order type
$k=\hbar/{\xi^2}=1$ for chosen typical nonlinear parameters
$K=2.2,2.3,2.4,2.5,2.6,2.7,2.8,2.9,3$. The perturbation $\sigma=0.1$
is used to see the difference numerically.}\label{quantum_fir_sec2}
\end{figure}

It can be seen in the figure 31 and 32 that the non-linear parameters
$K=1.65,1.7,1.85,1.9,2.1,2.3,2.6,2.7,2.8,2.9,3$ all have some
distinguished variations and in particular for the cases of
$K=1.65,1.9,2.1,2.6$ having obvious large variations. For other
parameters basically the fidelity decay of first version and second
version have same decay process. How to understand this numerical
result from our theory? The different effect should be related to
the function of $D$. When the value of $D$ change very little in
terms of different $p_0$ for some given time,the function of $D$
just change the integral window with Gaussian weight,collaborated
with the factor $1/D$ to give the basic same result for the
situation without $D$.But this condition can be changed and if $D$
is changed sharply in terms of the variation of $p_0$ for some given
time,we can not reasonably expect the same result.So this is our
explanation for the different expressions.

Based on the numerical observation above,we separate them as two
groups,the group having the basic same decay behaviors(called the
first group) and the other group lacking of this situation(called
the second group).Therefore,we could use the first-type uniform
semi-classical approach to treat the second type quantum fidelity
decay in terms of the first group,but should add the supplement of
first type quantum fidelity to see clearly what is accounted for the
difference between the first-type uniform semi-classical approach and
the second-type quantum fidelity decay.What we use to explain the
numerical result is not the direct uniform semi-classical approach
but revised version-we can call statistical semi-classical
method,which was introduced firstly in the paper\cite{WgWang_1},and
afterwards applied in the quantum map and cold
atoms\cite{WgWang_2,Zheng}.But there have some controversial idea
for the effectiveness about this method and the direct contrast
between the quantum decay process and corresponding semi-classical
one is deserved much attention.From our research in this paper,this
method is efficient and promising for explaining quantum results.
Now we want to explicitly clarify this theoretical method with some
likely misunderstanding points although the basic idea have existed
in the previous papers.

\subsection{B.The study of effect of semi-classical evaluation from the numerical realization}

The first work in terms of theoretical interpretation is to testify
this Levy distribution.We use the ensemble with the initial momentum
got in terms of Gaussian weight $ e^{{-({\bf p}_0-\tilde{\bf
p}_0)^2}/{(\hbar/\xi)^2}}$ just from the Eq.~(\ref{first_order}) and
choose a very narrow spanning field for the $r_0$,here we use the
uncertainty relation to give a Gaussian weight for the distribution
of $r_0$.Actually the exact corresponding semi-classical method we
should hire is the second-order semi-classical formula\cite{WgWang_2}
but with the argument for the issue of how to make a classical and
quantum contrast,we then begin to study the good extent for
describing the $P(s)$ with Levy distribution. For giving convincing
and extend evidence,we use the typical non-linear parameters with
some time discontinued,we can see directly the fitting
results.Meanwhile there also has a indirect evidence that can be
used from the formula Eq.~(\ref{Levy})and Eq.~(\ref{Flz}),if the
$P(s)$ can be taken as the Levy distribution quite closely,we can
make a Fourier transform and should find the near linear relation
for $ln(-ln|F(z)|)$ versus $lnz$ in terms of $z$ as the
frequency,for simplicity we can call this kind of relation as
frequency relation without the consideration of negligible
contribution from the frequencies with very small value of
$ln(-ln|F(z)|)$ that can be thought to be set in the cut-off region.
Therefore,we could use the direct contrast with probability density
distribution as well as the assumed linear relation in terms of
Fourier transform as a whole to extract the information that can
help to understand the approaching extent for the semi-classical
evaluation of quantum fidelity.For the considerations of brevity and
representativeness,from many numerical observation,we choose the
case of $K=1.5$ to give a typical and also convincing illustration
for these numerical evidences.The method we use in detail is to
extract the investigated statistical data for some evolution time we
choose run in the computer for one time,that is a dynamic
evolution,and then we can make a contrast study in terms of a given
specific time and observe very clearly about the tendency of
variation for different statistical density distributions for $P(s)$
which is a key point in our study for semi-classical interpretation.

Before we make a study for the comparison between the numerical
distribution of $P(s)$ and assumed Levy distribution,with the
knowledge of binary random variable in terms of initial Gaussian
ensemble we take,we can obtain an analytical formula for the initial
probability density distribution $P(s)$ as:\be P(s)=\frac {1}{(\pi
\xi^2)^{1/2} (1-s^2)^{1/2}}{\rm exp}\left[-\frac{(s-s_0)^2}{\xi^2
(1-s_0^2)}\right].\label{check_initial}\ee Here,$s_0$ is related to
the original center of position $r$ that can be written as $r_0$,for
the particular case of our study for standard map we have the simple
relation $s_0=\rm cos(r_0)$.From the analytical formula we obtain,we
can predict this distribution as Gaussian distribution with high
expectation as the original wave-packet for the initial ensemble is
so small and thus the variation of the term $(1-s^2)^{1/2}$ can be
approximately negligible compared to the latter term ${\rm
exp}\left[-\frac{(s-s_0)^2}{\xi^2 (1-s_0^2)}\right]$.To check our
theoretical estimation,we compare the numerical result and
theoretical curve for the distribution in the figure33 and also do a
Fourier transform to get the slope with the linear part,the fitting
slope is 2.01083 with Origin software that is close to 2 equivalent
to Gaussian distribution.Therefore,we can be loosely to say that the
initial distribution of $P(s)$ can be seen as Gaussian distribution.

\begin{figure}
\includegraphics[width=8cm,height=8cm]{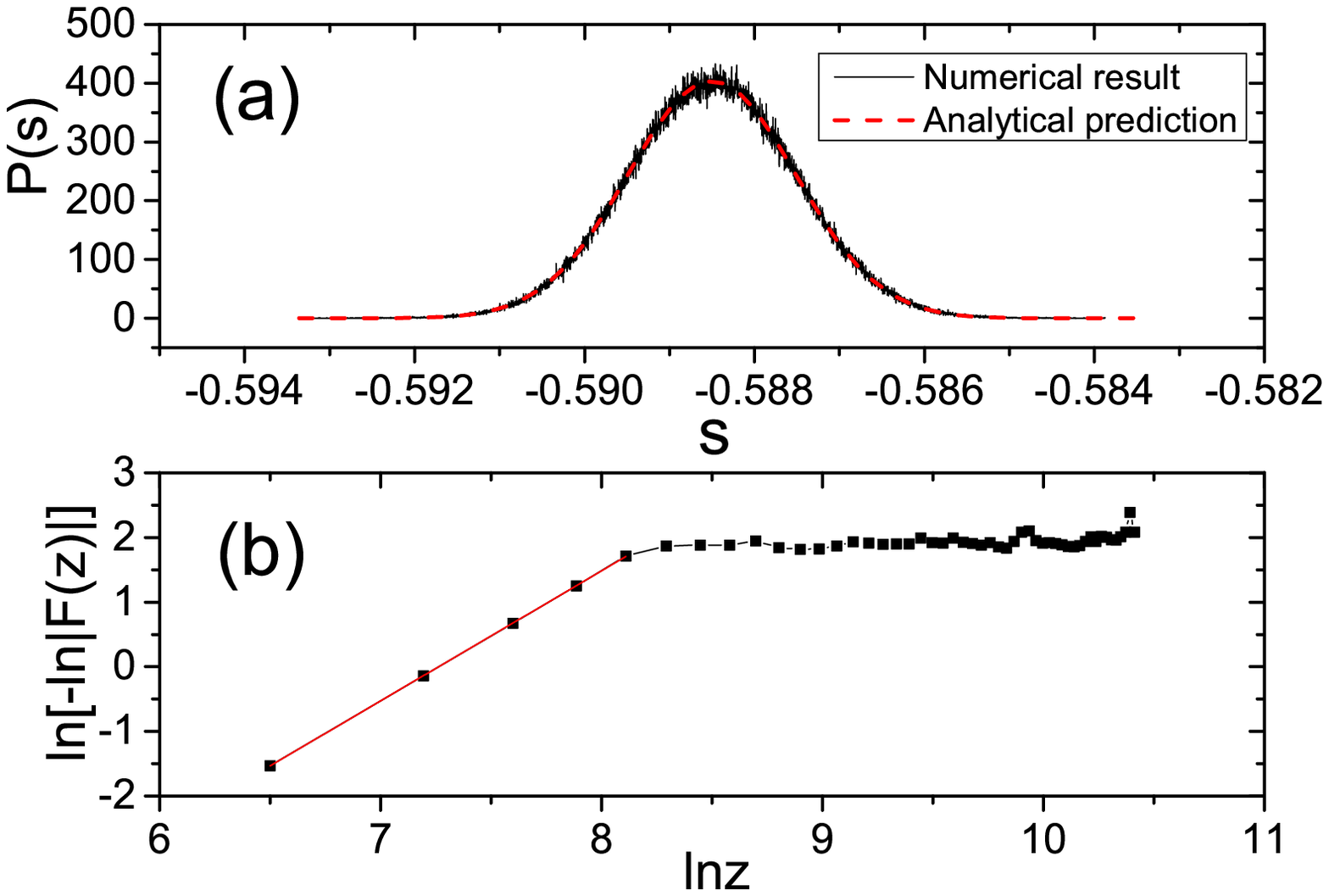}
\vspace{-0.5cm}\caption{The Panel(a)shows the comparison between the
initial distribution of $P(s)$ with direct numerical computation and
the theoretical curve,Panel(b)shows the frequency relation and the
frequencies with efficient contribution form the linear relation
connected using red line and fitting slope is around 2.01 close to 2
in terms of Gaussian distribution mutually confirmed with the
panel(a).}\label{check_initial_distribution}
\end{figure}

Now we want to study the variations of the distribution of $P(s)$
which is the central variable in our research.Obviously,the
distribution will deviate the Gaussian distribution with the
dynamical evolution for the initial ensemble and what we care about
is the tendency for the variations of distribution.By now,we still
can not find a systematically analytical method to treat the
problem,so we should appeal to the numerical method.We want to find
the general connection with the distribution and corresponding
frequency relation,Levy distribution is expected with the
corresponding linear relation formed.We have done a lot of numerical
observations and our expectations indeed exist for some evolutive
time with a given system parameter $K$.Obviously it is reasonable to
choose some represented case showing our observed results
typically,thus we choose the case of $K=1.5$ which corresponds a
highly mixed-type phase structure,and just need to study the revised
version $P(s-\la s \ra)$ without the variable $g$ presented in the
frequency formula Eq.~(\ref{Flz}).For the special case of $K=1.5$,we
find a interesting and common variation that the initial regular
distribution without evolution can be changed into a very wild
distribution having many peaks and then the peaks can be smoothed
out during the evolutive time passing and meanwhile the frequency
relation also have the variations reflected in the degree of linear
relation,and gradually the frequency relation tends to have the
shape of linearity formed with the frequencies giving significant
contribution in terms of $|F(z)|$ corresponding approximated Levy
distribution expected.Hence we can expect an accurate evaluation in
terms of using the formula Eq.~(\ref{semiclass})for a good linear
extent for frequency relation.We show this typical process in the
figure34 with some selected time using the comparison between the
distribution numerically computed and the theoretical curve came
from the formula Eq.~(\ref{levycompute1})and
Eq.~(\ref{levycompute2})although with the numerically fitting
variables $D_L$ and $\eta$.For the purpose of comparing the
approximating degree using the assumed Levy distribution,we also
depicted the frequency relation as a whole to comprehand.For the
case of $K=1.5$,it can be seen from the figure34 to show basically
monotonous variation for quite a long time.But we also find some
non-monotonous variations in some specific cases.The case of $K=1.5$
is just a typical one,thus we need to have a whole comprehension
about the transitions in terms of different parameters $K$,based on
the investigation we have before,it is necessary further to
investigate the tendency of the variations of frequency relations
with a broad parameter field.

\begin{figure}
\includegraphics[width=16cm,height=16cm]{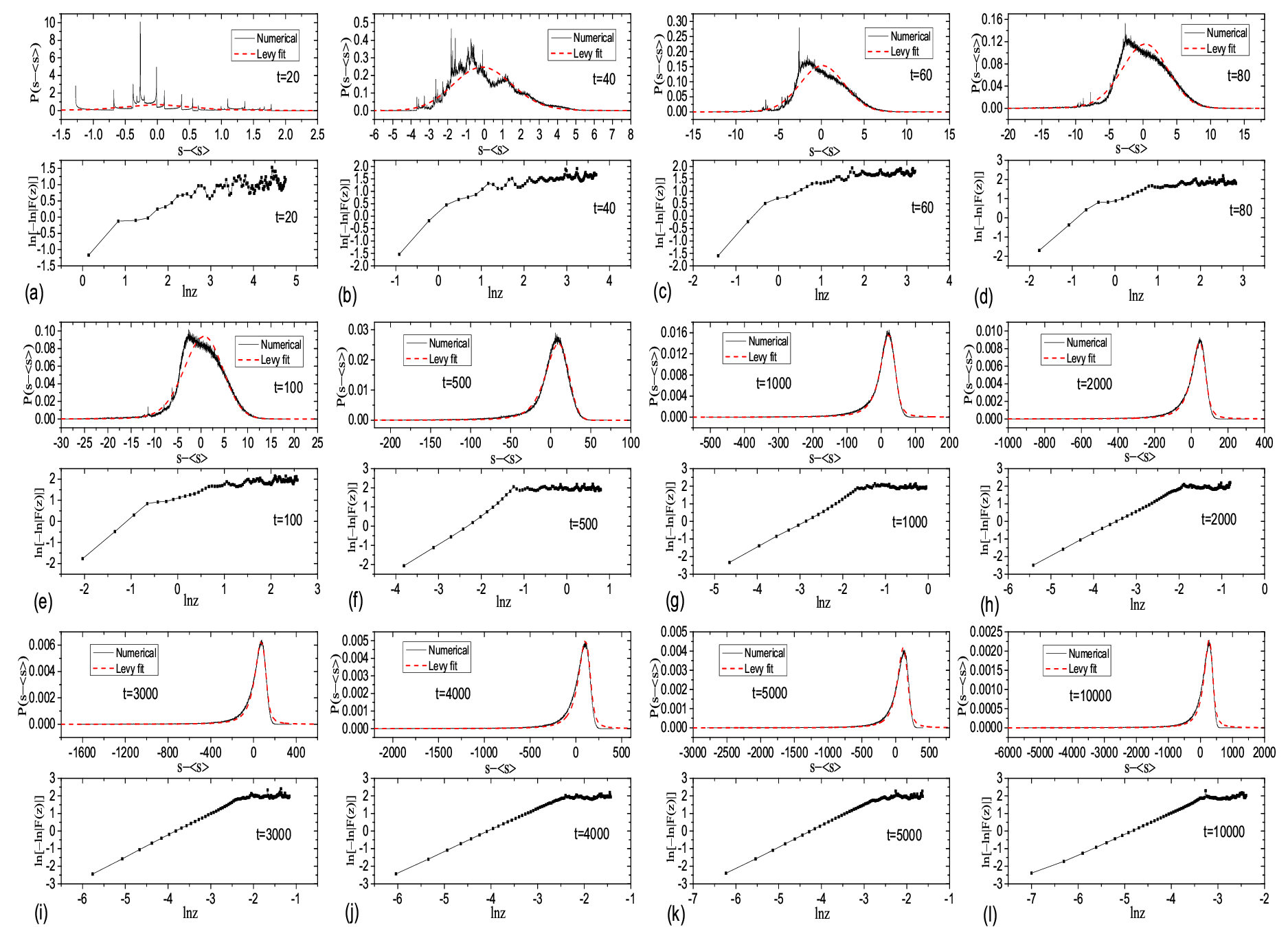}
\vspace{-0.1cm} \caption{Use the system $K=1.5$ to show the typical variations of the distribution $P(s-\la s \ra)$ vs $s-\la s
\ra$ for different selective time and corresponding the
relationship $ln(-ln|F(z)|)$ versus $lnz$ came from the Fourier
transform of $P(s-\la s \ra)$,$s$ comes from $\Delta S/\epsilon$.It
can be shown that Levy distribution can be approximately used for
the distribution $P(s-\la s \ra)$ after some evolutive time and
related the relationship $ln(-ln|F(z)|)$ versus $lnz$ can be turned
into linear in particular for the initial low frequency although
that can not be maintained for a some quite long time seen the
figure(l)in terms of evolutive time $t=10000$.Levy fitting technique
used in detail is to find the frequencies fitted to give the good
fitting curve for the numerical distribution joined with the most
near semiclassical value to the quantum fidelity in terms of formula
deduced as $M_{sc}(t)=\rm exp(-2(\epsilon/\hbar)^\eta D_L)$.The
frequencies we use counted without the zero frequency in terms of
the ascending order for the time we select with the same ascending
order are 2,2,2,3,4,13,4,2,3,4,6,17.}\label{levy_shape}
\end{figure}

\begin{figure}
\includegraphics[width=16cm,height=22cm]{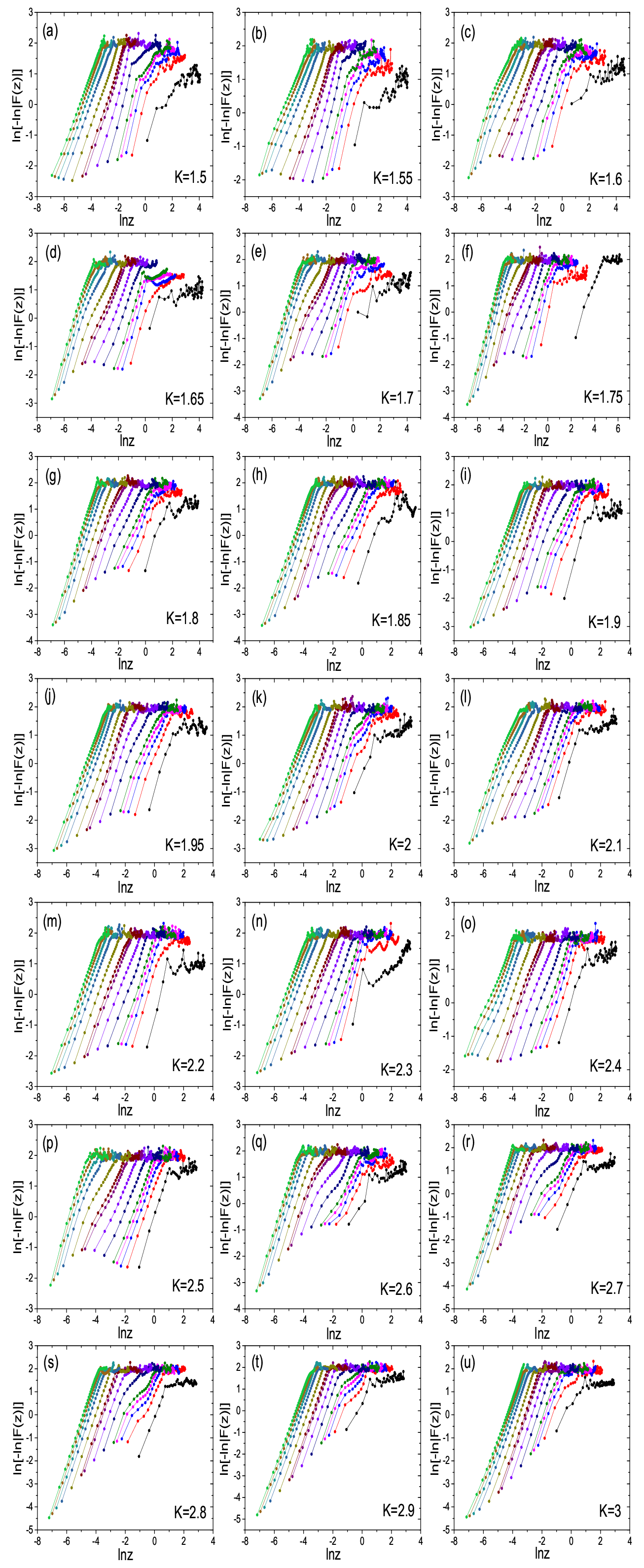}
\vspace{-0.1cm} \caption{Frequency relation as $ln(-ln|F(z)|)$
versus $lnz$ for selected nonlinear parameters $K$ from 1.5 to 3 in
terms of the typical evolutive time
20,40,60,80,100,200,400,800,1000,2000,4000,6000,8000,10000 we
choose.As the frequencies considered are inversely proportional to
the value range for the distribution of $P(s)$ and the value range
will be increased with the diffusion process and hence the
frequencies will be decreased for increasing evolutive time.As a
result,every single frequency relation numerically showed in the
figure is shifted towards the Horizontal direction to the left with
the given evolutive time enlarged.The whole frequency relation in
some time can be seen as linear relation corresponding to Levy
distribution we expect,and meanwhile the global pattern for the
variations of frequency relation can be distinguished from its
asymptotic expression belonging to linear relation class or not
depicted in each sub-figure.}\label{fft_all}
\end{figure}

To show the transitive pattern for the frequency relations,we
express them with different represented time put into together
depicted in every sub-panel of figure35 corresponding to a specific
system parameter $K$ and one can find the transitive pattern is
quite individually but typically have the tendency to have a linear
relation,and we can find the frequency relation can be changed into
the shape of concave or convex besides the expected good linearity
in terms of the consideration of the long-time expression.It is
obvious that the Hamiltonian dynamics determine a specific diffusion
for a initial ensemble,in other words,a stochastic process
corresponds to a complex Hamiltonian evolution for an ensemble,and
we have found some distributions of $P(s)$ are more complicated than
Levy distribution,hence the deeper mathematical understanding of
this kind of stochastic process is a high challenging question,in
particular for what kind of exact condition to form a specific
distribution of $P(s)$.Based on our investigation by now,the
behavior of long jump for an evolution of $s(t)$ can be explained
for the Levy distribution but how to extract useful and concise
mathematical formula is a open mathematical problem beyond our
research in this paper.

For giving the evaluation of fidelity decay in terms of
semi-classical formula,a certain number of selected frequencies
should be used,and what we care about is to consider the errors with
different frequencies selected as there have some arbitrariness that
can not be disregarded in our study.The basic consideration is to
check the idea that the positive correlation for errors variations
in terms of the extent of linearity of frequency relation.The first
one coming in our mind is to consider this problem technically in
detail which means the investigation for every single group of
chosen frequencies with ascending order added.Along this thought,we
still need to define a variable for describing the extent of
linearity of corresponding chosen frequency relation in terms of our
chosen frequencies.We can borrow the idea of variance in probability
theory to characterize the degree of linearity,the idea in detail is
just to consider the fitting line firstly as the reference line and
aggregate the deviation for every single frequency's difference for
$ln(-ln|F(z)|)$ to the corresponding value on the reference
line,with the consideration of different interval between adjacent
frequencies and the whole number for the chosen frequencies as a
single group,the averaged and re-scaled handling should be used.
Therefore,we get a variable with the name that can be called
$L_{v}$.Mathematically describe it as:
\be\label{LIN}L_{v}=\frac{(\sum_{i=1}^{n} {[k
lnz(i)+b-ln(-ln|F(z(i))|)]}^2)/n}{(\sum_{i=1}^{n-1}d_{i,i+1})/{n-1}}\ee
,where $n$ is the number of frequencies chosen,$z(i)$ express the
frequency value in terms of the order $i$,$d_{i,i+1}$ shows the
distance of two adjacent points in terms of phase plane
$ln(-ln|F(z)|)$ versus $lnz$ and $k$ and $b$ are the corresponding
parameters in terms of fitting linear equation $ln(-ln|F(z)|)=k
lnz+b$.

For a given evolutive time,there have a statistical distribution
$P(s)$ and corresponding frequency expression.But the variable we
define have a feature that it tends to increase with the fitting
frequencies increased with the new frequency added one by one,and
obviously the corresponding error can not increase monotonously as
we expect.Actually the variations are complicated based on every
individual application of semi-classical formula in detail,explain
explicitly more,for every single fitting result two semi-classical
elements $D_{L}$ and $\eta$ can be obtained to get the semi-classical
result $M_{sc}(t)=\rm exp(-2(\epsilon/\hbar)^\eta D_L)$,thus we can
get to know the error in terms of the comparison with the direct
quantum fidelity.Meanwhile our attention here is also not to find
the fitting method to get the nearest value to the quantum
fidelity,such as set an artificial regulation for the choosing
procedure within a fitting changeable slope field through decerning
the difference of slope as the new frequencies increased for
fitting,that is to say,to find the good linear part to fit,but we
still find this method can not give the nearest value and it can not
reflect the internal structure of frequency expression.Therefore we
notice the global structure of the frequency expression with time
variation is the key point to find the positive correlation we
expect.For simplicity we can call the error as the mathematical
symbol $\delta$.

As the amplitude contribution is different from different frequency,
and we can find many frequencies just give very small contribution
to the related amplitude,thus here we just consider the frequencies
corresponding to the relative large amplitude with the standard
$ln(-ln|F(z)|)<1$ as the remain frequencies give small contribution
for the amplitude and then we can think they are not important to
give the basic shape of statistical distribution of $P(s)$ if having
a inverse Fourier transform.Thus we should check our idea here in
terms of the time evolution from the averaged treatment.We show the
results we get and basically conform to our expectation,for
simplicity we call the averaged $L_{v}$ as the symbol $\la L_{v}\ra$
and the averaged $\delta$ as the symbol $\la\delta\ra$. when $\la
L_{v}\ra$ is changed to be higher and we could reasonably expect a
higher value for $\la\delta\ra$.The variations of different system
parameter $K$ in terms of time sequence we selected are showed in
the figure average relation and the basic expectation hold.

\begin{figure}
\includegraphics[width=16cm,height=24cm]{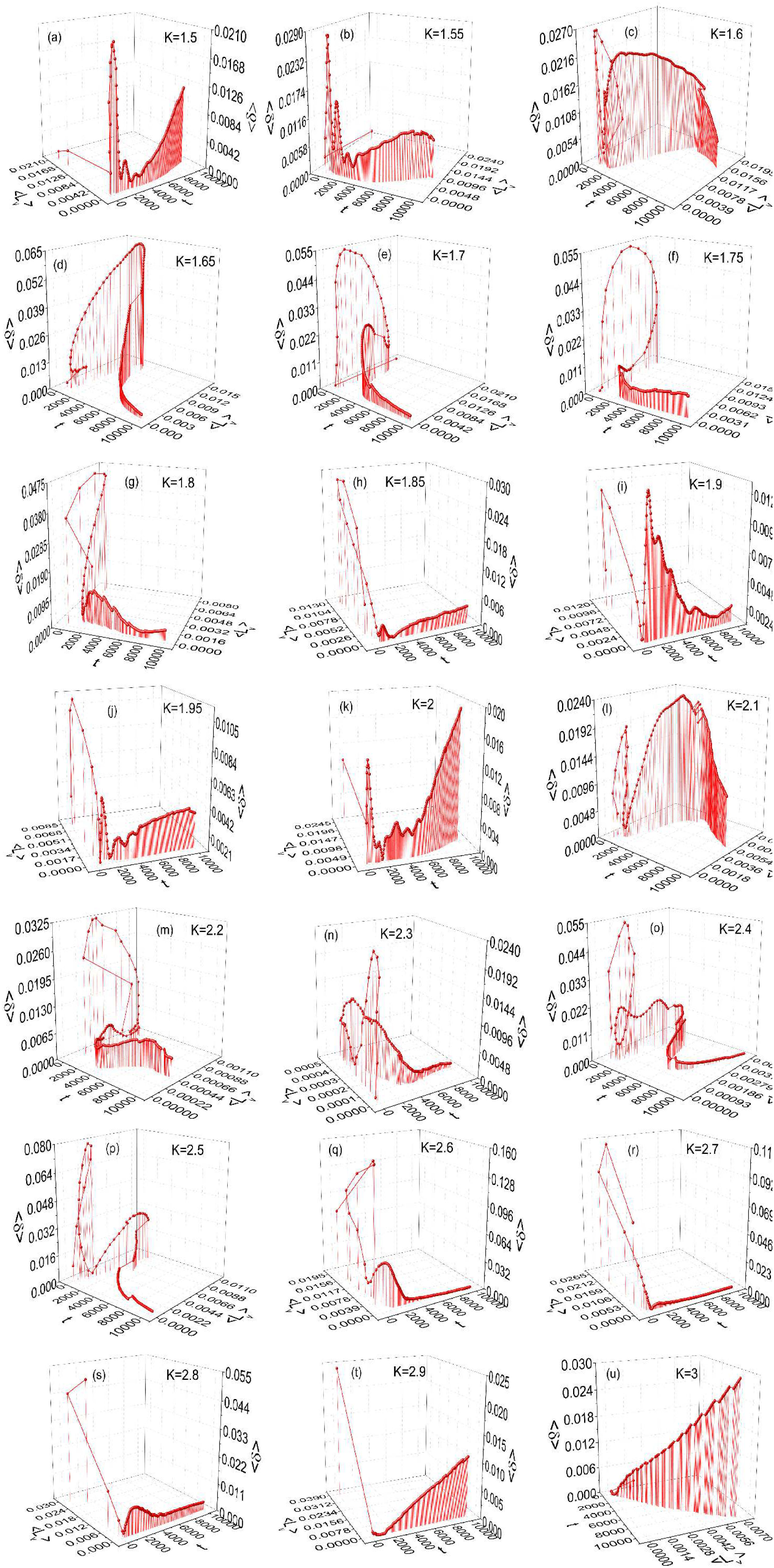}
\vspace{-0.5cm} \caption{The variations of averaged $\delta$ ($\la
\delta\ra$)to the averaged $L_{v}$ ($\la L_{v}\ra$)in terms of the
time interval 50 for different system parameters $K$.Averaging
procedure is performed for the frequencies with the contribution of
$ln(-ln|F(z)|)<1$ gradually added.The basic expectation of positive
relation for them is hold in terms of the tendency of variation
globally although we can find this relation can not be rigorously
correct in particular for the case of $K=1.6$ and $K=2.2$ as well as
some initial time with the order of magnitude
$10^2$.}\label{firgue_levy_time_LLE}
\end{figure}

Here we should give some notes for the result we get in terms of
figure average relation.Firstly we find the expression is highly
individually,there have not the common pattern for every
$K$.Secondly,in terms of the global tendency with the variation of
$\la L_{v}\ra$ via $\la\delta\ra$,the positive correlation actually
exist but not very rigorously in particular for the numerical
observations of the cases $K=1.6$ and $K=2.2$ as well as for some
initial time within the order of magnitude $10^2$.The likely
explanation is that positive correlation for the degree of linearity
to the error can not be hold monotonously and thus this kind of
variation can undergo some large deviation of expected
relation.Further more,the variable $L_{v}$ we define may be not a
very good indicator that have the coarse-grain property which means
to have the different configuration of frequency expression but with
the same value corresponding to the different errors.Thus we just
can have some reasonable expectation of the positive correlation for
typical situations,but can not guarantee exactly for the relation
happening sheer universally.

Surely,we also need to have the basic understanding of variations of
errors with different fitting in terms of different group of chosen
frequencies increased gradually.The most direct observation is to
see the comparison between fitted semi-classical result and related
quantum fidelity,then we can understand the errors in terms of
different fitting procedure.The extensively numerical results show
that it is not always true for the semi-classical value with initial
fitting frequencies more close to the corresponding quantum
value.This kind of situation just can be ascertained in the very
good linearity,thus we need to understand the variation based on the formula $M_{sc}(t)={\rm exp}(-2(\epsilon/\hbar)^\eta D_L)$ in
detail.

For simplicity,we can call the fitting number of corresponding
frequencies $N_{f}$.We can consider $D_L$ as the function of $\eta$
because every fitting procedure we can get a pair value of $\eta$
and $D_L$ with the knowledge of slope and intercept of fitting
line,it inspires us there have the interdependence with $\eta$ and
$D_L$ which means we can see $D_L$ as the function of $\eta$
mathematically.Then we can take the $\eta$ as the function of
$N_{f}$.Inserting into the semi-classical formula and subtracted by
the quantum fidelity,we can get to know the relation about how the
error is changed with increasing $N_{f}$.Further more,we make a lot
of numerical study about this question and find $D_{L}$ for a given
typical time can be seen approximately as the exponential function
of $\eta$ with the mathematical form:$D_{L}\approx c+a
e^{-b\eta}$,$a,b,c$ are the fitting parameters.Meanwhile we also can
predict the likely linear dependent relation for $D_{L}$ via $\eta$
as $D_L\approx a\eta+b$,$a,b$ are the fitting parameters,and get
numerical verification.(can be continued!)

From the variation of $\eta$ with $N_{f}$,we can understand the
variation of $M_{sc}$ with $N_{f}$ as well as with error $\delta$
further more.Thus we can see there have some competitions in the
expressions of $M_{sc}$ given in the appendix A in terms of specific
conditions that cause non-monotonous variation for the $\eta$ via
$N_{f}$ monotonously.We show the variations for different conditions
in the table below clearly,and then we also give the mathematical
analysis in the Appendix A.Therefore although it is a highly
individual manifestation for every single system parameter $K$ in
terms of a specific time,we can show there only have some basic
patterns for the likely changeable variations in the Table 1 and Table 2.

\begin{center}
\begin{table}[b]
{\footnotesize{\bf Table 1.}The monotonicity of $M_{sc}$ via $\eta$ in terms of the exponential relation for $D_{L}$ via $\eta$.\\
\vspace{2mm}
\begin{tabular}{@{}ccccccc}
\hline a&b&c&$\sigma$&Added condition&
 $\eta^{\ast}$&$M_{sc}$\\\hline
positive&negative&positive&$(0,1)$&Inexisence&Inexistence&Increasing function\\\hline
\multirow{3}*{Negative}&\multirow{3}*{Negative}&\multirow{3}*{Positive}&\multirow{3}*{$(0,1)$}&$\frac{c}{a}ln\sigma>|b+ln\sigma|$&Inexistence&Increasing function\\\cline{5-7}
&&&&\multirow{2}*{$\frac{c}{a}ln\sigma<|b+ln\sigma|$}&\multirow{2}*{Existence}&Decreasing function for $(0,\eta^{\ast})$\\&&&&&&Increasing function for $(\eta^{\ast},\infty)$\\\hline
\multirow{3}*{Positive}&\multirow{3}*{Positive}&\multirow{3}*{Positive}&\multirow{3}*{$(0,1)$}&$b+ln{\sigma}<0$&Inexistence&Increasing function\\\cline{5-7}
&&&&\multirow{2}*{$b+ln{\sigma}>0$,$|\frac{c}{a}ln\sigma|>b+ln\sigma$}&\multirow{2}*{Existence}&Increasing function for $(0,\eta^{\ast})$\\
&&&&&&Decreasing function for $(\eta^{\ast},\infty)$\\\hline
\multirow{3}*{Positive}&\multirow{3}*{Positive}&\multirow{3}*{Negative}&\multirow{3}*{$(0,1)$}&$b+ln{\sigma}>0$&Inexistence&Decreasing function\\\cline{5-7}
&&&&\multirow{2}*{$b+ln{\sigma}<0$,$\frac{c}{a}ln\sigma>|b+ln\sigma|$}&\multirow{2}*{Existence}&Decreasing function for $(0,\eta^{\ast})$\\&&&&&&Increasing function for $(\eta^{\ast},\infty)$\\\hline

Positive&Positive&Positive&$(1,\infty)$&Inexistence&Inexistence&Decreasing function\\
\hline

\multirow{4}*{Positive}&\multirow{4}*{Negative}&\multirow{4}*{Positive}&\multirow{4}*{$(1,\infty)$}&$b+ln{\sigma}>0$&Inexistence&Decreasing function\\\cline{5-7}
&&&&$b+ln{\sigma}<0$,$\frac{c}{a}ln{\sigma}>|b+ln{\sigma}|$&Inexistence&Decreasing function\\\cline{5-7}
&&&&\multirow{2}*{$b+ln{\sigma}<0$,$\frac{c}{a}ln{\sigma}<|b+ln{\sigma}|$}&\multirow{2}*{Existence}&Increasing function for $(0,\eta^{\ast})$\\
&&&&&&Decreasing function for $(\eta^{\ast},\infty)$\\\hline

\multirow{4}*{Negative}&\multirow{4}*{Negative}&\multirow{4}*{Positive}&\multirow{4}*{$(1,\infty)$}&$b+ln{\sigma}<0$&Inexistence&Decreasing function\\\cline{5-7}
&&&&$b+ln{\sigma}>0$,$|\frac{c}{a}ln\sigma|>b+ln\sigma$&Inexistence&Decreasing function\\\cline{5-7}
&&&&\multirow{2}*{$b+ln{\sigma}>0$,$|\frac{c}{a}ln\sigma|<b+ln\sigma$}&\multirow{2}*{Existence}&Increasing function for $(0,\eta^{\ast})$\\&&&&&&Decreasing function for $(\eta^{\ast},\infty)$\\\hline
\multirow{3}*{Positive}&\multirow{3}*{Positive}&\multirow{3}*{Negative}&\multirow{3}*{$(1,\infty)$}&$|\frac{c}{a}ln\sigma|<b+ln\sigma$&Inexistence&Decreasing function\\\cline{5-7}
&&&&\multirow{2}*{$|\frac{c}{a}ln\sigma|>b+ln\sigma$}&\multirow{2}*{Existence}&Increasing function for $(0,\eta^{\ast})$\\&&&&&&Decreasing function for $(\eta^{\ast},\infty)$\\\hline
\end{tabular}}
\end{table}
\end{center}

\begin{center}
\begin{table}[b]
{\footnotesize{\bf Table 2.}The monotonicity of $M_{sc}$ via $\eta$ in terms of the linear relation for $D_{L}$ via $\eta$.\\
\vspace{2mm}
\begin{tabular}{@{}cccccc}
\hline  a&b&$\sigma$&Added condition&
 $\eta^{\Delta}$&$M_{sc}$\\\hline
\multirow{2}*{Positive}&\multirow{2}*{Negative}&\multirow{2}*{$(0,1)$}&\multirow{2}*{Inexistence}&\multirow{2}*{Existence}&Decreasing function for $(0,\eta^{\Delta})$\\
&&&&&Increasing for $(\eta^{\Delta},\infty)$\\\hline
\multirow{2}*{Negative}&\multirow{2}*{Positive}&\multirow{2}*{$(0,1)$}&\multirow{2}*{Inexistence}&\multirow{2}*{Existence}&Increasing function for $(0,\eta^{\Delta})$\\
&&&&&Decreasing for $(\eta^{\Delta},\infty)$\\\hline
\multirow{3}*{Positive}&\multirow{3}*{Negative}&\multirow{3}*{$(1,\infty)$}&$\frac{a}{b}+ln{\sigma}<0$&Existence&Decreasing function\\\cline{4-6}
&&&\multirow{2}*{$\frac{a}{b}+ln{\sigma}>0$}&\multirow{2}*{Existence}&Increasing for $(0,\eta^{\Delta})$\\&&&&&Decreasing function for $(\eta^{\Delta},\infty)$\\\hline
\multirow{3}*{Negative}&\multirow{3}*{Positive}&\multirow{3}*{$(1,\infty)$}&$\frac{a}{b}+ln{\sigma}<0$&Inexistence&Increasing function\\\cline{4-6}
&&&\multirow{2}*{$\frac{a}{b}+ln{\sigma}>0$}&\multirow{2}*{Existence}&Decreasing function for $(0,\eta^{\Delta})$\\&&&&&Increasing function for $(\eta^{\Delta},\infty)$\\\hline
\end{tabular}}
\end{table}
\end{center}

For every single parameter $K$,we can find the common tendency for the variation of $D_{L}$ via $\eta$ with time increased,that is for the transitive pattern from decreasing function type rapidly to increasing function type.Now we need to show how to apply our analytical method to give the evaluation compatible with numerical results for the variation of $M_{sc}$ with fitting frequencies added,as their expressions depend heavily on every single situation with fitting parameters corresponding to specific system parameter $K$ as well as a given time,simply said just for a given specific distribution of $P(s)$,so our attention here is related to embody the analytical method itself with illustrative examples in particular for showing the likely existed critical value for $\eta^{\ast}$ and $\eta^{\Delta}$ in the table 1 and 2.Further more,the expression about $D_{L}$ via $\eta$ for some different $K$ in terms of a given specific time could be set in a very approximating curve in common although there generally have some deviated position for the same fitting order corresponding to increase fitting frequencies.This situation exists for the close expression for frequency relation that have been checked numerically and extensively which can be easily understood for the roughly similar distribution of $P(s)$ likely for some given time as the classical ensemble evolution from the same initial distribution although for different Hamiltonian in terms of different $K$.Therefore we want to have a demonstrative explanation about typical non-monotonous variation for $M_{sc}$ via $N_{f}$ from the knowledge of the tendency of variation about $D_{L}$ via $\eta$,the figure Mnf show our method using $K=1.9$ with the time for 4000.

\begin{center}
\begin{figure}
\includegraphics[width=14cm,height=18cm]{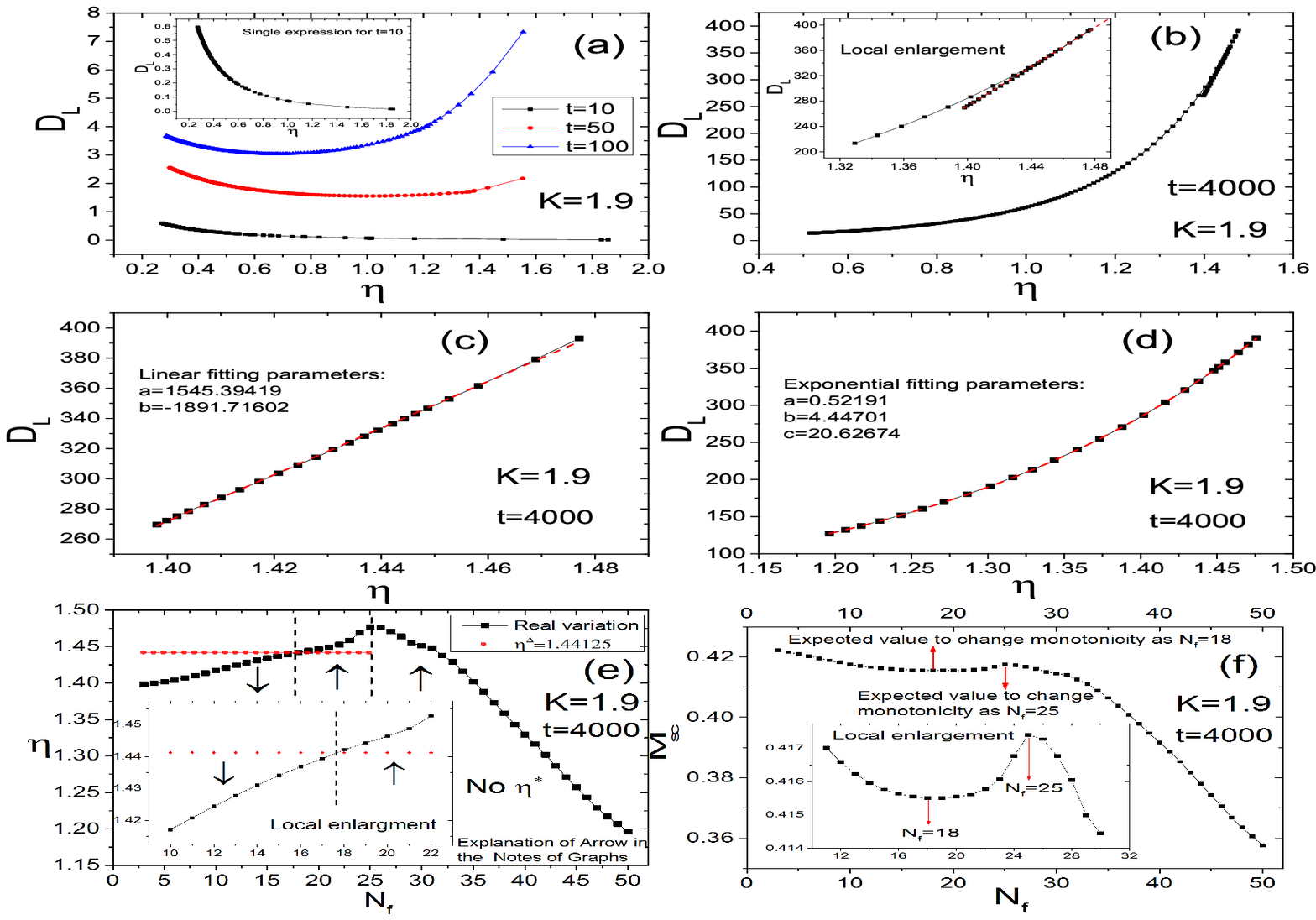}
\vspace{-0.5cm} \caption{A illustrative example to show how to use critical value of $\eta$ to understand the monotonicity of the variation of corresponding semiclassical fidelity value $M_{sc}$.We choose the system parameter as $K=1.9$ and the evolutive time for 4000.Based on the standard procedure to do a transformation of the distribution of $P(s)$ to get the module of spectrum $|F(z)|$ with $z$ as the frequency.Then we fit the initial three frequencies (without the zero frequency) to get the related $\eta$ and $D_{L}$,and then continuesly get the new $\eta$ and $D_{L}$ with a new nearest higher frequency added to form a new group for fitting,here $N_{f}$ the fitting number initially from three.Figure(a)shows the tendency of variation of $D_{L}$ via $\eta$ typically with time increased and firgure(b)shows the variation for time as 4000 of which firgure(c)and(d)are two subsections corresponding to  typical exponential and linear approximation.The arrows in the figure(e)show the property of monotonicity with $\uparrow$ for increasing function and $\downarrow$ for decreasing function in terms of the relationship between $M_{sc}$ and $\eta$,and $\eta^{\Delta}=1.44125$ is the critical value of $\eta$ caculated by the linear approximation from the variation of $D_{L}$ via $\eta$ in terms of a choosen perturbation $\sigma=0.01$.Correspondingly,through recursive relation,we show the property of monotonicity for $M_{sc}$ via $N_{f}$ through the relation of $\eta$ via $N_{f}$ in the Figure(f) and its expression for the monotonicity agrees with our prediction from the relation of  $\eta$ via $N_{f}$,in particular with attention for the change in the near field of critical value $\eta^{\Delta}$,but if we make a very accurate observation,we find the value of $M_{sc}$ corresponding to $N_{f}$ for 18 and 19 are 0.415514762664042 and 0.415514490855192 with high precision that deviate the prediction for increasing as the fitting itself can not be good enough to give such a good prediction to grasp the tendency for a very small variation.Therefore,we give a illustrative example to show the efficiency of our method to understand the basic varaition for semiclassical value $M_{sc}$ via $N_{f}$ in terms of a given perturbation with likely its limitation.}\label{mnf}
\end{figure}
\end{center}

Now we want to calculate the critical values for typical strength of perturbation with commonly used Levenberg-Marquardt Algorithm,and actually the prediction of changing tendency for monotonicity should be independent for different approximations with different but enough good fitting models as it is a objective truth which means we can uniformly substitute the exponential approximation for linear approximation,in other words,the critical values calculated with different fitting models could be equivalent as the different fitting parameters corresponding to different fitting models can guarantee this conception.Meanwhile,we just need to investigate a limited number frequencies fitted to observe the changing tendency of semi-classical decay fidelity,only concentrating on the initial part of frequencies.For these considerations,we use the first fitting frequencies as the number three(without zero frequency) and then add up to the number twenty-three,and then fit these groups of different frequencies to get the pair of variables $D_{L}$ and $\eta$ further taken as the data used as to get the fitting parameters in terms of our assumed mathematical model and then further we can get the critical value $\eta^{\ast}$ although there have some existing conditions for different compositions of corresponding parameters.

First of all,for $\sigma<1$ we have three likely combinations of signs for the fitting parameters $a,b,c$ that give the possible $\eta^{\ast}$:$(+,+,+)$,$(+,+,-)$ and $(-,-,+)$ from the Table 1.The corresponding conditions of existence of the critical value $\eta^{\ast}$ are $(b+ln{\sigma}>0,|\frac{c}{a}ln{\sigma}|>b+ln{\sigma})$,$(b+ln{\sigma}<0,\frac{c}{a}ln{\sigma}>|b+ln{\sigma}|)$,and $(\frac{c}{a}ln{\sigma}<|b+ln{\sigma}|)$.It is very easy to combine the two inequalities to the one inequality in terms of the reasonable meaning,such as we can not consider $ln{\sigma}$ larger than something that is positive as the $ln{\sigma}$ itself just is negative,so that kind of situation should not be included and what we get for constructing the one equality with some combination of the fitting parameters should have considered this kind of problem and thus guarantee the validity of the one inequality.Therefore,taken the the situation of $(+,+,+)$ as a case to illustrate,the condition $b+ln{\sigma}>0$ leads to $\sigma>e^{-b}$,and the condition of $|\frac{c}{a}ln{\sigma}|>b+ln{\sigma}$ leads to $\sigma<e^{-\frac{ab}{a+c}}$,so we can get the one inequality as $e^{-b}<\sigma<e^{\frac{-ab}{a+c}}$.For the case of combination of $(+,+,-)$,the existing condition is $b+ln{\sigma}<0$ and $\frac{c}{a}ln{\sigma}>|b+ln{\sigma}|$,so $b+ln{\sigma}<0$ leads to $\sigma<e^{-b}$,$\frac{c}{a}ln{\sigma}>|b+ln{\sigma}|$ leads to $ln\sigma(1+\frac{c}{a})>-b$,according to the positive or negative of $1+\frac{c}{a}$,we can get the two different results $\sigma>e^{-\frac{ab}{a+c}}$ and $\sigma<e^{-\frac{ab}{a+c}}$,combined with the previous result $\sigma<e^{-b}$,we can get $e^{-\frac{ab}{a+c}}<\sigma<e^{-b}$ and $0<\sigma<e^{-b}$.Then we can also use the same method to the combination of $(-,-,+)$,the existing condition is $\frac{c}{a}ln\sigma<|b+ln\sigma|$,this condition can lead to $ln\sigma(1+\frac{c}{a})<-b$,according to the positive or negative of $1+\frac{c}{a}$,we can get the two different results $\sigma<e^{-\frac{ab}{a+c}}$ and $\sigma>e^{-\frac{ab}{a+c}}$,if we pay attention to the field of $\sigma$ as $(0,1)$,                                                                                                                                                                                       so we can accordingly get $0<\sigma<e^{-\frac{ab}{a+c}}$ and $e^{-\frac{ab}{a+c}}<\sigma<1$.

Using the very same method,we can also treat the situation for $\sigma>1$,and we can find there have three kinds of combinations of fitting parameters in our numerical study as $(+,+,-),(+,-,+)(-,-,+)$ and show the likely existence of critical value $\eta^{\ast}$.We also can give a similar analysis,the corresponding conditions for existence are $|\frac{c}{a}ln\sigma|>b+ln\sigma$,$(b+ln\sigma<0,\frac{c}{a}ln\sigma<|b+ln\sigma|)$,and $(b+ln\sigma>0,|\frac{c}{a}ln\sigma|<b+ln\sigma)$ also given in the Table 1.For the combination of $(+,+,-)$,the condition $|\frac{c}{a}ln\sigma|>b+ln\sigma$ leads to $ln\sigma(1+\frac{c}{a})<-b$,according to the positive or negative of $1+\frac{c}{a}$,we can get two different results $\sigma<e^{-\frac{ab}{a+c}}<1$ and $\sigma>e^{-\frac{ab}{a+c}}$ ,for the first one is actually impossible as it contradicts with the precondition $\sigma>1$ and should be deleted.Then the available field for $\sigma$ is $\sigma>e^{-\frac{ab}{a+c}}$.For the combination of $+,-,+$,the condition $b+ln\sigma<0$ leads to $\sigma<e^{-b}$ and the condition $\frac{c}{a}ln\sigma<|b+ln\sigma|$ leads to $\sigma<e^{-\frac{ab}{a+c}}$,we combine the two results above and finally get the available field is $1<\sigma<e^{-\frac{ab}{a+c}}$.For the combination of $(-,-,+)$,the condition $b+ln\sigma>0$ leads to $\sigma>e^{-b}$ and the condition $|\frac{c}{a}ln\sigma|<b+ln\sigma$ leads to $ln\sigma(1+\frac{c}{a})>-b$,according to the positive or negative of $1+\frac{c}{a}$,we can get two different results $\sigma>e^{-\frac{ab}{a+c}}$ and $\sigma<e^{-\frac{ab}{a+c}}$,combined with $\sigma>e^{-b}$,the available field is $\sigma>e^{-\frac{ab}{a+c}}$.Actually the situation of $-,-,+$ is very few,so we don't need to consider it as a main fitting pattern rather than a uncertainty for fitting.We compare two algorithms with the basic framework of L-M method,and the algorithm we use finally can not have the situation of $(-,-,+)$,so we will don't consider this kind of pattern for fitting parameters.

Now we choose $K=1.95$ to give a typical case showing the variation of fitted parameters and for strictness we depict two kinds of parameters using different algorithms although in the basic framework of Levenberg-Marquardt method.From the figure,we can find the differences existing in the some kind of bad situations for fitting in terms of the exponential model used,the so-called term bad we use can be characterized with running steps,and we use the maximal running steps as 2000,and for most cases,the actual steps are far below this setting value,but for the situations with the obvious differences of the parameters the time steps are 2000 which show the outcome can not obtain the precision we set in the computer program but have a basic good fitting effect with sufficient long steps,so the maximal running step 2000 can be seen reasonably.The main algorithm we use from author Timothy Sauer in terms of the popular book Numerical Analysis and the other version is from the original article\cite{Denmarklm} which are all called in the figure 38 with the simple name as algorithm 1 and algorithm 2.In this paper,we hire the version of Timothy Sauer in the book of Numerical Analysis which show a better fitting effect.

\begin{center}
\begin{figure}
\includegraphics[width=18cm,height=8cm]{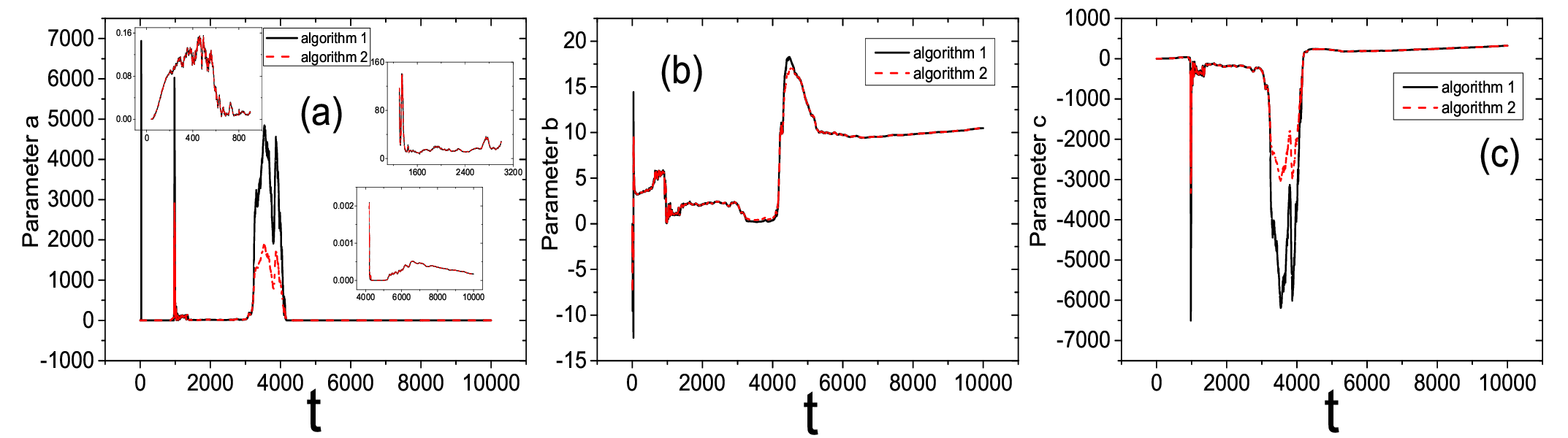}
\vspace{-0.5cm} \caption{The typical variations of fitted parameters  to the time for $K=1.95$,(a),(b)and (c)are for the three parameters $a,b,c$ fitted with the model $D_L=c+a{\rm exp}(b {\eta})$.We use the continual curve to express the variation rather than the points,and our emphasis is to show the basic tendency for the variations,the three sub-figures of (a)shows the variations in detail for the part which can not be seen in the origianl (a).The suddently rising or falling part have some obvious differences and the intermediate part of (a)and (c) also have very obvious differences which correspond to the expected precision can not be satisfied as the fitting data itself is not very good for expected exponential model.For very serious check,in terms of initial part,not just in this case of $K=1.95$,the fitting results of the algorithm 1 (The version of Timothy Sauer in the book of Numerical Analysis)have better fitting effect which maybe show a more flexible adaptability.}\label{parameters}
\end{figure}
\end{center}

The fitted parameters $a,b,c$ for the fitting model $D_L=c+a{\rm exp}(b {\eta})$ have basic features in our numerical computations,for $a$ is expected small but can suddenly rising for some time interval long or short and the variation of $b$ is some kind of complicated with the changing field among the order of $10$ and $c$ basically have the gradual tendency to increase to the maximal degree of the order of $10^{3}$ but often have some very large negative fluctuations to the negative value at large to the order of $-10^{4}$.Through our very rigorous check repeatedly,we have to claim that the parameters fitted can not be determined uniquely for all the time but we find our fitting parameters can have a good fitting effect which means the fitting points are very close to the original points in terms of $D_{L}$ versus $\eta$,and actually we use some alternative algorithm to fit to get the different parameters but with the same fitting effect which prove for some cases the fitting results are not unique but do not affect the basic variation of the parameters as a whole with our real interest here.

Then we also can make a simple mathematical analysis to predict the numerical results for the monotonicity of $\eta^{\ast}$ as $-\frac{1}{b}ln[-\frac{a}{c ln{\sigma}}(b+ln{\sigma})]$ assumed for existence.In terms of the available field of $\sigma$,we can make a differential of $-\frac{1}{b}ln[-\frac{a}{c ln{\sigma}}(b+ln{\sigma})]$,we can get the result as:
\be \frac{d\eta^{\ast}}{d\sigma}=\frac{1}{(b+ln{\sigma})(ln{\sigma})\sigma}.\label{dif_crieta}\ee So we can find that the parameters $a$ and $c$ are deleted and just the crucial part $b+ln{\sigma}$ determines the sign of the differential.For $\sigma<1$,the differential can be positive for the parameters patterns as $(+,+,-)$ and $(-,-,+)$,negative for the parameter pattern as $(+,+,+)$.For $\sigma>1$,the differential can be positive for the parameters patterns as $(+,+,-)$ and $(-,-,+)$,negative for the parameters pattern $(+,-,+)$.Actually we need not consider the case of $(-,-,+)$ for the practical numerical computation.

Based on the parameters we fit,we can decide the likely changing field for $\sigma$ separately considered with $0<\sigma<1$ and $\sigma>1$ as $\sigma=1$ being the boundary in terms of the above theoretical analysis.The main interest here is to consider the situation of $0<\sigma<1$ and we also get the changeable field for existence of critical value of $\eta$ for different combination of the parameters fitted.As for one time,there have two bounds for the maximal and minimal value for the field of $\sigma$ corresponding to existence of critical value $\eta^{\ast}$,and with time variation,we can obtain the changing tendency for different time.Actually the tendency of changing itself also can change just like the relation between velocity and acceleration and we just make a derivation of $\frac{1}{(b+ln{\sigma})(ln{\sigma})\sigma}$ and can find the term as:\be\label{second_order_eta}\frac{d^{2}\eta^{\ast}}{d{\sigma}^{2}}=-\frac{1}{(b+ln{\sigma}){\sigma}ln{\sigma}}\frac{d}{d{\sigma}}[(b+ln{\sigma}){\sigma}ln{\sigma}]=-\frac{1}{(b+ln{\sigma}){\sigma}ln{\sigma}}[{ln{\sigma}}^2+(b+2)ln{{\sigma}}+b]\ee 
   For practical computation,we can find there only have three combinations of fitting parameters $a,b,c$ with the patterns in terms of the sign of positive or negative as $(+,-,+)$,$(+,+,+)$,and $(+,+,-)$,actually the pattern $(-,-,+)$ basically can not be found that we do not consider actually in this paper.As the previous analysis of the bounds,for $\sigma<1$,there actually only have the bounds for the two patterns $(+,+,+)$,$(+,+,-)$,and corresponding bounds are $e^{-b}<\sigma<e^{-\frac{ab}{a+c}}$,$e^{-\frac{ab}{a+c}}<\sigma<e^{-b}$ with the additional condition $1+\frac{c}{a}>0$,and $0<\sigma<e^{-b}$ with the additional condition $1+\frac{c}{a}<0$,for $\sigma>1$,there also just have two patterns $(+,-,+)$,$(+,+,-)$ that have the possibilities for having the bounds and corresponding bounds are $1<\sigma<e^{-\frac{ab}{a+c}}$ and $\sigma>e^{-\frac{ab}{a+c}}$ with the addition condition $1+\frac{c}{a}<0$.For the consideration of second order derivative $\frac{d^{2}\eta^{\ast}}{d{\sigma}^{2}}$,we can introduce the transition value to decide its positive or negative  connected with two roots of ${ln{\sigma}}^2+(b+2)ln{{\sigma}}+b$ as $e^{[-(b+2)-\sqrt{b^2+4}]/2}$ and $e^{[-(b+2)+\sqrt{b^2+4}]/2}$.For simplicity,we can call the smaller root as the $\sigma_1$ and the bigger one as the $\sigma_2$.In terms of $\sigma<1$ corresponding to $b>0$,$[-(b+2)-\sqrt{b^2+4}]/2<[-(b+2)-\sqrt{b^2}]/2=[-(b+2)-b]/2=[-2{b}-2]/2=-b-1$,so $\sigma_1<e^{-b-1}=e^{-b}e^{-1}$,as $b>0$,thus we can get $\sigma_1<e^{-b}$ and $\sigma_1<e^{-1}$.For consideration of $\sigma_2$,$[-(b+2)+\sqrt{b^2+4}]/2=-\frac{1}{2}[(b+2)-\sqrt{b^2+4}]>-\frac{1}{2}[(b+2)-b]=-1$ and $[-(b+2)+\sqrt{b^2+4}]/2>[-(b+2)+2]/2=-\frac{b}{2}>-b$,thus we can get $\sigma_2>e^{-1}$ and $\sigma_2>e^{-b}$.In terms of $\sigma>1$,only the pattern $(+,-,+)$ have the possible transition value,and we find $|-(b+2)|<\sqrt{b^2+4}$ which means $-(b+2)-\sqrt{b^2+4}<0$ whatever $b$ is positive or negative,thus we can get $\sigma_1<1$ and $-(b+2)+\sqrt{b^2+4}>0$,and easily find $\sigma_2>e^{-b-1}$ and $\sigma_2>e^{-\frac{b}{2}}$.   
   
   Therefore,the transition value just is one of roots,and we can get the changing fields with bounds and possible transition value.In terms of $\sigma<1$,$e^{-b}<\sigma_2<e^{-\frac{ab}{a+c}}$ for $(+,+,+)$,$e^{-\frac{ab}{a+c}}<\sigma_1<e^{-b}$ with the additional condition $1+\frac{c}{a}>0$ for $(+,+,-)$,$0<\sigma_1<e^{-b}$ with the additional condition $1+\frac{c}{a}<0$ for $(+,+,-)$.In terms of $\sigma>1$,$1<\sigma_2<e^{-\frac{ab}{a+c}}$ for $(+,-,+)$.
As the field between $\sigma_1$ and $\sigma_2$,${ln{\sigma}}^2+(b+2)ln{{\sigma}}+b<0$ and beyond this field for the contrary.

Through the numerical computation based on the guide of theoretical judgement for how to determine the field of $\sigma$ for existence of critical value $\eta$,we can get the variation of lower bound and upper bound of $\sigma$.It also can be seen as a prediction and need to calculate the critical value $\eta^{\ast}$ with some different perturbations to testify our prediction and meanwhile to see directly the fields of the critical value as to what extent they are truly set in the changeable field of $\eta$ for just initial limited fitting,loosely to say roughly between 1 and 2.Further more,we also can test the changing tendency of critical value from our theoretical prediction in terms of the formula Eq.~(\ref{dif_crieta}) as well as the effect of the transition value originated from the formula Eq.~(\ref{second_order_eta}).
That is to say,we can compute the available fields numerically from the fitting parameters with all the time added with the likely transition value and study the variations of critical values in terms of some typical perturbations.Based on this spirit,we depict the figure 39 and 40 to show the basic numerical results agreed with our analysis in terms of the situation for $\sigma<1$,and use the figure 41 to show the corresponding expressions for $\sigma>1$ with a specific system parameter $K=1.75$ as well as the typical variations of the $\eta^{\ast}$ to $\sigma$ with the time fixed in terms of situations for $\sigma<1$ and $\sigma>1$ which mainly show the effect of the transition value.

Actually we also can find the tendency of the variation of the transition value is reconciled with the term $e^{-b}$ in terms of the situation for $\sigma<1$ without the consideration of the pattern $(-,-,+)$ and it is easy to know from simple mathematical derivation.For $b>0$,in terms of the pattern $(+,+,+)$,the transition value is $e^{[-(b+2)+\sqrt{b^2+4}]/2}$,if we make a derivation of $(b+2)-\sqrt{b^2+4}$,the result is $1-\frac{b}{\sqrt{b^2+4}}>0$ which show a increasing function with the same tendency of $b$ which is the simplest case of increasing function.For the pattern $(+,+,-)$ as well as $b>0$,the corresponding part of the transition value is $(b+2)+\sqrt{b^2+4}$ which is a obvious increasing function the same with $b$,thus we can expect the same tendency of the variation of the transition value to the lower bound of the pattern $(+,+,+)$ and the same tendency to the upper bound of the pattern $(+,+,-)$,and we can find our numerical results completely support our prediction showed clearly in the figure 39.For the situation of $\sigma>1$,in terms of the initial part of time,the pattern is dominated by $(+,-,+)$ and with the assumed small effect of $\frac{a}{a+c}$,the transition value is $e^{[-(b+2)+\sqrt{b^2+4}]/2}$,and use the same method,we can know the tendency of the variation of the transition value is close to the tendency of the upper bound $e^{-\frac{ab}{a+c}}$ of $(+,-,+)$,and basically it is true as the typical case for the fitting parameter $c$ often small and actually the term $\frac{a}{a+c}$ is close to 1 that could be neglected corresponding the bound $e^{-\frac{ab}{a+c}}$ dominated by the term $e^{-b}$.This kind of expression can be seen clearly in the figure 41(a).

\begin{center}
\begin{figure}
\includegraphics[width=18cm,height=20cm]{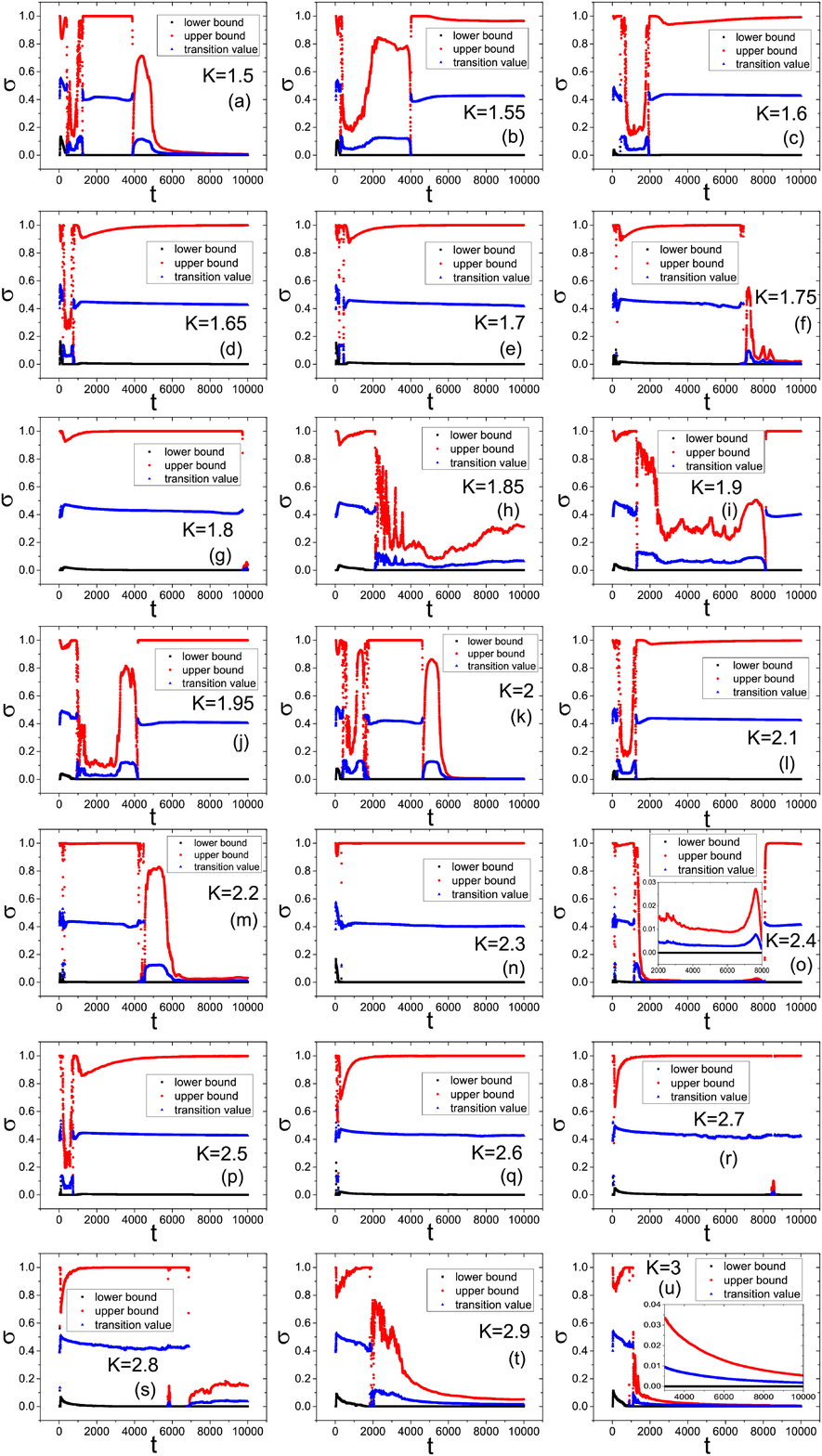}
\vspace{-0.5cm} \caption{The variations of the lower bound,upper bound and transition value of perturbation $\sigma$ to the time in terms of the situation for existing of critical value of $\eta$ within the consideration of $0<\sigma<1$,caculated with the different combination of fitted parameters in terms of $D_{L}$ versus $\eta$ with the number of fitting frequencies increased gradually from the initial three frequencies to the maxium twenty-two.For all the system parameters $K$ considered here,there show some basic patterns as the changing order for the pattern of parameters can be similar for some cases that can be seen easily but not entirely same expression for any two $K$,and also find there basically have the intermittency between two different continuous variations for a given $K$.The existence of transition value is very common and the tendency of the variation is reconciled with the term $e^{-b}$ corresponding to the lower bound of the pattern $(+,+,+)$ and the upper bound of the pattern $(+,+,-)$ that can be derivated mathematically and proved numerically here.}\label{sigma_fieldbelow1_transition_value}
\end{figure}
\end{center}

\begin{center}
\begin{figure}
\includegraphics[width=18cm,height=22cm]{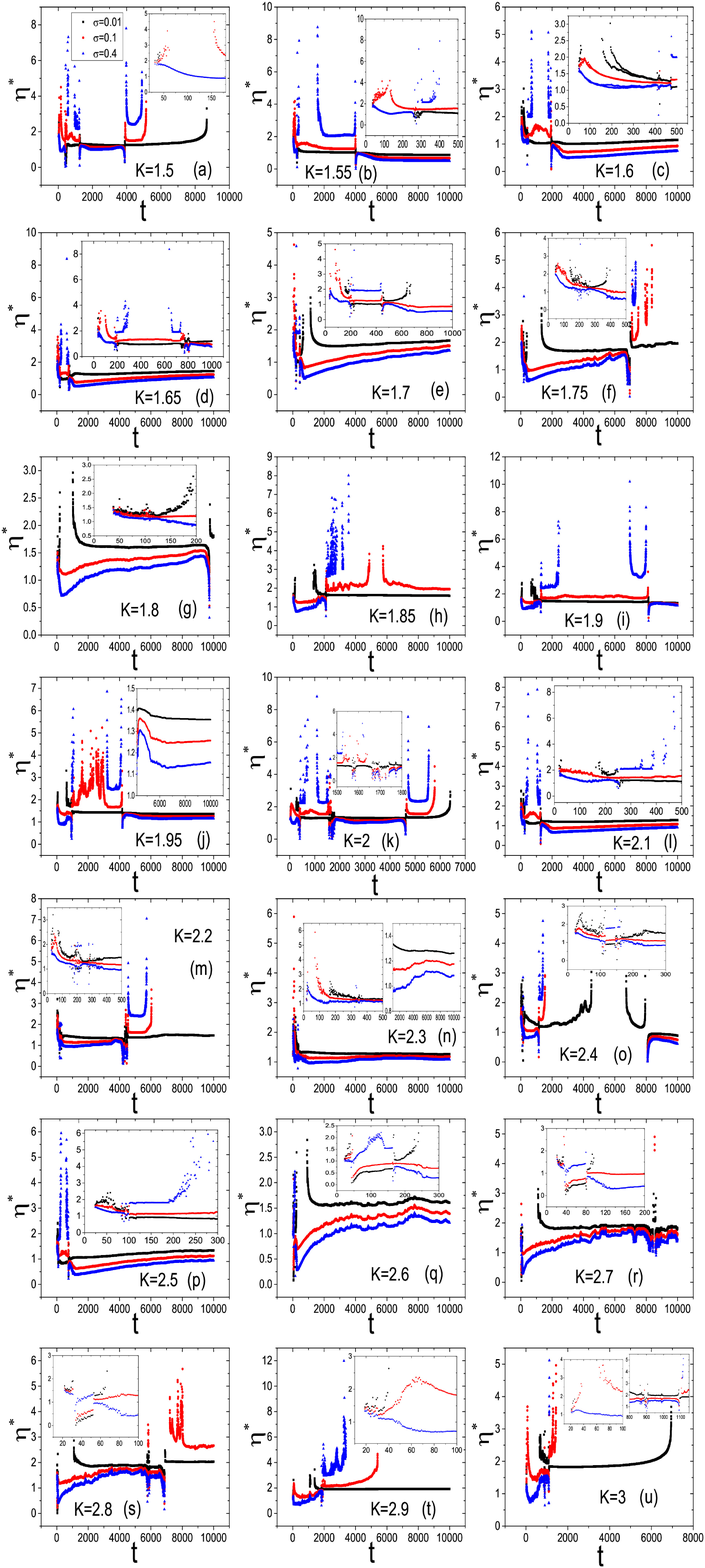}
\vspace{-0.5cm} \caption{The variations of critical value of $\eta^{\ast}$ to the time with three typical perturbation $\sigma=0.01,0.1,0.4$ used for the contrast,and the expressiones are agreed entirely with the changeable field of $\sigma$ we obtain in the figure 39 and the changing tendencies are also agreed with our theoretical analysis in terms of the formula Eq.~(\ref{dif_crieta}).One can find the U-shaped pattern for the variations of $\eta^{\ast}$ is very special but also quite common and  also find $\eta^{\ast}$ exists with a high expectation in the field between one and two corresponding the practical field for $\eta$,so it shows the non-monotonic variations of $M_{sc}$ for increasing fitting frequencies are common obsvered in a lot of numerical computation.}\label{cri_eta}
\end{figure}
\end{center}

\begin{center}
\begin{figure}
\includegraphics[width=18cm,height=18cm]{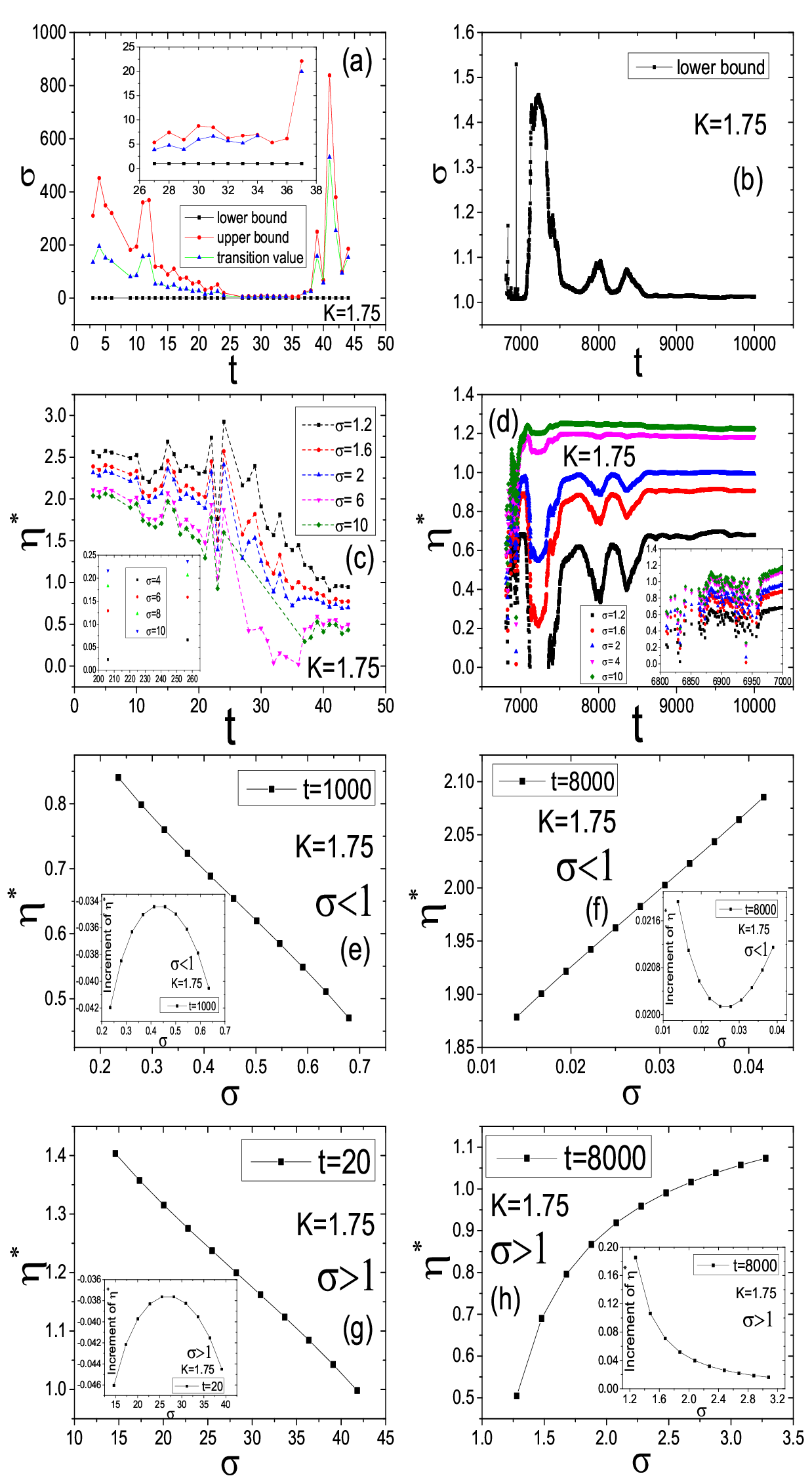}
\vspace{-0.5cm} \caption{A typical illustration for the variations of bounds for increasing time as well as corresponding variations of $\eta^{\ast}$ in terms of different available perturbations $\sigma$ with the category of $\sigma>1$ corresponding to the system parameter $K=1.75$ are showed in the figure(a),(b),(c)and(d).Besides the bounds,there also have the possible transition values to change the tendency of amplitude of variations just like the relation between the physical quantity velocity and acceleration,for the contrast,all the four basic patterns for the variations of $\eta^{\ast}$ to available $\sigma$ are illustrated using four typical cases in the figure(e),(f),(g)and(h) in terms of the situations for $\sigma<1$ and $\sigma>1$.The expressions of these variations are all agreed with the theoretical analysis and the sub-figures show the variations of the increment of $\eta^{\ast}$ computed with the symmetrical distribution around the transition value when there have the transition values and clearly show the changing situation happened as we expect.The basic linear relation in the figure (e),(f)and(g) can be explained as the transition value actually is a extremal point for $\frac{d\eta^{\ast}}{d\sigma}$,thus one can expect almost constant for it with the vincinity of the transition value which means basically the same slope for the variation of $\eta^{\ast}$ to $\sigma$,the closer the better showed clearly in the figure(f).}\label{typical_sigma_large1}
\end{figure}
\end{center}

One can find the existence and changing tendency for $\eta^{\ast}$ coming from numerical results typically showed in the figure 40 are agreed with our analysis and further more we can find the U-shaped pattern for the variations versus time is quite common that can be explained as the the variations of parameters $a,b,c$ for inverted U-shape,not large variation and U-shape accordingly combined with the formula of $\eta^{\ast}=-\frac{1}{b}ln[-\frac{a}{cln{\sigma}}(b+ln{\sigma})].$

In terms of the case of $\sigma>1$ which is not our main consideration here as the large perturbation leading to the very rapidly decay in a very short time making the semi-classical prediction with Levy distribution out of work as it requires relatively long time to form.For detail,the combination of parameters $a,b,c$ for $+,+,+$ can not have the existence of $\eta^{\ast}$ from our analysis,and we find the upper bound for the situation of $+,-,+$ corresponding to the initial time less than 50 and the values can be span from the order of $10^{1}$ to $10^{8}$,for the situation of $+,+,-$,there only have the lower bounds which span basically from 1 to 3 although very few case of some very large value sudden emergent with the order of $10^{3}$ or even larger.Why there have the very large fluctuation for the changeable field for $\sigma>1$?the reason is the term $-\frac{ab}{a+c}$ is positive and for some specific combination of the parameters the bound value $e^{-\frac{ab}{a+c}}$ can be very large even the variation of single parameter is some kind of small.So even there have a very large fluctuations for the variations of parameters,the variation of $-\frac{ab}{a+c}$ still could change small.For typical cases,$a+c$ or $ab$ become much smaller or larger with $ab$ or $a+c$ changing not so much,then the very large value of $e^{-\frac{ab}{a+c}}$ can be expected.We find the variations of the value of the bounds are more complicated than the situation for $\sigma<1$,the basic explanation here is the complexity of the variations in terms of $\sigma<1$ can be covered to some extent as the bound value tends to 1 or 0 when $\frac{ab}{a+c}$ is sufficient small or large but this situation can not applied to $\sigma>1$ as the bound value is taken as the increase function $e^{x}$ with $x>0$.here $x=-\frac{ab}{a+c}>0$,so the variations of parameters $a,b,c$ can have a more clear effect leading to a high individual feature for the variations of the bound value of $\sigma>1$.For practical consideration,the bounds just give the existence of $\eta^{\ast}$ but we need to know whether they can be set in the real fitting field basically between 1 and 2,and we find numerically this kind of situation exists but much fewer than the case of $\sigma<1$.

It is worth noting that the variations of $M_{sc}$ in terms of a very small variation of $\eta$ can be seen as a very simple Tylor expansion.For the exponential approximation $D_L=c+a{\rm exp}(b {\eta})$,we have
$ M_{sc}\approx {\rm
exp}\{-2a[\frac {c}{a}e^{\eta ln{\sigma}}+e^{\eta
(b+ln{\sigma})}]\}.$With $\eta=\eta^{0}+\Delta \eta$,and use the symbol $M_{sc}^{0}\equiv{\rm
exp}\{-2a[\frac {c}{a}e^{\eta^{0} ln{\sigma}}+e^{\eta^{0}
(b+ln{\sigma})}]\}$ as the approximated semi-classical value corresponding to the variable $\eta^{0}$,then we get $\Delta M_{sc}$ with a simplistic version as:
\be \Delta M_{sc}\approx -M_{sc}^{0}[2c\,e^{\eta^{0}ln{\sigma}}ln{\sigma}+2a\,e^{\eta^{0}(b+ln{\sigma})}(b+ln{\sigma})]\Delta \eta.\label{delt_Msc1}\ee
For linear approximation with the same Tylor expansion,and define here $M_{sc}^{0}\equiv {\rm
exp}\{-2b[\frac{a}{b}e^{ln{\eta^{0}}+\eta^{0}ln{\sigma}}+e^{\eta^{0}ln{\sigma}}]\}$,we have the similar expression for the relation of $\Delta M_{sc}$ via $\Delta \eta$ as:
\be \Delta M_{sc}\approx -M_{sc}^{0}[2a\,e^{ln{\eta^{0}}+\eta^{0}ln{\sigma}}(\frac{1}{\eta^{0}}+ln{\sigma})+2b\,e^{\eta^{0}ln{\sigma}}ln{\sigma}]\Delta \eta.\label{delt_Msc2}\ee
The numerical results we observe basically agree with the theoretical prediction at least in a very short field in terms of $\eta$,here we can give a typical case with $k=1.5$ for the time step as 2800 to show our prediction in the figure 41.

\begin{center}
\begin{figure}
\includegraphics[width=16cm,height=6cm]{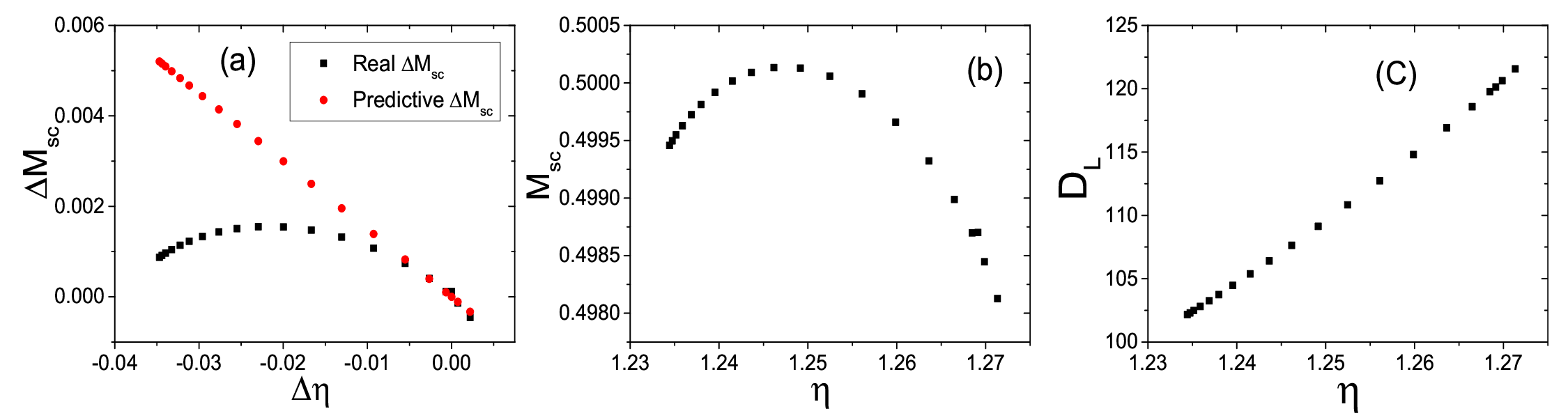}
\vspace{-0.5cm} \caption{A illustrative case for the Tylor expansion of $M_{sc}$ in terms of a short field of $\eta$,$K=1.5$ and $t=2800$.(a)The variations of $\Delta M_{sc}$ to the $\Delta \eta$ in terms of the direct numerical result and the other result for corresponding Tylor expansion.In the initial part,there are basically very close and then seperate quickly.(b)the original variation for $M_{sc}$ versus $\eta$ which shows a non-monotonic curve agreed well with our critical value caculated for $\eta{\ast}=1.24814$.For a more clear illustration,(c)shows the variation of $D_{L}$ to $\eta$.For a good fitting,we just use the initial twenty values of $\eta$ to obtain the fitting parameters $a,b,c$ in terms of the model $D_L=c+a{\rm exp}(b {\eta})$.}\label{deltm}
\end{figure}
\end{center}

Therefore,it seems some kind of wasting so much energy to study the fitting detail,but for seriousness,we consider this study is helpful to clarify the accuracy of the semi-classical prediction with the fitting procedure in detail,and the wide existence of $\eta^{\ast}$ and its monotonous variation versus increasing perturbation suggest us to use a fewer frequencies for fitting to get the $\eta$ and $D_{L}$ to decrease the chance to approach the $\eta^{\ast}$ to get the accuracy of semi-classical evaluation towards quantum fidelity mainly from corresponding statistical similarity to Levy distribution .Thus,we uniformly choose four frequencies to make a fitting for giving the semi-classical evaluations to the direct quantum fidelity decay in this paper.

\subsection{C.Typical properties for quantum fidelity joined with Semi-classical evaluation}

Now it is the time to use our semi-classical method to understand the quantum fidelity and the real important factor just the probability density distribution of $P(s)$ that is $s=cos(r)$ in our study.Further more,one with careful reading can find our method actually is based on the fitting technique we have discussed a lot,but can treat all the available perturbations just from the one thing $P(s)$,for more detail,the two variables $D_{L}$ and $\eta$.

So firstly we want to show our numerical calculations about the variations of $D_L$ and $\eta$ with the time in terms of all the systems with different $K$ taken consideration,the focus here is to find some basic patterns for the variations.As the initial time corresponds to the situation obviously without Levy distribution of $P(s)$,we do not consider the initial time with the scale as 50 and show the variations of $D_L$ and $\eta$ for the rest of time.These results are depicted in the figure 42 and 43.

\begin{center}
\begin{figure}
\includegraphics[width=18cm,height=20cm]{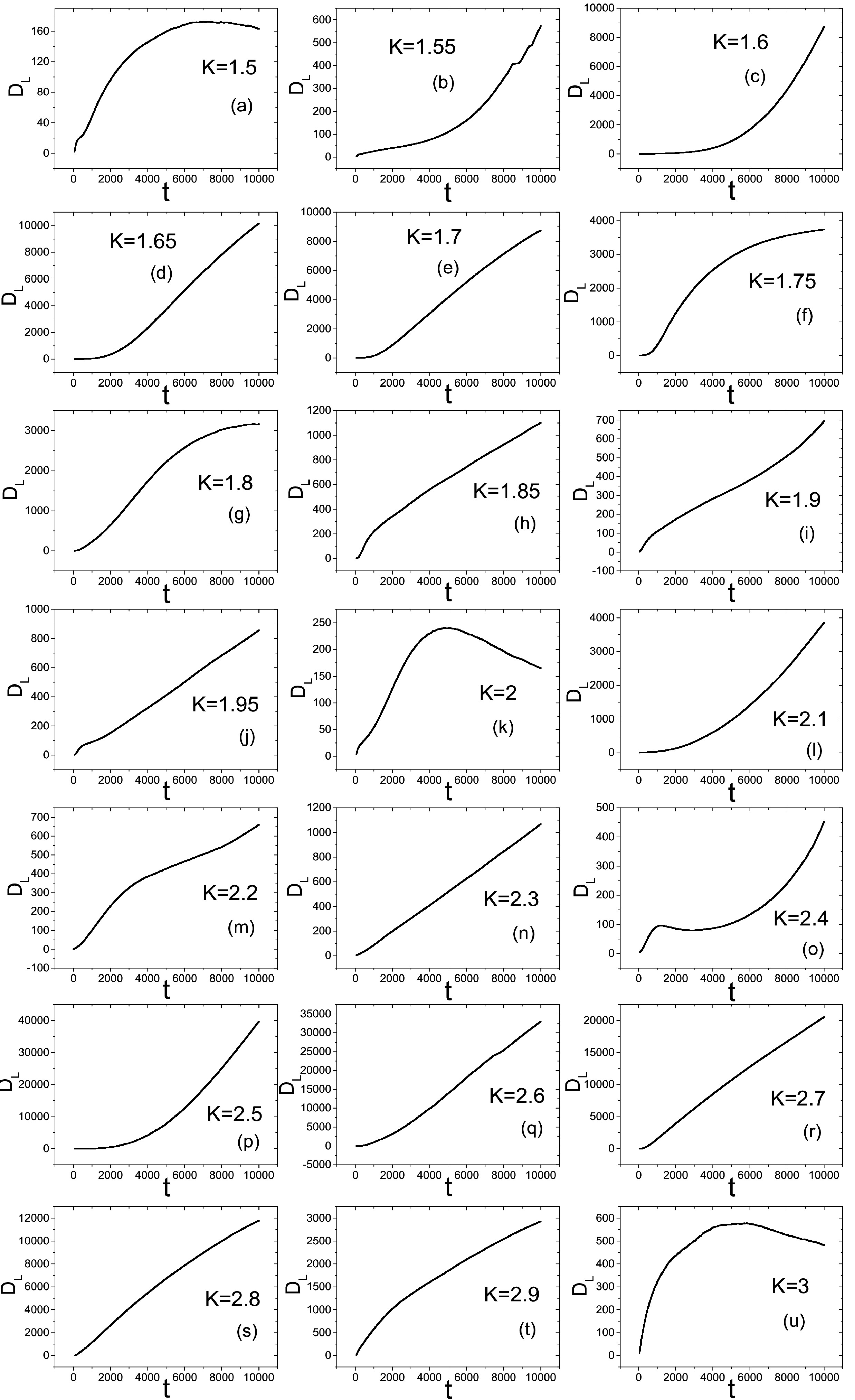}
\vspace{-0.5cm} \caption{The variations of $D_L$ to the time without initial 50 taken consideration for oviously deviating the assumation of Levy distribution of $P(s)$ corresponding to choosen system parameters $K$.In detail,$D_L$ is calculated with initial non-zero four frequencies corresponding to Fourier transform of $P(s)$ in terms of assumed Levy distribution.Basically $D_L$ increase with the time increasing but there also have decreasing situations within some time scale.These different expressions come from the diffusion of the initial classical ensemble corresponding to the initial quantum Gaussian wavepacket in terms of different choosen parameters $K$.}\label{DL}
\end{figure}
\end{center}

\begin{center}
\begin{figure}
\includegraphics[width=18cm,height=20cm]{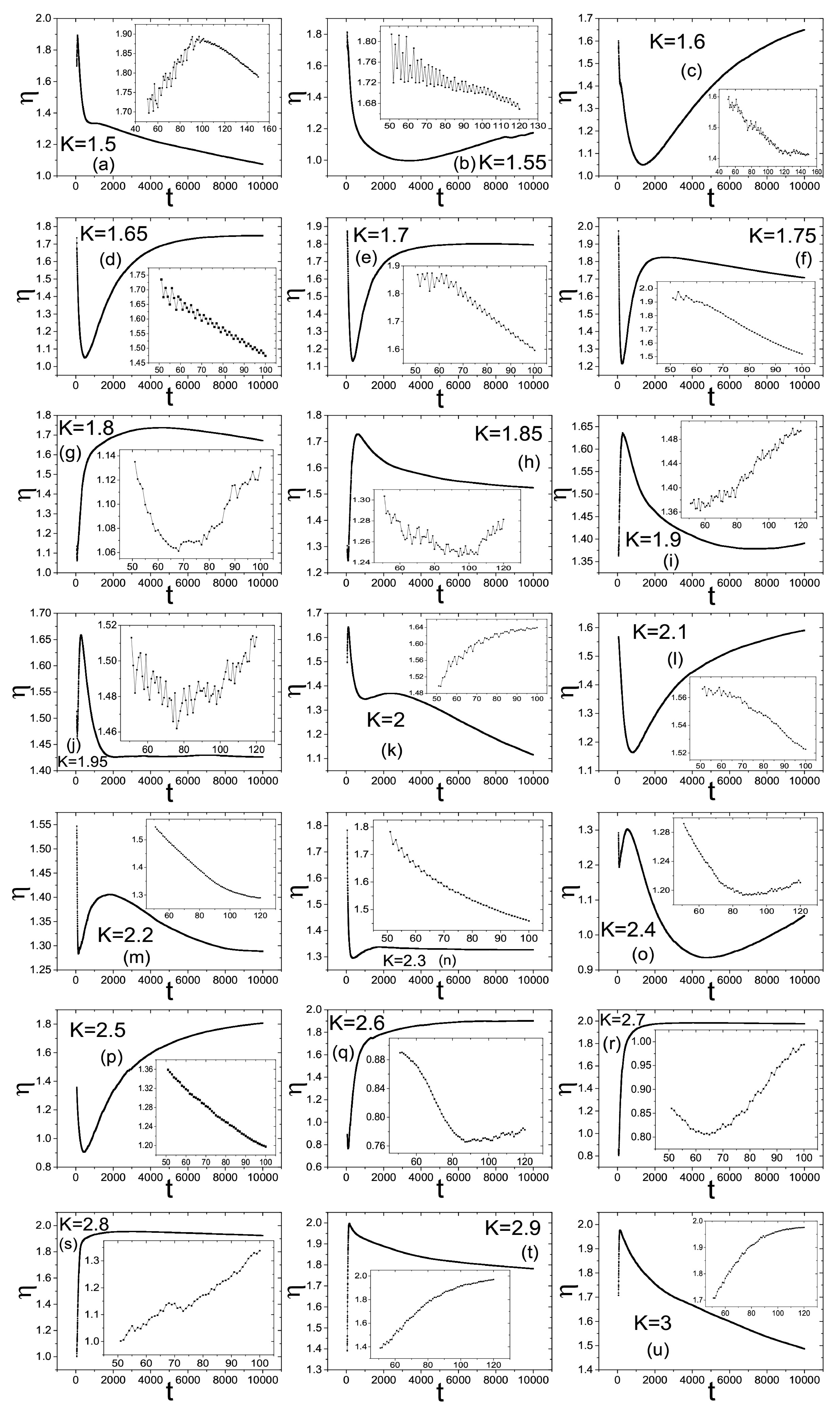}
\vspace{-0.5cm} \caption{The variations of $\eta$ to the time with the same consideration for $D_L$ as they are fitted at the same time with common fitting technique,the detail can be seen in the figure above for the variations of $D_L$.As the value of $\eta$ is a important index for the statistical feature of $P(s)$,the dependence of $s$ on the time can be seen as non-Gaussian random process with $\eta=2$ equivalent to Gaussian distribution of $P(s)$ reflecting the mixed struction of phase space in terms of the parameters $K$ we study.The variation of $\eta$ to the time as a whole shows a transition very similar to successive pattern from smaller $K$ changed to the larger which actually exists to some extent in the variation of $D_L$ as a weak version,thus we can expect this successive pattern can also hold obviously for $\eta$ with other study models taken as a universal expression with the system $K$ changed monotonously.The subfigures we depict show the likely initial oscillations of the variations of $\eta$ corresponding to some wild distributions of $P(s)$ which have been illustrated previouly for checking Levy distribution.}\label{eta}
\end{figure}
\end{center}

Now we show the numerical results of the dependence of $D_{L}$ and $\eta$ on the time and show the obvious non-Gaussian random process as the basic features of Gaussian random process is $\eta=2$ and the linear increasing function of time for $D_{L}$ which lead to exponential decay of fidelity decay satisfied in the strong chaos have been studied extensively.But for our study here,the random process we study actually even is not a standard random process from rigorous mathematical idea as the asymptotic value of $\eta$ seems to not be applied universally from our numerical computation,which implicit the random attractor can not always hold at least for quite a long time.From the numerical results we get showed in the figure 43 and 44,we can find there have successive pattern for the variation as a whole obviously for $\eta$ and we expect this expression is universal to some extent for other study models when the system parameters can change monotonously.Actually this expression also can be seen here as a simple and whole characterization of a very complicated phase space with chaotic and regular field in coexistence through the consideration of dynamical evolution of a ensemble.

Actually $D_{L}$ and $\eta$ are the seeds to get all the important information with our interest in fidelity decay as decay rate and decay exponent in terms of the good evaluation of real quantum fidelity,thus the direct contrast between quantum fidelity and corresponding semi-classical one is one of main task in our study.Further more,the effectiveness of our semi-classical method should be considered,and we want to use the contrast between quantum fidelity and direct semi-classical integral to investigate.Now firstly we should consider the decay rate and decay exponent with $D_{L}$ and $\eta$ separately and then hold together. 
      If we take the formula $M_{sc}(t)=\rm exp(-2(\epsilon/\hbar)^\eta D_L)$ to give the assuming decay law $M(t)\simeq e^{-ct^\alpha}$,it
is obviously found that the terms $(\epsilon/\hbar)^\eta$ and $D_L$
should give the $\alpha$ together,which are divided into two parts that
we can call as $\alpha_{\eta}$ and $a_{D_L}$ in terms of the relation
$(\epsilon/\hbar)^\eta\propto t^{\alpha_{\eta}}$ and $D_L\propto
t^{\alpha_{D_L}}$.Therefore we can use the relation $\eta ln{\sigma}$
via $lnt$ to give $\alpha_{\eta}$ and the relation $ln{D_L}$ via $lnt$ to
give $\alpha_{D_L}$.For simplicity,$\alpha_{D_{L}}$ can be expressed with $\alpha_{D}$.Now we also can do a easy analysis about the relative tendency for the variations of $\alpha_{\eta}$ and $\alpha_{D}$,as $D_{L}(t){\sigma}^{\eta(t)}$ can be expressed as $c_{\eta}t^{\alpha_{\eta}}c_{D}t^{\alpha_{D}}$ for a given $t$,then there will have some variations for the $\eta$ and $D_{L}$ taken as the function of time corresponding to $\alpha_{\eta}+\Delta \alpha_{\eta}$ and $\alpha_{D}+\Delta \alpha_{D}$.Meanwhile we can reasonably assume the variation of $c_{\eta}c_{D}$ can change not so sharp for every fitting time step and actually the $c_{\eta}$ and $c_{D}$ have great disparity in quantity stemming from basically $D_{L}$ much larger than $\sigma^{\eta}$.From the discussion above,we can approximately obtain the relation $t^{\Delta \alpha_{\eta}+\Delta \alpha_{D}}>1$ as the semi-classical evaluation is basically a decreasing function versus time which leads to $\Delta \alpha_{\eta}+\Delta \alpha_{D}>0$.Therefore there should have the positive value for one of $\Delta \alpha_{\eta}$ and $\Delta \alpha_{D}$ only for the corresponding different one 
as the negative value,and meanwhile we can get the variations of $c_{\eta}$ and $c_{D}$ are quite opposite.Actually there still have the possibility for $\Delta \alpha_{\eta}$ and $\Delta \alpha_{D}$ can all be positive which basically can not seen in practical numerical observations.

For clearly illustrating our study results,we firstly show our numerical results for the decay exponents with our great interest with $\sigma=0.01$ as a typical case in terms of the different selective parameters $K$.For a better comparison,we depict the fitting decay exponents from direct quantum fidelity,semi-classical evaluations from direct numerical integral and Levy assumption as well as the separated $\alpha_{\eta}$ and $\alpha_{D}$.The expressions in detail can be seen clearly in the figure 45.The small strength like $\sigma=0.01$ we use basically can have a good correspondence for a long time as the quantum fidelity can be expected to decay slowly,but for the situation of much faster decay,such as the case of $K=2.5$,the direct fitting method have some problem as the numerical fitting precision and the large fluctuation of $\alpha_{q}$ and $\alpha_{SC}$ can not reflect the real decay law and this problem can be escaped from $\alpha_{SCL}$.Our prediction of possibility of existing entirely opposite tendency for $\alpha_{\eta}$ and $\alpha_{D}$ is showed obviously and many gradual decay exponents showed are between 1 and 1.1 corresponding to slight stretched exponential decay which is different from the intense study of strong chaos.As it is just one case about $\sigma=0.01$ and we can expect this kind of new decay law can be even strengthened as $\alpha_{\eta}$ can be changed versus the variation of the perturbation $\sigma$.Here we also pay attention to the variation of $\alpha_{\eta}$ and $\alpha_{D}$ as a whole form some successive pattern which is previously expressed in the variation of $\eta$.

\begin{center}
\begin{figure}
\includegraphics[width=18cm,height=20cm]{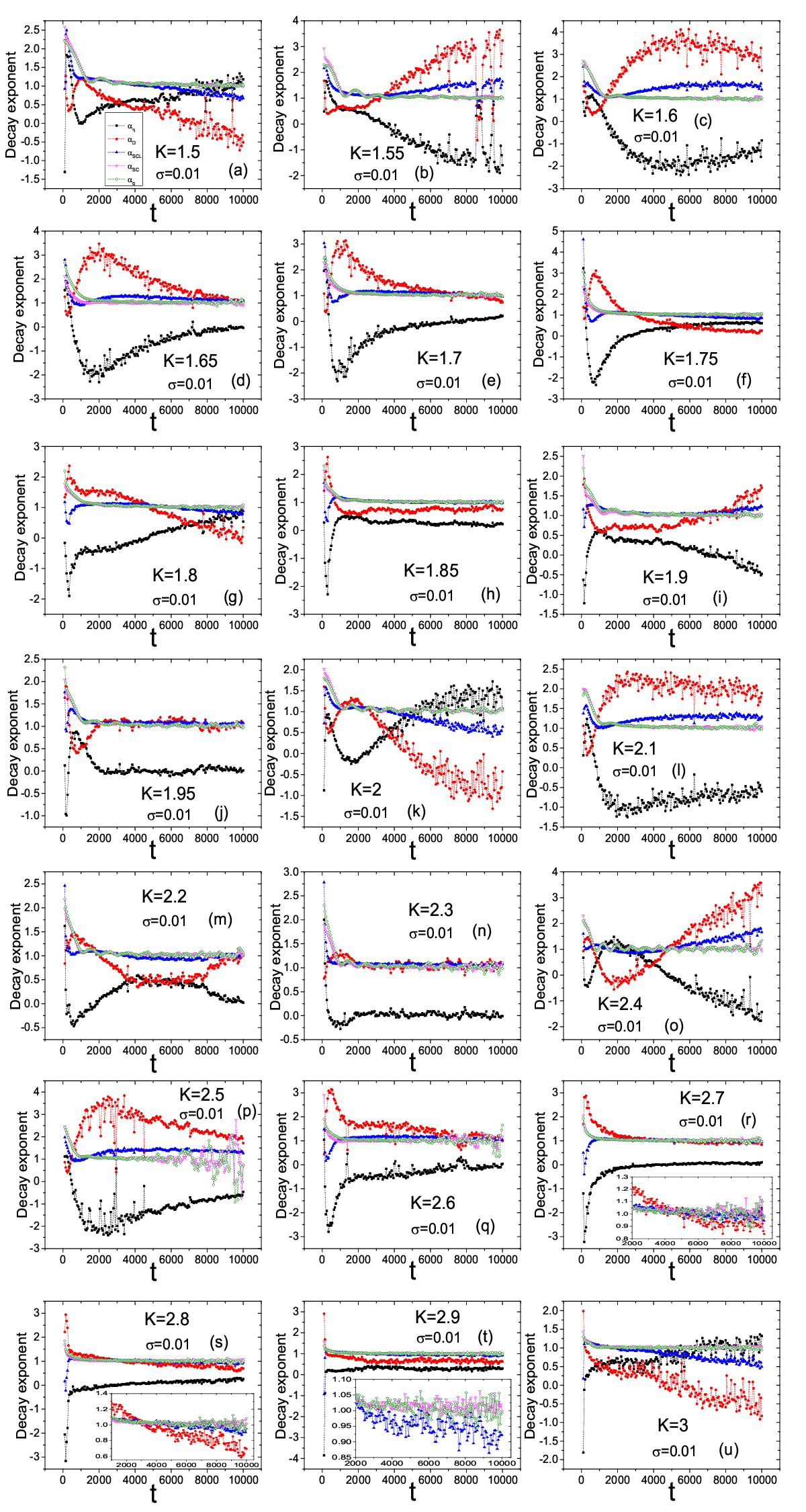}
\vspace{-0.5cm} \caption{We show the variation of decay exponents with $\sigma=0.01$,$\alpha_{q}$,$\alpha_{SC}$ and $\alpha_{SCL}$ are the decay exponents fitted according to direct quantum fidelity,semi-classical integral and Levy assumption for different system parameters $K$ with original 50 time steps as a whole to get the fitting value but without the initial 50 time steps taken consideration.The separated parts of decay exponent with Levy assumption $\alpha_{\eta}$ and $\alpha_{D}$ are also depicted here for comparison and the tendency of quite opposite expression have been predicted in our analysis which leads to basically slow variation of $\alpha_{SCL}$ as we expect.$\alpha_{SC}$ basically can have a good agreement with the direct quantum fitting but $\alpha_{SCL}$ can have some large deviation when the Levy assumption is not a good appximation.There have a transient process when $\alpha_{q}$ and $\alpha_{SC}$ change decreasingly and then a basic stable field can be expected although decreased very slowly,the decay exponents in terms of stable field can be seen between 1 and 1.1 which means a slight stretched exponential decay.As the reason of fitting accuracy,somehow large fluctuation of $\alpha_{q}$ and $\alpha_{SC}$ for the system parameter $K=2.5$ can not reflect the real decay law but $\alpha_{SCL}$ not yet beyond the diret fitting limitation.}\label{fit_exp_rate}
\end{figure}
\end{center}

Now we should pay attention to the difference of decay rate which have been studied very intensively,for $FGR$ field in terms of strong chaos the decay rate can be expressed as $c\approx 2.2\sigma^{2}$ and obviously we should observe the likely deviation in weak chaotic sea.As the separated parts $c_{\eta}$ and $c_{D}$ have quite different scale in value and what we really care about is the decay rate which means the whole $c_{\eta}c_{D}$,thus we illustrate the variation of decay rates fitted using the same fitting method we mentioned above and make a comparison among them corresponding to direct quantum fidelity,semi-classical evaluations with direct integral and Levy assumption.We still use $\sigma=0.01$ to show the typical illustration depicted in the figure 46.From the numerical results in terms of $\sigma=0.01$,we can find that decay rate fitted shows the basic regular as the temporary decay rate increase gradually to saturation with some oscillation and the expected decay rate for strong chaos as $2.2\times10^{-4}$ can have some deviation more or less corresponding to individual expression for a particular system parameter $K$.For the previous study of completely chaotic sawtooth map\cite{WgWang_2},even the Lyapunov exponent is small,the decay rate all below the $FGR$ expected decay rate,and our result here give the hints that the decay law of weak chaos is not a subordinate version of strong chaos,on the contrary,there could be highly non-trivial.

\begin{center}
\begin{figure}
\includegraphics[width=18cm,height=20cm]{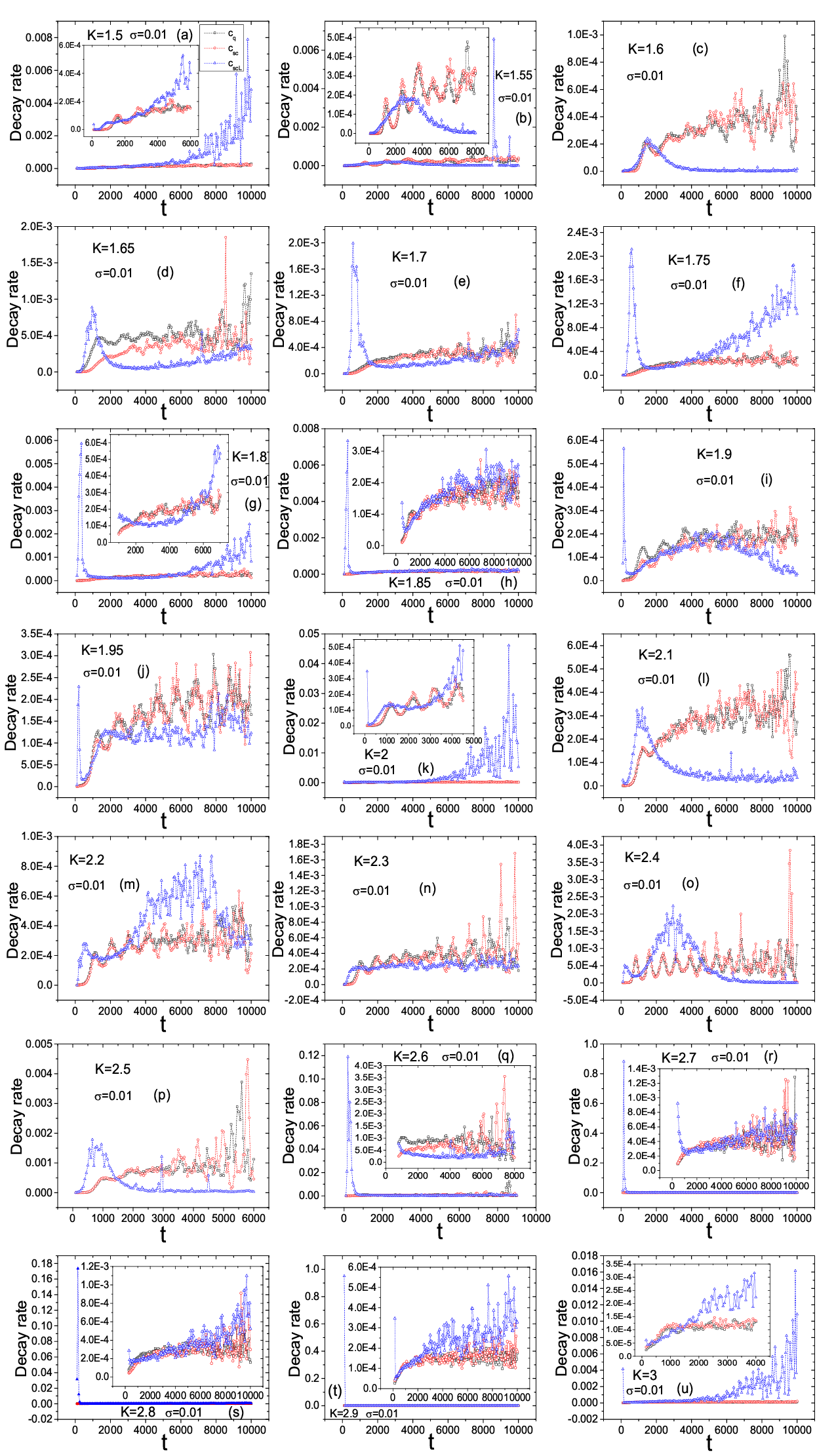}
\vspace{-0.5cm} \caption{The variation of decay rate fitted with the same procedure as the fitted decay exponents previously.One can find the initial decay rate is small and then gradually increase to some saturated field although having some oscillation obvious or not,besides the special cases of $K=1.65$ and $2.6$,$c_{sc}$ basically is agreed with $c_{q}$ very well in terms of time variation but $c_{scL}$ can have some obvious deviation when the Levy distribution is a not good assumption.The expected deviation from the $FRG$ decay rate as $2.2\times10^{-4}$ can be observed and larger or smaller situation really does exist which is heavily dependent on the individual expression.When time is long and decay value is quite small,somewhat large decay rate is the result of limitation of fitting precision.}\label{fit_rate}
\end{figure}
\end{center}

It seems that the semi-classical method we use is a quite good approximation although the case of $K=1.65,2.6$ showing some large deviation,so we should consider the effectiveness of our theoretical method for different perturbation,in particular for the likely limitation.Then we should consider the comparison between the quantum fidelity and the direct semi-classical integral and want to find the likely variation of the difference between them.Obviously,the semi-classical decay with the assumption of Levy distribution should also be considered here to see the accuracy approaching the direct integral.As the initial time the distribution of $P(s)$ is far away from the Levy distribution from our numerical computation,so for expression of our results appropriately showed,we do not consider the very short time for the semi-classical evaluation with Levy assumption in terms of the comparison for a long time using the relative weak perturbation,but for the comparison for a relative short time using some kind of strong perturbation,we can show the some big fluctuation clearly.Based on this idea,we use the perturbations $\sigma=0.01,0.05,0.4$ separately to see the variation of the difference of the evaluation.The figure 47 and 48 give the typical expressions and we can find the big deviation for quite a few parameters $K$ in terms of strong perturbation $\sigma=0.4$.So how to understand this kind of difference?through very carefully check,even in terms of small perturbation,there still have some minor difference,so we want to guess if it is a effect of the fluctuation term that be omitted and vary for different perturbation?it is a open problem in our paper.

\begin{center}
\begin{figure}
\includegraphics[width=18cm,height=20cm]{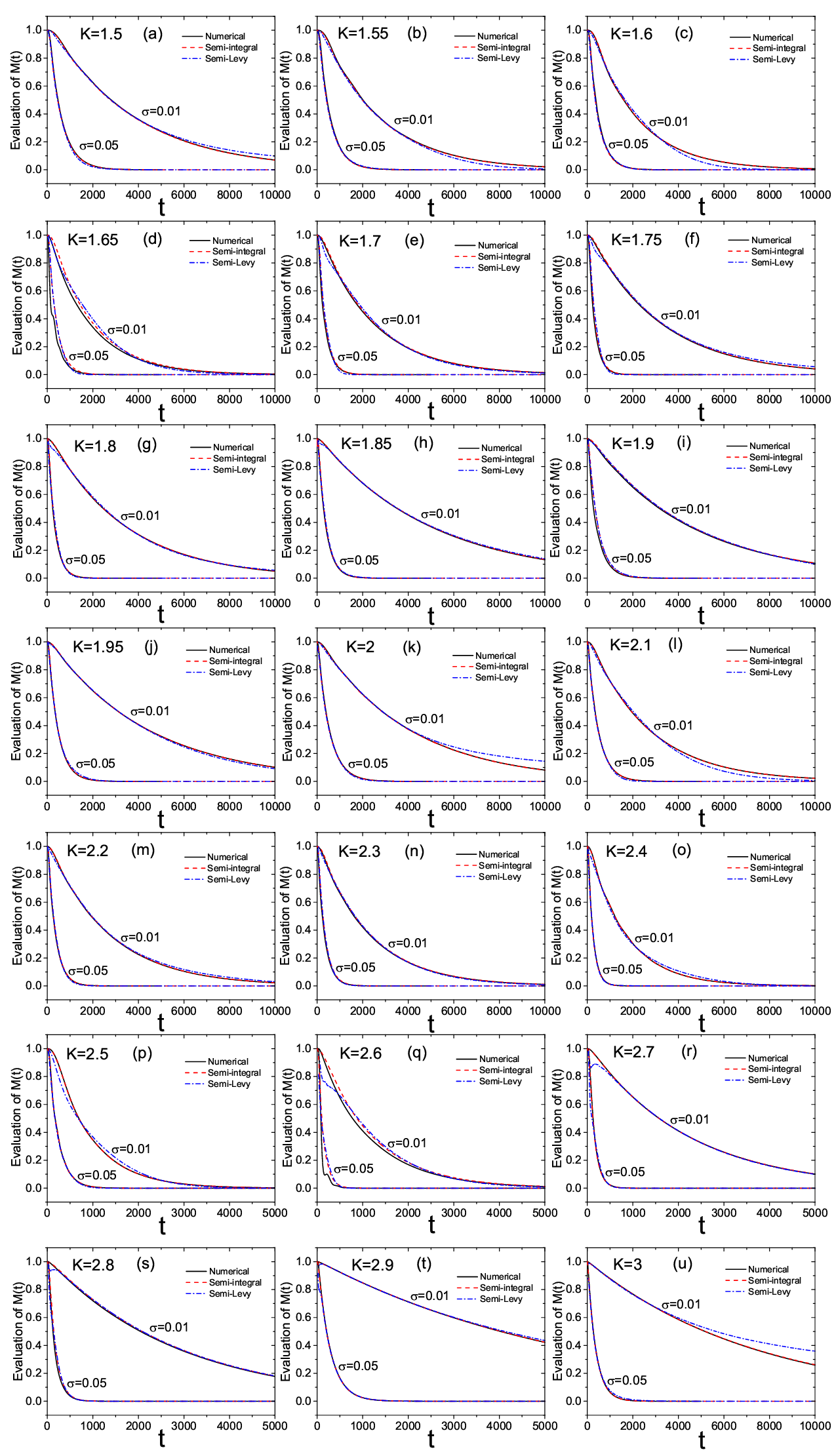}
\vspace{-0.5cm} \caption{we use the small perturbations $\sigma=0.01,0.05$ to show the long time evaluation of quantum fidelity with direct numerical computation and semi-classical caculation through direct integral and Levy assumption.One can find the decay expression from our theory basically is a good approximation as most semi-classical evaluation with direct integral agree well with the quantum fidelity except the cases of $K=1.65,2.6$ having some obvious deviations.Semi-classical evaluation with reasonable Levy assumption can also be a good approximation although the obvious deviation can be found when the Levy assumption is not good enough so the initial part with time scale around 30 can not be considered.From the figure,it seems that the deviation from quantum fidelity with the evaluation using Levy assumption can be smoothed out to some extent versus the perturbation increase as it is actually a pure mathematical effect from the semi-classical theory $M(t)\approx \rm exp(-2D_{L}\sigma^{\eta})$ with the variation of $\sigma$.The time scale we choose is to show the decay process clearly,and actually the decay time for the numerical evaluation just take 5000 for $\sigma=0.05$ when basically come to very small value.}\label{fidelity_contrast_small}
\end{figure}
\end{center}

\begin{center}
\begin{figure}
\includegraphics[width=18cm,height=20cm]{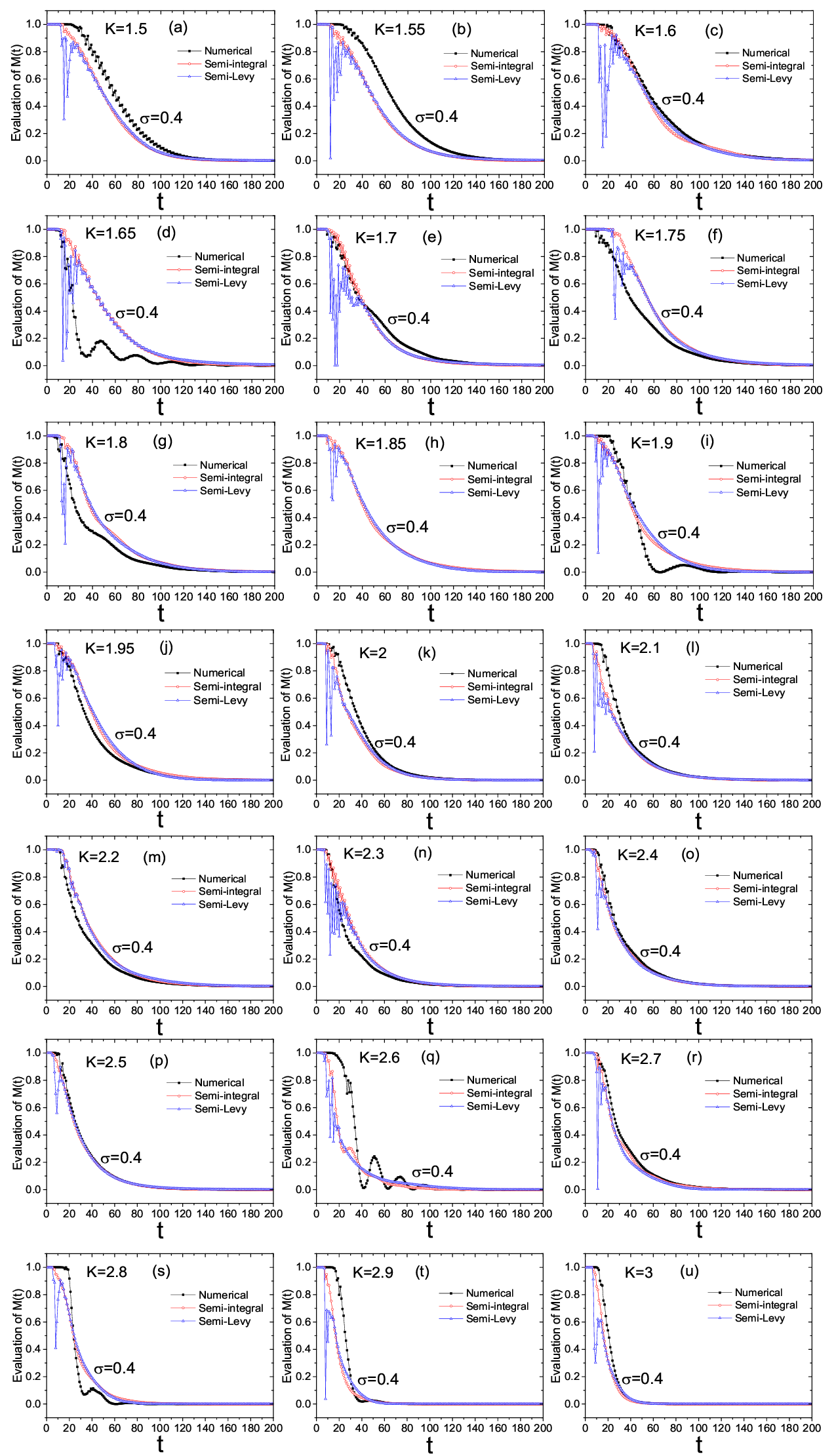}
\vspace{-0.5cm} \caption{We use the relative large perturbation $\sigma=0.4$ within well established FGR field to show the evaluation of quantum fidelity in terms of short time with the same procedure for the small perturbation above.As the relative short time for effective decay process,we show clearly the large fluctuation of the semi-classical evaluation with Levy assumption.The initial quantum frozen field predicted by Prosen etc\cite{Gorin} is clearly showed and basically can be captured by our semi-classical method even using the obviously poor Levy assumption in the initial stage.One can find the evaluations of semi-classical direct integral and Levy assumption are basically close each other but the difference between direct numerical computation have very individual expressions and there have quite a few cases corresponding to some large deviation which shows the limitation of our semi-classical method.}\label{fidelity_contrast_large}
\end{figure}
\end{center}

Therefore,we put forward a simple variable to investigate more,we consider the relative difference based on the difference of evaluations from direct numerical computation and semi-classical direct integral divided by the corresponding perturbation itself,and then averaged efficient decay time without consideration of the small value of $M(t)$ in terms of the order of $10^{-2}$ in practical.For simplicity,we can write semi-classical direct integral as $M_{int}$,and the mathematical formula can be written as:
\be \overline{\frac{\Delta M(t)}{\sigma}}=\frac{1}{N}\sum_{t=1}^N{\frac{M_{int}-M(t)}{\sigma}} \label{aver_dM}\ee

Now this variable is just the function of perturbation $\sigma$ in terms of a given system parameter $K$,and if the averaged difference $\overline{\Delta M(t)}$ can be seen as the linear relation to the perturbation,it is reasonable to expect the almost constant value for $\overline{\frac{\Delta M(t)}{\sigma}}$ in terms of increasing perturbation $\sigma$.Obviously,we should make a numerical study about the variable and the expressions in detail have been showed in the figure 49.We can find in the figure that the expression of $\overline{\frac{\Delta M(t)}{\sigma}}$ is quite individualized and the value around zero in particular for the large perturbation show that the theoretical approaching expectation basically can be hold,further more the field with some little change in the averaged value showed in figure correspond to the approximated proportional relationship for the dependence of the averaged difference of $M(t)$ on the increasing perturbation.

\begin{center}
\begin{figure}
\includegraphics[width=16cm,height=6cm]{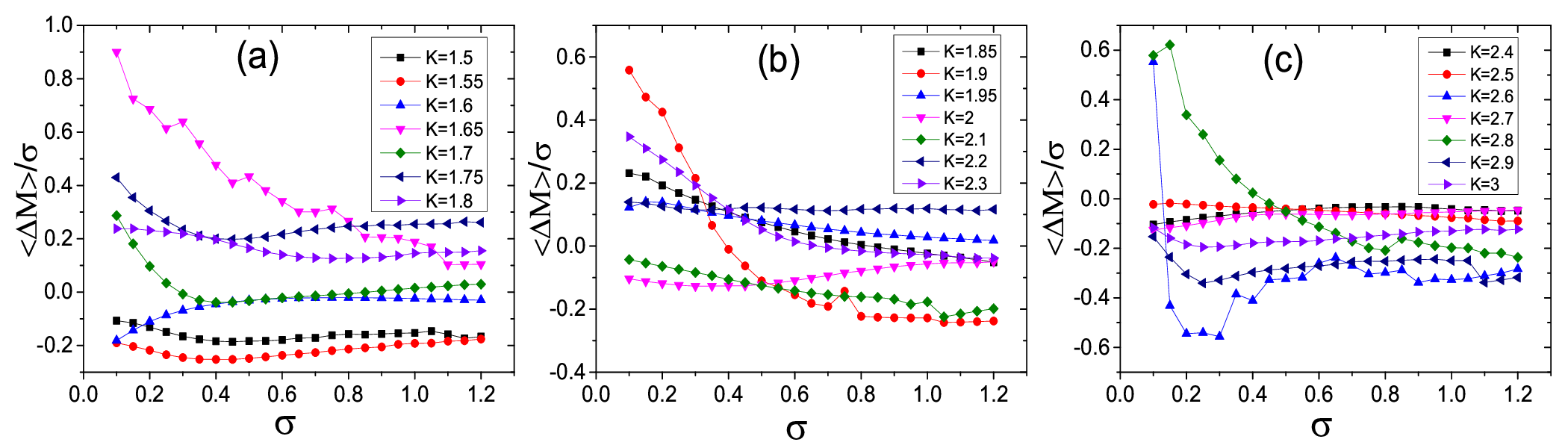}
\vspace{-0.5cm} \caption{The averaged difference of M(t) between direct numerical computation and semi-classical integral rescaled by the corresponding perturbation versus the perturbation itself,here the arrange of perturbation used is from $0.1$ to $1.2$ in terms of the interval taken $0.05$ and the averaged process is considered for the $M(t)$ larger than $0.01$.}\label{aver_diff_M_sc}
\end{figure}
\end{center}

Now we do know the accuracy of our semi-classical method have some limitation mainly for the large perturbation in terms of seeming common system parameters $K$,and in the practical numerical observation,the so-called decay rate $c$ defined rigorously in the strong chaos with the decay formula as $e^{-ct}$ is quite below the regular law about $c\approx 2.2\sigma^2$ with increasing perturbation,which means also for now,there have not a clear law governing the variation of $c$.Thus what we focus is the decay exponent as the sign for the likely stretched exponential decay different from the exponential decay in strong chaos.But we find there still have a little bit harder to do this work as there basically have two different main decay processes and how to distinguish the two decay processes is a essential step.From numerical viewpoint,this means we should pay attention to how to decide the time scale for doing the fitting procedure and we find there basically always have the situation about decreasing the fitting decay exponent if we continually use a carefully selected time interval to get a sequence of fitting exponents which smooth the fluctuation corresponding to the first decay process,then always have the situation to increase the decay exponent which can be seen to enter the second decay process taken as the transitive time.Further more,what we really care about is the main decay law,so we can combine these two different decay processes to compute the decay exponent.In detail,we can get a final time corresponding to a small number of $M(t)$ and we calculate the decay exponent of the main second decay process from the transitive time to the final time and then if the transitive time is more than final time we set which means the dominate decay process is the first decay process,and we can calculate the decay exponent of the first decay process as the main decay process.

Further more,we also want to check the accuracy of our theoretical method.In terms of our previous study of semi-classical method based on the Levy distribution,we can find it seem to be good to use fewer frequencies to hold the monotonicity of fitted semi-classical results but it doesn't mean it can be more accurate.Thus we use another ten frequencies to fit the semi-classical evaluation for the comparison and we can find the complicated expressions mainly based on the non-common monotonicity of fidelity evaluation have also been studied before and the variation of effective decay time in terms of changing perturbation corresponding to the different accuracy of Levy assumption.We consider the perturbations corresponding to the field from Fermi golden regime to independent regime in strong chaos and the numerical procedure is to use the 20 time interval to get a sequence of local decay exponents and then consider the situation about the turning time where fitting local exponent begin to increase beyond the threshold 0.01 we set,at last we can consider the effective final decay time numerically set as the decay value smaller than $10^{-4}$.As there have the initial quantum frozen time\cite{Gorin},so we do not consider the initial 25 time steps and just consider the fitting time  after that initial time.For simplicity,we call the three distinguished time as the $t_1,t_2,t_3$ and we fit the decay exponent with the time scale between $t_2$ and $t_3$ if $t_3>t_2+2$,and if this kind of condition can not be satisfied for some certain perturbation,we then fit the decay exponent with the time scale between $t_1$ and $t_2$ for the remain perturbation as we have taken the decay process for remain perturbation is dominated by the first decay process.In practical computation,$t_2$ should have the limitation below $t_3$,so we should change the situation for the time scale $[t_1,t_2]$ to $[t_1,t_3]$ when the limitation can not hold any more.At last,for small perturbation,there have some initial unexpected fluctuation leading to $t_2$ can be far smaller than actual reasonable value,so we particularly set a numerical condition that if $t_2$ is smaller than 100 for the perturbation smaller than 0.1,we do not consider and reconsider the new one to replace it.Here we should pay attention to $t_2$ is the times of $20$,and the study results are expressed in the figure 50 and we can find our theoretical method using direct integral is good for most cases in terms of not so large perturbation and the semi-classical evaluation with Levy assumption is basically better for small perturbation as the time for forming Levy distribution is somewhat quite long,and with the perturbation increased,the effective decay time become shorter and it seems basically harder to have a Levy distribution.

\begin{center}  
\begin{figure}
\includegraphics[width=18cm,height=20cm]{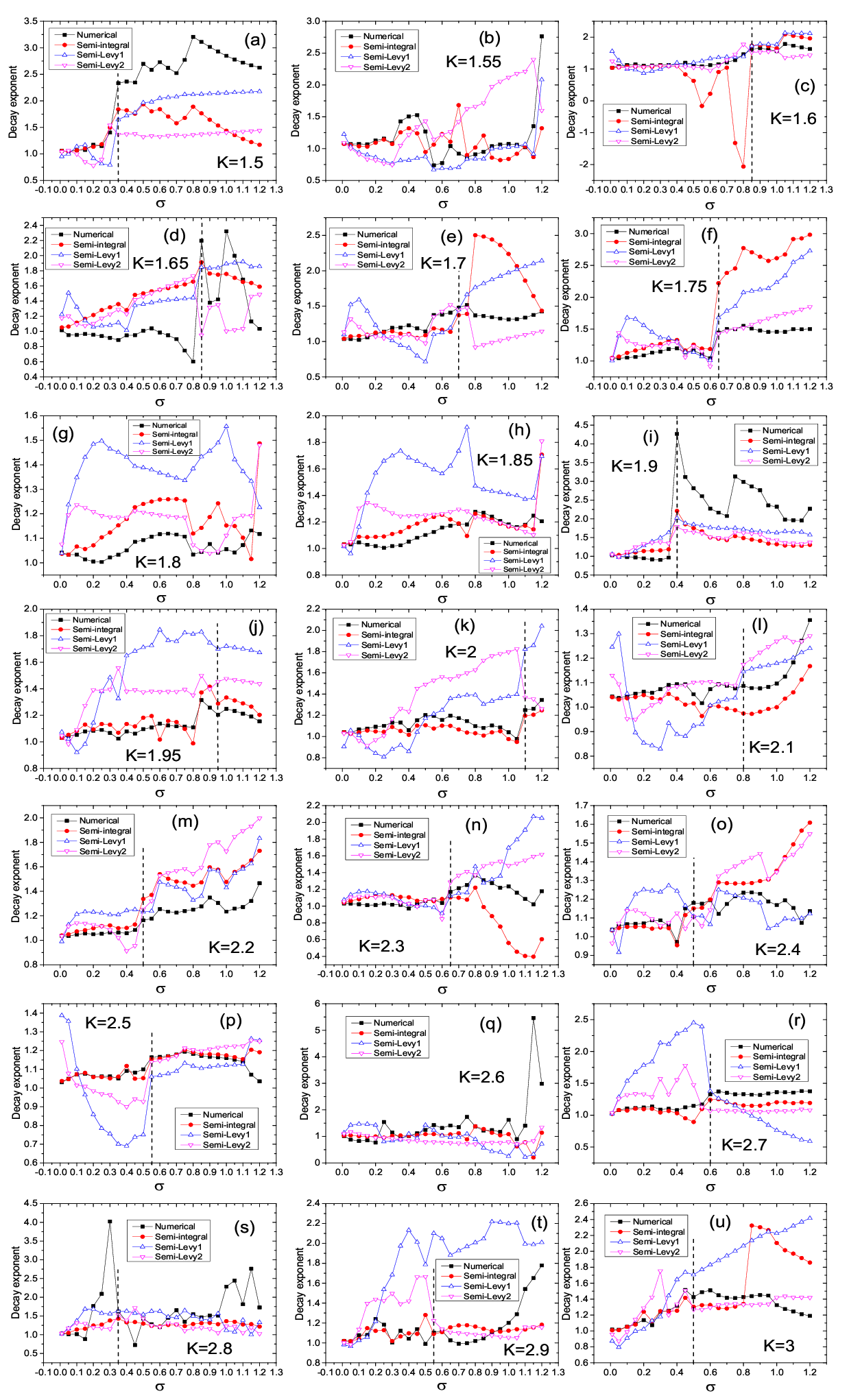}
\vspace{-0.5cm} \caption{Decay exponents fitted from direct quantum fidelity,direct semi-classical integral,semi-classical evaluation from four frequencies fitting and ten frequencies fitting based on the assumption of Levy distribution versus the changed perturbation from $\sigma=0.01$ to
 $\sigma=1.2$ we choose.The dash line is to indicate the alternation for the fitting time scale from $[t_2,t_3]$ to $[t_1,t_2]$ where $t_1$,$t_2$ and $t_3$ are the initial staring time corresponding basically to without the quantum frozen situation,turning time from the first decay process to the sencond decay process and the effective final time corresponding to the decay value smaller than $10^{-4}$,all the time scale should have the limitation smaller than the final time we set.The decay exponents fitted from our semi-classical integral can be basically close to the quantum decay exponent for some field of perturbation corresponding to most cases,and the fitted results from Levy assumption with initial small perturbation is basically better than the large perturbation afterwards.The somewhat unsatisfied expression of semi-classical evaluation from Levy distribution show the Levy assumption is not good for some large perturbation as the time to form a Levy distribution need quite a long time and the theory itself also can not be a entirely good approximation studied before.The different and somewhat complicated expressiones for four and ten frequencies fitting reflect the non-common monotonicity of the evaluation.}\label{sea_decayexp}
\end{figure}
\end{center}

\begin{center}  
\begin{figure}
\includegraphics[width=16cm,height=6cm]{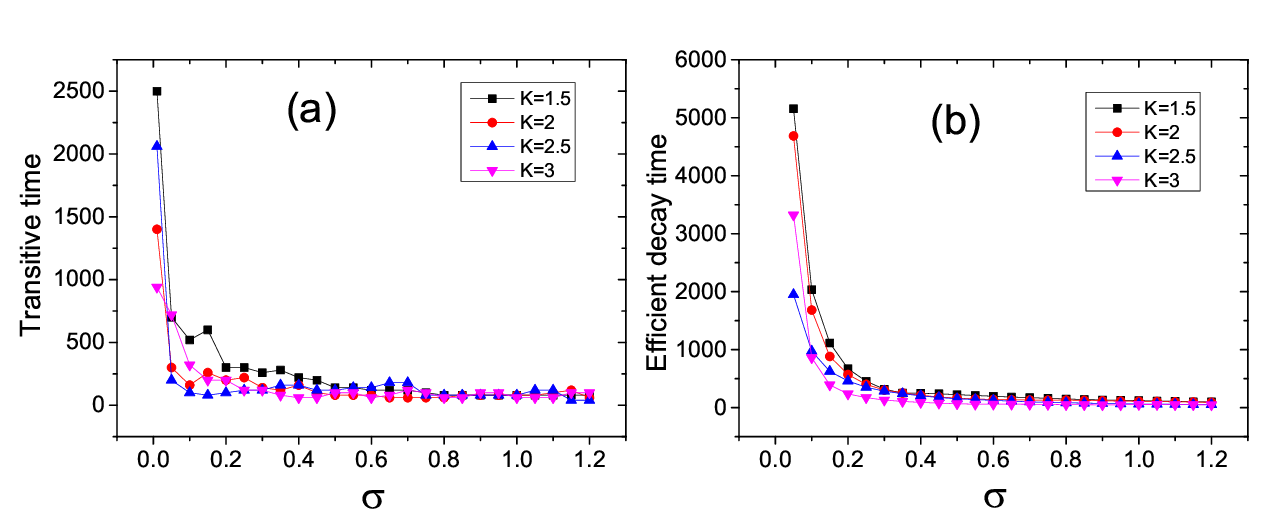}
\vspace{-0.5cm} \caption{The variation of transitive time and efficient decay time within the decay value for $10^{-4}$ via perturbations we choose.As the rapidly decreasing tendency for efficient decay time via perturbation leads to it is very large for small perturbation,so we do not consider $\sigma=0.01$ for the variation of efficient decay time but included in the variation of transitive time.The tendency to decrease for the transitive time is comformed to our expectation but not with entirely monotonicity from our numerical procedure to check the variational monotonicity of the local exponents fitted with $20$ time step illustrated in detail in the previous study,therefore the time we show here is the integer multiples of $20$.}\label{time_scale}
\end{figure}
\end{center}

   Now we naturally want to think about the issue of time scale which is a very important question but there still have a very few theoretical knowledge about it.From our numerical investigation,we can not find a general formula to summarize the time scale via perturbation.So now we can consider two situations numerically,one is the transitive time and the other time is the efficient decay time within the decay value not below $10^{-4}$ we set sufficiently larger than the saturation value basically with the order $10^{-6}$ numerically observed.Therefore we show our main numerical expressions with some cases that we choose evenly distributed for system parameter $K$ and depict them in the figure below.

   Then we should consider the effectiveness of our semi-classical theory to edge of chaos,but unfortunately we find there basically have not the Levy distribution from a lot of numerical study and should check the direct semi-integral from our theory and surprisingly find it can partly agree with the direct quantum result but with limited decay process,and the tendency is very clear that the agreement can become better as the quantum wave-packet gradually leaves the edge.It is a highly non-trivial result challenging our understanding the semi-classical theory in a more deep level.If the dephasing representation put forwarded by J.Van\'{i}\v{c}ek and Eric J. Heller is effective,then variation of the importance of fluctuation term should be considered and remains a open problem in the future research.Now we give the numerical result in the figure below and find there have obvious corresponding relationship between quantum expression and the dynamics of classical ensemble.

   From the numerical study,we can find the approaching degree with the semi-classical theory to evaluate the quantum fidelity tends to have a obvious change for much less accuracy during the critical variation of escaping to the chaotic sea or not for the classic correspondence,but we still find even under this kind of situation,the accuracy can be improved with the perturbation increased illustrated clearly in the figure using the case of $p_{center}=1.868$ and $p_{center}=1.872$.We have found the negative effect of the accuracy of semi-classical theory with the perturbation increased but there have basically positive effect.We also pay high attention to the revival situation from the semi-theory and weakened during the process of leaving the edge as well as the process for escaping to the chaotic sea.

  Therefore,there have a very important question about the effectiveness of our semi-classical theory with the variation of different perturbation and related induced result such as the revival situation existing in the edge of chaos.Obviously we need study more about the semi-classical theory and fluctuation term maybe need to be investigated more.Now it remains a open problem needing more models and theoretical investigation to find the precisely condition to apply our semi-classical theory as a future work.Based our study,we also find the Levy distribution is not a good approximation for $P(s)$ and it is a great interest to investigate what kind of typical distributions in terms of time variation.Then we show our numerical results in the very first time to see the typical distributions and find the variations of the contribution from the long tail of $P(s)$ show the transition of the situation gradually escaping to chaotic sea from the edge of chaos,meanwhile maybe the most feature here is the peak-like shape of $P(s)$ for the typical distribution for the edge of chaos clearly depicted in the figure below.

Now there remain a important question that can not be investigated carefully yet but it is vital to understand the decay process of Loschimidt echo,and this property is the time scale. Thus we can check to find whether there have a extension about the decay formula as $M(t)\approx e^{-{c_0}\sigma^{\nu}t^{\alpha(\sigma)}}$ for the common expression as $M(t)\approx e^{-2.2\sigma^{2}t}$ corresponding to the classical limit of strong chaos.Thus we need to figure out the expression of                                                                                                                                                                                                                                                                                                                                                                               $\nu$ and $c_0$ as well as the $\alpha$ discussed above but now investigated more within the semi-classical comparison,further more we need to find a new decay law as far as we could.

Firstly we want to check our idea directly which means we should find the relationship from numerical support.Thus we can consider a efficient procedure like this:if we consider $M(t)\approx e^{-{c_0}\sigma^{\nu}t^{\alpha(\sigma)}}$,then \be \label{decay_law}\ln(-\ln{M(t)})\approx \ln{c_0}+{\nu}ln{\sigma}+{\alpha}\ln{t}\ee
For a given time with different perturbation,we expect to observe the linear relationship for $ln(-ln(M(\sigma)))$ versus $ln(\sigma)$ and also can obtain the fitting slope as variable $\nu$.Then with the variation of time,we can obtain the information of the change of $\nu$ which is a key point here as we expect there will be some obvious deviation from the strong chaos with the $\nu$ always be 2.Then we also have a great interest in the variable $c_{0}$ and also want to investigate the variation to see the likely different expression.Based on the widely numerical experience for the study of variation of fitting $\nu$ that there basically always is the good linear relationship for $ln(-ln(M(\sigma)))$ versus $ln(\sigma)$ within a quite long time for sufficient small field of perturbation.Therefore we can use the small perturbation $\sigma=0.01$ to extract the variables $c_{0}$ joined force with the variable $\alpha$ fitted with time in terms of the equality  $ln(-lnM(t))=ln{c_0}+{\nu}ln{\sigma}+{\alpha}lnt$,therefore we can get the variation of $c_{0}$ with different time.With this method,we can escape the large fluctuation of the obtained variable if the two successive fitting procedures could be used which make the practical usage out of value.

Besides the corresponding comparison of direct semi-classical formula $M(t)\approx \left | \int ds e^{is\sigma} P(s)\right |^2$,we also should consider the corresponding treatment with the semi-classical formula as $M_{sc}(t)=\rm exp(-2(\epsilon/\hbar)^\eta D_L)$ in terms of Levy distribution consideration.Therefore we can assume $\sigma^{\eta(t)}\approx c_{\eta}{\sigma}^{\nu_{\eta}}t^{\alpha_{\eta}}$,and $D_{L}(t)\approx c_{D_{L}}t^{\alpha_{D_L}}$,for simplicity we call $c_{D_{L}}$ and $\alpha_{D_{L}}$ as $c_{D}$ and $\alpha_{D}$.Actually $c_{\eta}$ should be one for every fitting result in terms of a time step fixed and we do not consider this variable afterwards.One can find these considerations are basically same for previous study besides the explicit expression of perturbation $\sigma$ added.Actually the expression $\sigma^{\eta}$ accounts for the term $\sigma^{\nu}$ and the fitting result is just the $\eta$ itself if fitted using the $\eta ln{\sigma}$ versus $ln{\sigma}$.So we alternately consider the issue from the time fitting procedure and if we fit $\eta$ via $ln{t}$ and can get $\eta(t)\approx Aln{t}+B$,thus it leads to the expression as $\sigma^{\eta(t)}\approx \sigma^{B(t)}t^{A(t)ln{\sigma}}$ with $\nu$ as $B(t)$ and $\alpha_{\eta}$ as $A(t)ln{\sigma}$.Combining $\alpha_{D}$,we can get the decay exponent $\alpha$ which can be taken for the comparison.During the previous study,we find the semi-classical formula $M_{sc}(t)=\rm exp(-2(\sigma)^\eta D_L)$ can not be good for the perturbation increased and corresponding $c_D$ and $\nu_{\eta}$ basically can not agree well with the direct quantum parts and also have some very large fluctuation,thus we do not show in this paper.

Now we show the relationship for $ln(-ln(M(\sigma)))$ versus $ln(\sigma)$ for different typical time corresponding the system parameters $K$ we study,and the related figure illustrates the universal good linear relationship for initial time and then basically have some deviation from the linear relationship with the time increasing stemmed from the large perturbation field.If one want to fit the slope for the very initial time as $t=1$ and can find it definitely is very close to 2 with the precision smaller than $10^{-6}$.As the extensive numerical observation,we can expect it as a universal rule in the chaotic sea having mixed-type phase space.Here we also give the result of semi-integral as the reference to find the proximity to the direct numerical computation and the result can have a quite good agreement except the initial relative short time within the order $10^{2}$ and tend to better with the time increasing.This expression is consistent with the previous study as the good agreement of semi-classical evaluation with the fidelity in terms of absolute difference consists in the small perturbation field as well as relative long time without the very initial time,then for a long time only small perturbation can give the effective and also accurate contribution without entering the saturation field for other relative large perturbation which also can be seen clearly in the figure corresponding basically a platform around $2.5$ for the value of $\rm ln(-\rm ln(M(\sigma)))$.Although a little bit distorted by the time increasing according to all the perturbation field,the very good linear relationship still can hold for sufficient small perturbation even corresponding to a quite long time basically independent of different system parameter $K$,and it gives us the indication that our assumed decay law exists to some extent and thus it guarantees to get the reliable variables $c_{0}$ and $\nu$ from the fitting technique described before.Last but not least,the tendency of increasing the absolute difference in the comparison of semi-integral and direct numerical computation versus perturbation $\sigma$ in particular for a fixed short time can not be reflected well in the variation of $\rm ln(-\rm ln(M(\sigma)))$ as the high non-linearity of double logarithmic function.

\begin{center}  
\begin{figure}
\includegraphics[width=18cm,height=20cm]{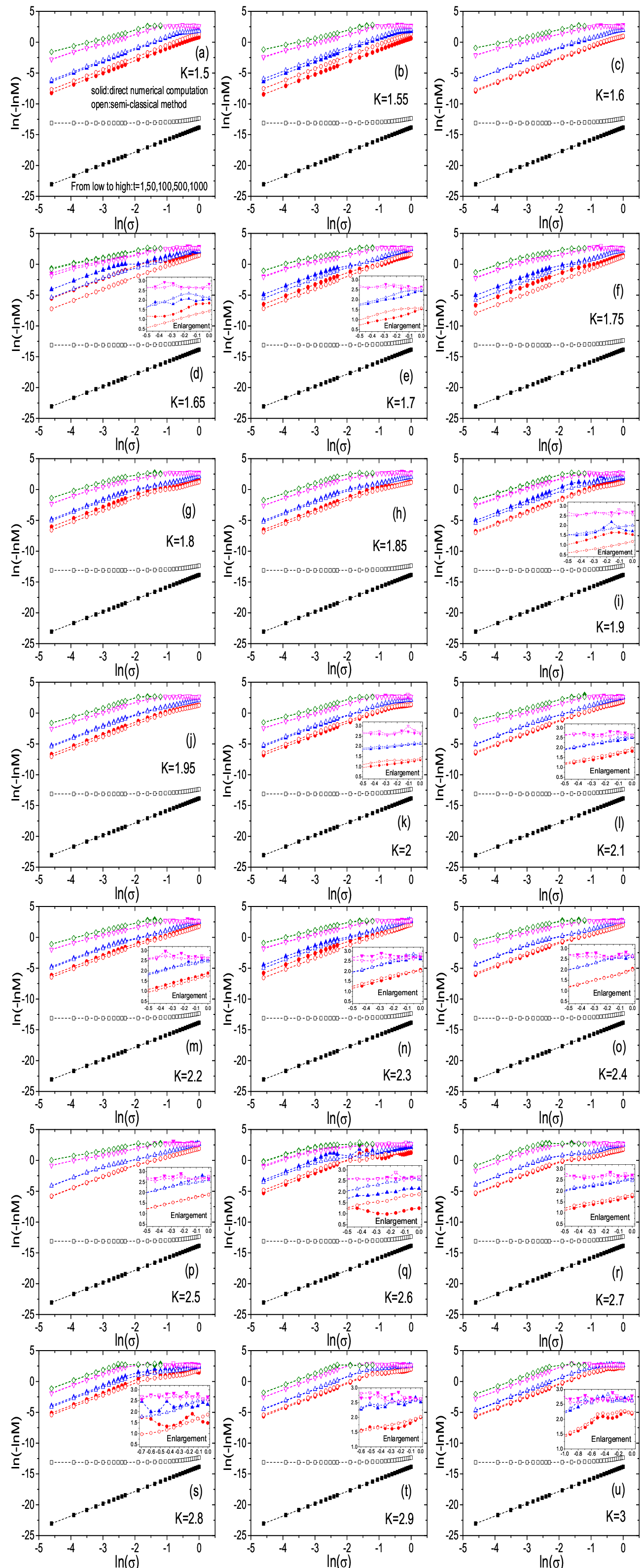}
\vspace{-0.5cm} \caption{Comparison of the variation of $\rm ln(-\rm ln{M})$ as a function of $\rm ln(\sigma)$ for direct numerical computation and semi-classical integral is represented with different time $t=1,50,100,500,1000$ corresponding to different system parameter $K$.To clearly show the variation in detail,$\sigma$ is taken from $0.1$ to $1$ with the interval as $0.05$ but $0.01$ for the field from $0.1$ to $1$.From direct numerical result,the good linear relationship can not hold with the time increasing and always is destructed from the field of large perturbation and gradually extends to the small perturbation field.The result from semi-classical integral agrees well with direct numerical computation for the long time and have some deviation for short time and this expression is consistent with the previous study as the better agreement of semi-classical integral with the fidelity consists in the smaller perturbation field and the effective contributions in the long time decay come from small perturbation.}\label{linear_relation_sigma}
\end{figure}
\end{center}

As $\rm ln(-\rm ln{M})$ can loosely be taken as a linear function of $\rm ln{\sigma}$ although there is the very good linear relationship existing only in the sufficient small perturbation field for a long time for most cases one can find,we can get the value of $\nu$ from the fitted slope.Obviously we need to have some reasonable conditions to effectively get the expected fitted $\nu$ and here is the two conditions we consider.One condition is the value $M(\sigma)$ should not too small approaching the saturation field for a given time and thus we just consider it should not smaller than $10^{-5}$ as the value of $10^{-6}$ can be seen as entering the saturation field.Another condition we set is related to the extent of tolerance for the distortion of the linear relationship and only increasing order for $\rm ln(-\rm ln{M})$ versus $\rm ln{\sigma}$ can be put into our consideration which means in detail we can choose from the contrast of two adjacent values of $\rm ln(-\rm ln{M})$ and the second one is taken as the reference making actually the last $\rm ln(-\rm ln{M})$ can not be included for fitting and thus we extend the study field to $\sigma=1.2$. What we want to find is the variation of $\nu$ which is expected to different from the $\nu=2$ and try to find some universal expression with extensive study for different system parameter $K$.Then we show our study result in the related figure.To show some kind of big fluctuation of the variation of fitted $\nu$ in terms of short time,we use the logarithmic coordinate to illustrate it in detail.

\begin{center}  
\begin{figure}
\includegraphics[width=18cm,height=22cm]{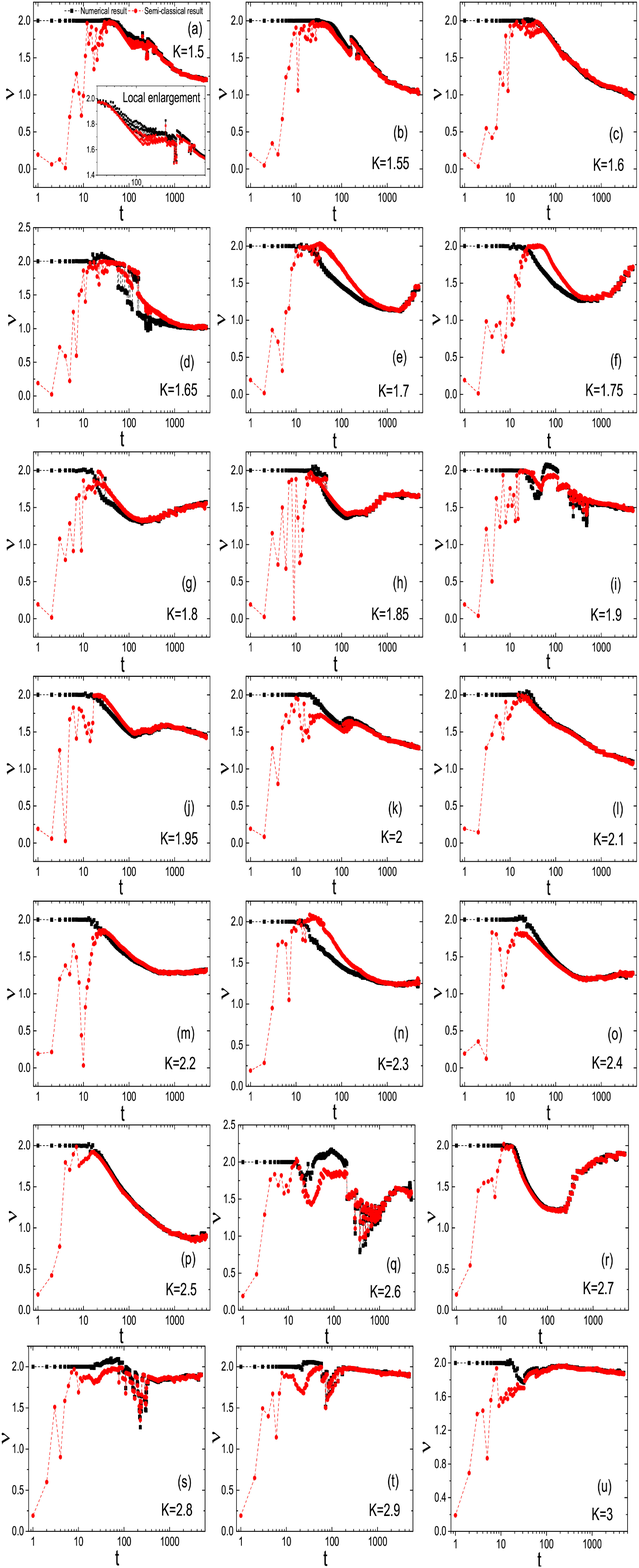}
\vspace{-0.5cm} \caption{Fitted $\nu$ as a function of time from direct numerical computation and semi-classical integral corresponding to different system parameters $K$.As there are some large discrepancy between them as well as some large fluctuation within the initial time up to the order $10^{2}$,we use log coordinate to express it in detail.One can find the variation of $\nu$ basically is put into the field between the numerical values 1 and 2 which are related to the typical indicators of time scale as $\sigma^{-1}$ and $\sigma^{-2}$ for the fields of stable dynamics and strong chaos and it is a universal expression for the very initial time having the exactly value $2$ independent of different parameters $K$.One also can find the pattern of variation can have some good continuation with $K$ increasing.Some discreteness in the process of variation of $\nu$ one can find dues to the large deviation of good linearity for $\rm ln(-\rm ln{M})$ versus $\rm ln(\sigma)$.Without initial time,there are basically good agreements for the $\nu$ fitted from numerical value and semi-classical integral.}\label{nu_t}
\end{figure}
\end{center}

    From our study,we find the variation of $\nu$ basically is between the value $1$ and $2$,and it is non-trivial to find there is the exact value $\nu=2$ for the very initial time independent of system parameters $K$ we choose.As we have pointed out that the time scales for the region of stable dynamics and strong chaos are $\sigma^{-1}$ and $\sigma^{-2}$ correspondingly and the fitted $\nu$ between them reasonably conform to our expectation and we can find there are some complicated change for the pattern of variation of $\nu$ but having some good continuation as well as the final tendency to enlarge the value towards $2$ corresponding to strong chaos as system parameter $K$ increases.As the linear relationship for $\rm ln(-\rm ln{M})$ versus $\rm ln(\sigma)$ is just our rough approximation from our numerical study and its cost consists in the some discreteness happened during the variation of fitted $\nu$ when the large deviation of good linearity happens.As time goes by,the perturbations for fitting will be smaller and a good agreement for the comparison between the quantum fidelity and our semi-classical integral can be expected generally for small perturbation,and thus we also can expect a good agreement for the comparison of fitted $\nu$ for a relative long time and it is indeed right from our study excepts the rule described above has to be broken obviously for $K=2.6$ illustrated in out figure with some poor efficiency of our semi-classical method even for small perturbation.

From the study of the variation of $\nu$,it seems that there is the approximation for the linear relationship for $\rm ln(-\rm ln{M})$ versus $\rm ln(\sigma)$ and hence the local decay exponent could independent of the perturbation $\sigma$ if only $c_{0}$ can hold the same.From our extensive numerical study,we indeed find the similar decay exponent for different perturbations existing for a relative long time with the order about $10^{2}$ and the value is between $1$ and $1.1$ mostly.As there is two different main decay processes and the first one can gradually evolve into the second one and the variation of $\alpha$ shall be obtained from the first decay part to the second decay part with perturbation $\sigma$ increasing for a give time which can show the decreasing pattern.For the very initial short time,we find similar common decay and from our study here the decay exponent is almost same with quite large value as even more than $6$ which surely depends on the fitting time we take.The initial large decay exponent is universal from our study and we expect it is the decay feature for the mixed phase space and for a sufficient strong perturbation,one can predict a very fast decay within this decay time and get extensive numerical observation.The origination for the quite large decay exponent even more than the Cubic-exponential decay\cite{vanicek_arxiv}is not clear by now but the classical-quantum correspondence could work which can be taken as the starting point for future research.

Based on the previous analysis of the decay exponent from semi-classical treatment with the consideration of Levy distribution,it can be divided into two parts and one of them including the factor $\ln(\sigma)$ tends to zero when $\sigma$ approach $1$,and the variation of local decay exponent $\alpha$ can be monotonously increased or decreased in terms of Logarithmic law relying on the positive or negative value of $A(t)$.To express the variation clearly also for the field of small perturbation,we use the Logarithmic coordinate with the base as $10$ to illustrate it with a additional factor as $\ln(10)\approx 2.3026$ multiplied by $A(t)$.Therefore,we show the result with the numerical and semi-classical method to treat the decay exponent together in the figure we give below and pay a high attention to the discrepancy among them to illustrate the typical decay features with different time scale we choose.For simplicity,we use the symbol $\alpha$ to represent the local decay exponent fitted in terms of a given time step which is different from commonly used name as decay exponent fitted with a whole time scale we choose.To guide anyone having interest in the comparison,there are two things we should consider.One is the closeness for $\alpha$ calculated from the semi-integral method to the direct numerical result and we should pay attention to the effectiveness of our semi-classical method as the good agreement can not reach out too far for most cases to the field of large perturbation shown clearly in the previous study,and it means the semi-classical method we use itself have the limitation.The other one we should consider is the assumption of Levy distribution of $P(s)$ and we also check it and find we could have a good expectation for a long time but it is not always the case with the implication that the variation is complicated originating form the classical dynamics in terms of mixed phase space.

To give a explicit illustration as there are quite a few system parameters $K$ we consider,we use the three figures to show the variation of $\alpha$ corresponding to different time we choose.As we know,the decay exponent is a key point to understand the decay laws and the variation of $\alpha$ we study here obviously can give some important information.Firstly the local fitted decay exponent $\alpha$ for all different perturbation we choose below $\sigma=1.2$ can have the value much larger than $3$ within the initial time,the time scale is mainly not more than $t=30$ numerically observed with the time step as $10$.To clarify the fact is that we can not say all the $\alpha$ numerically fitted within this time scale definitely have this feature described above but always most $\alpha$ in this time scale can have this feature.Meanwhile the shorter of the time,the smaller difference of the variation of $\alpha$ for the perturbation from small to large.One can find this feature easily by calculating the absolute value of the difference of $\alpha$ corresponding to the largest perturbation and smallest perturbation for a given time,in our study,one can use the variable $\vert \alpha(\sigma=1.2)-\alpha(\sigma=0.01)\vert$ to find it.This variable can be small at least with the order $10^{-2}$ corresponding to the time at $t=10$ with the fitting time step also as $10$.The expression of $\alpha$ in a short time also depends on the fitting time step we choose but the basic features of $\alpha$ we described above hold and we find the difference can be very small below the order of $10^{-4}$ for the time $t=5$ if fitted using the time step as $5$.

As it is a universal expression for $\alpha$ tending to be same for a quite short time independent of the system parameters we choose,thus it shows there is a common mechanism accounting for it.We guess this feature have something with the Ehrenfest time $\tau_{E}\propto \ln(1/\hbar)$ having the order of time as $10^{1}$ and one can find there is indeed somewhat platform emergent in our semi-classical integral method to treat $\alpha$ for the time as $t=20$ we choose in the first panel of every sub-figure with a specific $K$ although the agreement basically can not be good.It is a open problem,classical-quantum correspondence could work taken as the future study.Although there is a quite small variation of the $\alpha$ for the very initial time,with careful study we also find the linear dependence of $\alpha$ versus perturbation $\sigma$ within the time scale as $t=15$ in terms of using the fitting time step as $5$ and it is similar for the fitting time step taken as $10$.For the time as $t=5$,the linear relationship indeed is common for all the system parameters,here we use the fitting time step as $5$.But the linear dependence is poor according to the small perturbation field from $\sigma=0.01$ to $0.1$ and then we just illustrate this linear relationship with the range from $\sigma=0.1$ to $1.2$ and it is a universal situation independent of selected parameter $K$ although the extent of linearity varies.We show the linearity in the figure below,but it is still not clear for the very reason to have this expression.During this linear dependence,we also find some fluctuation for some parameter $K$.

Based on our analysis,the variation of $\alpha$ versus $\sigma$ should obey the logarithmic law if the Levy distribution can give a very good approximation for $P(s)$ and the forming process of Levy distribution is not a simple process with time going.Further more the corresponding semi-classical result based on the Levy distribution relies on how many frequencies are used in terms of Fourier transform of $P(s)$,and here we use the $4$ and $10$ frequencies to get $\alpha$ with two elements coming from the $D_{L}$ and $\eta$ and the difference between them for $\alpha$ can be expected to be small when the Levy distribution can be a good approximation of $P(s)$ which is quite efficient for a relative long time as there are sufficient large number of frequencies to give the effective contribution.For simplicity,we call the different methods relying on taking different frequencies the names as Semi-Levy1 and Semi-Levy2 used in the figure below.Therefore,to understand our result carefully,one also need to retrospect our study of frequency relation showing the variation of linearity to represent the good approximation of $P(s)$ with time increasing.Effective frequencies are increased with the time going,this means there always is some difference between them leading us to care about mainly the result from the fewer frequencies used equivalent to the main consideration of Semi-Levy1.Even some good agreement of Levy distribution of $P(s)$,there still is some small difference that can be analysed using our method developed with the assumed exponential relation for $D_{L}$ versus $\eta$ but it is not important in our study with the focus on the regulars of the variation of $\alpha$ and Semi-Levy1 and Semi-Levy2 taking as the references.

For the time $t=100$,there is a typical crossover from the first transitive decay process to the second stable decay process characterized by $\alpha$ with the perturbation $\sigma$ increasing,this feature is also illustrated directly in our previous study with the relation for $\ln(-\ln{M(t}))$ versus $\ln{t}$ giving a basic fact that the first transitive decay process is gradually shortened by $\sigma$ increasing.One can indeed find the variation of $\alpha$ shows a typical pattern from some larger value to smaller value expected to above $1$ with some small fluctuation,but the large deviation from this typical pattern also can be seen in our study as for the situation of $K=1.65,1.9,2.6,2.8$ with large fluctuation after the dropping process of the larger value of $\alpha$ as well as the situation of $K=1.6,2.5$ with obviously decreasing tendency all along.The agreement of $\alpha$ fitted by the Semi-classical integral to the direct numerical result varies corresponding a specific $K$,and for most cases they are not very close but basically share the common tendency for the variation without the situation happened for the large fluctuation happened.It means so-called increasing inaccuracy to treat fidelity from the semi-classical method with the perturbation $\sigma$ increasing has some bound which also is reflected in the previous study of difference of the fidelity.Here as a special case,the expression of $K=2.5$ have a good agreement deserving us to attention.The relative difference of $\alpha$ obtained by the method of Semi-Levy1 and Semi-Levy2 indeed has the tight connection with the approximation of $P(s)$ using Levy distribution.If the effective frequencies is similar to the number $4$,the result from Semi-Levy1 is more closer to $\alpha$ fitted from direct numerical result and when the effective frequencies is more than the number $10$ but the linearity is not hold,we can expect some obvious difference between them,and further more the number of effective frequencies are not only enough compared to $10$ but also the linearity can be hold for the frequency relation,then the relative difference will be small.This connection can help us to understand the variation of $\alpha$ based on the Levy distribution of $P(s)$ for all the time we consider not limited to the time $t=20$.The some large fluctuation closer to the large perturbation during the process of approaching $\sigma=1$ happened for some parameters $K$ have not the same origination from the so-called large fluctuation just after the dropping process of the initial larger $\alpha$ as the previous one is due to entering the saturation field.

Then we use the special cases of $t=500$ and $t=2000$ to show the transition from the short time to long time for the variation of $\alpha$ versus $\sigma$.From our extensive numerical work,we can find $\alpha$ can not more than $1.2$ basically in the stable decay process even it is likely to have a very slow decreasing tendency for some parameters $K$.Thus the process of the crossover between two typical decay processes will be shortened with the time going and the sign to characterize the crossover is the initial value of $\alpha$ compared to the value $1.2$,if smaller than it we can acknowledge the variation of $\alpha$ just is set in the stable decay process.Based on our semi-classical analysis with Levy distribution of $P(s)$,the variation of $\alpha$ conforms to the Logarithmic law and it can be seen as the same as $A(t)$ from the variable $\eta(t)$ is very small.We numerically check this idea and find it is not always the case which means $\alpha$ is not necessary for same and show the linear dependence versus $\sigma$ with the slope taken as $A(t)\ln(10)$ in terms of using the Logarithmic coordinate.During this time scale,we can find the agreement of $\alpha$ fitted from the treatment of semi-classical integral is basically good as the perturbation here is relatively small and some large divergence with the perturbation increasing mainly consists in approaching the saturation field.We find agreement of $\alpha$ obtained from the methods of Semi-Levy1 and Semi-Levy2 can be good if the approximation of Levy distribution
for $P(s)$ works well in terms of long time for the time as $t=2000$ and it can not work very well for the time as $t=500$ with the situation of crossover of different typical decay processes dominating in the initial variation of $\alpha$ although the basic tendency can be described to some extent.The basic judgement of the good approximation of Levy distribution for $P(s)$ can simply be obtained from the relative difference of the variation of $\alpha$ in terms of using the methods from Semi-Levy1 and Semi-Levy2.The variation of $\alpha$ as a Logarithmic law indeed exists not only for the good agreement with $\alpha$ obtained from Semi-Levy1 and Semi-Levy2,even there is a some large difference among them,we still could find the good linearity for $\alpha$ versus $\sigma$ in terms of Logarithmic coordinate we use.One can easily find the linear relationship is a common situation for most cases at least in some field of $\sigma$ whatever the methods from Semi-Levy1 and Semi-Levy2 can work well or not such as for the cases of $K=1.75,1.8,2.7$ of the time as $t=500$,to name a few.This fact actually is beyond our semi-classical method used in this paper and indicate a deeper theoretical analysis in the future.

\begin{center}  
\begin{figure}
\includegraphics[width=18cm,height=18cm]{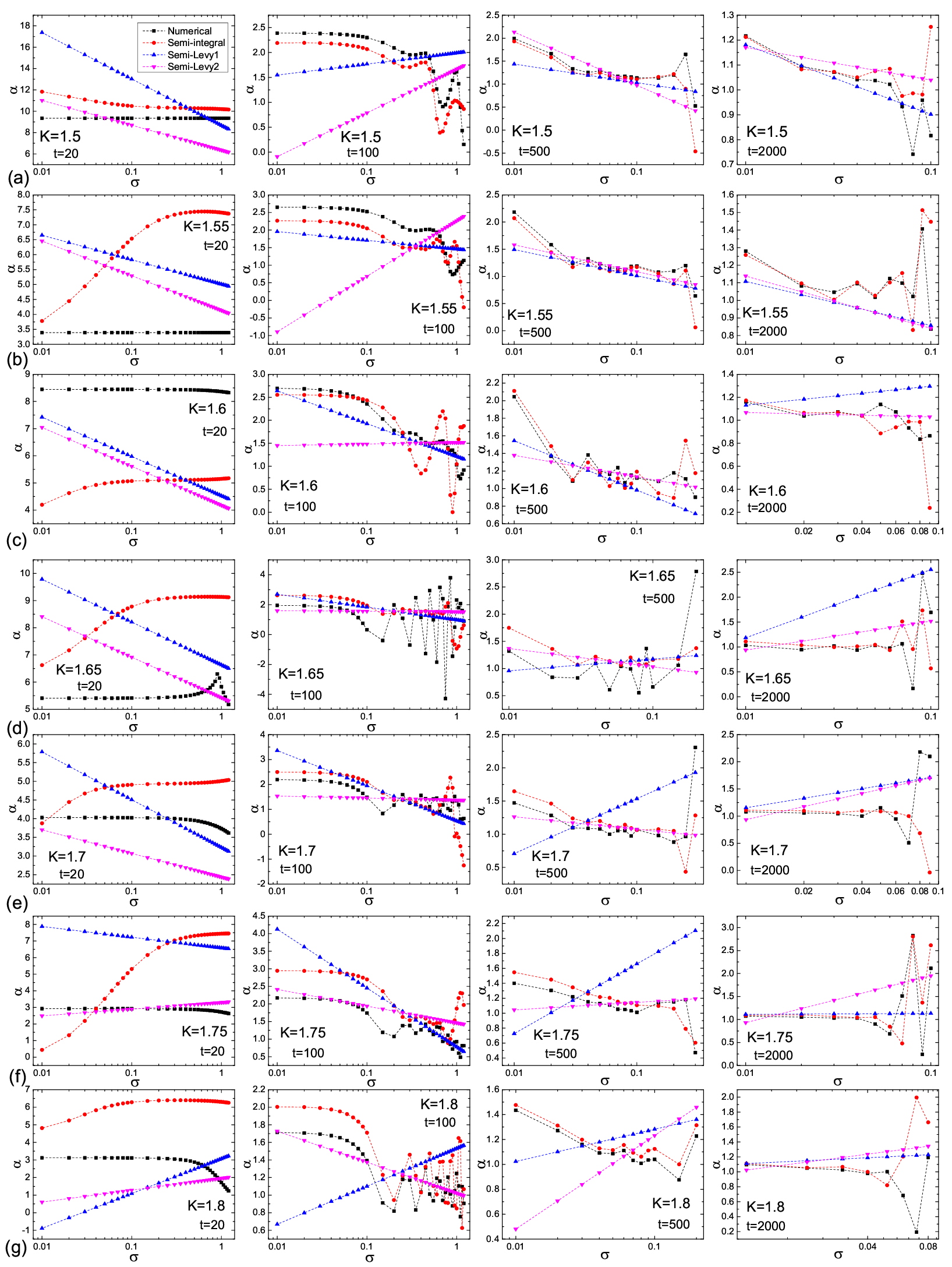}
\vspace{-0.5cm} \caption{Variation of the local fitted decay exponent $\alpha$ versus the perturbation $\sigma$.The four results of $\alpha$ are obtained respectively from the direct numerical computation,semi-classical integral,and the assumption of Levy distribution of $P(s)$ with $4$ and $10$ frequencies(call the names as Semi-Levy1 and Semi-Levy2) taken for the comparison of agreement.In this figure,we consider the parameter $K$ from $1.5$ to $1.8$.For a very short time as $t=20$,$\alpha$ from the direct numerical result can hold the very close value from small perturbation to some large perturbation or even all the field of perturbation we choose.The agreement is not good but we find there is always the plat region forming gradually for the $\alpha$ obtained by semi-classical integral.As there are typically two main decay processes as the transitive and stable decay processes along with time going,then the change of time means the different crossover from the former one to the latter one with $\sigma$ increasing.When time increases,this crossover feature shall be weakened as the initial value of $\alpha$ will decrease as well as the connected dropping process of $\alpha$ from the initial value will be also shortened.During the change from short time to long time,there is indeed such a transition seen clearly from every sub-figure of a specific $K$ for the variation of $\alpha$ and also the agreement of $\alpha$ from semi-classical integral to the corresponding direct numerical result show a better tendency.The agreement of $\alpha$ obtained from Semi-Levy1 and Semi-Levy2 is based on how good is for the approximation of $P(s)$ using Levy distribution and can be reflected by the relative difference of $\alpha$ between them and if the number of effective frequencies corresponding linear part in terms of frequency relation is below $10$ and $\alpha$ obtained from Semi-Levy1 have a better agreement such as for the case of $K=1.55$ with time fixed as $t=100$.The Logarithmic law represented by the linear relationship seems to not only exists for the good agreement of $\alpha$ obtained from Semi-Levy1 and Semi-Levy2 such as the cases of $K=1.75,1.8$ for $t=500$.}\label{alpha_sigma1}
\end{figure}
\end{center}

\begin{center}  
\begin{figure}
\includegraphics[width=18cm,height=20cm]{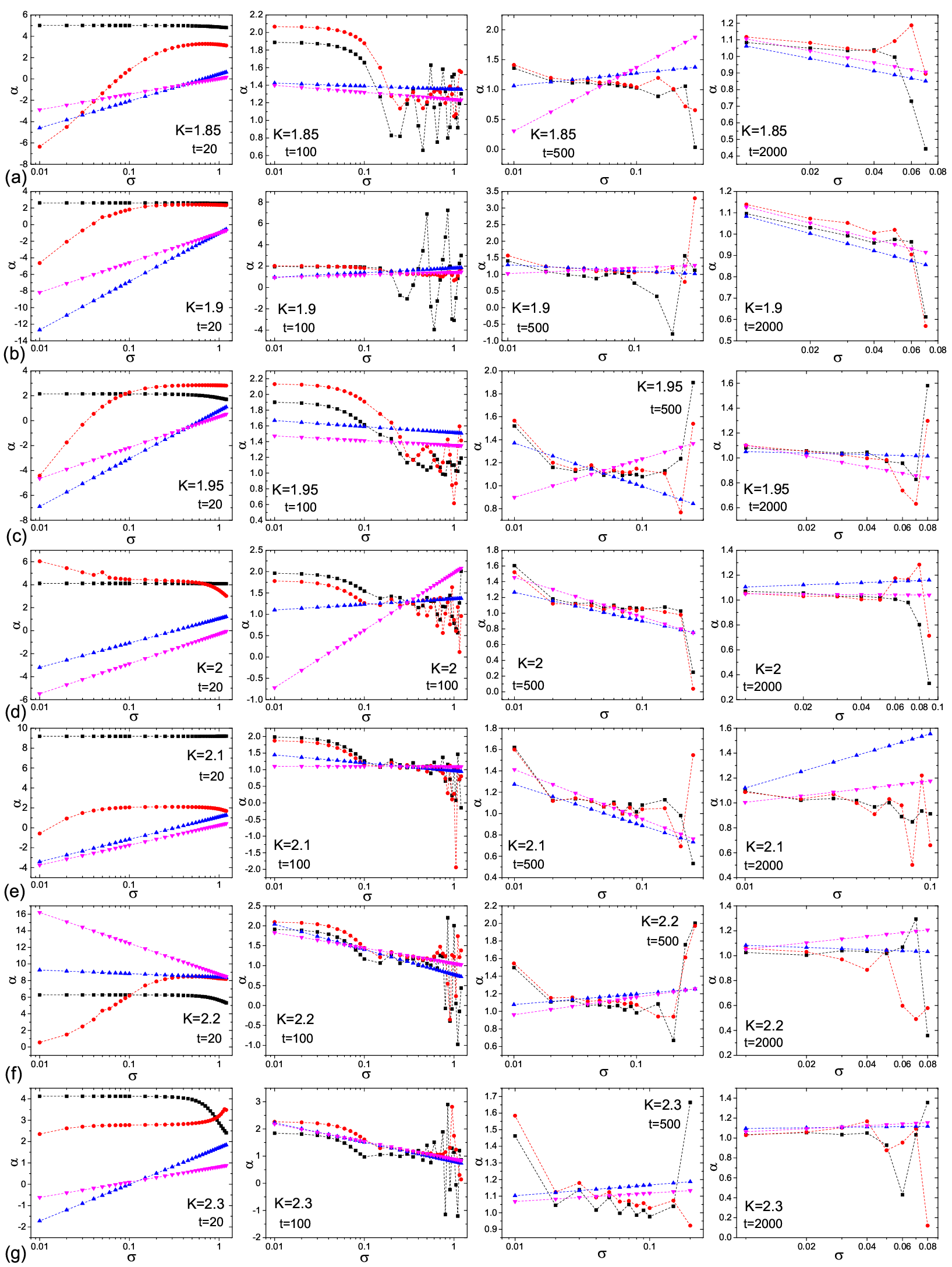}
\vspace{-0.5cm} \caption{Similar to the last figure,and here we consider the parameter $K$ from $1.85$ to $2.3$.There is the similar features for the variation of $\alpha$ and we want to concentrate some specific expression deserving attention.When time goes,the changing amplitude of the variation of $\alpha$ have a clear tendency to decrease as the crossover is weakened explained before in the last figure,but even for the time deeply setting into the so-called stable decay process corresponding to a long time,there still is some cognizable variation that can not be simply neglected clearly seen for some parameters $K$ with $t=2000$.In this figure we can find the typical pattern of the variation of $\alpha$ with the crossover from the initial larger value to the smaller value with some small fluctuation can not hold obviously for the case of $K=1.9$ for $t=100$ similar to the expression for the case of $K=1.65$ for $t=100$ in the last figure,thus it is not a exceptional one.The agreement of $\alpha$ obtained from semi-classical integral is far from good when such a large fluctuation happens from our study.}\label{alpha_sigma2}
\end{figure}
\end{center}

\begin{center}  
\begin{figure}
\includegraphics[width=18cm,height=20cm]{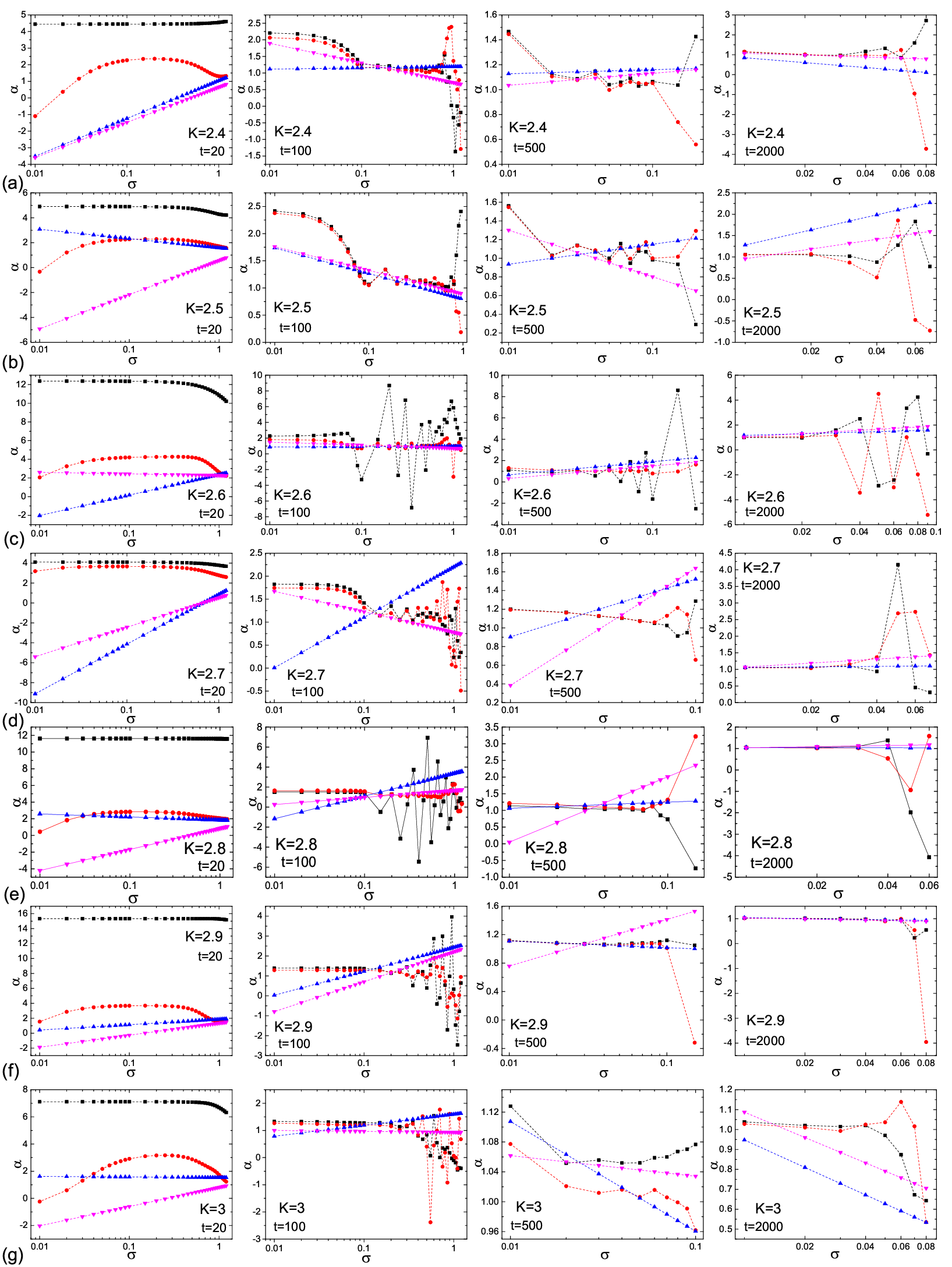}
\vspace{-0.5cm} \caption{Similar to the last two figures,and here we consider the parameter $K$ from $2.4$ to $3$.There is the similar features for the variation of $\alpha$ and there is not any other important features that can not be pointed out in the last two figures and once again we can find the big fluctuation after the dropping of initial larger value of $\alpha$ for the cases of $K=2.6,2.8$ for $t=100$ where the good agreement of $\alpha$ obtained from the semi-classical integral to the corresponding numerical result can not be seen even some similar tendency of variation of $\alpha$ compared with the direct numerical result also lost.}\label{alpha_sigma3}
\end{figure}
\end{center}

\begin{center}  
\begin{figure}
\includegraphics[width=18cm,height=20cm]{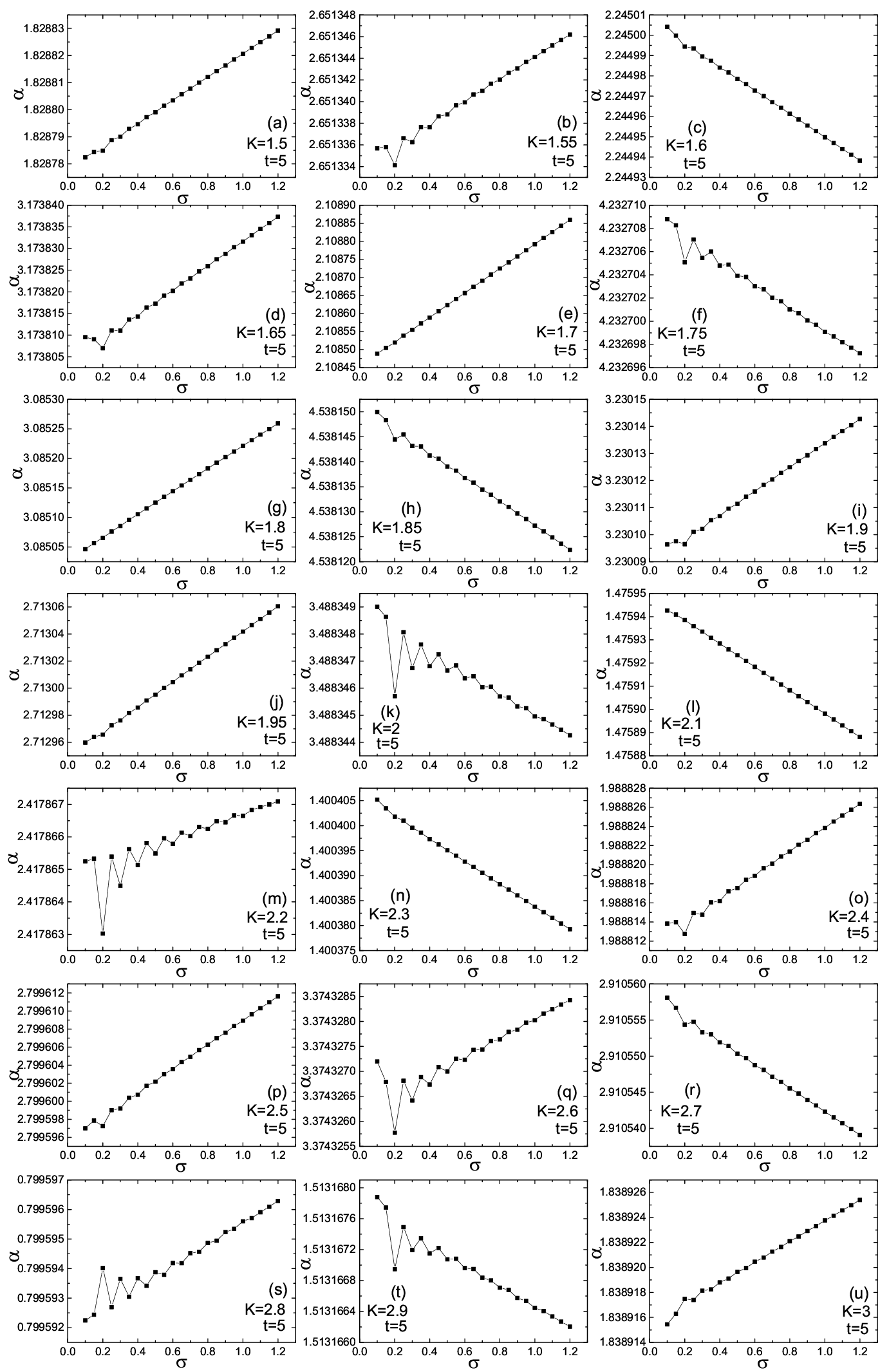}
\vspace{-0.5cm} \caption{The linear dependence of $\alpha$ on perturbation $\sigma$ fixed with the time as $t=5$.As there is some large deviation from the linear relationship for small perturbation and we use the field of perturbation from $\sigma=0.1$ to $\sigma=1.2$ to illustrate it.Some fluctuation small or large stemming from the initial part can be seen for most cases in the figure,but the basic linearity is very clear need to be explained theoretically yet.}\label{alpha_sigma_5step}
\end{figure}
\end{center}

  Based on the study of decay laws,we still need to get the information about the variable $c_{0}$ which is a constant for strong chaos in terms of classical limit.As the rough linear relationship for $\rm ln(-\rm ln{M})$ versus $\rm ln(\sigma)$,we can approximately take $c_{0}$ as a function of time $t$.Then we can get $c_{0}$ from Eq.~(\ref{decay_law}) with a specific small perturbation $\sigma=0.01$ used.We expect $c_{0}$ can be stable at some value although with some possible fluctuation.Therefore,we illustrate our study results in the figure below obtained from direct numerical computation and semi-classical integral as the comparison.To escape the likely fluctuation,we use some large time step as $t=50$ to get $c_{0}(t)$.As the longer time that we consider,the smaller perturbation that should be used to get the fitted $\nu$ which is the base to obtain the variable $c_{0}$.The smallest perturbation in this paper we take is $\sigma=0.01$ as the smaller perturbation will make numerical computation specially time-consuming as well as the basic tendency of the variation of $c_{0}$ is our main consideration here,then we use the whole time as $t=5000$ making the effective perturbation basically under the field we consider.

  From the figure we illustrate during the time,we can indeed find there is a clear transition from very small value to asymptotic stable value although having some large oscillation for quite a few parameters $K$.$c_{0}$ actually can be seen as to be stripped from the decay rate originally defined in the strong chaos in terms of classical limit and they could share some similarity between them.From our study here,one can find this similarity from the previous study of decay rate with the perturbation fixed as $\sigma=0.01$.As our calculation is based on the good linear dependence of $\ln(-ln{M(t)})$ on $ln{\sigma}$ with every time fixed,and it can be particularly valid for the field of small perturbation but there is some exceptional cases obviously for the variation of $c_{0}$ of $K=1.7,1.75$ having some large discontinuity corresponding to a long time larger than $t=2000$,although the considered perturbation for fitting is small for a long time but the linearity for these two cases is lacking from our careful check.We emphasize that it is different from the situation about the large fluctuation obviously for $K=2.5,2.6,2.7,2.8$ that also can not be trusted as the perturbation for fitting is quite few for quite a long time.  With parameter $K$ increasing,we can find the order of magnitude of $c_{0}$ undergoes a complicated change.For the parameter $K$ not more than $2.5$,the stable value is basically below $0.2$ and even smaller than $0.1$ or around for most cases.Above $K=2.5$,we can find there is a big jump to some value larger than $1$ and then the large oscillation just like before can not be observed.The value of $c_{0}$ is clearly has a tight connection with the variation of $\nu$,and we can find there also is a big change from $K=2.5$ to $K=2.6$ for the variation of $\nu$ in the previous study,for a simply analysis from Eq.~(\ref{decay_law}),obviously they are positive correlated for $\nu$ and $c_{0}$,the enlargement of $\nu$ makes $c_{0}$ increases in a high non-linear way.

\begin{center}  
\begin{figure}
\includegraphics[width=18cm,height=20cm]{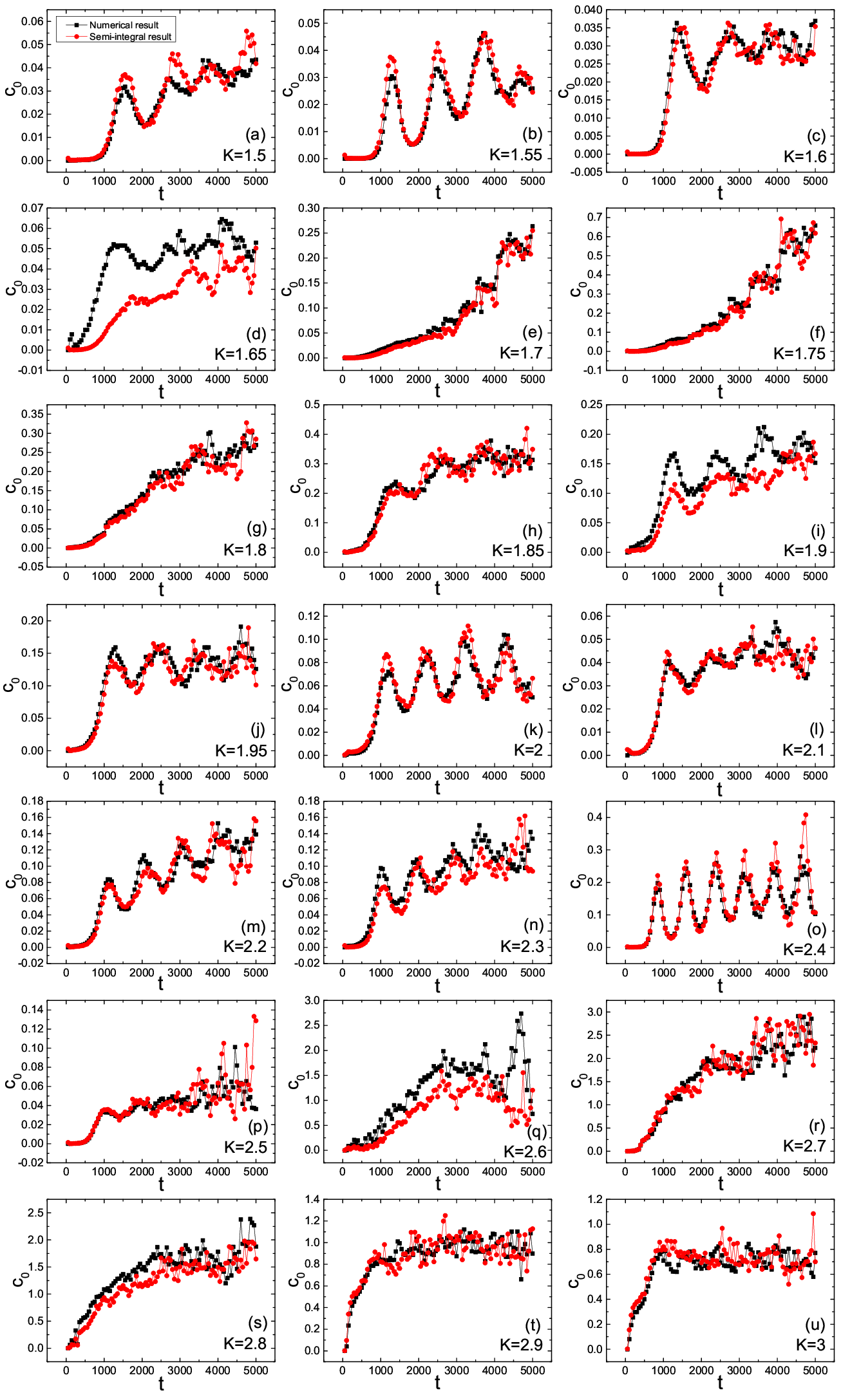}
\vspace{-0.5cm}\caption{Variation of $c_{0}$ versus the time $t$.The time step is taken as $50$ to escape some large fluctuation and the variable $c_{0}$ is calculated based on the good linear dependence of $\ln(-ln{M(t)})$ on $ln{\sigma}$ with every time fixed and thus can use the specific small perturbation $\sigma=0.01$ to obtain $c_{0}$ from the assumed decay law as $M(t)\approx e^{-c_{0}(t)\sigma^{\nu(t)}t^{\alpha(\sigma,t)}}$.The variation show there is a clear transition from very small value to some value gradually stabilized but with periodically oscillation for quite a few parameters $K$.One can find the tendency of variation of $c_{0}$ is similar to the decay rate with small perturbation as $\sigma=0.01$ in the previous study and thus it also gives some support for the decay law assumed at least as a effective approximation.One can find the cases of $K=1.7,1.75$ have some large discontinuity for $c_{0}$ corresponding to a long time larger than $t=2000$ as our calculation is based on the the good linear dependence of $\ln(-ln{M(t)})$ on $ln{\sigma}$ which means the term $\sigma^{\nu(t)}$ is effective in our assumed decay law,although the considered perturbation is small for a long time but the linearity for these two cases is lacking and then it can be reflected with this discontinuity that we can not trust.It is different from the situation about the large fluctuation obviously for $K=2.5,2.6,2.7,2.8$ that also can not be trusted as the perturbation for fitting is quite few for quite a long time.Meanwhile the order of magnitude of stabilized $c_{0}$ varies with parameter $K$ changing and has a tight connection with the variation of $\nu$ and thus we can find there is some large value around $1$ larger or smaller above $K=2.5$ and besides the value is basically below $0.2$ even smaller than $0.1$ for most cases.}\label{c_0}
\end{figure}
\end{center}

  Now we get the required information of basic components for studying the decay law that can be expected to useful for the study of time scale which is our interest here.As the fitted $\nu$ varies for the time going,we can not say there is a fixed rule for time scale summarising general perturbations but we can make a simple although rough estimate for the comparison of the extent of typical decay speed to the fields of stable dynamics and strong chaos.The fidelity could be taken explicitly as $M(t)\approx e^{-{c_0(t)}\sigma^{\nu(t)}t^{\alpha(\sigma,t)}}$ for not too large perturbation and the variation of $\alpha$ can be seen as the different decay laws and $\alpha=1$ and $\alpha=2$ with $c_{0}$ and $\nu$ respectively fixed represent exponential decay and Gaussian decay.For strong chaos,above the very small perturbation proportional to $1/\sqrt{N}$ smaller than $10^{-3}$,the typical decay laws as FRG and Lyapunov decay with the formulas as $M(t)\approx e^{-c_{0}\sigma^{2}t}$ and $M(t)\approx e^{\lambda t}$.$c_{0}$ corresponds to $2K(E)$ where K(E)$\int_{0}^{\infty}dt\la V[{\bf r}(t)]V[{\bf r}(0)]\ra$ is the classical action diffusion constant\cite{Cerruti} also with the name $\sigma_{cl}$\cite{Prosen_supp,Gorin} derived from the quantum correlation function in terms interaction picture when $\hbar$ is small.The symbol $\sigma_{cl}$ in the reference\cite{Prosen_supp,Gorin} have not any meaning with the perturbation $\sigma$ used in this paper.To get the knowledge of $K(E)$ and $\lambda$,we even have not the need to calculate it using original definition from classical dynamics but just use the established decay law with some typical quantum decay.One can find $c(t)$ of small perturbation $\sigma=0.01$ for $K=3$ is stable above $10^{-4}$ although the system have the mixed phase space and then obviously the value of expected $K(E)$ have the order of $10^{0}$ also for the $\lambda$ as the common knowledge of strong chaos in kicked rotator\cite{Ott}.

The typical decay process can be mainly divided into two decay processes besides the initial almost frozen and last saturation parts and then we try to use the equivalent idea to consider the time scale.We use the averaged method to consider the formula as $M(t)\approx e^{-\overline{c_{0}}\sigma^{\bar{\nu}}t^{\overline{\alpha(\sigma)}}}$,now we can consider the bound for these variables.The averaged decay exponent $\bar{\alpha}$ can be seen below the value $2$ as the first typical decay process basically can be seen to have the decay exponent more than $2$ but a little bit larger than $1$ for the second typical decay process,with the consideration all the decay process,one can numerically fit the result basically below the value $2$.$c_{0}$ is some kind of hard to predict and with the numerical computation and we will find $c_{0}$ is gradually evolve into a stable field illustrated in the study afterwards and we find the typically stable value around the order $10^{-2}$ which is quite different from the corresponding value $2K(E)$ and $\lambda$ with the order around $10^{0}$,then we also approximately consider the $\bar{\nu}\approx 1.5$ typically as the variation is between $1$ and $2$ with basic continuity.Then we can give a simple estimate as the time scale can be written as:
     \be \label{time_scale}\tau=e^{[-\frac{1}{\bar{\alpha}}\ln(\overline{c_{0}}\sigma^{\bar{\nu}})]}\ee
Thus we can make a comparison and get the condition as $\sigma^{2-\frac{\bar{\nu}}{\bar \alpha}}>\frac{\overline{c_{0}}^{\frac{1}{\bar{\alpha}}}}{2K(E)}$ to have the faster decay for strong chaos,
and $K(E)$ rely on the system $K$ in detail and here we can set the $K$ as $7$ which corresponds to the strong chaos used in previous study extensively.From the fitting results in terms of the quantum decay with the typical perturbations $\sigma=0.01,0.1$ one can get $2K(E)\approx 0.3$ and $\overline{c_{0}}\propto 10^{-2},\bar{\nu}\approx 1.5,\overline{\alpha}\approx 2$ based on our argument before,hence we can get the clear condition as $\sigma>0.4152$.Although we only make a rough estimate,but we expect the condition indeed show there is a transition for the quantum fidelity from the lower speed to faster speed for strong chaos compared to the mixed phase space in terms of classical limit.

Then we illustrate the transition for the comparison of decay speed using the system parameters as $K=1.5,2,3,7$ as well as the selected perturbations and one can find the transition indeed happen and our prediction basically is right,after the value as $\sigma=0.4$,we can basically take the decay process of $K=7$ corresponding strong chaos faster than other decay processes.There is a interesting situation one can find in our illustrated figure if we concentrate on the decay processes for the system parameters $K=1.5,2,3$ in terms the classical limit of mixed phase space,we still find the transition from the slower to faster for the system parameter $K=3$ compared to other parameters $K=1.5,2$ although this transition field is much smaller,it gives the hints that this transition may be universal at least for the study model kicked rotator for the comparison of decay speed in terms of any two system with sufficient large difference for the dynamical stability.Further more we want to illustrate possible turning value of perturbation numerically from the variation of time scale as the function of perturbation,we show our numerical results and obviously support the theoretical analysis.

\begin{center}  
\begin{figure}
\includegraphics[width=18cm,height=16cm]{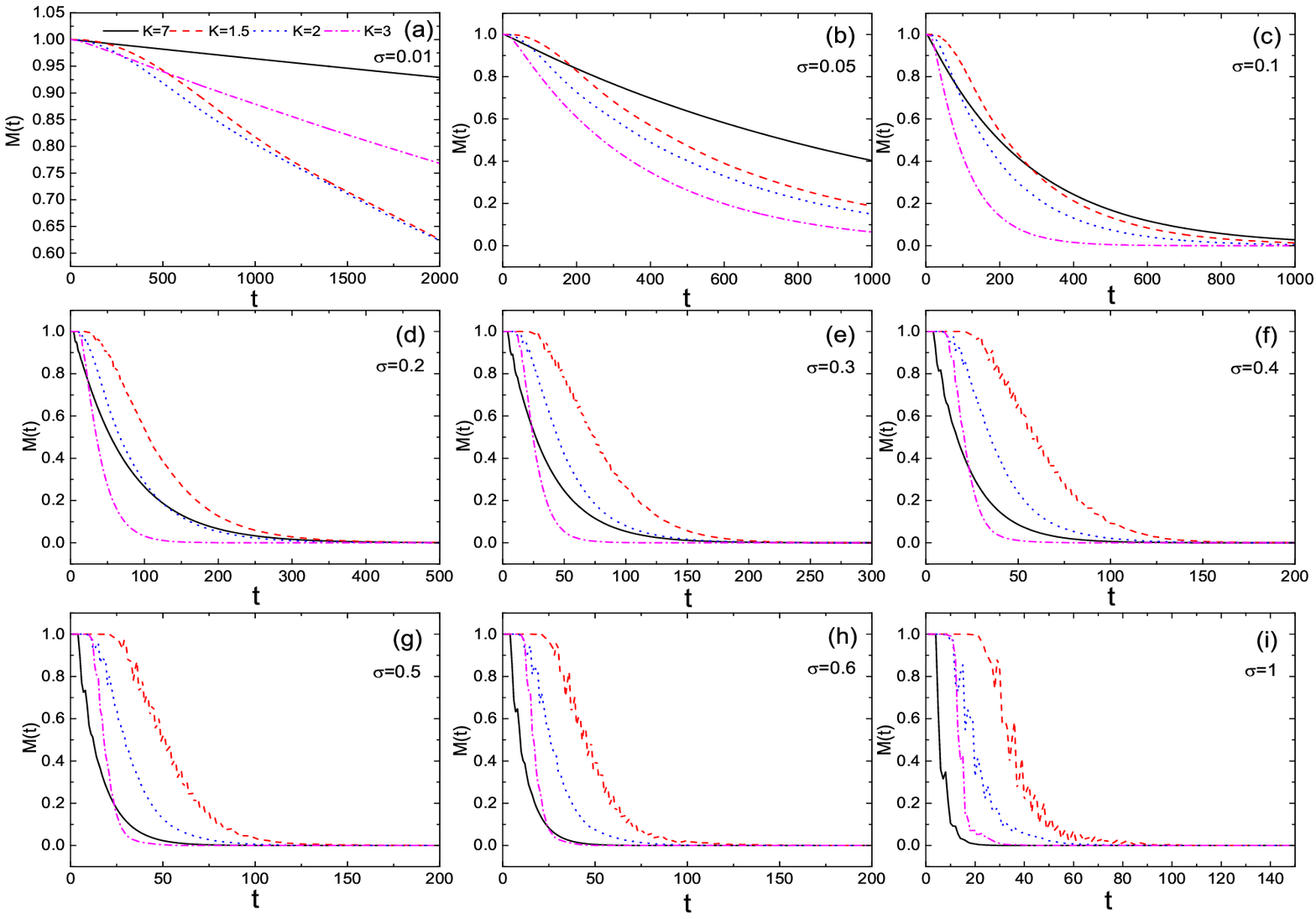}
\vspace{-0.5cm} \caption{The comparison of decay speed for different system parameters $K=1.5,2,3,7$ corresponding to increasing perturbation we select.There is a transition from slower to faster decay for the case of strong chaos with the parameter as $K=7$ compared to other system parameters $K=1.5,2,3$ corresponding to mixed phase space in terms of classical limit.The faster situation happened basically agree with the theoretical analysis as the estimated critical condition is $\sigma$ larger than $0.4$.Similar transition for the comparison is even also found for the decay process of $K=3$ compared to other two parameters $K=2,3$ in terms of small perturbation field illustrated clearly in the figure(a) and (b).}\label{scale_decay}
\end{figure}
\end{center}

Further more we need to show the comparison of time scales of quantum fidelity related to the strong chaos and mixed phase space in terms of classical limit and give a clear illustration about the existence of the critical perturbation accounting for the transition of decay speed among them.To strengthen convincingness,we add one parameter $K$ as $10$ corresponding to strong chaos and the critical perturbation is calculated as $0.1585$ with the same method we introduce to treat the case of $K=7$,thus we show them together to compare the time scales with the corresponding one for the mixed phase space and we find it is indeed a universal expression.Actually the study is a natural expansion of the work\cite{Gorin} to compare the time scales of quantum fidelity of strong chaos and stable dynamics.Besides the comparison of decay speed,we also want to find the basic law to govern time scale just like 
the typical established law as $\sigma^{-1}$ and $\sigma^{-2}$ for stable dynamics and strong chaos.Therefore we can approach these two subjects through the relationship for decay time $\tau$ versus perturbation $\sigma$ with the logarithmic coordinates,and $\tau$ can be numerically determined by the decay process within the value of quantum fidelity as $e^{-1}$,then we can assure the clear law to govern time scale if the linear dependence between $\tau$ and $\sigma$ can be found.Meanwhile the existence of critical perturbation can be showed as the intersection point for the variation of $\tau$ as a function of $\sigma$.Based on the consideration described before,we illustrate the variation of $\tau$ versus $\sigma$ in the figure below.

\begin{center}  
\begin{figure}
\includegraphics[width=18cm,height=20cm]{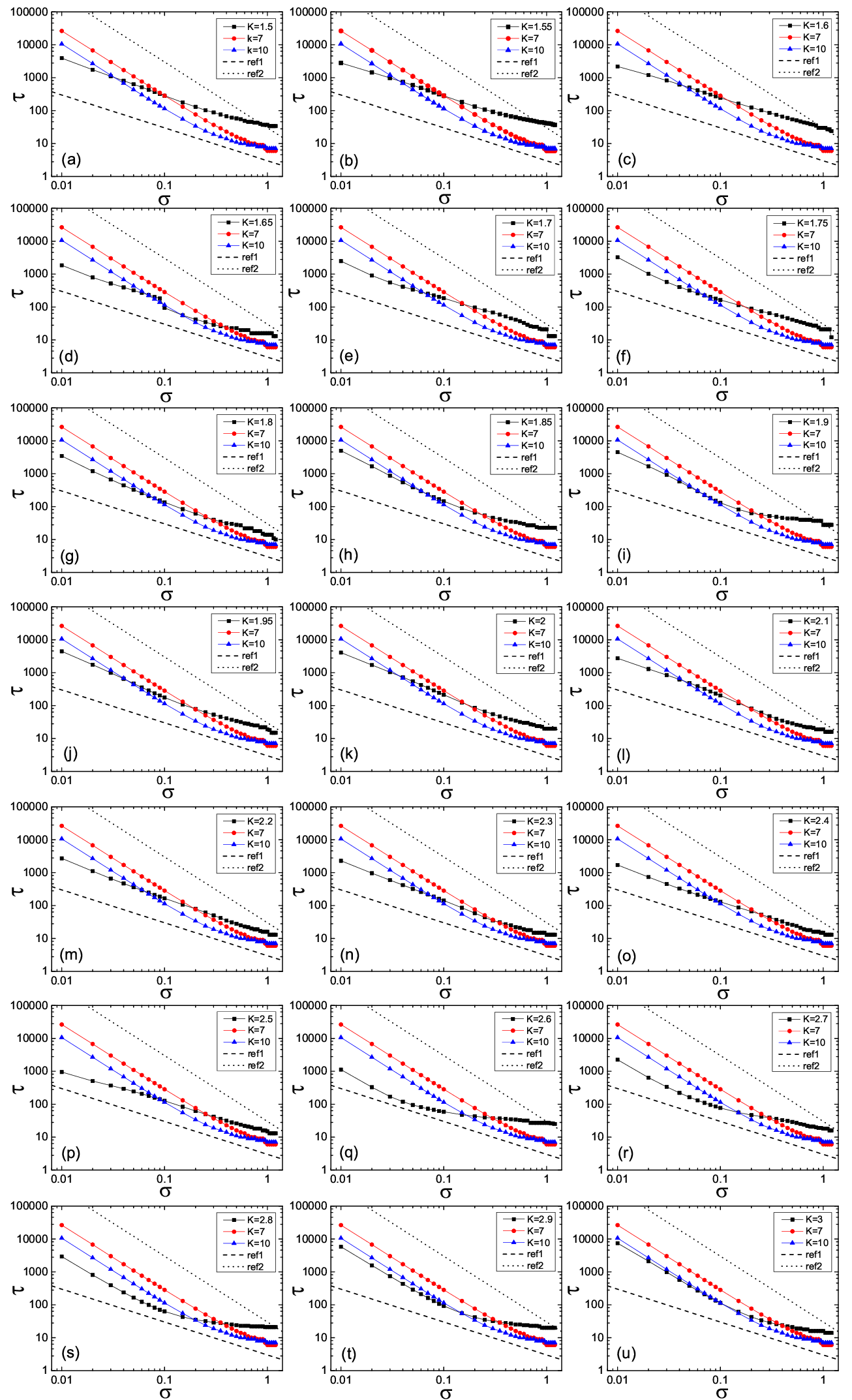}
\vspace{-0.5cm} \caption{Variation of decay time of quantum fidelity numerically calculated versus the perturbation $\sigma$ with logarithmic coordinates.$\tau$ can be numerically determined by the decay process within the value of quantum fidelity as $e^{-1}$ and two lines with the names as $ref{1}$ and $ref{2}$ are illustrated to guide the eyes for variation of slope with respective slopes as $1$ and $2$ from low to high.System parameters $K=7,10$ are used to testify our theoretical prediction of existence of critical perturbation characterised by the intersection points illustrated clearly in every sub-figure,there is a universal expression for the transition from the slower decay speed to faster decay speed for the quantum fidelity of strong chaos compared with the decay process of mixed phase space with the parameter $K$ from $1.5$ to $3$.The good linear dependence of $\tau $ on $\sigma$ indicates there is a clear rule to govern time scale such as $\sigma^{-1}$ and $\sigma^{-2}$ for stable dynamics and strong chaos,and we can find this good expression also can be seen in the mixed phase space that can not be valid very well for all the field of perturbation we study,but the good linear relationship can be seen separately for different part linked by some transition which means it still be valid for the rule of time scale but only can be applied to some specific field of perturbation for mixed phase space basically.} \label{time_scale_critical}
\end{figure}
\end{center}

From the figure we show,the existence of intersection point is a universal situation that can be seen in every sub-figure agreeing with our analysis.The predictive critical perturbation is a rough estimate and varies for different system $K$ in detail,but the basic prediction is valid at least taken as a basic bound.Meanwhile we can find the critical perturbation of $K=7$ is always smaller than corresponding value of $K=10$ for whatever comparative parameter $K$ for mixed phase space can be,and this numerical result is compatible with our calculation of the critical perturbation as $\sigma_{critical}=0.4152$ for $K=7$ is larger than $\sigma_{critical}=0.1585$ for $K=10$.Further more,we can find the linear relationship indeed can be seen clearly but basically can not hold for all the effective field of perturbation,one can easily find this feature with the contrast of the linear relationship for the cases of $K=7,10$.With very careful check for fitting,we can find the variation of $\tau$ versus $\sigma$ can be divided into different parts with the linear relationship held approximately.The main typical situation is to have two parts which show the linear relationship,the initial part with small perturbation always can be seen to have the linear relationship which can last for longer or shorter depending on the individual expression for a given specific $K$,and then we can get another part having the linear relationship with some possible transitive part.But the variation of $\tau$ also can be divided into three basic parts if the initial part with small perturbation basically below $\sigma=0.1$ can be divided into two parts which can be seen obviously in the cases for $K=1.7,1.75,1.9,2.1,2.7$.Here we use the dash line and dot line respectively to show exactly the slope as $1$ and $2$ to guide the eyes to have a intuitive impression of the variation of the slope which represent the rule of time scale.If we take a viewpoint for the continued variation,the slope always tends to decrease and the pattern for the variation undergoes some complicated change with the slope tending to $2$ for small perturbation finally.There is some sharp change from the case of $K=2.5$ to $K=2.6$ and this similar change is also seen in the previous study such as for $c_{0}$.

    As a bonus,we find the linear relationship is better for the case of $K=7$ than the case of $K=10$,it gives the hint that the well established rule of the time scale of strong chaos as $\tau \propto \sigma^{-2}$ still can not govern all the effective perturbation without saturation,and it is not better for more chaotic and it is a very interesting subject which shows that we still need more investigation in the study of time scale for strong chaos.The last but not the least,there is still the need for us to give the value of slope at least for the very initial part with promising linear relationship and the latter part with approximately linear relationship although with some fluctuation.Based on the careful check,it is reasonable to use the very first three perturbations as $\sigma=0.01,0.02,0.03$ to get a fitted slope as the first slope and then use the field of perturbation from $\sigma=0.3$ to $\sigma=0.6$ to get the second slope.Thus we call the field of perturbation for fitting as the field of initial perturbation and the field of latter perturbation with specifically determination.As $M(t)\approx e^{-{c_0(t)}\sigma^{\nu(t)}t^{\alpha(\sigma,t)}}$ can be written as $M(t)\approx e^{-(t/\tau)^{\alpha}}$,and further more time scale $\tau$ can be expressed as $\tau \approx c_{\gamma}\sigma^{-\gamma}$ with good linear relationship for $\ln{\tau}$ versus $\ln{\sigma}$.The slope described above actually is $-\gamma$ and we have to use the numerical calculation to find the variables $c_{\gamma}$ and $\gamma$ for every specific $K$ and the study result is illustrated in the figure below.The variation of $\gamma$ is not just between $1$ and $2$ but have the common value smaller than $1$ obtained from the field of latter perturbation.$\gamma$ and $c_{\gamma}$ undergoes some complicated change during the variation of system parameter $K$,and the closeness of $\gamma$ implicates there is some good rule as a whole to govern time scale which can be seen specially obvious for the case of $K=1.6$.From the relationship for $\ln \tau$ versus $\ln{\sigma}$ we show in the last figure,here we can easily understand the values of $\gamma$ from the field of initial perturbation are basically larger than the corresponding values from the field of latter perturbation and also the contrast of $c_{\gamma}$ is mainly contrary to the previous expression of $\gamma$.Meanwhile,there is a trend opposite to the variations of $c_{\gamma}$ and $\gamma$ obtained from the same field of perturbation,and the very reason accounting for it is the correlation between them in terms of expression as $\tau \approx c_{\gamma}\sigma^{-\gamma}$ if the corresponding $\tau$ does not change so much with different $K$.

\begin{center}  
\begin{figure}
\includegraphics[width=12cm,height=5cm]{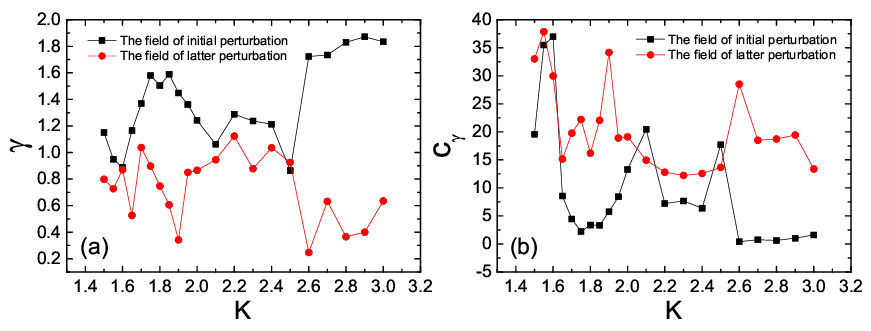}
\vspace{-0.2cm} \caption{Variation of $\gamma$ and $c_{\gamma}$ versus system parameter $K$,and $\gamma$ and $c_{\gamma}$ are the variables to characterise time scale expressed as $\tau \approx c_{\gamma}\sigma^{-\gamma}$.The field of initial perturbation and the field of latter perturbation in the figure are related to the specific perturbations as $\sigma=0.01,0.02,0.03$ and $\sigma=0.3,0.35,0.4,0.45,0.5,0.55,0.6$ for getting $\gamma$ and $c_{\gamma}$ respectively from the relationship for $\ln{\tau}$ versus $\ln{\sigma}$.$\gamma$ of the field of initial perturbation undergoes some complicated change in particular for the sharp change from $K=2.5$ to $K=2.6$ basically between $1$ and $2$ with the final tendency towards $2$ and $\gamma$ of the field of latter perturbation also undergoes some complicated change but basically below $1$ and the situation for their closeness implicates there is some good rule as a whole to govern time scale which can be seen clearly for the case of $K=1.6$.The contrast of $\gamma$ shows the value from the field of initial perturbation is basically larger than corresponding value from the field of latter perturbation and the contrast of $c_{\gamma}$ is mainly contrary to the previous expression of $\gamma$ easily understood from the last figure for the variation of the intercept with expected linear relationship for $\ln{\tau}$ versus $\sigma$.Further more,the tendencies of variation of $\gamma$ and $c_{\gamma}$ obtained from the same field of perturbation are contrary to each other basically and it can be understood by the correlation between them in terms of using the relationship as $\tau \approx c_{\gamma}\sigma^{-\gamma}$ if the corresponding $\tau$ does not change so much with different $K$.}  \label{chaotic_slope_fit}
\end{figure}
\end{center}

Now we can reasonable to consider the transition of decay speed among different dynamics in terms of classical limit is a universal situation,and thus the similar critical perturbation to characterise the transition may also exist for the contrast for quantum fidelity in terms of stable dynamics and mixed phase space.Actually we can make a simple analysis similar to the previous study for the contrast of time scales for strong chaos and mixed dynamics.We can take a special case as the initial position of wave packet for $r_{centre}=2.2,p_{centre}=1.5$ with $K=3$ to guide our understanding of the existence of possible transition for the decay speed.This special case is related to stable field that actually has been studied in previous section about the numerical study of edge of chaos.Firstly time scale for quantum fidelity in terms of stable field can be expressed typically as $\tau=c_{s}\sigma^{-1}$,and $c_{s}$ is the name we call as the coefficient.Obviously if we can get the value of $c_{s}$,we can make a prediction of the possible critical perturbation.From the simple fitting technique with two perturbations as $\sigma=0.01,0.1$,we can extract the value of $c_{s}$ basically as $3.20\times{10^{3}}$ averaged and actually $c_{s}\approx 3.2752\times{10^{3}}$ from $\sigma=0.01$ and $c_{s}\approx 3.1235\times{10^{3}}$ from $\sigma=0.1$.For simplicity,we call the time scale from the strong chaos,stable dynamics and mixed dynamics as $\tau_{strong},\tau_{stable},\tau_{mixed}$.Then we can consider the condition for $\tau_{mixed}>\tau_{stable}$,it means $\frac{(\overline{c_{0}}\sigma^{\bar{\nu}})^{-\frac{1}{\overline{\alpha}}}}{c_{s}\sigma^{-1}}>1$.Through a simple mathematical calculation with the values of variables put into,we can get the term as $\ln{\sigma}>23.0732$,and further more we have a result as $\sigma>1.0485\times{10^{10}}$ which is useless in the application as it is surprisingly large.So we can safely say that $\tau_{mixed}$ is always smaller than $\tau_{stable}$ and there is not critical perturbation for the special case we consider.Then we consider another case for the quantum fidelity of strong chaos as the initial position of wave packet for $r_{centre}=2.2,p_{centre}=3$ with $K=7$.For similar calculation based on the condition for $\tau_{strong}>\tau_{stable}$ which means $(2K(E)\sigma^{2})^{-1}>c_{s}\sigma^{-1}$,we can get the result as $\sigma<1.0417\times{10^{-3}}$.This result is reconciled with the important conclusion as faster fidelity decay for mixing than for regular dynamics for sufficient small perturbation\cite{Gorin}.

From the above analysis from the typical cases,we can find there are two intersection points if the decay time of quantum fidelity of strong chaos,mixed dynamics and stable dynamics can be illustrated together as a function of perturbation $\sigma$,we expect this situation can typically hold for the general $K$ varied corresponding to mixed phase space and stable dynamics.As it is a special time-consuming work to calculate the decay time for small perturbation such as below $\sigma=0.01$,we do not attempt to illustrate it in this paper to show the intersection point and one can find this kind of expression for much smaller Hilbert space with the study model as kicked top in the previous work\cite{Gorin}.Actually the study of quantum fidelity with the classical limit as stable dynamics still have something unclear and we find a special situation deserving us attention for quantum freeze which had been seriously studied for a vanishing time-averaged perturbation operator in terms of classical limit as stable dynamics and strong chaos\cite{Prosen_NJP03,prosen2005}.Now we find a important situation that can not be clearly studied although it is not a main consideration in this paper,which is the long-time quantum freeze after Gaussian decay which have been involved in the previous study but have not a systematically study in particular for lack of a theoretical explanation.In terms of comparison of decay speed,the quantum freeze obviously makes it more complicated except the value of freeze is very small that can not be seen as a important decay process happened for some very large perturbation which is useless to find the critical perturbation.

We find the quantum freeze can exist for a broad field of perturbation in particular for a very strong perturbation which is quite different from a vanishing time-averaged perturbation.Further more it is interesting to find the time scale $t_{1}$ for the initial Gaussian decay is independent of perturbation $\sigma$ as $t_{1}\sim \hbar^{-1/2}$ agreed well with the previous work considering the vanishing time-averaged perturbation but then we can find the time of quantum freeze as $t_{2}$ is so long which seems to be the same for all the perturbation we consider at least with our limited observation that can not be used for $t_{2}\sim \rm{min}\{\hbar^{1/2}\epsilon^{-2},\hbar^{-1/2}\epsilon^{-1}\}$ such as $t_{2}$ of $\sigma=10$ can also last for the order more than $10^{3}$.To illustrate this non-trivial decay features,we use the cases of $K=0.2,0.5$ with the initial position of wave packet for $r_{centre}=2.2,p_{centre}=3$ fixed belonging to stable dynamics for all the phase space and a satisfied explanation in theory is still yet to be known.

\section{5 DISCUSSION AND CONCLUSION}
In this paper we consider Loschimidt echo with initial quantum Gaussian wave-packet for the two typical fields as edge of chaos and chaotic sea belonging to classical mixed phase with well-known kick rotator as the study model.Besides the extensive numerical study,the baisic analytical method we use is to construct the classical and quantum correspondence,in detail is that we constructed a classical ensemble with quantum uncertainty relation corresponding to a initial quantum state we want to investigate and then we can study the dynamics of this ensemble naturally reconciled with the statistical-type semi-classical method initiated by the one of author\cite{WgWang_1} to study so called weak chaos beyond the Gaussian approximation\cite{Jalabert}.During the dynamical process for a ensemble evolution alongside with the related time-dependent density of probability distribution of action differences in terms of shadowing theorem\cite{vanicek},we can consider corresponding features of extensive numerical computations of quantum decay of Loschmidt echo with the central semi-classical formula(\ref{Ma_new})reconciled with the diffusion process of the classical ensemble.This is indeed the whole story about out method to treat the problem we put forward in the part of introduction for searching for some kind of new decay laws as $M(t)\approx e^{-ct^{\alpha}}$ beyond the situations about strong chaos or regular field,where $c$ can be called the decay rate and $\alpha$ as the decay exponent.

For extensive numerical study,we fixed the average position $\tilde r_{0}$ and change the average momentum $\tilde p_{0}$ for initial quantum state with the system parameter as $K=3$ in terms of edge of chaos and from the classical dynamical diffusion with direct tracking and also abnormal diffusion,one can find there have a transitive process from stable field to chaotic sea not smoothly as the multi-fractal boundary\cite{Ott}although has been coarse grained here and in terms of this process,the quantum decay process can be basically divided into three processes in our study without the situation coming back to stable field with power law decay\cite{Gorin}.The first one is the initial Cubic-exponential decay predicted theoretically by J.Van\'{i}\v{c}ek \cite{vanicek_arxiv} for quasi-integrable field with short time using the argument of correlation of action differences and numerically observed firstly in our study with universal same time scale.The second intermediated and also complicated decay process has the merging tendency to form a similar decay process within some perturbation field basically enlarged with the setting of initial quantum state gradually going towards the chaotic sea and then have a rough expression for the decay law at least for some time scale as $M(t)\approx e^{-ct^{\alpha}}$ with $c$ mainly varied.As the complicated variation they have,we use the reference lines to show it.The third one is basically the saturation field without obvious decay process or likely have a small decay process linked with the second decay process mainly relying on the initial setting of quantum state we study with some perturbation field.

Besides numerical study of the decay features of the edge of chaos,we also find a positive connection between the fluctuation of Loschimidt echo and corresponding perturbation and further more the fluctuation can be universally characterized by the measures as fractal dimension and direct average fluctuation we define sharing the same variation together and this variation pattern versus perturbation is independently  with different setting of initial quantum state.It is a first discovery and it could be taken as a unique feature for edge of chaos in the study of Loschimidt echo although we need more study models to check,but our study strongly support the suggestion.We also pay attention very recent work in the related fluctuation for the mixed-type phase space with the information flow as the main method\cite{Mata},but our finding for the universal expression in the fluctuation of Loschimidt echo is the first time.

For the numerical study of chaotic sea,we can find there also have three decay processes as the first one is the quantum frozen process with a short time firstly pointed out by T.Prosen\cite{Prosen_supp} using the so-called classical correlation integral and then the second decay process generally have two decay sub-processes mostly with some kind of good relation that can be described using the formula $M(t)\approx e^{-ct^{\alpha}}$,further more with the perturbation increased this kind decay process can gradually evolve into a single decay process described with $M(t)\approx e^{-ct^{\alpha}}$ or not relying on the initial setting of quantum state corresponding to different system parameter $K$,and finally the saturation field without obvious decay can be found.We find that it is more obvious for the chaotic sea to show the clear non-exponential decay law with the $\alpha$ basically larger than 1 varied with the perturbation.By contrast,we find the saturation values of loschimidt echo for the edge of chaos and chaotic sea are quite different,for the previous one the value can be decreased basically during the process of leaving out of the edge of chaos for the initial setting of quantum state and at last converges to the saturation value for the theoretical prediction as $1/N$ that currently is the common sense in this research field\cite{Scholarpedia},here $N$ is the dimension of Hilbert space of the system taken as $2^{21}$ in this paper and thus the predictive saturation value have the order about $10^{-7}$ verified indeed by the latter expression for chaotic sea from our study.

Thus our finding challenge the general paradigm of the saturation value and it seems to be treated universally with our simple statistical semi-classical formula attributed to the variation of the density of probability of action differences with the mathematical expression as $P(\Delta S)$ actually taken the simper vision as $P(s)$ with $s$ from $s=\Delta S/\epsilon$ as $\epsilon$ for direct classic perturbation although there still have some open problem for the applied condition.Here we should say that actually our analytical work is based on the established result as the Dephasing representation\cite{vanicek} with the advantage that can treat single quantum decay and many work have to treat the average Loschimidt echo with some argument to get the clear decay laws mainly in strong chaos but actually Dephasing representation is also a approximation and some condition to apply is not very clear and very recently a more serious analytical work was done\cite{vanicek_exact}that may join the force of our work and will be likely a important consideration in the future.

Actually in this paper we seriously discuss the likely problem of the statistical-type semi-classical 
method and consider the difference of so-called first order and second order approximation\cite{WgWang_2} in terms of the variation of $k=\hbar/{\xi^2}$ for $\hbar$ as the effective Pluck constant and $\xi$ as the effective width and actually the quantum state we study with the equal uncertainty for the position $r$ and momentum $p$ belonging to the second order expression with $k=1$ for most work in this research field.Then we have the second-order semi-classical formula to treat the quantum Loshimidt echo and afterwards there have a subtle procedure that have not been clarified  with rigorous mathematical ground yet as to use the first-order statistical semi-classical formula with the classical ensemble from second-ordered quantum state for evaluating the quantum Loschimidt echo.Rigorously to say,it is a trick that seems to have not exact mathematical base but it seems that this idea has been applied to study Loschimidt echo in previous work \cite{WgWang_2,Zheng} to get good results to some extent,hence we still in this paper use the first-ordered type semi-classical method to approach Loschimidt echo although we pay a high attention to the differences and the likely reasons.

In this paper for chaotic sea with the classical limit,we used the probability theory to get a analytical formula(\ref{check_initial}) for the initial $P(s)$ with the prediction as approximately Gaussian distribution and get numerical agreement.Further more,we considered the simple expectation for the positive connection for the accuracy of evaluation of semi-classical method using Levy distribution of $P(s)$ and the extent for the linearity of the frequency relation corresponding to the Fourier transform of $P(s)$,and from the variable we define(\ref{LIN})with some average procedure,the expectation basically can be seen hold.With some example,we also show the direct comparison of $P(s)$ with corresponding Levy fit in terms of some given time as well as the likely linear frequency relation(\ref{Flz}),and indeed find it is a common tendency to have approximately Levy distribution for relatively long time with the order at least about $10^2$.Then we have some kind of solid background to seriously consider the fitting affect in the semi-classical formula(\ref{semiclass})with Levy assumption,and it is a question easily neglected but should be considered in practice.Therefore,in this paper we seriously consider the variation of semi-classical evaluation with different fitting procedure which means to use some group of given frequencies together to fit and find the possible situation of existing the critical value of $\eta$ generally taken the name as $\eta^{\ast}$(A.4)accounting for the non-monotonic variation of semi-classical evaluation of Loschimidt echo and $\eta$ is a index characterizing $P(s)$ with $\eta=2$ for Gaussian distribution and $0<\eta<2$ for Levy distribution.

The basic study idea for investigating the non-monotonic variation of semi-classical evaluation seems complicated but actually it is very logic.Firstly,we fitted the parameters mainly for the so-called exponential model taken as the $D_L=c+a{\rm exp}(b {\eta})$ with $D_L$ as the width of $P(s)$ in terms of levy distribution assumption,and $a,b,c$ are the fitting parameters corresponding the assumed function for $D_{L}$ versus $\eta$ as there have the interdependence for the fitting with different group of frequencies which is a starting point to understand our basic treatment.Then we can apply this idea to which group of frequencies for fitting likely corresponds to change the monotonicity of the variation of semi-classical evaluation.Anyone who still have some puzzlement about how to apply the critical value $\eta^{\ast}$ can have a look at our illustrative example in the FIG 37 and also a very detailed derivation in the part of appendix can help you for the understanding.Then we can find the field of perturbation for existing $\eta^{\ast}$ corresponding to given fitted parameters $a,b,c$ with different combination of positive or negative values to set the lower bound and upper bound from the study of condition of existence of the critical value $\eta^{\ast}$ summarized in the Table 1 and Table 2 and in this available field of perturbation we also can know the variation of $\eta^{\ast}$ itself versus perturbation with a simple formula(\ref{dif_crieta})numerically checked.Further more,the tendency of this variation also can be studied with the formula(\ref{second_order_eta})and so-called transition value can be obtained.

From our study,we can find the variational field of $\eta^{\ast}$ can have the intersection of the effective changing field for the fitting $\eta$ corresponding to some field of perturbation with different expression in terms of different system parameter $K$.For escaping the possibility to include the $\eta^{\ast}$,it seems to be reasonable to use fewer frequencies to fit but actually this can not always guarantee the better accuracy for the semi-classical evaluation as this depends on the monotonicity for $M_{sc}$ versus $\eta$ connected with the combination of fitted parameters $a,b,c$ for a given time as well as the value of $\eta^{\ast}$ as the function of $a,b,c$ and perturbation.Obviously we have developed a unified method to treat the non-variation of $M_{sc}$ joined the force of numerical computation and relatively easy mathematical analysis.In a word,if we can get the information about the fitted parameters $a,b,c$ with the time variation,we can get to know the non-monotonic variation of semi-classical evaluation in detail for any more careful investigation.

Then we use the few frequencies as four unified to treat semi-classical evaluation of quantum fidelity which is another name used in other papers rather than with the name Loschimidt echo.For chaotic sea,we have find Levy distribution of $P(s)$ can be hired for some evolution time and also was applied in previous study\cite{Zheng}but we find there have some important factors missing.The most important thing is the $\eta$ taken as the function of time numerically realized for a fixed fitting procedure.Thus the decay exponent in the semi-classical treatment should be divided into two parts from the variables $\eta$ and $D_{L}$ together that can be called as $\alpha_{\eta}$ and $\alpha_{D}$ and find there likely have a entirely opposite tendency for variation versus time using some easy mathematical analysis and also get indeed numerical conformation which means there have a tight correlation between seeming separated expressions for $\eta$ and $D_{L}$.Further more,we consider the direct integral with our essential semi-classical formula(\ref{Ma_new}) to see the efficiency of our theory and this direct integral can treat all the situations of chaotic sea and edge of chaos together without some assumption of $P(s)$ although the price we should pay for is it entirely depends on numerical computation.

As the decay process in terms of chaotic sea is quite different from the strong chaos but the related decay laws can give some reference for our study,so in this paper we basically have two steps to treat decay laws systematically.Meanwhile during this kind of study,we always have the comparison among the expressions from quantum Loshimidt echo,semi-classical direct integral and semi-classical evaluation from Levy distribution assumption for $P(s)$.

Firstly,we assume the local very limited decay process the decay law as $M(t)\approx e^{-ct^{\alpha}}$ hold and then for a given time interval we can make a fitting to get related local decay rate $c$ and decay exponent $\alpha$ as well,thus we can study the variation of $c$ and $\alpha$ versus time closely to a continual variation with the time interval for fitting sufficient small.If the variation of fitting local $c$ basically around a constant and $\alpha$ around 1,this situation actually belong to the typical exponential decay in strong chaos,otherwise we can expect the new decay laws.From our extensive numerical study,this idea is quite good for small perturbations and corresponding decay law in strong chaos can be basically expressed as Fermi Golden rule decay\cite{Scholarpedia} with the main feature as $c\approx 2.2\sigma^{2}$ that can be taken for the comparison.Based on this study method,we find the local decay rate can enter a stable field with some fluctuation around the value deviating from $2.2\sigma^{2}$ more or less relying on the system parameter $K$ but with the same order,and the local decay exponent $\alpha$ correspondingly also undergo a transitive process from a higher value to a basically stable value although actually having some tendency to slowly decrease that basically is a universal feature that could be explained by the variational relation for $\eta$ and $D_{L}$ to some extent in our study.The fitting results from direct semi-integral have a good agreement with numerical fitting of Loschimidt echo for most system parameters $K$ and the fitting results coming from the Levy distribution can agree well with direct integral in particular for $\alpha$ in terms of some time scale corresponding to a good assumption of $P(s)$ as our expectation.Secondly for large perturbation,the variation of local decay rate have some kind of large fluctuation even without the similar order in the quantity and we just need another method to treat the decay law as a whole,further more as the complicated situation then what we focus now is the decay exponent,along this thought we can fit the decay exponent $\alpha$ from some procedure to judge the related time scale if it could be taken as the stable decay process actually also relied on the local fitting technique.Based on this numerical method,we got the variation of related decay exponent $\alpha$ with different perturbation.From our study,non-exponential decay is very common in chaotic sea as well as the edge of chaos.

Here,we find our semi-classical method tends to be good to approach Loschimidt echo for a effective field of perturbation relying on the system parameter $K$ but with the good level for small perturbation universally in chaotic sea.This kind of expression can be reflected in the study of decay exponent as well as the direct comparison for decay process which never seriously was investigated.This finding have some hints for us that there have some reason in common for the condition for applying our statistical-type semi-classical method,but until now we do not know.With the average difference between the semi-integral and Loschimidt echo rescaled by perturbation,we can find if there have a tendency to evolve into some small value around zero near field with perturbation increased,the good extent for approaching can be expected,but for our study this kind of expectation is not common.Actually it is a very important issue but not easy to treat in our paper,which is the time scale for the decay process.It is a seeming confusing question for a new comer in the research field of Loshimidt echo,as quite a few researchers did the research from their own research background and some kind of complicated with many mathematical discussion in detail,actually the basic idea is simple.Loshimidt echo is very similar to relaxation process in statistical physics,and the typical decay law for strong chaos is $M(t)\approx e^{-ct}$ that can be rewrite as $M(t)\approx e^{-t/(1/c)}$,and we can replace $1/c$ for $\gamma$ to have the regular style as $M(t)\approx e^{-t/\gamma}$,and $\gamma$ actually is the time scale characterizing the decay time.As $c\approx 2.2 \sigma^{2}$,thus the order of $1/c$ is about $\sigma^{-2}$ as the typical time scale which is the one of main results in the seminal work done by T.Prosen\cite{Gorin}.

Thus we can rewrite $M(t)\approx e^{-ct^{\alpha}}$ as $M(t)\approx e^{-(t/\gamma)^{\alpha}}$,and here  
$\gamma$ still can be defined as time scale as the extension of the definition in strong chaos.For edge of chaos,the decay process is quite complicated and it seems to be hard to treat.To point it out as the focus is that the decay process even with good expression of $M(t)\approx e^{-ct^{\alpha}}$ also have related $\alpha$ commonly varying versus the perturbation and this make the outcome even challenge the general idea that the time scale can be shorten by the perturbation increased\cite{Scholarpedia}.For the decay process that can not have a good approximation with the formula $M(t)\approx e^{-ct^{\alpha}}$,the situation is more complicated and how to develop new concept connected with efficient decay time equivalent to time scale deserves a serious attention in the future work.For chaotic sea,the situation seems easier as the decay process can be basically divided into two clear decay sub-processes that can be described separately the decay law as $M(t)\approx e^{-ct^{\alpha}}$ and generally the corresponding decay rate $c$ and decay exponent $\alpha$ vary with the perturbation,thus the decay law can be written as $M(t)\approx e^{-c(\sigma)t^{\alpha(\sigma)}}$ and time scale $\gamma$ could be written in terms of $M(t)\approx e^{(-t/\gamma)^{\alpha}}$,with some easy math one can get $\gamma=e^{{lnc(\sigma)}/{\alpha(\sigma)}}$.As there have very few theoretical understanding,this problem should have to be relied on the extensive numerical work added with some smart trick,we are working on it and have show some variation of efficient decay time as well as the transitive time for some special cases,we may summarize related results in other paper if we can find some clear regulars.

In this paper we also want to use statistical-type semi-classical method to approach Loschimidt echo in the edge of chaos and we can find there generally have not the clearly Levy distribution but have typical peak-like shape of $P(s)$ which means the distribution is highly localized and during the process for the initial quantum state coming out of the edge of chaos,the contribution from chaotic sea become more and more significant corresponding to the original long tail of distribution in $P(s)$ that initially can be neglected become more and more important with the time evolution,at last the Levy distribution can be expected.This variation of $P(s)$ obviously agree with the ensemble dynamic as there have a transition from the edge of chaos towards chaotic sea.For the multi-fractal dimensional edge,we should point it out that there have extend distributions of $P(s)$ in the stable fields without the situation of escaping to chaotic sea initially that actually we also can take as the mark for the edge of chaos.

Then in this paper,we seriously use the semi-integral with this special distribution of $P(s)$,and find some non-trivial results.For stable field,we can find the basically good agreement for the evaluation of our semi-classical theory entirely inside into the stable torus and otherwise also have a good agreement as a whole but with some large fluctuation.For the situation of firstly escaping to chaotic sea,the some good agreement just have a limited time scale for small perturbation and can be become good in all the decay process with the large perturbation but have the revive process again that can not make sure whether it is just a numerical computation issue simply from the variation from $P(s)$ or have some more deep root for this expression.Then with the situation of hindering again for escaping to the chaotic sea,the evaluation of semi-integral can be good for a limited time scale and have a large fluctuation afterwards.After the situation of obviously escaping to the chaotic sea again,the evaluation of semi-integral approach to Loschimidt echo can be gradually better at least for the main decay process for the initial quantum state coming out the edge of chaos correspondingly as well as the perturbation increased,it is a basic experienced rule only just taken as the basic guideline to some expectation.Further more,there have some burst and large revival as a universal situation splitting as the perturbation doubled obviously observed for large perturbation larger than 1.Therefore,we can find a clearly different evaluation of efficiency for semi-classical method we used in this paper,and the very reason for the difference need more theoretical understanding added with more study models in the future work.

At last,we want to search for the correspondence between the abnormal diffusion and the feature of quantum decay and it is just a first try,but we indeed find there have some clear connection in particular from the variation of diffusion exponent as the value one for the normal diffusion.To connect the abnormal diffusion to the distribution of $P(s)$ is a interesting question but some hard for physicists although there have some important work in it from the complicated dynamics to the statistical formula\cite{Levydistr}.Actually we pay attention to the study with non-extensive statistical mechanics\cite{WLT02} to approach Loshimidt echo for the edge of chaos,and the evaluation of decay process need to fit for every perturbation and also not very carefully for the decay process in detail,and in our study,we can treat the quantum decay with a effective perturbation field with the so-called seed $P(s)$ although our semi-classical method need to improved to have more clearer applied condition.Nowadays the experimental technique have been improved to single atom level,and our study show there have a clear correspondence between a classical ensemble and quantum state in terms of uncertainty relation,thus we hope our study in this paper can access to the interest of broad research field.

From more analytical viewpoint,the semi-classical method with so called pair of orbits developed firstly by Martin Sieber and Klaus Richter\cite{Martin,Muller}can be taken great attention as it indeed is a reflection of shadowing theorem accounting for the dephasing representation to Loschmidt echo that is the starting point for us to develop the statistical-type semi-classical method in our study\cite{Heller}.As the initial important paper\cite{Jalabert}stimulated wide interest for fidelity which just considered the short correlation of classical orbits with some approximation using the diagonal contribution and non-diagonal contribution in terms of semi-classical profile,but the long range correlation corresponding to weak chaos can not be applied.Our work obviously have some intersection with this research direction,we hope to develop a more deeper work mathematically in the future with more study models under investigation.At last but not least,the transition from weak chaos to strong chaos is deserved much attention,the variation of mathematical description of decay laws could be the next work in the future.

\section{ACKNOWLEDGMENTS}
Wenjun Shi is grateful to Professor Wenge Wang of University of Science and Technology of China for many valuable discussions and the suggestion and encouragement for this study should be owned to him. Wenjun Shi thanks Dr.A. Goussev for useful discussion about semi-classical physics.This work is supported by the project of visiting scholar for excellent teacher from colleges and universities(2009.9-2010.7).
%\begin{appendices}

\section{Appendix}
\subsection{Appendix A. Monotonicity of $M_{sc}$ for assumed
variations of $D_{L}$ via $\eta$ with frequency relation}

\subsubsection{1.Exponential variations of $D_{L}$ via $\eta$}

For the starting point,$M_{sc}(t)={\rm exp}(-2(\epsilon/\hbar)^\eta
D_L)$.We use the assumption for the $D_L=c+a{\rm exp}(b {\eta})$ and
$a,b,c$ are the fitting parameters.Inserting this expression about
$D_L$,we can get the expression \be\tag{A.1} M_{sc}\approx {\rm
exp}\{-2a[\frac {c}{a}e^{\eta ln{\sigma}}+e^{\eta
(b+ln{\sigma})}]\}.\ee There are two terms having the likely
competitive relation for $\frac {c}{a}e^{\eta ln{\sigma}}$ and
$e^{\eta (b+ln{\sigma})}$,so we want to figure out the monotonicity
and have to make a derivation to $\frac {c}{a}e^{\eta
ln{\sigma}}+e^{\eta (b+ln{\sigma})}$,thus we can get the essential
term \be\tag{A.2}{\rm exp}[\eta (b+ln{\sigma}][\frac
{c}{a}e^{-b\eta}ln{\sigma}+b+ln{\sigma}].\ee Finally the positive or
negative of the term $\frac {c}{a}e^{-b\eta}ln{\sigma}+b+ln{\sigma}$
decides the monotonicity of $M_{sc}$.In some situations,the
derivation of the term  $\frac
{c}{a}e^{-b\eta}ln{\sigma}+b+ln{\sigma}$ also should be
considered,and it is very easy to find the result $(\frac
{c}{a})(-b)ln{\sigma}e^{-b\eta}$.For simplicity,we can call the term
$\frac {c}{a}e^{\eta ln{\sigma}}+e^{\eta (b+ln{\sigma})}$ as
$M_{0}$,$\frac {c}{a}e^{-b\eta}ln{\sigma}+b+ln{\sigma}$ as
$M_{1}$,and $(\frac {c}{a})(-b)ln{\sigma}e^{-b\eta}$ as $M_{2}$.

Here we do not use purified mathematical classification
discussion,what we care about are the situations existing in the
very extensive numerical observation,but our analysis can be very
easily to apply the new possibilities,actually we show the basic
patterns are very limited.Therefore we need to consider different
typical cases,and the first classified factor is $\sigma=1$ as the
sign of $ln{\sigma}$ can be changed with $\sigma>1$ or
$0<\sigma<1$,the other factors are the different combination of
$a,b,c$.

$1)$ $0<\sigma<1$ and $b<0$

a)$a>0,c>0$

As here $b+ln{\sigma}<0$,and $\frac {c}{a}>0$,thus the essential
term $M_1<0$.That means $M_0$ is a decreasing function and $M_{sc}$
is a increasing function for $a>0$.Now we can visually and also
conveniently illustrate the decreasing and increasing tendency using
symbol $\downarrow$ $\uparrow$,and get the logic sequence clearly in
this situation:$\eta \uparrow$ $\to$ $M_0 \downarrow$ $\to$ $M_{sc}
\uparrow$, $\eta \downarrow$ $\to$ $M_0 \uparrow$ $\to$ $M_{sc}
\downarrow$.

b)$a<0,c>0$

Here still $b+ln{\sigma}<0$,but $\frac
{c}{a}e^{-b\eta}ln{\sigma}>0$,so we can not know right now for the
monotonicity of $M_{0}$.We need to consider the term $M_{2}=(\frac
{c}{a})(-b)ln{\sigma}e^{-b\eta}$,and find $M_{2}>0$ which means
$M_{1}$ is a increasing function.Thus we can investigate the value
of $M_{1}$ in terms of $\eta=0$ and find
$\frac{c}{a}ln\sigma+b+ln\sigma$.For general cases with
$|\frac{c}{a}|\gg1$,so we can expect
$\frac{c}{a}ln\sigma+b+ln\sigma>0$ added with the condition about
increasing function of $M_{1}$,$M_{1}$ will always be positive with
$\eta\ge0$.Therefore,$M_{0}$ is a increasing function with
$\eta$,and that leads to $M_{sc}$ is also a increasing function with
$a<0$ here.With the logic sequence,we have:$\eta \uparrow$ $\to$
$M_{0} \uparrow$ $\to$ $M_{sc}\uparrow$,$\eta \downarrow$ $\to$
$M_{0} \downarrow$ $\to$ $M_{sc}\downarrow$.

For some situation,if $\frac{c}{a}\ln\sigma+b+\ln\sigma<0$ as the
condition $|\frac{c}{a}|\gg1$ can not be satisfied,then we can find
there have one critical point for $\eta$ that can be called
$\eta^{\ast}$ dividing the $\eta$ as two fields for
$[0,\eta^{\ast})$ and $(\eta^{\ast},\infty]$ corresponding to
$M_{1}<0$ or $M_{1}>0$.The $\eta^{\ast}$ can be determined by the
equation\be\tag{A.3}\frac{c}{a}ln{\sigma}{\rm
exp}(-b\eta)+b+ln{\sigma}=0,\ee and get
\be\tag{A.4}\eta^{\ast}=-\frac{1}{b}ln[-\frac{a}{c
ln{\sigma}}(b+ln{\sigma})].\ee Therefore $M_{0}$ is a decreasing
function with the field $[0,\eta^{\ast})$ and is a increasing
function with $(\eta^{\ast},\infty]$.As $a<0$,$M_{sc}$ share the
same monotonicity of $M_{0}$.With the logic sequence,we have:in
terms of $[0,\eta^{\ast})$,$\eta \uparrow(\downarrow)$ $\to$
$M_{0}\downarrow(\uparrow)$ $\to$ $M_{sc}\downarrow(\uparrow)$,and
in terms of $(\eta^{\ast},\infty]$,$\eta \uparrow(\downarrow)$ $\to$
$M_{0}\uparrow(\downarrow)$ $\to$ $M_{sc}\uparrow(\downarrow)$.

Actually $\eta$ is not more than 2 in our numerical observation,and
there always have the global tendency of decreasing $\eta$ with
fitting number of frequencies continually added,thus the $\infty$
for $\eta$ just only be seen as the mathematical seriousness in
form,it is enough to consider $\eta$ in the field of $[0,2]$.To
point out further more,the situation for
$\frac{c}{a}\ln\sigma+b+\ln\sigma<0$ in terms of $a<0,c>0$ is very
few,then we can reasonably expect the monotonous increasing function
for $M_{sc}$ with $\eta$ varied in terms of $0<\sigma<1$ and $b<0$.

$2)$ $0<\sigma<1$ and $b>0$

a)$a>0,c>0$

We need to investigate the essential term $\frac
{c}{a}e^{-b\eta}ln{\sigma}+b+ln{\sigma}$,and there have likely
competitive relation for $b+ln{\sigma}>0$,with the derivation we
know $M_{2}>0$,therefore $M_{1}$ is a increasing function.For
general cases,$\frac {c}{a}\gg1$,so the term $\frac
{c}{a}ln{\sigma}+b+ln{\sigma}$ could be negative with $b$ not very
large that is reasonable from numerical observation.As $M_{1}$ is an
increasing function,then there have a transition from negative to
positive in terms of increasing $\eta$ and we can get the same
expression of critical value $\eta^{\ast}$ for $\rm {A.4}$ with the
equation $\rm {A.3}$.As $a>0$,the monotonicity of $M_{sc}$ is
opposite to $M_{0}$,with logic sequences,we
have:$[0,\eta^{\ast})$,$\eta \uparrow(\downarrow)$ $\to$
$M_{0}\downarrow(\uparrow)$ $\to$ $M_{sc}\uparrow(\downarrow)$,and
in terms of $(\eta^{\ast},\infty]$,$\eta \uparrow(\downarrow)$ $\to$
$M_{0}\uparrow(\downarrow)$ $\to$ $M_{sc}\downarrow(\uparrow)$.

If $b+ln{\sigma}<0$,$M_{1}<0$.Thus $M_{0}$ is a decreasing function
and it is very easy to get the tendency of variation:$\eta \uparrow$
$\to$ $M_{0}\downarrow$ $\to$ $M_{sc}\uparrow$,$\eta \downarrow$
$\to$ $M_{0}\uparrow$ $\to$ $M_{sc}\downarrow$.

b)$a>0,c<0$

Consider the essential term $\frac
{c}{a}e^{-b\eta}ln{\sigma}+b+ln{\sigma}$,if $b+ln{\sigma}>0$,the
term $M_{1}>0$,thus $M_{0}$ is a increasing function.For $a>0$,the
monotonicity of $M_{sc}$ is opposite to $M_{0}$.With the logic
sequences,we have:$\eta \uparrow$ $\to$ $M_{0}\uparrow$ $\to$
$M_{sc}\downarrow$,$\eta \downarrow$ $\to$ $M_{0}\downarrow$ $\to$
$M_{sc}\uparrow$.

If $b+ln{\sigma}<0$,as $\frac {c}{a}e^{-b\eta}ln{\sigma}>0$,we need
to consider the monotonicity of $M_{1}$ which means to consider
$M_{2}=(\frac {c}{a})(-b)ln{\sigma}e^{-b\eta}$.It is easy to find
$M_{2}<0$ which means $M_{1}$ is a decreasing function.Thus we want
to investigate the maximum value of $M_{1}$ with $\eta=0$,that is
$\frac{c}{a}ln\sigma+b+ln\sigma$.For our very limited numerical
observation,$|\frac{c}{a}|\gg1$ can be found and pay attention to
$M_{1}$ as a decreasing function,we can reasonably to expect there
have a transition from positive value to negative value for
$M_{1}$,and using the same equation $\rm {A.3}$ to get the critical
value as $\eta^{\ast}$ with the form $\rm {A.4}$.With the logic
sequences,we have:$[0,\eta^{\ast})$,$\eta \uparrow(\downarrow)$
$\to$ $M_{0}\uparrow(\downarrow)$ $\to$
$M_{sc}\downarrow(\uparrow)$,and in terms of
$(\eta^{\ast},\infty]$,$\eta \uparrow(\downarrow)$ $\to$
$M_{0}\downarrow(\uparrow)$ $\to$ $M_{sc}\uparrow(\downarrow)$.

$3)$ $\sigma>1$ and $b>0$

a) $a>0,c>0$

Starting point still is the essential term $\frac
{c}{a}e^{-b\eta}ln{\sigma}+b+ln{\sigma}$,then it is easy to find
this term $M_{1}>0$ and that means $M_{0}$ is a increasing
function.As $a>0$,the monotonicity of $M_{sc}$ is opposite to
$M_{0}$,with logic sequences,we have:$\eta \uparrow$ $\to$
$M_{0}\uparrow$ $\to$ $M_{sc}\downarrow$,$\eta \downarrow$ $\to$
$M_{0}\downarrow$ $\to$ $M_{sc}\uparrow$.

b) $a>0,c<0$

The starting point is $\frac
{c}{a}e^{-b\eta}ln{\sigma}+b+ln{\sigma}$,and the sub-term $\frac
{c}{a}e^{-b\eta}ln{\sigma}<0$ and $b+ln{\sigma}>0$,so we need to
consider the $M_{2}$ as the derivative of $M_{1}$ and find $(\frac
{c}{a})(-b)ln{\sigma}e^{-b\eta}>0$ that means $M_{1}$ is a
increasing function.With the condition of $|\frac {c}{a}|\gg1$ for
most cases,the value of $M_{1}$ in terms of $\eta=0$ is
negative.Thus there have a transition from negative to positive for
$M_{1}$ and the critical value $\eta^{\ast}$ for $M_{1}=0$ can be
obtained with the equation $\rm A.3$ in the form $\rm A.4$.With the
logic sequences,we have:$[0,\eta^{\ast})$,$\eta
\uparrow(\downarrow)$ $\to$ $M_{0}\downarrow(\uparrow)$ $\to$
$M_{sc}\uparrow(\downarrow)$,and in terms of
$(\eta^{\ast},\infty]$,$\eta \uparrow(\downarrow)$ $\to$
$M_{0}\uparrow(\downarrow)$ $\to$ $M_{sc}\downarrow(\uparrow)$.

$4)$ $\sigma>1$ and $b<0$

a) $a>0,c>0$

If $b+ln{\sigma}>0$,the essential term $\frac
{c}{a}e^{-b\eta}ln{\sigma}+b+ln{\sigma}>0$,it is easy to know
$M_{0}$ is a increasing function.Thus with logic sequences,we
have:$\eta \uparrow$ $\to$ $M_{0}\uparrow$ $\to$
$M_{sc}\downarrow$,$\eta \downarrow$ $\to$ $M_{0}\downarrow$ $\to$
$M_{sc}\uparrow$.

If $b+ln{\sigma}<0$,we need to consider $M_{2}=(\frac
{c}{a})(-b)ln{\sigma}e^{-b\eta}$ and find easily $M_{2}>0$ that
means $M_{1}$ is a increasing function.Thus we can consider the
value of $M_{1}$ in terms of $\eta=0$,and with the condition for
$\frac {c}{a}\gg1$,we can find $M_{1}$ always is positive for all
the value of $\eta$ choose.So we have the logic sequences for
tendency of $M_{sc}$:$\eta \uparrow$ $\to$ $M_{0}\uparrow$ $\to$
$M_{sc}\downarrow$,$\eta \downarrow$ $\to$ $M_{0}\downarrow$ $\to$
$M_{sc}\uparrow$.But with condition for $\frac {c}{a}ln{\sigma}$
smaller than $|b+ln{\sigma}|$,there have a transition from negative
to positive for $M_{1}$ and the critical value $\eta^{\ast}$ can be
got with the equation $\rm A.3$ in the form $\rm A.4$.Thus the logic
sequences we have are:$[0,\eta^{\ast})$,$\eta \uparrow(\downarrow)$
$\to$ $M_{0}\downarrow(\uparrow)$ $\to$
$M_{sc}\uparrow(\downarrow)$,and in terms of
$(\eta^{\ast},\infty]$,$\eta \uparrow(\downarrow)$ $\to$
$M_{0}\uparrow(\downarrow)$ $\to$ $M_{sc}\downarrow(\uparrow)$.

b) $a<0,c>0$

If $b+ln{\sigma}<0$,thus the term $M_{1}=\frac
{c}{a}e^{-b\eta}ln{\sigma}+b+ln{\sigma}<0$,then we can know $M_{0}$
is a decreasing function.With the logic sequences,we have:$\eta
\uparrow$ $\to$ $M_{0}\downarrow$ $\to$ $M_{sc}\downarrow$,$\eta
\downarrow$ $\to$ $M_{0}\uparrow$ $\to$ $M_{sc}\uparrow$.If
$b+ln{\sigma}>0$,we should consider $M_{2}=(\frac
{c}{a})(-b)ln{\sigma}e^{-b\eta}$ and find it is negative.Thus
$M_{1}$ is a decreasing function and we check the value of $M_{1}$
in terms of $\eta=0$ and it can be negative if
$|\frac{c}{a}|\gg1$.Therefore $M_{0}$ is a decreasing function and
we can get the monotonicity of $M_{sc}$ with logic sequences as
follows:$\eta \uparrow$ $\to$ $M_{0}\downarrow$ $\to$
$M_{sc}\downarrow$,$\eta \downarrow$ $\to$ $M_{0}\uparrow$ $\to$
$M_{sc}\uparrow$.

\subsubsection{2.Linear variations of $D_{L}$ via $\eta$}

We find there have a approximated linear relation for $D_{L}$ with
$\eta$ changed in terms of some initial frequencies,mathematically
describe it as:$D_L=a\eta+b$,$a,b$ are the fitting parameters.The
staring point still is $M_{sc}={\rm exp}(-2(\epsilon/\hbar)^\eta
D_L)$,with the linear expression of $D_{L}$,we can get
\be\tag{A.5}M_{sc}\approx {\rm exp}\{-2b[\frac{a}{b}e^{ln{\eta}+\eta
ln{\sigma}}+e^{\eta ln{\sigma}}]\}.\ee

For sake of getting the monotonicity of $M_{sc}$,we need to
differentiate $\frac{a}{b}e^{ln{\eta}+\eta ln{\sigma}}+e^{\eta
ln{\sigma}}$,and easily to get the expression \be\tag{A.6}{\rm
exp}[ln{\eta}+\eta
ln{\sigma}][\frac{a}{b}(\frac{1}{\eta}+ln{\sigma})+ln{\sigma}e^{-ln{\eta}}].\ee
To get the information about value of the term
$\frac{a}{b}(\frac{1}{\eta}+ln{\sigma})+ln{\sigma}e^{-ln{\eta}}$ for
positive or negative,we still need to consider the derivative of
this term,and we can get the new
term:\be\tag{A.7}-[\frac{a}{b}\frac{1}{{\eta}^2}+\frac{ln{\sigma}}{\eta}e^{-ln{\eta}}].\ee
Therefore,we can find there have three terms determining the
monotonicity of $M_{sc}$,and for simplicity,we use simple name for
these terms:$\frac{a}{b}e^{ln{\eta}+\eta ln{\sigma}}+e^{\eta
ln{\sigma}}$ as $M^{'}$
,$\frac{a}{b}(\frac{1}{\eta}+ln{\sigma})+ln{\sigma}e^{-ln{\eta}}$ as
$M^{''}$ and
$-[\frac{a}{b}\frac{1}{{\eta}^2}+\frac{ln{\sigma}}{\eta}e^{-ln{\eta}}]$
as $M^{'''}$.

We here consider two typical linear relations for $D_{L}$
corresponding to different fields of $\sigma$ for $0<\sigma<1$ and
$\sigma>1$,and discuss them separately.

$1)$ $0<\sigma<1$

a) $a>0,b<0$

We firstly consider the essential term
$M^{''}=\frac{a}{b}(\frac{1}{\eta}+ln{\sigma})+ln{\sigma}e^{-ln{\eta}}$
and find $M^{''}$ will be negative if
$\frac{1}{\eta}+ln{\sigma}>0$.As $\frac{1}{\eta}+ln{\sigma}$ is a
decreasing function and there is a transitive value for $\eta$,and
$\frac{1}{\eta}+ln{\sigma}$ can be positive below that critical
value given name as $\eta^{'}$.The $\eta^{'}$ is given with the
equation very easily:\be\tag{A.8}\frac{1}{\eta}+ln{\sigma}=0,\ee
\be\tag{A.9}\eta^{'}=-\frac{1}{ln{\sigma}}.\ee

But we still find it is uncertain for the condition about
$\eta>\eta^{'}$,thus we still need to consider further for $M^{'''}$
which is the derivative of $M^{''}$.With the condition about
$\frac{a}{b}<0$,$M^{'''}$ is positive and the value of $M^{''}$ in
terms of $\eta=\eta^{'}$ is negative,therefore there has a critical
value for $\eta$ larger than $\eta^{'}$ that we can call $\eta^{''}$
determined by the
equation:\be\tag{A.10}\frac{a}{b}(\frac{1}{\eta}+ln{\sigma})+ln{\sigma}\frac{1}{\eta}=0,\ee
\be\tag{A.11}\eta^{''}=-\frac{b}{a}-\frac{1}{ln{\sigma}}\ee

Thus we can combine $\eta^{'}$ and $\eta^{''}$ to find the
$\eta^{''}$ is the only deciding factor for monotonicity of
$M_{sc}$.With logic sequences,we have:$[0,\eta^{''})$,$\eta
\uparrow(\downarrow)$ $\to$ $M^{'}\downarrow(\uparrow)$ $\to$
$M_{sc}\downarrow(\uparrow)$,and in terms of
$(\eta^{''},\infty]$,$\eta \uparrow(\downarrow)$ $\to$
$M^{'}\uparrow(\downarrow)$ $\to$ $M_{sc}\uparrow(\downarrow)$.

b) $a<0,b>0$

We use the same method but more convenient as we find the deciding
factor about the monotonicity of $M_{sc}$ can come from directly
from $M^{''}$ and $M^{'''}$.As $\frac{a}{b}<0$ and $0<\sigma<1$,we
get $M^{'''}>0$ which means $M^{''}$ is a increasing function.It is
very easy to find the value of $M^{''}$ with $\eta=0$ is
negative,thus we can find there has a transitive value of $\eta$
that can be called $\eta^{\ast}$ dividing the $\eta$ as two fields
$[0,\eta^{\ast})$ and $(\eta^{\ast},\infty]$ corresponding to
$M^{''}<0$ and $M^{''}>0$.Obviously $\eta^{\ast}$ is decided by $\rm
A.10$ with the form as $\rm A.11$.Pay attention to $b>0$,with logic
sequences,we have:$[0,\eta^{\ast})$,$\eta \uparrow(\downarrow)$
$\to$ $M^{'}\downarrow(\uparrow)$ $\to$
$M_{sc}\uparrow(\downarrow)$,and in terms of
$(\eta^{\ast},\infty]$,$\eta \uparrow(\downarrow)$ $\to$
$M^{'}\uparrow(\downarrow)$ $\to$ $M_{sc}\downarrow(\uparrow)$.

$2)$ $\sigma>1$

a) $a>0,b<0$

The starting point is still for the essential term
$\frac{a}{b}(\frac{1}{\eta}+ln{\sigma})+ln{\sigma}e^{-ln{\eta}}$,and
it will be convenient to study the monotonicity of $M_{sc}$ if we
can rewrite $M^{''}$ as
$(\frac{a}{b}+ln{\sigma})\frac{1}{\eta}+\frac{a}{b}ln{\sigma}$.$M^{'''}$
rewritten as $-\frac{1}{{\eta}^2}(\frac{a}{b}+ln{\sigma})$ can be
negative,if $\frac{a}{b}+ln{\sigma}>0$.Thus $M^{''}$ is a decreasing
function.Actually what we care about is $\eta>0$ even in the field
between 0 and 2,but it is better to understand the monotonicity with
all the field put in consideration.One can notice $\eta=0$ is a
discontinuity point for $M^{''}$ and it is decreased continuously
for $[-\infty,0]$ and $(0,\infty]$ and have the same value of
$M^{''}=\frac{a}{b}ln{\sigma}$ in terms of
$\eta=-\infty,\infty$.Therefore we can find $\eta$ has a critical
value $\eta^{\ast}$ with form $\rm A.11$ corresponding to
$M^{''}=0$,and it is positive that can be deduced by
$\frac{a}{b}+ln{\sigma}>0$.With the logic sequences,we
have:$[0,\eta^{\ast})$,$\eta \uparrow(\downarrow)$ $\to$
$M^{'}\uparrow(\downarrow)$ $\to$ $M_{sc}\uparrow(\downarrow)$,and
in terms of $(\eta^{\ast},\infty]$,$\eta \uparrow(\downarrow)$ $\to$
$M^{'}\downarrow(\uparrow)$ $\to$ $M_{sc}\downarrow(\uparrow)$.

Now we consider the condition about $\frac{a}{b}+ln{\sigma}<0$ and
$M^{''}=(\frac{a}{b}+ln{\sigma})\frac{1}{\eta}+\frac{a}{b}ln{\sigma}$
is negative for $\eta>0$.But we need to deepen our understanding for
all the field of $\eta$,and
$M^{'''}=-\frac{1}{{\eta}^2}(\frac{a}{b}+ln{\sigma})>0$ which means
$M^{''}$ is a increasing function.As $\eta=0$ is a discontinuity
point for $M^{''}$,and it is increased continuously for
$[-\infty,0]$ and $(0,\infty]$ and have the same value of
$M^{''}=\frac{a}{b}ln{\sigma}$ in terms of $\eta=-\infty,\infty$.So
it is easy to find $\eta$ has a critical value
$\eta^{\ast}=-\frac{b}{a}-\frac{1}{ln{\sigma}}<0$ that can deduced
directly by $\frac{a}{b}+ln{\sigma}<0$.Thus this critical value does
not exist in our basic caring field $(0,\infty]$ in which $M^{''}$
is always negative.In terms of $(0,\infty]$,with the logic
sequences,we have:$\eta \uparrow$ $\to$ $M^{'} \downarrow$ $\to$
$M_{sc} \downarrow$, $\eta \downarrow$ $\to$ $M^{'} \uparrow$ $\to$
$M_{sc} \uparrow$.

b) $a<0,b>0$

We firstly consider the essential term $M^{''}$ as
$(\frac{a}{b}+ln{\sigma})\frac{1}{\eta}+\frac{a}{b}ln{\sigma}$ with
condition $\frac{a}{b}+ln{\sigma}>0$.As we can not make sure about
the value of $M^{''}$ for positive or negative,but $M^{'''}$ as
$-\frac{1}{{\eta}^2}(\frac{a}{b}+ln{\sigma})$ is negative which
means $M^{''}$ is a decreasing function disconnected by
$\eta=0$.From the value of $M^{''}=\infty$ with $\eta=0+$ and the
value of $M^{''}=\frac{a}{b}ln{\sigma}<0$ with $\eta=\infty$,with
continuously decreasing property,there has a critical value
$\eta^{\ast}$ between $0+$ and $\infty$ corresponding to
$M^{''}=0$,and have the form with $\rm A.11$ which is positive
equivalent to the condition $\frac{a}{b}+ln{\sigma}>0$.Pay attention
to $b>0$ in the term $\rm A.5$,with logic sequences,we
have:$[0,\eta^{\ast})$,$\eta \uparrow(\downarrow)$ $\to$
$M^{'}\uparrow(\downarrow)$ $\to$ $M_{sc}\downarrow(\uparrow)$,and
in terms of $(\eta^{\ast},\infty]$,$\eta \uparrow(\downarrow)$ $\to$
$M^{'}\downarrow(\uparrow)$ $\to$ $M_{sc}\uparrow(\downarrow)$.

For the condition about $\frac{a}{b}+ln{\sigma}<0$,we find the
processing method is exactly same to the situation of $a>0,b<0$
taken as a whole to treat in the terms $M^{'}$,$M^{''}$ and
$M^{'''}$ except the monotonicity of $M_{sc}$ is opposite to $M^{'}$
as $b>0$ in the form $\rm A.5$.In terms of $(0,\infty]$,with the
logic sequences,we have:$\eta \uparrow$ $\to$ $M^{'}\downarrow$
$\to$ $M_{sc} \uparrow$, $\eta \downarrow$ $\to$ $M^{'}\uparrow$
$\to$ $M_{sc} \downarrow$.Actually we surly can treat the condition
for $\frac{a}{b}+ln{\sigma}>0$ using the same concise way,but we
want to investigate in detail for giving a deep impression
previously to at last enlighten the universal idea we use
repeatedly.

%\end{appendices}


\begin{thebibliography}{0}
\expandafter\ifx\csname natexlab\endcsname\relax\def\natexlab#1{#1}\fi
\expandafter\ifx\csname bibnamefont\endcsname\relax
  \def\bibnamefont#1{#1}\fi
\expandafter\ifx\csname bibfnamefont\endcsname\relax
  \def\bibfnamefont#1{#1}\fi
\expandafter\ifx\csname citenamefont\endcsname\relax
  \def\citenamefont#1{#1}\fi
\expandafter\ifx\csname url\endcsname\relax
  \def\url#1{\texttt{#1}}\fi
\expandafter\ifx\csname urlprefix\endcsname\relax\def\urlprefix{URL }\fi
\providecommand{\bibinfo}[2]{#2}
\providecommand{\eprint}[2][]{\url{#2}}

\end{thebibliography}


\begin{thebibliography}{99}
\bibitem{vanicek_arxiv}J.Van\'{i}\v{c}ek,arXiv:quant-ph/0410205v1.
\bibitem{Prosen_supp}T.~Prosen,T.~Seligman,M.~\v{Z}nidari\v{c},Prog.Theor.Phys.Supp.
{\bf 150},200(2003).
\bibitem{WgWang_1}W.G.~Wang,G.~Castati,and B.W.~Li,Phys.Rev.E
{\bf 69},025201(R)(2004).
\bibitem{Zheng}Q.~Zheng, W.G.~Wang,P.P.~Qin,P.~Wang,X.P.~Zhang,and
Z.Z.~Ren,Phys.Rev.E.{\bf80},016214 2009).
\bibitem{Nielsen}M.A.Nielsen and I.L.Chuang,Quantum Computation and
Quantum Information(Cambridge University Press,Cambridge,2000).
\bibitem{Datta}S.Datta,Quantum Transport:Atom to Transistor(Cambridge
University Press,Cambridge,2005).
\bibitem{Gorin} T.~Gorin,T.~Prosen,T.~Seligman,M.~\v{Z}nidari\v{c},Phy.~Rep.
{\bf 435},33(2006);
\bibitem{Jacquod}P.~Jacquod,C.Petitjean,Advance.~in.~Physics.~{\bf 58},67(2009)
\bibitem{Raimond}J.M.~Raimond,M.~Brune,and S.Haroche,Rev.Mod.Phys.
{\bf 73},565(2001).
\bibitem{Sohn}L.L.~Sohn,L.P.~Kouwenhoven,and G.~Sch\"{o}n,Mesoscopic
Electron Transport(Spring,New York,1997).
\bibitem{Casabone}B.~Casabone,I.~Garc\'{i}a-Mata,and D.A.~Wisniacki,
Europhysics Letters {\bf 89},50009(2010).
\bibitem{peres}A.~Peres,Phys.Rev.A.{\bf 30},1610(1984).
\bibitem{Jalabert}R.A.~Jalabert and H.M.~Pastawski,Phys.Rev.Lett.{\bf 86},
2490(2001).
\bibitem{You}W.L.~You,Y.W.~Li,S.J.Gu,Phys.Rev.E.{\bf 76},
022101(2007).
\bibitem{Quan}H.T.~Quan,Z.~Song,X.F.~Liu,P.~Zanardi,and C.P.~Sun,
Phys.Rev.Lett.{\bf 96},140604(2006).
\bibitem{Wang}W.~Wang,P.~Qin,L.~He,and P.~Wang,Phys.Rev.E.{\bf 81},
016214(2010).
\bibitem{Jacquod_01}P.~Jacquod,P.G.~Silvestrov,and C.W.J.~Beenakker,
Phys.Rev.E.{\bf 64},055203(2001).
\bibitem{Wisniacki}D.A.~Wisniaki and D.~Cohen,Phys.Rev.E.{\bf 66},
046209(2002).
\bibitem{Wisniacki_03}D.A.~Wisniaki,Phys.Rev.E.{\bf 67},
016205(2003).
\bibitem{Guti}M.~Guti\'{e}rrez and A.~Goussev,Phys.Rev.E.{\bf 79},
046211(2009).
\bibitem{Andersen}M.F.~Andersen,A.~Kaplan,T.~Grunzweig,and N.~Davidson,
Phys.Rev.Lett.{\bf 97},104102(2006).
\bibitem{WgWang_2} W.G.~Wang and B.W.Li,Phys.Rev.E{\bf 71},066203(2005).
\bibitem{Pozzo}E.N.~Pozzo and D.~Dominguez,Phys.Rev.Lett.
{\bf 98},057006(2007).
\bibitem{Goussev}A.~Goussev and D.~Waltner,and R.A.~Jalabert,
New J.Phys.{\bf10},093010(2008).
\bibitem{WgWang_3}W.G.~Wang,Phys.Rev.E{\bf77},036206(2008).
\bibitem{WLT02} Y.~Weinstein,S.~Lloyd,C.~Tsallis,Phys.Rev.Lett
{\bf 89},214101(2002).
\bibitem{Casati05}G.~Casati,C.~Tsallis,and F.~Baldovin. Europhysics
Letters {\bf 72}, 355 (2005).
\bibitem{Hannay}J.H.~Hannay and M.V.~Berry,Physica D{\bf1},267(1980).
\bibitem{Ford}J.~Ford,G.~Mantica,and G.H.~Ristow,Physica D{\bf50},493(1991).
\bibitem{Wilkie}J.~Wilkie and P.~Brumer,Phys.Rev.E{\bf49},1968(1994).
\bibitem{Haake}Haake,Quantum Signatures of Chaos,2nd ed.(Springer-Verlag,
Berlin,2001).
\bibitem{WLT02} Y.~Weinstein,S.~Lloyd,C.~Tsallis,Phys.Rev.Lett {\bf 89},214101(2002).
\bibitem{Ott}E.~Ott, Chaos in Dynamical Syatems, 2nd ed.
(Cambridge University Press, Cambridge, England, 2001).
\bibitem{Casati05}G.~Casati,C.~Tsallis,and F.~Baldovin.Europhysics
Letters {\bf 72},355 (2005).
\bibitem{CLMPV02}F.M.~Cucchietti{\it et al.},Phys.~Rev.~E {\bf 65},046209 (2002).
\bibitem{RS96} S.~Ruffo, D.~Shepelyansky, Phys.~Rev.~Lett {\bf 76},3300(1996).
\bibitem{Levydistr}K.~Umeno,Phys.Rev.E {\bf 58} 2644(1998).
\bibitem{Robledo}A.~Robledo,L.G.~Moyano,Phys.Rev.E {\bf 77},036213(2008).
\bibitem{Liujie}J.~Liu,W.G.~Wang,C.W.~Zhang,Q.~Niu,B.W.~Li.
Phys.Rev.A{\bf 72}063623(2005).
\bibitem{Adamov}Y.~Adamov,I.V.~Gornyi and A.D.~Mirlin.
Phys.Rev.E{\bf 67},056217(2003).
\bibitem{Pellegrini}F.~Pellegrini,s.~Montangero.Phys.Rev.A{\bf
76},052327(2007).
\bibitem{prosen2005}T.~Prosen and M.~\v{Z}nidaric,Phys.Rev.Lett{\bf 94},044101(2005).
\bibitem{Wwang}W.~Wang,Z.~ Liu and B.~Hu,Phys.Rev.Lett
{\bf 84},2610(2000).
\bibitem{Korus}L.~Korus,Int.J.Appl.Math.Comput.Sci.{\bf 21},149(2011).
\bibitem{Martin}M.Sieber and K.Richter,Phys.Scr.{\bf
T90},128(2001);
\bibitem{Muller}S.Muller,S.Heusler,P.Braun,F.Haake,and
A.Altland,Phys.Rev.Lett.{\bf 93},014103(2004).
\bibitem{Heller}J.Van\'{i}\v{c}ek,Phys.Rev.E
{\bf 73},046204(2006).
\bibitem{vanicek}J.Van\'{i}\v{c}ek and Eric J.~Heller,Phys.Rev.E
{\bf 68},056208(2003).
\bibitem{Mata}I.~Garc\'{l}a-Mata,C.~Pineda and Diego A.~Wisniacki,J.~Phys.~A{\bf
47},115301(2014).
\bibitem{Denmarklm}K.Madsen,H.B.Nielsen,and O.Tingleff. Method for Non-Linear Least Squares Problems.Technical University of Denmark,2004.
\bibitem{Scholarpedia}A.~Goussev,R.A.~ Jalabert,H.M.~Pastawski and D.~Wisniacki,Scholarpedia{\bf
7},11687(2012).
\bibitem{vanicek_exact}J.Van\'{i}\v{c}ek,D.~Cohen,Phil.Trans.R.Soc.A{\bf 374},20150164(2016).
\bibitem{Levydistr}K.~Umeno,Phys.Rev.E {\bf 58} 2644 (1998).
\bibitem{Cerruti}N.R.~Cerruti and S.~Tomsovic,Phys.Rev.Lett {\bf 88},054103(2002);J.Phys.A {\bf 36},3451(2003).
\bibitem{Prosen_NJP03}T.~Prosen and M.\v{Z}nidari,New J.Phys.{\bf5},109(2003).
\bibitem{moving_average}D.~Jansen,J.~Stolpp,L.~Vidmar,and F.~Heidrich-Meisner,Phys.Rev.B.{\bf 99},155130(2019).
\end{thebibliography}
\end{document}